\documentstyle[eqsecnum,aps,epsf,
feynmp]{revtex}
\newcommand{\nc}{\newcommand}
\newcommand{\scs}{\scriptstyle}
\nc{\fulla}{\mbox{\boldmath $A$}^{\rm c}}
\nc{\fulld}{\mbox{\boldmath $D$}^{\rm c}}
\nc{\fulls}{\mbox{\boldmath $S$}^{\rm c}}
\nc{\fullg}{\mbox{\boldmath $G$}^{\rm c}}
\nc{\fullgs}{\mbox{\scriptsize \boldmath $G^{\rm c}$}}
\nc{\fullas}{\mbox{\scriptsize \boldmath $A^{\rm c}$}}
\nc{\fullds}{\mbox{\scriptsize \boldmath $D^{\rm c}$}}
\nc{\fullss}{\mbox{\scriptsize \boldmath $S^{\rm c}$}}
\nc{\beq}{\begin{eqnarray}}
\nc{\eeq}{\end{eqnarray}}
\nc{\la}{\label}
\nc{\r}{\ref}
\nc{\no}{\nonumber}
\nc{\ci}{\cite}
\begin{document}
\nc{\setval}{\fmfset{wiggly_len}{1.5mm}\fmfset{arrow_len}{1.5mm}\fmfset{arrow_ang}{20}\fmfset{dash_len}{1.5mm}\fmfpen{0.125mm}\fmfset{dot_size}{1thick}}
\setlength{\unitlength}{1mm}
\newcommand{\ddfermi}[5]{\frac{\delta^2 #1}{\delta
\parbox{10mm}{\centerline{
\begin{fmfgraph*}(5,3)
\setval
\fmfleft{v1}
\fmfright{v2}
\fmf{fermion}{v2,v1}
\fmfv{decor.size=0,label={\footnotesize #2},l.dist=0.5mm}{v1}
\fmfv{decor.size=0,label={\footnotesize #3},l.dist=0.5mm}{v2}
\end{fmfgraph*}
}}\,\delta
\parbox{10mm}{\centerline{
\begin{fmfgraph*}(5,3)
\setval
\fmfleft{v1}
\fmfright{v2}
\fmf{fermion}{v2,v1}
\fmfv{decor.size=0,label={\footnotesize #4},l.dist=0.5mm}{v1}
\fmfv{decor.size=0,label={\footnotesize #5},l.dist=0.5mm}{v2}
\end{fmfgraph*}
}}
}}
\nc{\dvertex}[4]{\frac{\delta #1}{\rule[0pt]{0pt}{15pt}\delta
\parbox{10mm}{\centerline{
\begin{fmfgraph*}(4,3.464)
\setval
\fmfforce{1w,0h}{v1}
\fmfforce{0w,0h}{v2}
\fmfforce{0.5w,1h}{v3}
\fmfforce{0.5w,0.2886h}{vm}
\fmf{fermion}{v1,vm}
\fmf{fermion}{vm,v2}
\fmf{photon}{v3,vm}
\fmfv{decor.size=0,label=${\scs #3}$,l.dist=0.5mm}{v1}
\fmfv{decor.size=0,label=${\scs #4}$,l.dist=0.5mm}{v2}
\fmfv{decor.size=0,label=${\scs #2}$,l.dist=0.5mm}{v3}
\fmfdot{vm}
\end{fmfgraph*}}}\rule[0pt]{0pt}{15pt}}}
\nc{\dbphi}[3]{\frac{\delta #1}{\delta
\parbox{10mm}{\centerline{
\begin{fmfgraph*}(5,3)
\setval
\fmfleft{v1}
\fmfright{v2}
\fmf{photon}{v2,v1}
\fmfv{decor.size=0,label=${\scs #2}$,l.dist=0.5mm}{v1}
\fmfv{decor.size=0,label=${\scs #3}$,l.dist=0.5mm}{v2}
\end{fmfgraph*}
}}}}
\nc{\dephi}[3]{\frac{\delta #1}{\delta
\parbox{10mm}{\centerline{
\begin{fmfgraph*}(5,3)
\setval
\fmfleft{v1}
\fmfright{v2}
\fmf{electron}{v2,v1}
\fmfv{decor.size=0,label=${\scs #2}$,l.dist=0.5mm}{v1}
\fmfv{decor.size=0,label=${\scs #3}$,l.dist=0.5mm}{v2}
\end{fmfgraph*}
}}}}
\nc{\cdphi}[2]{\frac{\delta #1}{\delta
\parbox{11mm}{\centerline{  
\begin{fmfgraph*}(7,7)
\setval
\fmfforce{0w,0h}{i1}
\fmfforce{0w,6/7h}{i2}
\fmfforce{1w,3/7h}{o}
\fmfforce{0.5w,3/7h}{v}
\fmf{double,width=0.2mm}{i1,v,i2}
\fmf{boson}{v,o}
\fmfdot{v}
\fmfv{decor.size=0, label=${\scs 1}$, l.dist=1mm, l.angle=0}{o}
\end{fmfgraph*}
}}}}
\nc{\deephi}[3]{\frac{\delta #1}{\delta
\parbox{10mm}{\centerline{
\begin{fmfgraph*}(5,3)
\setval
\fmfleft{v1}
\fmfright{v2}
\fmf{heavy,width=0.2mm}{v2,v1}
\fmfv{decor.size=0,label=${\scs #2}$,l.dist=0.5mm}{v1}
\fmfv{decor.size=0,label=${\scs #3}$,l.dist=0.5mm}{v2}
\end{fmfgraph*}
}}}}
\nc{\dbbphi}[3]{\frac{\delta #1}{\delta
\parbox{10mm}{\centerline{
\begin{fmfgraph*}(5,3)
\setval
\fmfleft{v1}
\fmfright{v2}
\fmf{dbl_wiggly,width=0.2mm}{v2,v1}
\fmfv{decor.size=0,label=${\scs #2}$,l.dist=0.5mm}{v1}
\fmfv{decor.size=0,label=${\scs #3}$,l.dist=0.5mm}{v2}\end{fmfgraph*}
}}}}
\nc{\dddvertex}[4]{\frac{\delta #1}{\rule[0pt]{0pt}{15pt}\delta
\parbox{10mm}{\centerline{
\begin{fmfgraph*}(4,3.464)
\setval
\fmfforce{1w,0h}{v1}
\fmfforce{0w,0h}{v2}
\fmfforce{0.5w,1h}{v3}
\fmfforce{0.5w,0.2886h}{vm}
\fmf{heavy,width=0.2mm}{v1,vm}
\fmf{heavy,width=0.2mm}{vm,v2}
\fmf{photon,width=0.2mm}{v3,vm}
\fmfv{decor.size=0,label=${\scs #3}$,l.dist=0.5mm}{v1}
\fmfv{decor.size=0,label=${\scs #4}$,l.dist=0.5mm}{v2}
\fmfv{decor.size=0,label=${\scs #2}$,l.dist=0.5mm}{v3}
\fmfdot{vm}
\end{fmfgraph*}}}\rule[0pt]{0pt}{15pt}}}
\nc{\ddvertex}[4]{\frac{\delta #1}{\rule[0pt]{0pt}{15pt}\delta
\parbox{13mm}{\centerline{
\begin{fmfgraph*}(6.93,6)
\setval
\fmfforce{1w,-2.5/6h}{v1}
\fmfforce{0w,-2.7/6h}{v2}
\fmfforce{0.5w,3.5/6h}{v3}
\fmfforce{0.5w,0.5/6h}{v4}
\fmfforce{0.5w,0/6h}{v5}
\fmfforce{0.5w,0/6h}{v6}
\fmfforce{0.5w,-1.5/6h}{v7}
\fmfforce{0.333w,-1/6h}{v8}
\fmfforce{0.667w,-1/6h}{v9}
\fmf{dbl_wiggly,width=0.2mm}{v3,v4}
\fmf{heavy,width=0.2mm}{v8,v2}
\fmf{heavy,width=0.2mm}{v1,v9}
\fmf{double,left=1,width=0.2mm}{v4,v7,v4}
\fmfv{decor.size=0, label=${\scs #3 }$, l.dist=1mm, l.angle=-150}{v2}
\fmfv{decor.size=0, label=${\scs #4 }$, l.dist=1mm, l.angle=-30}{v1}
\fmfv{decor.size=0, label=${\scs #2 }$, l.dist=1mm, l.angle=90}{v3}
\end{fmfgraph*}}}\rule[0pt]{0pt}{15pt}}}
\begin{fmffile}{sd11}
\title{Functional Closure of Schwinger-Dyson Equations in Quantum Electrodynamics\\
Part 1: Generation of Connected and One-Particle Irreducible Feynman Diagrams}
\author{Axel Pelster, Hagen Kleinert}
\address{Institut f\"ur Theoretische Physik, Freie Universit\"at Berlin, 
Arnimallee 14, D-14195 Berlin, Germany\\
E-mails: pelster@physik.fu-berlin.de, kleinert@physik.fu-berlin.de}
\author{Michael Bachmann}
\address{Institut f\"ur Theoretische Physik, Universit\"at Leipzig,  Augustusplatz 10/11, D-04109 Leipzig, Germany\\
E-mail: michael.bachmann@itp.uni-leipzig.de\\ $\mbox{}$}
\date{\today}
\maketitle
\begin{abstract}
Using functional derivatives with respect to free propagators and interactions
we derive a closed set of Schwinger-Dyson equations in quantum electrodynamics.
Its conversion to graphical recursion relations allows us to systematically generate all connected and
one-particle irreducible Feynman diagrams for the $n$-point functions and the vacuum 
energy together with their correct weights.
\end{abstract}
\section{Introduction}
In quantum field theory, the calculation of physical quantities usually relies on evaluating Feynman integrals which are pictured by
diagrams. Each diagram is associated with a certain weight depending on its topology. 
There exist various convenient computer programs, for instance {\it FeynArts} \ci{FeynArts1,FeynArts2,FeynArts3} 
or {\it QGRAF} \ci{QGRAF1,QGRAF2}, for constructing these diagrams and for determining their weights in different field theories, 
Some of them are based on a combinatorial enumeration of all possible ways of
connecting vertices by lines according to Feynman's rules. Others use a systematic generation of homeomorphically
irreducible star graphs \cite{Heap,Nagle}. The latter approach is quite efficient and popular 
at higher orders, it has, however,  the conceptual disadvantage that it renders at an intermediate stage 
numerous diagrams with different vertex degrees which have to be discarded at the end.\\

A more systematic and physical approach to construct all Feynman diagrams of a quantum field theory was proposed a long time
ago \cite{Kleinert1,Vasiliev}. It is based on the observation that the complete knowledge of the vacuum energy implies the
knowledge of the entire theory (``the vacuum is the world'') \cite{Streater,Schwinger}. In this spirit, all vacuum diagrams are 
initially generated by a recursive graphical procedure. This procedure is derived from a functional differential equation
involving functional derivatives with respect to free propagators and interactions.
In a subsequent step, the $n$-point functions are found graphically by applying the functional derivatives to the vacuum energy.
Recently, this approach was used to systematically generate all connected and one-particle irreducible Feynman diagrams of the
euclidean multicomponent scalar $\phi^4$-theory both in the disordered, symmetric phase \cite{PHI4} and in the ordered, 
spontaneously broken-symmetry phase \cite{Boris,asym} (see also the related work in Ref.~\ci{Verschelde}). 
The approach was also applied to QED \cite{QED} and scalar QED \cite{gl} 
to construct the connected Feynman diagrams. In contrast
to the conventional generating functional technique \cite{Drell,Amit,Zuber,Bellac,Zinn,Peskin}, 
no external currents coupled to single fields are used, such that there is no need for introducing
Grassmann sources for fermion fields. An additional advantage is that the number
of derivatives necessary to generate a certain correlation function is half as big as with external sources.\\

The purpose of the present paper is to develop a modification of the approach in Ref. \cite{Glaum} for QED. Rather than starting from 
vacuum diagrams as elaborated in Ref. \cite{QED}, we generate the Feynman diagrams of $n$-point functions directly 
and find that they obey an infinite hierarchy of coupled Schwinger-Dyson equations
\cite{Drell,Amit,Zuber,Bellac,Zinn,Peskin}. We show that using functional derivatives with respect to the
free propagators and the interaction allows to close these Schwinger-Dyson equations functionally. 
In this way we obtain in Section \r{CONFD}
a closed set of equations determining the connected electron and photon two-point function as well as the connected 
three-point function. Analogously, we derive in Section \r{IRRFD} a closed set of equations for the electron and photon
self-energy as well as the one-particle irreducible three-point function. In both cases, the closed set of Schwinger-Dyson
equations can be converted into graphical recursion relations for the connected and one-particle irreducible Feynman
diagrams. From these follow the corresponding vacuum diagrams by short-circuiting external legs.
\section{Connected Feynman Diagrams}
\la{CONFD}
Following the short-hand notation introduced in Ref. \ci{QED}, the action
of QED in euclidean spacetime reads
\beq
\la{AC}
{\cal A} [ \bar{\psi} , \psi , A ] = \int_{12} S^{-1}_{12} \bar{\psi}_1
\psi_2 + \frac{1}{2} \int_{12}
D^{-1}_{12} A_1 A_2 + \int_{123} V_{123}
\bar{\psi}_1 \psi_2 A_3 - \int_1 J_1 A_1 \, ,
\eeq
where $\bar{\psi},\psi$ denote the electron fields and $A$ stands for the
photon field. For brevity, we omit all spinor or vector indices of the fields
and indicate their spacetime arguments by simple number indices, i.e.,
$1 = x_1, 2 = x_2, \ldots$ and $\int_1 = \int d^4 x_1$. Throughout
the paper we assume that the current $J$, the electron kernel $S^{-1}$, the photon kernel $D^{-1}$, and 
the interaction $V$ are completely general non-singular functional matrices
and their physical values for QED are inserted only at the end. By doing so,
we regard the action (\r{AC}) as the functional 
\beq
{\cal A} [ \bar{\psi} , \psi , A ] = 
{\cal A} [ \bar{\psi} , \psi , A ; J,S^{-1},D^{-1},V] \, .
\eeq
The same functional dependences are inherited by all field-theoretic
quantitites derived from it. In particular, we are interested in studying
the functional dependence of the partition function, which is defined by
a functional integral over a Boltzmann weight in natural units
\beq
\la{PF}
Z [J,S^{-1},D^{-1},V] = \oint {\cal D} \bar{\psi} {\cal D} \psi 
{\cal D} A \, e^{-{\cal A} [ \bar{\psi} , \psi , A ; J,S^{-1},D^{-1},V]} \, ,
\eeq
and of the $n$-point functions
\beq
&& \langle \psi_{n-2} \ldots \psi_4 \psi_1 
\bar{\psi}_2 \bar{\psi}_5  \ldots \bar{\psi}_{n-1} A_3 A_6 
\ldots A_{n} \rangle [ J,S^{-1},D^{-1},V] \nonumber \\
\la{EXP}
&& \hspace*{2cm}= \frac{1}{Z} \oint {\cal D} \bar{\psi} {\cal D} \psi 
{\cal D} A  \, \psi_{n-2} \ldots \psi_4 \psi_1 
\bar{\psi}_2 \bar{\psi}_5  \ldots \bar{\psi}_{n-1} A_3 A_6 
\ldots A_{n} 
\, e^{-{\cal A} [ \bar{\psi} , \psi , A ; J,S^{-1},D^{-1},V]} \, .
\eeq
By expanding the functional integrals (\r{PF}) and (\r{EXP})
in powers of the interaction $V$, the expansion coefficients
of the partition function and the $n$-point functions
consist of free-field expectation values. These are evaluated
with the help of Wick's rule as a sum of Feynman integrals, which are pictured
as diagrams constructed from lines and vertices. To illustrate the current we use the symbol
\beq
\la{CR}
\parbox{11mm}{\centerline{  
\begin{fmfgraph*}(7,6)
\setval
\fmfleft{i1,i2}
\fmfright{o}
\fmfforce{0.5w,0.5h}{v}
\fmf{double,width=0.2mm}{i1,v,i2}
\fmf{boson}{v,o}
\fmfdot{v}
\fmfv{decor.size=0, label=${\scs 1}$, l.dist=1mm, l.angle=0}{o}
\end{fmfgraph*}}}
\quad \equiv \quad \hspace*{1mm} J_1 \, .
\eeq
As usual, a straight line with an arrow represents a free electron propagator
\beq
\la{vac07}
\parbox{20mm}{\centerline{
\begin{fmfgraph*}(7,3)
\setval
\fmfleft{v1}
\fmfright{v2}
\fmf{fermion}{v2,v1}
\fmfv{decor.size=0, label=${\scs 1}$, l.dist=1mm, l.angle=-180}{v1}
\fmfv{decor.size=0, label=${\scs 2}$, l.dist=1mm, l.angle=0}{v2}
\end{fmfgraph*}}}  
\equiv \quad S_{12},
\eeq
and a wiggly line indicates a free photon propagator
\beq
\la{vac08}
\parbox{20mm}{\centerline{
\begin{fmfgraph*}(7,3)
\setval
\fmfleft{v1}
\fmfright{v2}
\fmf{boson}{v1,v2}
\fmfv{decor.size=0, label=${\scs 1}$, l.dist=1mm, l.angle=-180}{v1}
\fmfv{decor.size=0, label=${\scs 2}$, l.dist=1mm, l.angle=0}{v2}
\end{fmfgraph*}}}
\equiv \quad D_{12}.
\eeq
Both propagators are inverse functional matrices of the corresponding kernels in the action (\r{AC}):
\beq
\int_{2} S_{12} S^{-1}_{23} &= &\delta_{13} \, , \la{FI1} \\
\int_{2} D_{12} D^{-1}_{23} &= & \delta_{13} \, .\la{FI2}
\eeq
A three-vertex represents the interaction potential:
\beq
\la{vac09}
\mbox{} \no \\
\parbox{15mm}{\centerline{
\begin{fmfgraph*}(5,4.33)
\setval
\fmfforce{1w,0h}{v1}
\fmfforce{0w,0h}{v2}
\fmfforce{0.5w,1h}{v3}
\fmfforce{0.5w,0.2886h}{vm}
\fmf{fermion}{v1,vm}
\fmf{fermion}{vm,v2}
\fmf{photon}{v3,vm}
\fmfv{decor.size=0,label=${\scs 2}$,l.dist=0.5mm}{v1}
\fmfv{decor.size=0,label=${\scs 1}$,l.dist=0.5mm}{v2}
\fmfv{decor.size=0,label=${\scs 3}$,l.dist=0.5mm}{v3}
\fmfdot{vm}
\end{fmfgraph*}
}}
\equiv \quad - V_{123}\, . \\ \no 
\eeq
In this section we generate the subset of connected Feynman diagrams contributing to the partition function 
(\ref{PF}) and to the $n$-point functions (\r{EXP}) together with their
weights. To this end we introduce in Subsection \r{FDD} 
functional derivatives with respect to the graphical elements $J$, $S^{-1}$,
$D^{-1}$, $V$ of the Feynman diagrams. With these we derive in Subsection
\r{CSCD} a closed set of Schwinger-Dyson equations determining the connected $n$-point
functions. In Subsection \r{GRR} they are converted
into graphical recursion relations for the
corresponding connected Feynman diagrams.
Finally the connected vacuum diagrams contributing to the vacuum energy are constructed in a graphical way
in Subsection \r{CONNVAC}.
\subsection{Functional Derivatives}
\la{FDD}
Each Feynman diagram of QED will be considered as a {\it functional} of the quantities characterizing the underlying field theory,
the electron kernel $S^{-1}$, the photon kernel $D^{-1}$, the interaction $V$, and
an  current $J$. Following Refs.~\cite{PHI4,Boris,asym,gl,QED} we introduce in this subsection functional derivatives
with respect to these, identify their associated
graphical operation and derive fundamental field-theoretic relations between them.
\subsubsection{Graphical Representation}
We start by studying the functional derivative with respect to the
current $J$, which fulfills the identity
\beq
\la{DR1}
\frac{\delta J_2}{\delta J_1} = \delta_{12} \, .
\eeq
To represent this graphically, we represent the $\delta$-function on the right-side by
extending the elements of Feynman diagrams by an open dot with two labeled wiggly line ends 
\beq
\label{DELTA}
\parbox{8mm}{\centerline{
\begin{fmfgraph*}(4,3)
\setval
\fmfforce{0w,0.5h}{i1}
\fmfforce{1w,0.5h}{o1}
\fmfforce{0.5w,0.5h}{v1}
\fmf{photon}{i1,v1}
\fmf{photon}{v1,o1}
\fmfv{decor.size=0, label=${\scs 1}$, l.dist=1mm, l.angle=-180}{i1}
\fmfv{decor.size=0, label=${\scs 2}$, l.dist=1mm, l.angle=0}{o1}
\fmfv{decor.shape=circle,decor.filled=empty,decor.size=0.6mm}{v1}
\end{fmfgraph*}
}} 
\hspace*{0.5cm} = \quad \delta_{12} \, ,
\eeq
and picture the identity (\r{DR1}) graphically as follows
\beq
\cdphi{}{1} 
\parbox{11mm}{\centerline{  
\begin{fmfgraph*}(7,6)
\setval
\fmfleft{i1,i2}
\fmfright{o}
\fmfforce{0.5w,0.5h}{v}
\fmf{double,width=0.2mm}{i1,v,i2}
\fmf{boson}{v,o}
\fmfdot{v}
\fmfv{decor.size=0, label=${\scs 2}$, l.dist=1mm, l.angle=0}{o}
\end{fmfgraph*}}}
\quad = \quad
\parbox{6mm}{\centerline{
\begin{fmfgraph*}(4,3)
\setval
\fmfforce{0w,0.5h}{i1}
\fmfforce{1w,0.5h}{o1}
\fmfforce{0.5w,0.5h}{v1}
\fmf{photon}{i1,v1}
\fmf{photon}{v1,o1}
\fmfv{decor.size=0, label=${\scs 1}$, l.dist=1mm, l.angle=-180}{i1}
\fmfv{decor.size=0, label=${\scs 2}$, l.dist=1mm, l.angle=0}{o1}
\fmfv{decor.shape=circle,decor.filled=empty,decor.size=0.6mm}{v1}
\end{fmfgraph*}}}  
\hspace*{0.5cm} ,
\eeq
which leaves the spatial index at the line end to which the current was connected.\\

Since the photon field $A$ is bosonic, the kernel $D^{-1}$ 
is a symmetric functional matrix obeying $D^{-1}_{12}=D^{-1}_{21}$.  This property is taken into account when performing
functional derivatives with respect to the photon kernel $D^{-1}$ with the basic rule  
\beq
\la{DR2}
\frac{\delta D^{-1}_{12}}{\delta D^{-1}_{34}} = \frac{1}{2} \left\{ 
\delta_{13} \delta_{42} + \delta_{14} \delta_{32} \right\} \, .
\eeq
From the identity (\r{FI2}) and the functional chain rule of differentiation we find the derivative of the free propagator:
\beq
\la{ACT}
- 2 \, \frac{\delta D_{12}}{\delta D^{-1}_{34}} = 
D_{13} D_{42} + D_{14} D_{32} \, .
\eeq
This has the graphical representation
\beq
- 2 \, \frac{\delta}{\delta D^{-1}_{34}}
\parbox{20mm}{\centerline{
\begin{fmfgraph*}(7,3)
\setval
\fmfleft{v1}
\fmfright{v2}
\fmf{photon}{v1,v2}
\fmfv{decor.size=0, label=${\scs 1}$, l.dist=1mm, l.angle=-180}{v1}
\fmfv{decor.size=0, label=${\scs 2}$, l.dist=1mm, l.angle=0}{v2}
\end{fmfgraph*}
}}
=
\parbox{18mm}{\centerline{
\begin{fmfgraph*}(7,3)
\setval
\fmfleft{v1}
\fmfright{v2}
\fmf{photon}{v1,v2}
\fmfv{decor.size=0, label=${\scs 1}$, l.dist=1mm, l.angle=-180}{v1}
\fmfv{decor.size=0, label=${\scs 3}$, l.dist=1mm, l.angle=0}{v2}
\end{fmfgraph*}
}} \hspace*{-0.4cm}
\parbox{18mm}{\centerline{
\begin{fmfgraph*}(7,3)
\setval
\fmfleft{v1}
\fmfright{v2}
\fmf{photon}{v1,v2}
\fmfv{decor.size=0, label=${\scs 4}$, l.dist=1mm, l.angle=-180}{v1}
\fmfv{decor.size=0, label=${\scs 2}$, l.dist=1mm, l.angle=0}{v2}
\end{fmfgraph*}
}}
+
\parbox{18mm}{\centerline{
\begin{fmfgraph*}(7,3)
\setval
\fmfleft{v1}
\fmfright{v2}
\fmf{photon}{v1,v2}
\fmfv{decor.size=0, label=${\scs 1}$, l.dist=1mm, l.angle=-180}{v1}
\fmfv{decor.size=0, label=${\scs 4}$, l.dist=1mm, l.angle=0}{v2}
\end{fmfgraph*}
}}\hspace*{-0.4cm}
\parbox{18mm}{\centerline{
\begin{fmfgraph*}(7,3)
\setval
\fmfleft{v1}
\fmfright{v2}
\fmf{photon}{v1,v2}
\fmfv{decor.size=0, label=${\scs 3}$, l.dist=1mm, l.angle=-180}{v1}
\fmfv{decor.size=0, label=${\scs 2}$, l.dist=1mm, l.angle=0}{v2}
\end{fmfgraph*}
}} \, .
\eeq
Thus, differentiating a photon propagator
with respect to the kernel $D^{-1}$
amounts to cutting the associated wiggly line
into two pieces. 
The differentiation rule (\r{DR2}) ensures that the spatial indices 
of the kernel are symmetrically attached to the newly created line ends. When differentiating a general
Feynman integral 
with respect to $D^{-1}$, the product rule of functional
differentiation leads to a sum of diagrams in which
each photon line is treated in this way.\\

We now study the graphical effect of functional derivatives
with respect to the photon propagator $D$,
where the basic differentiation rule reads
\beq
\la{DR2b}
\frac{\delta D_{12}}{\delta D_{34}} = \frac{1}{2} \left\{ 
\delta_{13} \delta_{42} + \delta_{14} \delta_{32} \right\} \, .
\eeq
This is graphically written as follows:
\beq
\dbphi{}{3}{4} 
\parbox{20mm}{\centerline{
\begin{fmfgraph*}(7,3)
\setval
\fmfleft{v1}
\fmfright{v2}
\fmf{photon}{v1,v2}
\fmfv{decor.size=0, label=${\scs 1}$, l.dist=1mm, l.angle=-180}{v1}
\fmfv{decor.size=0, label=${\scs 2}$, l.dist=1mm, l.angle=0}{v2}
\end{fmfgraph*}
}} = \quad \frac{1}{2} \,\Bigg\{ \hspace*{0.2cm}
\parbox{7mm}{\centerline{
\begin{fmfgraph*}(4,3)
\setval
\fmfforce{0w,0.5h}{i1}
\fmfforce{1w,0.5h}{o1}
\fmfforce{0.5w,0.5h}{v1}
\fmf{photon}{i1,v1}
\fmf{photon}{v1,o1}
\fmfv{decor.size=0, label=${\scs 1}$, l.dist=1mm, l.angle=-180}{i1}
\fmfv{decor.size=0, label=${\scs 3}$, l.dist=1mm, l.angle=0}{o1}
\fmfv{decor.shape=circle,decor.filled=empty,decor.size=0.6mm}{v1}
\end{fmfgraph*}
}}  \quad 
\parbox{7mm}{\centerline{
\begin{fmfgraph*}(4,3)
\setval
\fmfforce{0w,0.5h}{i1}
\fmfforce{1w,0.5h}{o1}
\fmfforce{0.5w,0.5h}{v1}
\fmf{photon}{i1,v1}
\fmf{photon}{v1,o1}
\fmfv{decor.size=0, label=${\scs 4}$, l.dist=1mm, l.angle=-180}{i1}
\fmfv{decor.size=0, label=${\scs 2}$, l.dist=1mm, l.angle=0}{o1}
\fmfv{decor.shape=circle,decor.filled=empty,decor.size=0.6mm}{v1}
\end{fmfgraph*}}}  \quad + \quad
\parbox{7mm}{\centerline{
\begin{fmfgraph*}(4,3)
\setval
\fmfforce{0w,0.5h}{i1}
\fmfforce{1w,0.5h}{o1}
\fmfforce{0.5w,0.5h}{v1}
\fmf{photon}{i1,v1}
\fmf{photon}{v1,o1}
\fmfv{decor.size=0, label=${\scs 1}$, l.dist=1mm, l.angle=-180}{i1}
\fmfv{decor.size=0, label=${\scs 4}$, l.dist=1mm, l.angle=0}{o1}
\fmfv{decor.shape=circle,decor.filled=empty,decor.size=0.6mm}{v1}
\end{fmfgraph*}}}  \quad 
\parbox{7mm}{\centerline{
\begin{fmfgraph*}(4,3)
\setval
\fmfforce{0w,0.5h}{i1}
\fmfforce{1w,0.5h}{o1}
\fmfforce{0.5w,0.5h}{v1}
\fmf{photon}{i1,v1}
\fmf{photon}{v1,o1}
\fmfv{decor.size=0, label=${\scs 3}$, l.dist=1mm, l.angle=-180}{i1}
\fmfv{decor.size=0, label=${\scs 2}$, l.dist=1mm, l.angle=0}{o1}
\fmfv{decor.shape=circle,decor.filled=empty,decor.size=0.6mm}{v1}
\end{fmfgraph*}}} 
\hspace*{0.3cm} \Bigg\} \, .
\eeq
Thus differentiating a wiggly
line with respect to the photon propagator
removes the wiggly line, leaving in a symmetrized way the spatial indices of the wiggly line and 
the  photon propagator.\\

Setting up functional derivatives for electrons is different from 
the photon case, since the electron kernel $S^{-1}$ is not symmetric. The functional derivative is simply
\beq
\la{DDE}
\frac{\delta S^{-1}_{12}}{\delta S^{-1}_{34}} = 
\delta_{13} \delta_{42} \, .
\eeq
From a differentiation of the identity (\r{FI1}), we find
\beq
- \frac{\delta S_{12}}{\delta S^{-1}_{34}} = S_{13} S_{42} \, .
\eeq
Its graphical representation 
\beq
- \, \frac{\delta}{\delta S^{-1}_{34}}
\parbox{20mm}{\centerline{
\begin{fmfgraph*}(7,3)
\setval
\fmfleft{v1}
\fmfright{v2}
\fmf{electron}{v2,v1}
\fmfv{decor.size=0, label=${\scs 1}$, l.dist=1mm, l.angle=-180}{v1}
\fmfv{decor.size=0, label=${\scs 2}$, l.dist=1mm, l.angle=0}{v2}
\end{fmfgraph*}
}}
=
\parbox{18mm}{\centerline{
\begin{fmfgraph*}(7,3)
\setval
\fmfleft{v1}
\fmfright{v2}
\fmf{electron}{v2,v1}
\fmfv{decor.size=0, label=${\scs 1}$, l.dist=1mm, l.angle=-180}{v1}
\fmfv{decor.size=0, label=${\scs 3}$, l.dist=1mm, l.angle=0}{v2}
\end{fmfgraph*}
}} \hspace*{-0.4cm}
\parbox{18mm}{\centerline{
\begin{fmfgraph*}(7,3)
\setval
\fmfleft{v1}
\fmfright{v2}
\fmf{electron}{v2,v1}
\fmfv{decor.size=0, label=${\scs 4}$, l.dist=1mm, l.angle=-180}{v1}
\fmfv{decor.size=0, label=${\scs 2}$, l.dist=1mm, l.angle=0}{v2}
\end{fmfgraph*}
}}
\eeq
states that differentiating an electron propagator with respect to the
kernel $S^{-1}$ amouts to cutting the associated straight line with an
arrow once.\\

As in (\r{DDE}) the functional derivative with respect to the electron propagator $S$ reads
\beq
\la{DE}
\frac{\delta S_{12}}{\delta S_{34}} = 
\delta_{13} \delta_{42} \, .
\eeq
By analogy with (\ref{DELTA}), we represent a $\delta$-function by an open dot with two
labeled straight line ends with arrows
\beq
\parbox{8mm}{\centerline{
\begin{fmfgraph*}(4,3)
\setval
\fmfforce{0w,0.5h}{i1}
\fmfforce{1w,0.5h}{o1}
\fmfforce{0.5w,0.5h}{v1}
\fmf{electron}{v1,i1}
\fmf{electron}{o1,v1}
\fmfv{decor.size=0, label=${\scs 1}$, l.dist=1mm, l.angle=-180}{i1}
\fmfv{decor.size=0, label=${\scs 2}$, l.dist=1mm, l.angle=0}{o1}
\fmfv{decor.shape=circle,decor.filled=empty,decor.size=0.6mm}{v1}
\end{fmfgraph*}
}} 
\hspace*{0.5cm} = \quad \delta_{12} \, ,
\eeq
so that the differentiation rule (\r{DE}) has the graphical form
\beq
\dephi{}{3}{4} 
\parbox{20mm}{\centerline{
\begin{fmfgraph*}(7,3)
\setval
\fmfleft{v1}
\fmfright{v2}
\fmf{electron}{v2,v1}
\fmfv{decor.size=0, label=${\scs 1}$, l.dist=1mm, l.angle=-180}{v1}
\fmfv{decor.size=0, label=${\scs 2}$, l.dist=1mm, l.angle=0}{v2}
\end{fmfgraph*}
}} = \quad 
\parbox{7mm}{\centerline{
\begin{fmfgraph*}(4,3)
\setval
\fmfforce{0w,0.5h}{i1}
\fmfforce{1w,0.5h}{o1}
\fmfforce{0.5w,0.5h}{v1}
\fmf{electron}{v1,i1}
\fmf{electron}{o1,v1}
\fmfv{decor.size=0, label=${\scs 1}$, l.dist=1mm, l.angle=-180}{i1}
\fmfv{decor.size=0, label=${\scs 3}$, l.dist=1mm, l.angle=0}{o1}
\fmfv{decor.shape=circle,decor.filled=empty,decor.size=0.6mm}{v1}
\end{fmfgraph*}}}  \quad 
\parbox{7mm}{\centerline{
\begin{fmfgraph*}(4,3)
\setval
\fmfforce{0w,0.5h}{i1}
\fmfforce{1w,0.5h}{o1}
\fmfforce{0.5w,0.5h}{v1}
\fmf{electron}{v1,i1}
\fmf{electron}{o1,v1}
\fmfv{decor.size=0, label=${\scs 4}$, l.dist=1mm, l.angle=-180}{i1}
\fmfv{decor.size=0, label=${\scs 2}$, l.dist=1mm, l.angle=0}{o1}
\fmfv{decor.shape=circle,decor.filled=empty,decor.size=0.6mm}{v1}
\end{fmfgraph*}}} 
\hspace*{0.5cm} .
\eeq
Thus differentiating an electron line with respect to the electron propagator
removes the line, leaving the spatial indices of
the  electron propagator at the vertices to which the straight line with
an arrow was connected.\\

The functional derivative with respect to the interaction $V$ is defined by
\beq
\frac{\delta V_{123}}{\delta V_{456}} & = & 
\delta_{14} \delta_{25} \delta_{36}
\, , \la{DRV}
\eeq
which has the graphical representation
\beq
\dvertex{}{6}{5}{4}
\parbox{15mm}{\centerline{
\begin{fmfgraph*}(5,4.33)
\setval
\fmfforce{1w,0h}{v1}
\fmfforce{0w,0h}{v2}
\fmfforce{0.5w,1h}{v3}
\fmfforce{0.5w,0.2886h}{vm}
\fmf{fermion}{v1,vm}
\fmf{fermion}{vm,v2}
\fmf{photon}{v3,vm}
\fmfv{decor.size=0,label=${\scs 2}$,l.dist=0.5mm}{v1}
\fmfv{decor.size=0,label=${\scs 1}$,l.dist=0.5mm}{v2}
\fmfv{decor.size=0,label=${\scs 3}$,l.dist=0.5mm}{v3}
\fmfdot{vm}
\end{fmfgraph*}
}}& = & \hspace*{2mm}
{\displaystyle \rule[-10pt]{0pt}{40pt}\hspace*{2mm}
\parbox{15mm}{\centerline{
\begin{fmfgraph*}(12,10.392)
\setval
\fmfforce{1w,0h}{v1}
\fmfforce{0.7w,0.2h}{v1b}
\fmfforce{0.85w,0.1h}{v1c}
\fmfforce{0w,0h}{v2}
\fmfforce{0.3w,0.2h}{v2b}
\fmfforce{0.15w,0.1h}{v2c}
\fmfforce{0.5w,1h}{v3}
\fmfforce{0.5w,0.639h}{v3b}
\fmfforce{0.5w,0.82h}{v3c}
\fmf{photon}{v3,v3c,v3b}
\fmf{electron}{v1,v1c,v1b}
\fmf{electron}{v2b,v2c,v2}
\fmfv{decor.size=0,label=${\scs 2}$,l.dist=0.5mm}{v1}
\fmfv{decor.size=0,label=${\scs 1}$,l.dist=0.5mm}{v2}
\fmfv{decor.size=0,label=${\scs 3}$,l.dist=0.5mm}{v3}
\fmfv{decor.size=0,label=${\scs 5}$,l.dist=0.5mm,l.angle=120}{v1b}
\fmfv{decor.size=0,label=${\scs 4}$,l.dist=0.5mm,l.angle=60}{v2b}
\fmfv{decor.size=0,label=${\scs 6}$,l.dist=0.5mm,l.angle=-90}{v3b}
\fmfv{decor.shape=circle,decor.filled=empty,decor.size=0.6mm}{v1c}
\fmfv{decor.shape=circle,decor.filled=empty,decor.size=0.6mm}{v2c}
\fmfv{decor.shape=circle,decor.filled=empty,decor.size=0.6mm}{v3c}
\end{fmfgraph*} }}} \, .  \\ \nonumber
\eeq
Thus, differentiating a 3-vertex with respect to the 
interaction removes this vertex, leaving
the spatial indices of the interaction at the line ends to which the
vertex was connected.
\subsubsection{Field-Theoretic Identities}
With the help of these graphical operations, products of fields can be 
rewritten as functional derivatives of the action (\r{AC}). Thus we obtain
\beq
A_1 & = & - \frac{\delta {\cal A}[ \bar{\psi} , \psi , A ]}{\delta J_1} \, ,
\la{SUB1}\\
\psi_1 \bar{\psi}_2 & = & - \frac{\delta {\cal A}[ \bar{\psi} , 
\psi , A ]}{\delta S^{-1}_{21}} \, ,\la{SUB3}\\
A_1 A_2 & = & 
2 \,\frac{\delta {\cal A}[ \bar{\psi} , \psi , A ]}{\delta 
D^{-1}_{12}} \, , \la{SUB2}\\
\psi_1 \bar{\psi}_2 A_3 & = & 
- \, \frac{\delta {\cal A}[ \bar{\psi} , 
\psi , A ]}{\delta V_{213}} \, ,\la{SUB4}
\eeq
as follows from (\r{DR1}), (\r{DR2}), (\r{DDE}), and (\r{DRV}). 
Applying these derivatives to the integrands of the functional integrals (\r{EXP}) for the $n$-point functions, 
they can be determined from functional derivatives of the
partition function (\r{PF}) or its logarithm, the vacuum energy
\beq
\la{W}
W [ J , S^{-1}, D^{-1}, V ] = \ln Z [ J , S^{-1}, D^{-1}, V ] \, .
\eeq
Thus we obtain the derivative rules
\beq 
\langle A_1 \rangle & = & \frac{\delta W}{\delta J_1} \, , \la{C1} \\
\langle \psi_1 \bar{\psi}_2 \rangle & = & \frac{\delta W}{\delta S^{-1}_{21}}
\, , \la{C3} \\
\langle A_1 A_2 \rangle & = & - 2 \, \frac{\delta W}{\delta D^{-1}_{12}} 
\, , \la{C2} \\
\langle \psi_1 \bar{\psi}_2 A_3 \rangle & = & 
\frac{\delta W}{\delta V_{213}}\, . \la{C4}
\eeq
By doing so, we have to take into account compatibility relations between the
different functional derivatives
\beq
\frac{\delta W}{\delta D^{-1}_{12}} & = & - \frac{1}{2} \left\{
\frac{\delta^2 W}{\delta J_1\delta J_2} + \frac{\delta W}{\delta J_1}
\frac{\delta W}{\delta J_2} \right\} \, , \la{CP1} \\ 
\frac{\delta W}{\delta V_{213}} & = & \frac{\delta^2 W}{\delta S^{-1}_{21}
\delta J_3} + 
\frac{\delta W}{\delta S^{-1}_{21}} \frac{\delta W}{\delta J_3} \, , \la{CP2}
\eeq
which follow from the functional integral (\r{PF}) for the partition function and (\r{SUB1})--(\r{W}). Thus there  
exist different ways of obtaining all diagrams of the $n$-point
functions from the connected vacuum diagrams. From (\r{C2}) and (\r{CP1})
we read off that, for instance, the diagrams of the photon two-point
function follow either from cutting a photon line or from removing two
currents of the connected vacuum diagrams in all possible ways.
Consider as an example the vacuum energy for a vanishing 
interaction $V$, which follows directly from the functional integral 
according to (\r{AC}), (\r{PF}), and (\r{W}) 
\beq
\la{W0}
W^{({\rm free})} = W [J,S^{-1},D^{-1},0] = \mbox{Tr} \, \ln S^{-1} - \frac{1}{2} 
\mbox{Tr} \, \ln D^{-1} + \frac{1}{2} \int_{12} D_{12} J_1 J_2 \, , 
\eeq
where the trace of the logarithm of a kernel $K^{-1}=S^{-1}, D^{-1}$ is defined by
the series \cite[p.~16]{Kleinert3}
\beq
\la{LOG}
\mbox{Tr} \ln K^{-1} = \sum_{n = 1}^{\infty} \frac{(-1)^{n + 1}}{n}
\int_{1 \ldots n} \left\{ K^{-1}_{12} - \delta_{12} \right\} \cdots
\left\{ K^{-1}_{n1} - \delta_{n1} \right\} \, .
\eeq
The free photon two-point function can be
determined by applying functional derivatives to (\r{W0}) either with 
respect to the photon kernel $D^{-1}$ or with respect to the current $J$. In both cases we obtain
\beq
\la{PH0}
\langle A_1 A_2 \rangle^{({\rm free})} = D_{12} + \int_{34} D_{13} D_{24} J_3 J_4 \,.
\eeq
Similar relations follow for the connected $n$-point functions which are
defined as
\beq
\fulla_1 &= &\langle A_1 \rangle \, , \la{A1}\\
\fulls_{12} &=& \langle \psi_1 \bar{\psi}_2 \rangle \, , \la{S1} \\
\fulld_{12} &=& \langle A_1 A_2 \rangle - \langle A_1 \rangle
\langle A_2 \rangle \, , \la{D1} \\
\fullg_{123}&=& \langle \psi_1 \bar{\psi}_2 A_3 \rangle -
\langle \psi_1 \bar{\psi}_2 \rangle \langle A_3 \rangle \,, \la{V1}
\eeq
resulting in the following derivative rules for $W$:
\beq
\la{NANA}
\fulla_1 &= &\frac{\delta W}{\delta J_1} \, , \la{A2}\\
\fulls_{12} &= &\frac{\delta W}{\delta S^{-1}_{21}} \, , \la{S2}\\
\fulld_{12} &= &- 2 \, \frac{\delta W}{\delta D^{-1}_{12}} - 
\fulla_1 \fulla_2 \, , \la{D2}\\
\fullg_{123} & = &
\frac{\delta W}{\delta V_{213}} - \fulls_{12} 
\fulla_3 \la{V2} \, .
\eeq
From (\r{D2}) we read off that, for instance, cutting a photon line of
the connected vacuum diagrams in all possible ways also leads to connected and 
disconnected pieces. The latter are removed by the term $\fulla_1 \fulla_2$,
leading to the diagrams contributing to the connected photon  
two-point function $\fulld_{12}$. For later purposes we note that the
connected three-point function (\r{V2}) may be rewritten as
\beq
\la{V3B}
\fullg_{123} = \frac{\delta^2 W}{\delta S^{-1}_{21} \delta J_3} \, .
\eeq
This follows from the compatibility relation (\r{CP2}) as well as from (\r{A2}) and (\r{S2}).
\subsection{Closed Set of Schwinger-Dyson Equations for Connected $n$-Point Functions}
\la{CSCD}
We now apply the above functional derivatives 
to certain trivial functional identities which immediately follow from the definition of
the functional integral. By doing so, we derive a closed set of functional equations
determining the connected electron and photon two-point function as
well as the connected three-point function. 
\subsubsection{Connected Electron Two-Point Function}
Consider the functional identity
\beq
\oint {\cal D} \bar{\psi} {\cal D} \psi {\cal D} A 
\frac{\delta}{\delta \bar{\psi}_1} \left\{ \bar{\psi}_2 \,
e^{-{\cal A} [ \bar{\psi} , \psi , A ; J,S^{-1},D^{-1},V]} 
\right\} = 0 \, ,
\eeq
which follows by functional integration from the vanishing of the
exponential at infinite fields. Performing the functional derivative in the
integrand and taking into account the action (\r{AC}) leads to
\beq
\oint {\cal D} \bar{\psi} {\cal D} \psi {\cal D} A 
\left\{ \delta_{12} + \int_3 S^{-1}_{13} \bar{\psi}_2 \psi_3 + 
\int_{34} V_{134} \bar{\psi}_2 \psi_3 A_4 \right\} \,
e^{-{\cal A} [ \bar{\psi} , \psi , A ; J,S^{-1},D^{-1},V]} = 0 \, .
\eeq
Replacing the fields $A_4$ and $\bar{\psi}_2 \psi_3$ by functional derivatives with respect to the current $J_4$  and
the electron kernel $S^{-1}_{23}$ using (\r{SUB1}) and (\r{SUB3}), respectively,
the equation can be expressed in terms of the vacuum energy $W$ by using (\r{PF}) and (\r{W}) as
\beq
\delta_{12} - \int_3 S^{-1}_{13} \frac{\delta W}{\delta S^{-1}_{23}}
- \int_{34} V_{134} \left\{ \frac{\delta^2 W}{\delta S^{-1}_{23} 
\delta J_4} + \frac{\delta W}{\delta S^{-1}_{23}} 
\frac{\delta W}{\delta J_4} \right\} = 0 \, .
\la{ID1}
\eeq
The functional derivative with respect to the current $J$ in the last term can be eliminated with the help of the 
second functional identity
\beq
\oint {\cal D} \bar{\psi} {\cal D} \psi {\cal D} A 
\frac{\delta}{\delta A_1} 
e^{-{\cal A} [ \bar{\psi} , \psi , A ; J,S^{-1},D^{-1},V]} = 0 \, .
\eeq
After differentiating the action (\r{AC}) in the exponential, we find
\beq
\oint {\cal D} \bar{\psi} {\cal D} \psi {\cal D} A  \left\{
- J_1 + \int_2 D^{-1}_{12} A_2 +  \int_{23} V_{231} \bar{\psi}_2
\psi_3 \right\} \,
e^{-{\cal A} [ \bar{\psi} , \psi , A ; J,S^{-1},D^{-1},V]} = 0 \, ,
\eeq
which leads to
\beq
\la{ID2}
\frac{\delta W}{\delta J_4} = \int_5 D_{45} J_5 + \int_{567} V_{567}
D_{47} \frac{\delta W}{\delta S^{-1}_{56}} \, .
\eeq
Inserting this into (\r{ID1}), we obtain
\beq
\delta_{12} - \int_3 S^{-1}_{13} \frac{\delta W}{\delta S^{-1}_{23}}
- \int_{34} V_{134} \frac{\delta^2 W}{\delta S^{-1}_{23} \delta J_4}
- \int_{345} V_{134} D_{45} J_5 \frac{\delta W}{\delta S^{-1}_{23}}
- \int_{34567} V_{134} V_{567} D_{47} 
\frac{\delta W}{\delta S^{-1}_{23}} \frac{\delta W}{\delta S^{-1}_{56}} 
= 0 \, .
\eeq
Taking into account the definitions of the connected electron 
two-point function (\r{S2}) and the connected
three-point function (\r{V3B}),
this equation reduces to the Schwinger-Dyson equation for $\fulls$:
\beq
\la{IES}
\fulls_{12} = S_{12} - \int_{345} V_{345} S_{13} \fullg_{425}
- \int_{345678} V_{345} V_{678} S_{13} D_{58} \fulls_{42} \fulls_{76}
- \int_{3456} V_{345} D_{56} J_6 \fulls_{42} S_{13} \, .
\eeq
In order to represent this graphically, we extend the elements of Feynman diagrams by a symbol
for the fully interacting connected electron two-point function
\beq
\la{FS}
\parbox{20mm}{\centerline{
\begin{fmfgraph*}(7,3)
\setval
\fmfleft{v1}
\fmfright{v2}
\fmf{heavy,width=0.2mm}{v2,v1}
\fmfv{decor.size=0, label=${\scs 1}$, l.dist=1mm, l.angle=-180}{v1}
\fmfv{decor.size=0, label=${\scs 2}$, l.dist=1mm, l.angle=0}{v2}
\end{fmfgraph*}}}  
\equiv \quad \fulls_{12}\,,
\eeq
and a three-vertex with an open dot representing the fully interacting connected three-point function
\beq
\mbox{} \nonumber \\
\parbox{13.39mm}{\centerline{
\begin{fmfgraph*}(10.39,9)
\setval
\fmfforce{1/2w,2/9h}{v1}
\fmfforce{1/2w,4/9h}{v2}
\fmfforce{1/2w,1h}{v3}
\fmfforce{0w,0h}{v4}
\fmfforce{1w,0h}{v5}
\fmfforce{0.5834w,2.5/9h}{v6}
\fmfforce{0.4166w,2.5/9h}{v7}
\fmf{plain,left=1}{v1,v2,v1}
\fmf{boson}{v2,v3}
\fmf{fermion}{v7,v4}
\fmf{fermion}{v5,v6}
\fmfv{decor.size=0, label=${\scs 1}$, l.dist=1mm, l.angle=-150}{v4}
\fmfv{decor.size=0, label=${\scs 2}$, l.dist=1mm, l.angle=-30}{v5}
\fmfv{decor.size=0, label=${\scs 3}$, l.dist=1mm, l.angle=90}{v3}
\end{fmfgraph*} }} 
\equiv \quad \fullg_{123} \, . \\ \nonumber
\eeq
With this, the Schwinger Dyson equation (\ref{IES}) reads graphically 
\beq
\parbox{20mm}{\centerline{
\begin{fmfgraph*}(7,3)
\setval
\fmfleft{v1}
\fmfright{v2}
\fmf{heavy,width=0.2mm}{v2,v1}
\fmfv{decor.size=0, label=${\scs 1}$, l.dist=1mm, l.angle=-180}{v1}
\fmfv{decor.size=0, label=${\scs 2}$, l.dist=1mm, l.angle=0}{v2}
\end{fmfgraph*}}}  
=
\parbox{20mm}{\centerline{
\begin{fmfgraph*}(7,3)
\setval
\fmfleft{v1}
\fmfright{v2}
\fmf{fermion}{v2,v1}
\fmfv{decor.size=0, label=${\scs 1}$, l.dist=1mm, l.angle=-180}{v1}
\fmfv{decor.size=0, label=${\scs 2}$, l.dist=1mm, l.angle=0}{v2}
\end{fmfgraph*}}}  
+ \hspace*{0.3cm}
\parbox{20mm}{\centerline{
\begin{fmfgraph*}(17,3)
\setval
\fmfforce{0w,0h}{v1}
\fmfforce{5/17w,0h}{v2}
\fmfforce{10/17w,0h}{v3}
\fmfforce{12/17w,0h}{v4}
\fmfforce{1w,0h}{v5}
\fmfforce{11/17w,0h}{v6}
\fmfforce{10.8/17w,0.33h}{v7}
\fmf{plain,left=1}{v3,v4,v3}
\fmf{boson,left=0.85}{v2,v7}
\fmf{fermion}{v5,v4}
\fmf{fermion}{v3,v2,v1}
\fmfv{decor.size=0, label=${\scs 1}$, l.dist=1mm, l.angle=-180}{v1}
\fmfv{decor.size=0, label=${\scs 2}$, l.dist=1mm, l.angle=0}{v5}
\fmfdot{v2}
\end{fmfgraph*} }} 
\hspace*{0.3cm} - \hspace*{0.3cm}
\parbox{13mm}{\begin{center}
\begin{fmfgraph*}(10,10)
\setval
\fmfforce{0w,0h}{v1}
\fmfforce{1/2w,0h}{v2}
\fmfforce{1/2w,0.5h}{v2b}
\fmfforce{1w,0h}{v3}
\fmf{heavy,width=0.2mm}{v3,v2}
\fmf{fermion}{v2,v1}
\fmf{boson}{v2,v2b}
\fmfi{heavy,width=0.2mm}{fullcircle rotated 270 
scaled 1/2w shifted (0.5w,0.75h)}
\fmfdot{v2,v2b}
\fmfv{decor.size=0, label=${\scs 1}$, l.dist=1mm, l.angle=-180}{v1}
\fmfv{decor.size=0, label=${\scs 2}$, l.dist=1mm, l.angle=0}{v3}
\end{fmfgraph*}
\end{center}}
\hspace*{0.3cm} + \hspace{0.3cm}
\parbox{13mm}{\begin{center}
\begin{fmfgraph*}(10,7)
\setval
\fmfforce{0w,0h}{v1}
\fmfforce{1/2w,0h}{v2}
\fmfforce{1/2w,4/7h}{v2b}
\fmfforce{1w,0h}{v3}
\fmfforce{2/10w,1h}{v4}
\fmfforce{8/10w,1h}{v5}
\fmf{double,width=0.2mm}{v4,v2b,v5}
\fmf{heavy,width=0.2mm}{v3,v2}
\fmf{fermion}{v2,v1}
\fmf{boson}{v2,v2b}
\fmfdot{v2,v2b}
\fmfv{decor.size=0, label=${\scs 1}$, l.dist=1mm, l.angle=-180}{v1}
\fmfv{decor.size=0, label=${\scs 2}$, l.dist=1mm, l.angle=0}{v3}
\end{fmfgraph*}
\end{center}} 
\la{REC1}\hspace*{0.4cm} .
\eeq
\subsubsection{Connected Photon Two-Point Function}
Now we determine in a similar way the connected photon two-point function. To this end we consider the third functional identity
\beq
\oint {\cal D} \bar{\psi} {\cal D} \psi {\cal D} A \, 
\frac{\delta}{\delta A_1} \left\{ A_2 e^{- 
{\cal A} [ \bar{\psi} , \psi , A ; J,S^{-1},D^{-1},V]}  \right\} = 0 \, ,
\eeq
which leads to
\beq
\oint {\cal D} \bar{\psi} {\cal D} \psi {\cal D} A \, 
\left\{ \delta_{12} + A_2 J_1 - \int_3 D^{-1}_{13} A_2 A_3 
- \int_{34} V_{341} A_2 \bar{\psi}_3 \psi_4 
\right\} e^{- {\cal A} [ \bar{\psi} , \psi , A ; J,S^{-1},D^{-1},V]} = 0 \, .
\eeq
Substituting products of fields according to the equations (\r{SUB1}), (\r{SUB2}), and
(\r{SUB4}), we obtain
\beq
\la{ID3}
\delta_{12} + J_1 \frac{\delta W}{\delta J_2} + 2 \int_3 D^{-1}_{13}
\frac{\delta W}{\delta D^{-1}_{23}} + \int_{34}
V_{341} \left\{ \frac{\delta^2 W}{\delta S^{-1}_{34} \delta J_2 }
+ \frac{\delta W}{\delta S^{-1}_{34}}
\frac{\delta W}{\delta J_2} \right\} = 0 \, .
\eeq
Taking into account the definitions of 
the connected photon two-point function (\r{D2}), 
the connected electron two-point function (\r{S2}) 
and the connected three-point function (\r{V3B}), 
the functional derivative
with respect to the current $J$ is eliminated by using (\r{ID2}). In this
way we result in the Schwinger-Dyson equation determining $\fulld$:
\beq
\la{FUD}
\fulld_{12} = D_{12} + \int_{345} V_{345} \fullg_{432} D_{15} \, .
\eeq
Extending the elements of Feynman diagrams by a symbold for the fully interacting connected photon two-point function 
\beq
\label{FD}
\parbox{20mm}{\centerline{
\begin{fmfgraph*}(7,3)
\setval
\fmfleft{v1}
\fmfright{v2}
\fmf{dbl_wiggly,width=0.2mm}{v2,v1}
\fmfv{decor.size=0, label=${\scs 1}$, l.dist=1mm, l.angle=-180}{v1}
\fmfv{decor.size=0, label=${\scs 2}$, l.dist=1mm, l.angle=0}{v2}
\end{fmfgraph*}}}  
\equiv \quad \fulld_{12}\,,
\eeq
this Schwinger-Dyson 
equation reads graphically
\beq
\parbox{20mm}{\centerline{
\begin{fmfgraph*}(7,3)
\setval
\fmfleft{v1}
\fmfright{v2}
\fmf{dbl_wiggly,width=0.2mm}{v2,v1}
\fmfv{decor.size=0, label=${\scs 1}$, l.dist=1mm, l.angle=-180}{v1}
\fmfv{decor.size=0, label=${\scs 2}$, l.dist=1mm, l.angle=0}{v2}
\end{fmfgraph*}}}  
=
\parbox{20mm}{\centerline{
\begin{fmfgraph*}(7,3)
\setval
\fmfleft{v1}
\fmfright{v2}
\fmf{photon}{v2,v1}
\fmfv{decor.size=0, label=${\scs 1}$, l.dist=1mm, l.angle=-180}{v1}
\fmfv{decor.size=0, label=${\scs 2}$, l.dist=1mm, l.angle=0}{v2}
\end{fmfgraph*}}}  
- \hspace*{0.5cm}
\parbox{20mm}{\centerline{
\begin{fmfgraph*}(17,6)
\setval
\fmfforce{0w,1/2h}{v1}
\fmfforce{5/17w,1/2h}{v2}
\fmfforce{10/17w,1/2h}{v3}
\fmfforce{12/17w,1/2h}{v4}
\fmfforce{1w,1/2h}{v5}
\fmfforce{11/17w,1/2h}{v6}
\fmfforce{10.8/17w,2/3h}{v7}
\fmfforce{10.8/17w,1/3h}{v8}
\fmf{plain,left=1}{v3,v4,v3}
\fmf{fermion,right=0.85}{v7,v2}
\fmf{fermion,right=0.85}{v2,v8}
\fmf{boson}{v5,v4}
\fmf{boson}{v2,v1}
\fmfv{decor.size=0, label=${\scs 1}$, l.dist=1mm, l.angle=-180}{v1}
\fmfv{decor.size=0, label=${\scs 2}$, l.dist=1mm, l.angle=0}{v5}
\fmfdot{v2}
\end{fmfgraph*} }} 
\la{REC2} \hspace*{0.4cm} .
\eeq
\subsubsection{Connected Three-Point Function}
The iteration of the integral equations (\r{REC1}) and (\r{REC2})
for the connected  electron and photon two-point function $\fulls$ and $\fulld$ 
requires the knowledge of the connected three-point function $\fullg$. 
Therefore we evaluate (\r{V3B}) further by inserting (\r{ID2}):
\beq
\fullg_{123} = \int_{456} V_{456} D_{36} \frac{\delta^2 W}{\delta 
S^{-1}_{21} \delta S^{-1}_{45}} \, .
\eeq
Taking into account the definition of the connected electron
two-point function (\r{S2}) and the functional chain rule
\beq
\la{FCR}
\frac{\delta}{\delta S^{-1}_{12}} = - \int_{34} S_{31} S_{24} \,
\frac{\delta}{\delta S_{34}} \, ,
\eeq
this equation leads
to a functional integrodifferential equation for the connected three-point
function
\beq
\la{FIDE}
\fullg_{123} = - \int_{45678} V_{456} D_{36} S_{74} S_{58} 
\frac{\delta \fulls_{12}}{\delta S_{78}} \, .
\eeq
Its graphical representation reads
\beq
\la{REC3}
\parbox{13.39mm}{\centerline{
\begin{fmfgraph*}(10.39,9)
\setval
\fmfforce{1/2w,2/9h}{v1}
\fmfforce{1/2w,4/9h}{v2}
\fmfforce{1/2w,1h}{v3}
\fmfforce{0w,0h}{v4}
\fmfforce{1w,0h}{v5}
\fmfforce{0.5834w,2.5/9h}{v6}
\fmfforce{0.4166w,2.5/9h}{v7}
\fmf{plain,left=1}{v1,v2,v1}
\fmf{boson}{v2,v3}
\fmf{fermion}{v7,v4}
\fmf{fermion}{v5,v6}
\fmfv{decor.size=0, label=${\scs 1}$, l.dist=1mm, l.angle=-150}{v4}
\fmfv{decor.size=0, label=${\scs 2}$, l.dist=1mm, l.angle=-30}{v5}
\fmfv{decor.size=0, label=${\scs 3}$, l.dist=1mm, l.angle=90}{v3}
\end{fmfgraph*} }} 
\hspace*{0.1cm} = \hspace*{0.3cm} 
\dephi{
\,\parbox{12mm}{\centerline{
\begin{fmfgraph*}(7,3)
\setval
\fmfleft{v1}
\fmfright{v2}
\fmf{heavy,width=0.2mm}{v2,v1}
\fmfv{decor.size=0, label=${\scs 1}$, l.dist=1mm, l.angle=-180}{v1}
\fmfv{decor.size=0, label=${\scs 2}$, l.dist=1mm, l.angle=0}{v2}
\end{fmfgraph*}}}  
}{4}{5}
\hspace*{0.5cm}
\parbox{10.5mm}{\begin{center}
\begin{fmfgraph*}(7.5,8.66)
\setval
\fmfforce{0w,0h}{v1}
\fmfforce{0w,1h}{v2}
\fmfforce{1/3w,1/2h}{v3}
\fmfforce{1w,1/2h}{v4}
\fmf{fermion}{v1,v3}
\fmf{fermion}{v3,v2}
\fmf{boson}{v3,v4}
\fmfdot{v3}
\fmfv{decor.size=0, label=${\scs 4}$, l.dist=1mm, l.angle=-180}{v2}
\fmfv{decor.size=0, label=${\scs 5}$, l.dist=1mm, l.angle=-180}{v1}
\fmfv{decor.size=0, label=${\scs 3}$, l.dist=1mm, l.angle=0}{v4}
\end{fmfgraph*}
\end{center}}
\hspace*{0.4cm} ,
\eeq
so that the diagrams of the connected three-point function follow from those
of the connected electron two-point function by inserting a three-vertex
in an electron line in all possible ways. Thus the closed set of Schwinger-Dyson equations is given by
(\r{REC1}), (\r{REC2}), and (\r{REC3}). 
\subsection{Graphical Recursion Relations} 
\la{GRR}
Now we demonstrate how the diagrams of the connected electron and photon two-point function
as well as of the connected three-point function
are recursively generated in a graphical way. To simplify
the discussion we restrict ourselves to the case of a vanishing external
current, so that we can neglect the last term in (\r{REC1}). Performing
a loop expansion of the connected electron and photon two-point function
\beq
\label{L1}
\parbox{20mm}{\centerline{
\begin{fmfgraph*}(7,3)
\setval
\fmfleft{v1}
\fmfright{v2}
\fmf{heavy,width=0.2mm}{v2,v1}
\fmfv{decor.size=0, label=${\scs 1}$, l.dist=1mm, l.angle=-180}{v1}
\fmfv{decor.size=0, label=${\scs 2}$, l.dist=1mm, l.angle=0}{v2}
\end{fmfgraph*}}}  
& = & \quad \sum_{l=0}^{\infty}
\parbox{15mm}{\centerline{
\begin{fmfgraph*}(7,3)
\setval
\fmfleft{v1}
\fmfright{v2}
\fmfforce{0.5w,2/3h}{v3}
\fmf{heavy,width=0.2mm}{v2,v1}
\fmfv{decor.size=0, label=${\scs 1}$, l.dist=1mm, l.angle=-180}{v1}
\fmfv{decor.size=0, label=${\scs 2}$, l.dist=1mm, l.angle=0}{v2}
\fmfv{decor.size=0, label=${\scs (l)}$, l.dist=1mm, l.angle=90}{v3}
\end{fmfgraph*}}}  \hspace*{0.4cm} , \\
\parbox{20mm}{\centerline{
\begin{fmfgraph*}(7,3)
\setval
\fmfleft{v1}
\fmfright{v2}
\fmf{dbl_wiggly,width=0.2mm}{v2,v1}
\fmfv{decor.size=0, label=${\scs 1}$, l.dist=1mm, l.angle=-180}{v1}
\fmfv{decor.size=0, label=${\scs 2}$, l.dist=1mm, l.angle=0}{v2}
\end{fmfgraph*}}}  
& = &\quad
\sum_{l=0}^{\infty}
\parbox{15mm}{\centerline{
\begin{fmfgraph*}(7,3)
\setval
\fmfleft{v1}
\fmfright{v2}
\fmfforce{0.5w,2/3h}{v3}
\fmf{dbl_wiggly,width=0.2mm}{v2,v1}
\fmfv{decor.size=0, label=${\scs 1}$, l.dist=1mm, l.angle=-180}{v1}
\fmfv{decor.size=0, label=${\scs 2}$, l.dist=1mm, l.angle=0}{v2}
\fmfv{decor.size=0, label=${\scs (l)}$, l.dist=1mm, l.angle=90}{v3}
\end{fmfgraph*}}}  
\hspace*{0.4cm} , 
\eeq
as well as for the connected three-point function
\beq
\mbox{} \no \\
\parbox{13.39mm}{\centerline{
\begin{fmfgraph*}(10.39,9)
\setval
\fmfforce{1/2w,2/9h}{v1}
\fmfforce{1/2w,4/9h}{v2}
\fmfforce{1/2w,1h}{v3}
\fmfforce{0w,0h}{v4}
\fmfforce{1w,0h}{v5}
\fmfforce{0.5834w,2.5/9h}{v6}
\fmfforce{0.4166w,2.5/9h}{v7}
\fmf{plain,left=1}{v1,v2,v1}
\fmf{boson}{v2,v3}
\fmf{fermion}{v7,v4}
\fmf{fermion}{v5,v6}
\fmfv{decor.size=0, label=${\scs 1}$, l.dist=1mm, l.angle=-150}{v4}
\fmfv{decor.size=0, label=${\scs 2}$, l.dist=1mm, l.angle=-30}{v5}
\fmfv{decor.size=0, label=${\scs 3}$, l.dist=1mm, l.angle=90}{v3}
\end{fmfgraph*} }} %
\hspace*{0.3cm} = \hspace*{0.3cm} \sum_{l=0}^{\infty}\hspace*{0.2cm}
\parbox{17mm}{\centerline{
\begin{fmfgraph*}(10,8.66)
\setval
\fmfforce{1.166w,-0.12h}{v1}
\fmfforce{-0.1666w,-0.12h}{v2}
\fmfforce{0.5w,1.177h}{v3}
\fmfforce{0.25w,0.18h}{v4}
\fmfforce{0.75w,0.18h}{v5}
\fmfforce{0.5w,0.6h}{v6}
\fmfforce{0.5w,-0.0114h}{vm1}
\fmfforce{0.5w,0.2886h}{vm}
\fmfforce{0.5w,0.5886h}{vm2}
\fmf{fermion}{v4,v2}
\fmf{fermion}{v1,v5}
\fmf{boson}{v3,v6}
\fmf{plain,left=1}{vm1,vm2,vm1}
\fmfv{decor.size=0,label=${\scs l}$, l.dist=0mm, l.angle=0}{vm}
\fmfv{decor.size=0,label=${\scs 2}$,l.dist=0.5mm}{v1}
\fmfv{decor.size=0,label=${\scs 1}$,l.dist=0.5mm}{v2}
\fmfv{decor.size=0,label=${\scs 3}$,l.dist=0.5mm}{v3}
\end{fmfgraph*}
}} \hspace*{0.4cm} ,
\\ \no
\eeq%
we obtain from (\r{REC1}), (\r{REC2}), and (\r{REC3}) the following closed set of graphical recursion relations:
\beq
\parbox{15mm}{\centerline{
\begin{fmfgraph*}(7,3)
\setval
\fmfleft{v1}
\fmfright{v2}
\fmfforce{0.5w,2/3h}{v3}
\fmf{heavy,width=0.2mm}{v2,v1}
\fmfv{decor.size=0, label=${\scs 1}$, l.dist=1mm, l.angle=-180}{v1}
\fmfv{decor.size=0, label=${\scs 2}$, l.dist=1mm, l.angle=0}{v2}
\fmfv{decor.size=0, label=${\scs (l)}$, l.dist=1mm, l.angle=90}{v3}
\end{fmfgraph*}}}
\quad & = & \quad   
\parbox{24mm}{\begin{center}
\begin{fmfgraph*}(21,6)
\setval
\fmfforce{0w,0.5h}{v1}
\fmfforce{5/21w,0.5h}{v2}
\fmfforce{10/21w,0.5h}{v3}
\fmfforce{13/21w,0h}{v4}
\fmfforce{13/21w,1/2h}{v5}
\fmfforce{13/21w,1h}{v6}
\fmfforce{1w,0.5h}{v7}
\fmfforce{12/21w,0.97h}{v8}
\fmfforce{16/21w,1/2h}{v9}
\fmf{fermion}{v3,v2,v1}
\fmf{fermion}{v7,v9}
\fmf{boson,left=0.7}{v2,v8}
\fmf{plain,left=1}{v4,v6,v4}
\fmfdot{v2}
\fmfv{decor.size=0, label=${\scs l-1}$, l.dist=0mm, l.angle=0}{v5}
\fmfv{decor.size=0, label=${\scs 1}$, l.dist=1mm, l.angle=-180}{v1}
\fmfv{decor.size=0, label=${\scs 2}$, l.dist=1mm, l.angle=0}{v7}
\end{fmfgraph*}
\end{center} }
\quad - \hspace*{0.2cm} \sum_{k=0}^{l-1} \quad
\parbox{13mm}{\begin{center}
\begin{fmfgraph*}(10,10)
\setval
\fmfforce{0w,0h}{v1}
\fmfforce{1/2w,0h}{v2}
\fmfforce{1/2w,0.5h}{v2b}
\fmfforce{1w,0h}{v3}
\fmfforce{0.85w,0.3h}{v4}
\fmfforce{0.5w,1.3h}{v5}
\fmf{heavy,width=0.2mm}{v3,v2}
\fmf{fermion}{v2,v1}
\fmf{boson}{v2,v2b}
\fmfi{heavy,width=0.2mm}{fullcircle rotated 270 
scaled 1/2w shifted (0.5w,0.75h)}
\fmfdot{v2,v2b}
\fmfv{decor.size=0, label=${\scs (k)}$, l.dist=0mm, l.angle=0}{v4}
\fmfv{decor.size=0, label=${\scs (l-k-1)}$, l.dist=0mm, l.angle=0}{v5}
\fmfv{decor.size=0, label=${\scs 1}$, l.dist=1mm, l.angle=-180}{v1}
\fmfv{decor.size=0, label=${\scs 2}$, l.dist=1mm, l.angle=0}{v3}
\end{fmfgraph*}
\end{center}} \hspace*{0.4cm} , 
\la{CSD1} \\
\parbox{15mm}{\centerline{
\begin{fmfgraph*}(7,3)
\setval
\fmfleft{v1}
\fmfright{v2}
\fmfforce{0.5w,2/3h}{v3}
\fmf{dbl_wiggly,width=0.2mm}{v2,v1}
\fmfv{decor.size=0, label=${\scs 1}$, l.dist=1mm, l.angle=-180}{v1}
\fmfv{decor.size=0, label=${\scs 2}$, l.dist=1mm, l.angle=0}{v2}
\fmfv{decor.size=0, label=${\scs (l)}$, l.dist=1mm, l.angle=90}{v3}
\end{fmfgraph*}}}
\quad & = & \quad
- \hspace*{0.2cm} 
\parbox{24mm}{\begin{center}
\begin{fmfgraph*}(21,6)
\setval
\fmfforce{0w,0.5h}{v1}
\fmfforce{5/21w,0.5h}{v2}
\fmfforce{10/21w,0.5h}{v3}
\fmfforce{13/21w,0h}{v4}
\fmfforce{13/21w,1/2h}{v5}
\fmfforce{13/21w,1h}{v6}
\fmfforce{1w,0.5h}{v7}
\fmfforce{12/21w,0.97h}{v8}
\fmfforce{16/21w,1/2h}{v9}
\fmfforce{12/21w,0.03h}{v10}
\fmf{boson}{v2,v1}
\fmf{boson}{v7,v9}
\fmf{fermion,right=0.7}{v8,v2}
\fmf{fermion,right=0.7}{v2,v10}
\fmf{plain,left=1}{v4,v6,v4}
\fmfdot{v2}
\fmfv{decor.size=0, label=${\scs l-1}$, l.dist=0mm, l.angle=0}{v5}
\fmfv{decor.size=0, label=${\scs 1}$, l.dist=1mm, l.angle=-180}{v1}
\fmfv{decor.size=0, label=${\scs 2}$, l.dist=1mm, l.angle=0}{v7}
\end{fmfgraph*}
\end{center} }
\hspace*{0.4cm} ,
\la{CSD2} \\
\parbox{17mm}{\centerline{
\begin{fmfgraph*}(10,8.66)
\setval
\fmfforce{1.166w,-0.12h}{v1}
\fmfforce{-0.1666w,-0.12h}{v2}
\fmfforce{0.5w,1.177h}{v3}
\fmfforce{0.25w,0.18h}{v4}
\fmfforce{0.75w,0.18h}{v5}
\fmfforce{0.5w,0.6h}{v6}
\fmfforce{0.5w,-0.0114h}{vm1}
\fmfforce{0.5w,0.2886h}{vm}
\fmfforce{0.5w,0.5886h}{vm2}
\fmf{fermion}{v4,v2}
\fmf{fermion}{v1,v5}
\fmf{boson}{v3,v6}
\fmf{plain,left=1}{vm1,vm2,vm1}
\fmfv{decor.size=0,label=${\scs l}$, l.dist=0mm, l.angle=0}{vm}
\fmfv{decor.size=0,label=${\scs 2}$,l.dist=0.5mm}{v1}
\fmfv{decor.size=0,label=${\scs 1}$,l.dist=0.5mm}{v2}
\fmfv{decor.size=0,label=${\scs 3}$,l.dist=0.5mm}{v3}
\end{fmfgraph*}
}}
\quad & = & \quad 
\dephi{\,\parbox{12mm}{\centerline{
\begin{fmfgraph*}(7,3)
\setval
\fmfleft{v1}
\fmfright{v2}
\fmfforce{0.5w,2/3h}{v3}
\fmf{heavy,width=0.2mm}{v2,v1}
\fmfv{decor.size=0, label=${\scs 1}$, l.dist=1mm, l.angle=-180}{v1}
\fmfv{decor.size=0, label=${\scs 2}$, l.dist=1mm, l.angle=0}{v2}
\fmfv{decor.size=0, label=${\scs (l)}$, l.dist=1mm, l.angle=90}{v3}
\end{fmfgraph*}}}  
}{4}{5}
\hspace*{0.5cm}
\parbox{10.5mm}{\begin{center}
\begin{fmfgraph*}(7.5,8.66)
\setval
\fmfforce{0w,0h}{v1}
\fmfforce{0w,1h}{v2}
\fmfforce{1/3w,1/2h}{v3}
\fmfforce{1w,1/2h}{v4}
\fmf{fermion}{v1,v3}
\fmf{fermion}{v3,v2}
\fmf{boson}{v3,v4}
\fmfdot{v3}
\fmfv{decor.size=0, label=${\scs 4}$, l.dist=1mm, l.angle=120}{v2}
\fmfv{decor.size=0, label=${\scs 5}$, l.dist=1mm, l.angle=-120}{v1}
\fmfv{decor.size=0, label=${\scs 3}$, l.dist=1mm, l.angle=0}{v4}
\end{fmfgraph*}
\end{center}}
\hspace*{0.4cm} .
\la{CSD3} 
\eeq
This is solved starting from
\beq
\la{EL1}
\parbox{15mm}{\centerline{
\begin{fmfgraph*}(7,3)
\setval
\fmfleft{v1}
\fmfright{v2}
\fmfforce{0.5w,2/3h}{v3}
\fmf{heavy,width=0.2mm}{v2,v1}
\fmfv{decor.size=0, label=${\scs 1}$, l.dist=1mm, l.angle=-180}{v1}
\fmfv{decor.size=0, label=${\scs 2}$, l.dist=1mm, l.angle=0}{v2}
\fmfv{decor.size=0, label=${\scs (0)}$, l.dist=1mm, l.angle=90}{v3}
\end{fmfgraph*}}}
\quad & = &  
\parbox{20mm}{\centerline{
\begin{fmfgraph*}(7,3)
\setval
\fmfleft{v1}
\fmfright{v2}
\fmf{fermion}{v2,v1}
\fmfv{decor.size=0, label=${\scs 1}$, l.dist=1mm, l.angle=-180}{v1}
\fmfv{decor.size=0, label=${\scs 2}$, l.dist=1mm, l.angle=0}{v2}
\end{fmfgraph*}}}
\hspace*{0.4cm} , \\
\la{EL2}
\parbox{15mm}{\centerline{
\begin{fmfgraph*}(7,3)
\setval
\fmfleft{v1}
\fmfright{v2}
\fmfforce{0.5w,2/3h}{v3}
\fmf{dbl_wiggly,width=0.2mm}{v2,v1}
\fmfv{decor.size=0, label=${\scs 1}$, l.dist=1mm, l.angle=-180}{v1}
\fmfv{decor.size=0, label=${\scs 2}$, l.dist=1mm, l.angle=0}{v2}
\fmfv{decor.size=0, label=${\scs (0)}$, l.dist=1mm, l.angle=90}{v3}
\end{fmfgraph*}}}
\quad & = &
\parbox{20mm}{\centerline{
\begin{fmfgraph*}(7,3)
\setval
\fmfleft{v1}
\fmfright{v2}
\fmf{boson}{v1,v2}
\fmfv{decor.size=0, label=${\scs 1}$, l.dist=1mm, l.angle=-180}{v1}
\fmfv{decor.size=0, label=${\scs 2}$, l.dist=1mm, l.angle=0}{v2}
\end{fmfgraph*}}}
\hspace*{0.4cm} .
\eeq
At first we evaluate the amputation of one electron line from (\r{EL1}),
\beq
\dephi{\parbox{12mm}{\centerline{
\begin{fmfgraph*}(7,3)
\setval
\fmfleft{v1}
\fmfright{v2}
\fmfforce{0.5w,2/3h}{v3}
\fmf{heavy,width=0.2mm}{v2,v1}
\fmfv{decor.size=0, label=${\scs 1}$, l.dist=1mm, l.angle=-180}{v1}
\fmfv{decor.size=0, label=${\scs 2}$, l.dist=1mm, l.angle=0}{v2}
\fmfv{decor.size=0, label=${\scs (0)}$, l.dist=1mm, l.angle=90}{v3}
\end{fmfgraph*}}}}{4}{5}
\quad = \quad
\parbox{7mm}{\centerline{
\begin{fmfgraph*}(4,3)
\setval
\fmfforce{0w,0.5h}{i1}
\fmfforce{1w,0.5h}{o1}
\fmfforce{0.5w,0.5h}{v1}
\fmf{electron}{v1,i1}
\fmf{electron}{o1,v1}
\fmfv{decor.size=0, label=${\scs 1}$, l.dist=1mm, l.angle=-180}{i1}
\fmfv{decor.size=0, label=${\scs 4}$, l.dist=1mm, l.angle=0}{o1}
\fmfv{decor.shape=circle,decor.filled=empty,decor.size=0.6mm}{v1}
\end{fmfgraph*}}}  \quad 
\parbox{7mm}{\centerline{
\begin{fmfgraph*}(4,3)
\setval
\fmfforce{0w,0.5h}{i1}
\fmfforce{1w,0.5h}{o1}
\fmfforce{0.5w,0.5h}{v1}
\fmf{electron}{v1,i1}
\fmf{electron}{o1,v1}
\fmfv{decor.size=0, label=${\scs 5}$, l.dist=1mm, l.angle=-180}{i1}
\fmfv{decor.size=0, label=${\scs 2}$, l.dist=1mm, l.angle=0}{o1}
\fmfv{decor.shape=circle,decor.filled=empty,decor.size=0.6mm}{v1}
\end{fmfgraph*}}} \hspace*{0.4cm} ,
\eeq
and insert this into (\r{CSD3}) to obtain the connected three-point
function for $l=0$:
\beq
\parbox{17mm}{\centerline{
\begin{fmfgraph*}(10,8.66)
\setval
\fmfforce{1.166w,-0.12h}{v1}
\fmfforce{-0.1666w,-0.12h}{v2}
\fmfforce{0.5w,1.177h}{v3}
\fmfforce{0.25w,0.18h}{v4}
\fmfforce{0.75w,0.18h}{v5}
\fmfforce{0.5w,0.6h}{v6}
\fmfforce{0.5w,-0.0114h}{vm1}
\fmfforce{0.5w,0.2886h}{vm}
\fmfforce{0.5w,0.5886h}{vm2}
\fmf{fermion}{v4,v2}
\fmf{fermion}{v1,v5}
\fmf{boson}{v3,v6}
\fmf{plain,left=1}{vm1,vm2,vm1}
\fmfv{decor.size=0,label=${\scs 0}$, l.dist=0mm, l.angle=0}{vm}
\fmfv{decor.size=0,label=${\scs 2}$,l.dist=0.5mm}{v1}
\fmfv{decor.size=0,label=${\scs 1}$,l.dist=0.5mm}{v2}
\fmfv{decor.size=0,label=${\scs 3}$,l.dist=0.5mm}{v3}
\end{fmfgraph*}
}}
\quad = \quad 
\parbox{12mm}{\begin{center}
\begin{fmfgraph*}(8.66,7.5)
\setval
\fmfforce{0w,0h}{v1}
\fmfforce{1w,0h}{v2}
\fmfforce{1/2w,1/3h}{v3}
\fmfforce{1/2w,1h}{v4}
\fmf{boson}{v3,v4}
\fmf{fermion}{v3,v1}
\fmf{fermion}{v2,v3}
\fmfv{decor.size=0, label=${\scs 3}$, l.dist=1mm, l.angle=90}{v4}
\fmfv{decor.size=0, label=${\scs 1}$, l.dist=1mm, l.angle=-150}{v1}
\fmfv{decor.size=0, label=${\scs 2}$, l.dist=1mm, l.angle=-30}{v2}
\fmfdot{v3}
\end{fmfgraph*}
\end{center} }
\hspace*{0.4cm} .
\eeq
With this we get from (\r{CSD1}) and (\r{CSD2}) the one-loop contribution to the connected electron and photon two-point function:
\beq
\parbox{15mm}{\centerline{
\begin{fmfgraph*}(7,3)
\setval
\fmfleft{v1}
\fmfright{v2}
\fmfforce{0.5w,2/3h}{v3}
\fmf{heavy,width=0.2mm}{v2,v1}
\fmfv{decor.size=0, label=${\scs 1}$, l.dist=1mm, l.angle=-180}{v1}
\fmfv{decor.size=0, label=${\scs 2}$, l.dist=1mm, l.angle=0}{v2}
\fmfv{decor.size=0, label=${\scs (1)}$, l.dist=1mm, l.angle=90}{v3}
\end{fmfgraph*}}}
\quad & = & \quad 
\parbox{18mm}{\begin{center}
\begin{fmfgraph*}(15,5)
\setval
\fmfforce{0w,0.5h}{v1}
\fmfforce{1/3w,0.5h}{v2}
\fmfforce{2/3w,0.5h}{v3}
\fmfforce{1w,0.5h}{v4}
\fmf{fermion}{v4,v3,v2,v1}
\fmf{boson,left=1}{v2,v3}
\fmfdot{v2,v3}
\fmfv{decor.size=0, label=${\scs 1}$, l.dist=1mm, l.angle=-180}{v1}
\fmfv{decor.size=0, label=${\scs 2}$, l.dist=1mm, l.angle=0}{v4}
\end{fmfgraph*}
\end{center} }
\hspace*{0.3cm} - \hspace*{0.3cm}
\parbox{13mm}{\begin{center}
\begin{fmfgraph*}(10,10)
\setval
\fmfforce{0w,0h}{v1}
\fmfforce{1/2w,0h}{v2}
\fmfforce{1/2w,0.5h}{v2b}
\fmfforce{1w,0h}{v3}
\fmf{fermion}{v3,v2}
\fmf{fermion}{v2,v1}
\fmf{boson}{v2,v2b}
\fmfi{fermion}{fullcircle rotated 270 
scaled 1/2w shifted (0.5w,0.75h)}
\fmfdot{v2,v2b}
\fmfv{decor.size=0, label=${\scs 1}$, l.dist=1mm, l.angle=-180}{v1}
\fmfv{decor.size=0, label=${\scs 2}$, l.dist=1mm, l.angle=0}{v3}
\end{fmfgraph*}
\end{center}} \hspace*{0.4cm}, \la{ELL2} \\
\parbox{15mm}{\centerline{
\begin{fmfgraph*}(7,3)
\setval
\fmfleft{v1}
\fmfright{v2}
\fmfforce{0.5w,2/3h}{v3}
\fmf{dbl_wiggly,width=0.2mm}{v2,v1}
\fmfv{decor.size=0, label=${\scs 1}$, l.dist=1mm, l.angle=-180}{v1}
\fmfv{decor.size=0, label=${\scs 2}$, l.dist=1mm, l.angle=0}{v2}
\fmfv{decor.size=0, label=${\scs (1)}$, l.dist=1mm, l.angle=90}{v3}
\end{fmfgraph*}}}
\quad & = & \quad - \hspace*{0.3cm}
\parbox{18mm}{\begin{center}
\begin{fmfgraph*}(15,5)
\setval
\fmfforce{0w,0.5h}{v1}
\fmfforce{1/3w,0.5h}{v2}
\fmfforce{2/3w,0.5h}{v3}
\fmfforce{1w,0.5h}{v4}
\fmf{boson}{v1,v2}
\fmf{fermion,right}{v2,v3,v2}
\fmf{boson}{v3,v4}
\fmfdot{v2,v3}
\fmfv{decor.size=0, label=${\scs 1}$, l.dist=1mm, l.angle=-180}{v1}
\fmfv{decor.size=0, label=${\scs 2}$, l.dist=1mm, l.angle=0}{v4}
\end{fmfgraph*}
\end{center}} \hspace*{0.4cm} .
\eeq
Amputating one electron line from (\r{ELL2}),
\beq
\dephi{\parbox{12mm}{\centerline{
\begin{fmfgraph*}(7,3)
\setval
\fmfleft{v1}
\fmfright{v2}
\fmfforce{0.5w,2/3h}{v3}
\fmf{heavy,width=0.2mm}{v2,v1}
\fmfv{decor.size=0, label=${\scs 1}$, l.dist=1mm, l.angle=-180}{v1}
\fmfv{decor.size=0, label=${\scs 2}$, l.dist=1mm, l.angle=0}{v2}
\fmfv{decor.size=0, label=${\scs (1)}$, l.dist=1mm, l.angle=90}{v3}
\end{fmfgraph*}}}}{4}{5}
\quad & = & \quad 
%
%
\parbox{12mm}{\begin{center}
\begin{fmfgraph*}(9.24,4.24)
\setval
\fmfforce{-1.42/9.24w,-1.42/4.24h}{v1}
\fmfforce{0w,1h}{v2}
\fmfforce{2.12/9.24w,2.12/4.24h}{v3}
\fmfforce{7.12/9.24w,1/2h}{v4}
\fmfforce{10.66/9.24w,-1.42/4.24h}{v5}
\fmfforce{1w,1h}{v6}
\fmf{boson}{v3,v4}
\fmf{fermion}{v3,v1}
\fmf{fermion}{v2,v3}
\fmf{fermion}{v5,v4}
\fmf{fermion}{v4,v6}
\fmfv{decor.size=0, label=${\scs 1}$, l.dist=1mm, l.angle=-180}{v1}
\fmfv{decor.size=0, label=${\scs 2}$, l.dist=1mm, l.angle=0}{v5}
\fmfv{decor.size=0, label=${\scs 4}$, l.dist=1mm, l.angle=-180}{v2}
\fmfv{decor.size=0, label=${\scs 5}$, l.dist=1mm, l.angle=0}{v6}
\fmfdot{v3,v4}
\end{fmfgraph*}
\end{center} }
%
%
\quad + \quad 
\parbox{17mm}{\begin{center}
\begin{fmfgraph*}(15,5)
\setval
\fmfforce{0w,0.5h}{v1}
\fmfforce{1/3w,0.5h}{v2}
\fmfforce{2/3w,0.5h}{v3}
\fmfforce{13/15w,0.5h}{v4}
\fmf{fermion}{v4,v3,v2,v1}
\fmf{boson,left=1}{v2,v3}
\fmfdot{v2,v3}
\fmfv{decor.size=0, label=${\scs 1}$, l.dist=1mm, l.angle=-180}{v1}
\fmfv{decor.size=0, label=${\scs 4}$, l.dist=1mm, l.angle=0}{v4}
\end{fmfgraph*}
\end{center} }
\hspace*{0.1cm}
\parbox{7mm}{\centerline{
\begin{fmfgraph*}(4,3)
\setval
\fmfforce{0w,0.5h}{i1}
\fmfforce{1w,0.5h}{o1}
\fmfforce{0.5w,0.5h}{v1}
\fmf{electron}{v1,i1}
\fmf{electron}{o1,v1}
\fmfv{decor.size=0, label=${\scs 5}$, l.dist=1mm, l.angle=-180}{i1}
\fmfv{decor.size=0, label=${\scs 2}$, l.dist=1mm, l.angle=0}{o1}
\fmfv{decor.shape=circle,decor.filled=empty,decor.size=0.6mm}{v1}
\end{fmfgraph*}}} 
%
%
\quad + \quad
\parbox{7mm}{\centerline{
\begin{fmfgraph*}(4,3)
\setval
\fmfforce{0w,0.5h}{i1}
\fmfforce{1w,0.5h}{o1}
\fmfforce{0.5w,0.5h}{v1}
\fmf{electron}{v1,i1}
\fmf{electron}{o1,v1}
\fmfv{decor.size=0, label=${\scs 1}$, l.dist=1mm, l.angle=-180}{i1}
\fmfv{decor.size=0, label=${\scs 4}$, l.dist=1mm, l.angle=0}{o1}
\fmfv{decor.shape=circle,decor.filled=empty,decor.size=0.6mm}{v1}
\end{fmfgraph*}}} 
\hspace*{0.2cm}  
\parbox{17mm}{\begin{center}
\begin{fmfgraph*}(15,5)
\setval
\fmfforce{2/15w,0.5h}{v1}
\fmfforce{1/3w,0.5h}{v2}
\fmfforce{2/3w,0.5h}{v3}
\fmfforce{1w,0.5h}{v4}
\fmf{fermion}{v4,v3,v2,v1}
\fmf{boson,left=1}{v2,v3}
\fmfdot{v2,v3}
\fmfv{decor.size=0, label=${\scs 5}$, l.dist=1mm, l.angle=-180}{v1}
\fmfv{decor.size=0, label=${\scs 2}$, l.dist=1mm, l.angle=0}{v4}
\end{fmfgraph*}
\end{center} }
\no \\
&& 
%
%
\quad - \quad 
\parbox{7mm}{\centerline{
\begin{fmfgraph*}(4,3)
\setval
\fmfforce{0w,0.5h}{i1}
\fmfforce{1w,0.5h}{o1}
\fmfforce{0.5w,0.5h}{v1}
\fmf{electron}{v1,i1}
\fmf{electron}{o1,v1}
\fmfv{decor.size=0, label=${\scs 1}$, l.dist=1mm, l.angle=-180}{i1}
\fmfv{decor.size=0, label=${\scs 4}$, l.dist=1mm, l.angle=0}{o1}
\fmfv{decor.shape=circle,decor.filled=empty,decor.size=0.6mm}{v1}
\end{fmfgraph*}}} 
\quad
\parbox{11mm}{\begin{center}
\begin{fmfgraph*}(8,10)
\setval
\fmfforce{0w,0h}{v1}
\fmfforce{3/8w,0h}{v2}
\fmfforce{3/8w,0.5h}{v2b}
\fmfforce{1w,0h}{v3}
\fmfforce{0.5/8w,0.75h}{v4}
\fmfforce{5.5/8w,0.75h}{v5}
\fmf{fermion}{v3,v2}
\fmf{fermion}{v2,v1}
\fmf{boson}{v2,v2b}
\fmf{plain,right=1}{v4,v5}
\fmf{fermion,right=1}{v5,v4}
\fmfdot{v2,v2b}
\fmfv{decor.size=0, label=${\scs 5}$, l.dist=1mm, l.angle=-180}{v1}
\fmfv{decor.size=0, label=${\scs 2}$, l.dist=1mm, l.angle=0}{v3}
\end{fmfgraph*}
\end{center}}
%
%
\quad - \quad 
\parbox{11mm}{\begin{center}
\begin{fmfgraph*}(8,10)
\setval
\fmfforce{0w,0h}{v1}
\fmfforce{5/8w,0h}{v2}
\fmfforce{5/8w,0.5h}{v2b}
\fmfforce{1w,0h}{v3}
\fmfforce{2.5/8w,0.75h}{v4}
\fmfforce{7.5/8w,0.75h}{v5}
\fmf{fermion}{v3,v2}
\fmf{fermion}{v2,v1}
\fmf{boson}{v2,v2b}
\fmfdot{v2,v2b}
\fmf{plain,right=1}{v4,v5}
\fmf{fermion,right=1}{v5,v4}
\fmfv{decor.size=0, label=${\scs 1}$, l.dist=1mm, l.angle=-180}{v1}
\fmfv{decor.size=0, label=${\scs 4}$, l.dist=1mm, l.angle=0}{v3}
\end{fmfgraph*}
\end{center}}
\quad
\parbox{7mm}{\centerline{
\begin{fmfgraph*}(4,3)
\setval
\fmfforce{0w,0.5h}{i1}
\fmfforce{1w,0.5h}{o1}
\fmfforce{0.5w,0.5h}{v1}
\fmf{electron}{v1,i1}
\fmf{electron}{o1,v1}
\fmfv{decor.size=0, label=${\scs 5}$, l.dist=1mm, l.angle=-180}{i1}
\fmfv{decor.size=0, label=${\scs 2}$, l.dist=1mm, l.angle=0}{o1}
\fmfv{decor.shape=circle,decor.filled=empty,decor.size=0.6mm}{v1}
\end{fmfgraph*}}} 
%
%
\quad - \quad
\parbox{11mm}{\centerline{
\begin{fmfgraph*}(6,5)
\setval
\fmfforce{-2/6w,0h}{v1}
\fmfforce{1/2w,0h}{v2}
\fmfforce{8/6w,0h}{v3}
\fmfforce{0w,1h}{v4}
\fmfforce{1/2w,1h}{v5}
\fmfforce{1w,1h}{v6}
\fmf{electron}{v3,v2,v1}
\fmf{electron}{v6,v5,v4}
\fmf{photon}{v2,v5}
\fmfdot{v2,v5}
\fmfv{decor.size=0, label=${\scs 1}$, l.dist=1mm, l.angle=-180}{v1}
\fmfv{decor.size=0, label=${\scs 2}$, l.dist=1mm, l.angle=0}{v3}
\fmfv{decor.size=0, label=${\scs 4}$, l.dist=1mm, l.angle=-180}{v4}
\fmfv{decor.size=0, label=${\scs 5}$, l.dist=1mm, l.angle=0}{v6}
\end{fmfgraph*}}} 
\hspace*{0.4cm},
\eeq
we find the one-loop contribution to the connected three-point function:
\beq
\parbox{17mm}{\centerline{
\begin{fmfgraph*}(10,8.66)
\setval
\fmfforce{1.166w,-0.12h}{v1}
\fmfforce{-0.1666w,-0.12h}{v2}
\fmfforce{0.5w,1.177h}{v3}
\fmfforce{0.25w,0.18h}{v4}
\fmfforce{0.75w,0.18h}{v5}
\fmfforce{0.5w,0.6h}{v6}
\fmfforce{0.5w,-0.0114h}{vm1}
\fmfforce{0.5w,0.2886h}{vm}
\fmfforce{0.5w,0.5886h}{vm2}
\fmf{fermion}{v4,v2}
\fmf{fermion}{v1,v5}
\fmf{boson}{v3,v6}
\fmf{plain,left=1}{vm1,vm2,vm1}
\fmfv{decor.size=0,label=${\scs 1}$, l.dist=0mm, l.angle=0}{vm}
\fmfv{decor.size=0,label=${\scs 2}$,l.dist=0.5mm}{v1}
\fmfv{decor.size=0,label=${\scs 1}$,l.dist=0.5mm}{v2}
\fmfv{decor.size=0,label=${\scs 3}$,l.dist=0.5mm}{v3}
\end{fmfgraph*}
}}
& \quad = \quad &
\parbox{23mm}{\begin{center}
\begin{fmfgraph*}(20,5)
\setval
\fmfforce{0w,0h}{v1}
\fmfforce{1/4w,0h}{v2}
\fmfforce{2/4w,0h}{v3}
\fmfforce{3/4w,0h}{v4}
\fmfforce{4/4w,0h}{v5}
\fmfforce{1/2w,1h}{v6}
\fmf{fermion}{v5,v4,v3,v2,v1}
\fmf{boson,right=0.75}{v2,v4}
\fmf{boson}{v3,v6}
\fmfdot{v2,v3,v4,v6}
\fmfv{decor.size=0, label=${\scs 1}$, l.dist=1mm, l.angle=-180}{v1}
\fmfv{decor.size=0, label=${\scs 2}$, l.dist=1mm, l.angle=0}{v5}
\fmfv{decor.size=0, label=${\scs 3}$, l.dist=1mm, l.angle=90}{v6}
\end{fmfgraph*}
\end{center} }
\hspace*{0.3cm} + \hspace*{0.3cm}
\parbox{23mm}{\begin{center}
\begin{fmfgraph*}(20,5)
\setval
\fmfforce{0w,0h}{v1}
\fmfforce{1/4w,0h}{v2}
\fmfforce{2/4w,0h}{v3}
\fmfforce{3/4w,0h}{v4}
\fmfforce{1w,0h}{v5}
\fmfforce{1/4w,1h}{v6}
\fmf{fermion}{v5,v4,v3,v2,v1}
\fmf{boson,left=1}{v3,v4}
\fmf{boson}{v2,v6}
\fmfdot{v2,v3,v4}
\fmfv{decor.size=0, label=${\scs 1}$, l.dist=1mm, l.angle=-180}{v1}
\fmfv{decor.size=0, label=${\scs 2}$, l.dist=1mm, l.angle=0}{v5}
\fmfv{decor.size=0, label=${\scs 3}$, l.dist=1mm, l.angle=90}{v6}
\end{fmfgraph*}
\end{center} }
\hspace*{0.3cm} + \hspace*{0.3cm}
\parbox{23mm}{\begin{center}
\begin{fmfgraph*}(20,5)
\setval
\fmfforce{0w,0h}{v1}
\fmfforce{1/4w,0h}{v2}
\fmfforce{2/4w,0h}{v3}
\fmfforce{3/4w,0h}{v4}
\fmfforce{1w,0h}{v5}
\fmfforce{3/4w,1h}{v6}
\fmf{fermion}{v5,v4,v3,v2,v1}
\fmf{boson,left=1}{v2,v3}
\fmf{boson}{v4,v6}
\fmfdot{v2,v3,v4}
\fmfv{decor.size=0, label=${\scs 1}$, l.dist=1mm, l.angle=-180}{v1}
\fmfv{decor.size=0, label=${\scs 2}$, l.dist=1mm, l.angle=0}{v5}
\fmfv{decor.size=0, label=${\scs 3}$, l.dist=1mm, l.angle=90}{v6}
\end{fmfgraph*}
\end{center} }
\no \\
 & & - \hspace*{0.3cm}
\parbox{18mm}{\begin{center}
\begin{fmfgraph*}(15,10)
\setval
\fmfforce{0w,0h}{v1}
\fmfforce{1/3w,0h}{v2}
\fmfforce{2/3w,0h}{v3}
\fmfforce{1w,0h}{v4}
\fmfforce{1/3w,1/2h}{v5}
\fmfforce{2/3w,1/2h}{v6}
\fmf{fermion}{v4,v3,v2,v1}
\fmfi{fermion}{fullcircle rotated 270 
scaled 1/3w shifted (2/3w,0.75h)}
\fmf{boson}{v2,v5}
\fmf{boson}{v3,v6}
\fmfdot{v2,v3,v6}
\fmfv{decor.size=0, label=${\scs 1}$, l.dist=1mm, l.angle=-180}{v1}
\fmfv{decor.size=0, label=${\scs 2}$, l.dist=1mm, l.angle=0}{v4}
\fmfv{decor.size=0, label=${\scs 3}$, l.dist=1mm, l.angle=90}{v5}
\end{fmfgraph*}
\end{center} }
\hspace*{0.3cm} - \hspace*{0.3cm}
\parbox{18mm}{\begin{center}
\begin{fmfgraph*}(15,10)
\setval
\fmfforce{0w,0h}{v1}
\fmfforce{1/3w,0h}{v2}
\fmfforce{2/3w,0h}{v3}
\fmfforce{1w,0h}{v4}
\fmfforce{1/3w,1/2h}{v5}
\fmfforce{2/3w,1/2h}{v6}
\fmf{fermion}{v4,v3,v2,v1}
\fmfi{fermion}{fullcircle rotated 270 
scaled 1/3w shifted (1/3w,0.75h)}
\fmf{boson}{v2,v5}
\fmf{boson}{v3,v6}
\fmfdot{v2,v3,v5}
\fmfv{decor.size=0, label=${\scs 1}$, l.dist=1mm, l.angle=-180}{v1}
\fmfv{decor.size=0, label=${\scs 2}$, l.dist=1mm, l.angle=0}{v4}
\fmfv{decor.size=0, label=${\scs 3}$, l.dist=1mm, l.angle=90}{v6}
\end{fmfgraph*}
\end{center} }
\hspace*{0.3cm} - \hspace*{0.3cm}
\parbox{13mm}{\begin{center}
\begin{fmfgraph*}(10,15)
\setval
\fmfforce{0w,0h}{v1}
\fmfforce{1/2w,0h}{v2}
\fmfforce{1w,0h}{v3}
\fmfforce{1/2w,1/3h}{v4}
\fmfforce{1/2w,2/3h}{v5}
\fmfforce{1/2w,1h}{v6}
\fmf{fermion}{v3,v2,v1}
\fmf{fermion,right=1}{v4,v5}
\fmf{fermion,right=1}{v5,v4}
\fmf{boson}{v2,v4}
\fmf{boson}{v5,v6}
\fmfdot{v2,v4,v5}
\fmfv{decor.size=0, label=${\scs 1}$, l.dist=1mm, l.angle=-180}{v1}
\fmfv{decor.size=0, label=${\scs 2}$, l.dist=1mm, l.angle=0}{v3}
\fmfv{decor.size=0, label=${\scs 3}$, l.dist=1mm, l.angle=90}{v6}
\end{fmfgraph*}
\end{center} }
\hspace*{0.4cm} \, .
\eeq
Thus we obtain from (\r{CSD1}) and (\r{CSD2}) the two-loop connected
two-point function of the electron
\beq
\parbox{15mm}{\centerline{
\begin{fmfgraph*}(7,3)
\setval
\fmfleft{v1}
\fmfright{v2}
\fmfforce{0.5w,2/3h}{v3}
\fmf{heavy,width=0.2mm}{v2,v1}
\fmfv{decor.size=0, label=${\scs 1}$, l.dist=1mm, l.angle=-180}{v1}
\fmfv{decor.size=0, label=${\scs 2}$, l.dist=1mm, l.angle=0}{v2}
\fmfv{decor.size=0, label=${\scs (2)}$, l.dist=1mm, l.angle=90}{v3}
\end{fmfgraph*}}}
\quad & = & \quad 
%
%
\parbox{28mm}{\begin{center}
\begin{fmfgraph*}(25,10)
\setval
\fmfforce{0w,1/2h}{v1}
\fmfforce{1/5w,1/2h}{v2}
\fmfforce{2/5w,1/2h}{v3}
\fmfforce{3/5w,1/2h}{v4}
\fmfforce{4/5w,1/2h}{v5}
\fmfforce{1w,1/2h}{v6}
\fmf{fermion}{v6,v5,v4,v3,v2,v1}
\fmf{boson,left=0.75}{v2,v4}
\fmf{boson,right=0.75}{v3,v5}
\fmfdot{v2,v3,v4,v5}
\fmfv{decor.size=0, label=${\scs 1}$, l.dist=1mm, l.angle=-180}{v1}
\fmfv{decor.size=0, label=${\scs 2}$, l.dist=1mm, l.angle=0}{v6}
\end{fmfgraph*}
\end{center} }
\hspace*{0.3cm} - \hspace*{0.3cm} 
%
%
\parbox{18mm}{\begin{center}
\begin{fmfgraph*}(15,7.5)
\setval
\fmfforce{0w,0h}{v1}
\fmfforce{1/3w,0h}{v2}
\fmfforce{2/3w,0h}{v3}
\fmfforce{3/3w,0h}{v4}
\fmfforce{1/3w,2/3h}{v5}
\fmfforce{2/3w,2/3h}{v6}
\fmf{fermion}{v4,v3,v2,v1}
\fmf{fermion,right=1}{v5,v6}
\fmf{fermion,right=1}{v6,v5}
\fmf{boson}{v2,v5}
\fmf{boson}{v3,v6}
\fmfdot{v2,v3,v5,v6}
\fmfv{decor.size=0, label=${\scs 1}$, l.dist=1mm, l.angle=-180}{v1}
\fmfv{decor.size=0, label=${\scs 2}$, l.dist=1mm, l.angle=0}{v4}
\end{fmfgraph*}
\end{center} }
\hspace*{0.3cm} + \hspace*{0.3cm} 
%
%
\parbox{28mm}{\begin{center}
\begin{fmfgraph*}(25,5)
\setval
\fmfforce{0w,0h}{v1}
\fmfforce{1/5w,0h}{v2}
\fmfforce{2/5w,0h}{v3}
\fmfforce{3/5w,0h}{v4}
\fmfforce{4/5w,0h}{v5}
\fmfforce{1w,0h}{v6}
\fmf{fermion}{v6,v5,v4,v3,v2,v1}
\fmf{boson,left=0.75}{v2,v5}
\fmf{boson,left=1}{v3,v4}
\fmfdot{v2,v3,v4,v5}
\fmfv{decor.size=0, label=${\scs 1}$, l.dist=1mm, l.angle=-180}{v1}
\fmfv{decor.size=0, label=${\scs 2}$, l.dist=1mm, l.angle=0}{v6}
\end{fmfgraph*}
\end{center} }
\no \\
&&
\quad + \hspace*{0.3cm} 
\parbox{28mm}{\begin{center}
\begin{fmfgraph*}(25,2.5)
\setval
\fmfforce{0w,0h}{v1}
\fmfforce{1/5w,0h}{v2}
\fmfforce{2/5w,0h}{v3}
\fmfforce{3/5w,0h}{v4}
\fmfforce{4/5w,0h}{v5}
\fmfforce{1w,0h}{v6}
\fmf{fermion}{v6,v5,v4,v3,v2,v1}
\fmf{boson,left=1}{v2,v3}
\fmf{boson,left=1}{v4,v5}
\fmfdot{v2,v3,v4,v5}
\fmfv{decor.size=0, label=${\scs 1}$, l.dist=1mm, l.angle=-180}{v1}
\fmfv{decor.size=0, label=${\scs 2}$, l.dist=1mm, l.angle=0}{v6}
\end{fmfgraph*}
\end{center} }
\hspace*{0.3cm} - \hspace*{0.3cm} 
\parbox{23mm}{\begin{center}
\begin{fmfgraph*}(20,10)
\setval
\fmfforce{0w,0h}{v1}
\fmfforce{1/4w,0h}{v2}
\fmfforce{2/4w,0h}{v3}
\fmfforce{3/4w,0h}{v4}
\fmfforce{4/4w,0h}{v5}
\fmfforce{3/4w,1/2h}{v6}
\fmfforce{3/4w,1h}{v7}
\fmf{fermion}{v5,v4,v3,v2,v1}
\fmfi{fermion}{fullcircle rotated 270 
scaled 1/4w shifted (3/4w,0.75h)}
\fmf{boson,left=1}{v2,v3}
\fmf{boson}{v4,v6}
\fmfdot{v2,v3,v4,v6}
\fmfv{decor.size=0, label=${\scs 1}$, l.dist=1mm, l.angle=-180}{v1}
\fmfv{decor.size=0, label=${\scs 2}$, l.dist=1mm, l.angle=0}{v5}
\end{fmfgraph*}
\end{center} }
\hspace*{0.3cm} - \hspace*{0.3cm} 
\parbox{23mm}{\begin{center}
\begin{fmfgraph*}(20,10)
\setval
\fmfforce{0w,0h}{v1}
\fmfforce{1/4w,0h}{v2}
\fmfforce{2/4w,0h}{v3}
\fmfforce{3/4w,0h}{v4}
\fmfforce{4/4w,0h}{v5}
\fmfforce{1/4w,1/2h}{v6}
\fmfforce{1/4w,1h}{v7}
\fmf{fermion}{v5,v4,v3,v2,v1}
\fmfi{fermion}{fullcircle rotated 270 
scaled 1/4w shifted (1/4w,0.75h)}
\fmf{boson,left=1}{v3,v4}
\fmf{boson}{v2,v6}
\fmfdot{v2,v3,v4,v6}
\fmfv{decor.size=0, label=${\scs 1}$, l.dist=1mm, l.angle=-180}{v1}
\fmfv{decor.size=0, label=${\scs 2}$, l.dist=1mm, l.angle=0}{v5}
\end{fmfgraph*}
\end{center} }
\no \\
&&
\quad - \hspace*{0.3cm} 
\parbox{23mm}{\begin{center}
\begin{fmfgraph*}(20,10)
\setval
\fmfforce{0w,0h}{v1}
\fmfforce{1/4w,0h}{v2}
\fmfforce{2/4w,0h}{v3}
\fmfforce{3/4w,0h}{v4}
\fmfforce{4/4w,0h}{v5}
\fmfforce{1/2w,1/2h}{v6}
\fmfforce{1/2w,1h}{v7}
\fmf{fermion}{v5,v4,v3,v2,v1}
\fmfi{fermion}{fullcircle rotated 270 
scaled 1/4w shifted (1/2w,0.75h)}
\fmf{boson,right=0.75}{v2,v4}
\fmf{boson}{v3,v6}
\fmfdot{v2,v3,v4,v6}
\fmfv{decor.size=0, label=${\scs 1}$, l.dist=1mm, l.angle=-180}{v1}
\fmfv{decor.size=0, label=${\scs 2}$, l.dist=1mm, l.angle=0}{v5}
\end{fmfgraph*}
\end{center} }
\hspace*{0.3cm}+ \hspace*{0.3cm}
\parbox{20.5mm}{\begin{center}
\begin{fmfgraph*}(17.5,10)
\setval
\fmfforce{0w,0h}{v1}
\fmfforce{2/7w,0h}{v2}
\fmfforce{5/7w,0h}{v3}
\fmfforce{1w,0h}{v4}
\fmfforce{2/7w,1/2h}{v5}
\fmfforce{5/7w,1/2h}{v6}
\fmf{fermion}{v4,v3,v2,v1}
\fmfi{fermion}{fullcircle rotated 270 
scaled 1/2h shifted (2/7w,0.75h)}
\fmfi{fermion}{fullcircle rotated 270 
scaled 1/2h shifted (5/7w,0.75h)}
\fmf{boson}{v2,v5}
\fmf{boson}{v3,v6}
\fmfdot{v2,v3,v5,v6}
\fmfv{decor.size=0, label=${\scs 1}$, l.dist=1mm, l.angle=-180}{v1}
\fmfv{decor.size=0, label=${\scs 2}$, l.dist=1mm, l.angle=0}{v4}
\end{fmfgraph*}
\end{center} }
\hspace*{0.3cm} + \hspace*{0.3cm} 
\parbox{13mm}{\begin{center}
\begin{fmfgraph*}(10,20)
\setval
\fmfforce{0w,0h}{v1}
\fmfforce{1/2w,0h}{v2}
\fmfforce{1w,0h}{v3}
\fmfforce{1/2w,1/4h}{v4}
\fmfforce{1/2w,2/4h}{v5}
\fmfforce{1/2w,3/4h}{v6}
\fmf{fermion}{v3,v2,v1}
\fmfi{fermion}{fullcircle rotated 270 
scaled 1/2w shifted (1/2w,0.875h)}
\fmf{fermion,right=1}{v4,v5}
\fmf{fermion,right=1}{v5,v4}
\fmf{boson}{v2,v4}
\fmf{boson}{v5,v6}
\fmfdot{v2,v4,v5,v6}
\fmfv{decor.size=0, label=${\scs 1}$, l.dist=1mm, l.angle=-180}{v1}
\fmfv{decor.size=0, label=${\scs 2}$, l.dist=1mm, l.angle=0}{v3}
\end{fmfgraph*}
\end{center} }
\hspace*{0.3cm} - \hspace*{0.3cm} 
\parbox{15.5mm}{\begin{center}
\begin{fmfgraph*}(10,12.5)
\setval
\fmfforce{0w,0h}{v1}
\fmfforce{1/2w,0h}{v2}
\fmfforce{1w,0h}{v3}
\fmfforce{1/2w,5/12.5h}{v4}
\fmfforce{0.81w,0.85h}{v5}
\fmfforce{0.19w,0.85h}{v6}
\fmf{fermion}{v3,v2,v1}
\fmf{boson}{v2,v4}
\fmf{fermion,right=0.6}{v4,v5}
\fmf{fermion,right=0.6}{v6,v4}
\fmf{fermion,right=0.6}{v5,v6}
\fmf{boson,left=0.6}{v5,v6}
\fmfdot{v2,v4,v5,v6}
\fmfv{decor.size=0, label=${\scs 1}$, l.dist=1mm, l.angle=-180}{v1}
\fmfv{decor.size=0, label=${\scs 2}$, l.dist=1mm, l.angle=0}{v3}
\end{fmfgraph*}
\end{center} }
\la{PHOA}
\eeq
and of the photon
\beq
\parbox{15mm}{\centerline{
\begin{fmfgraph*}(7,3)
\setval
\fmfleft{v1}
\fmfright{v2}
\fmfforce{0.5w,2/3h}{v3}
\fmf{dbl_wiggly,width=0.2mm}{v2,v1}
\fmfv{decor.size=0, label=${\scs 1}$, l.dist=1mm, l.angle=-180}{v1}
\fmfv{decor.size=0, label=${\scs 2}$, l.dist=1mm, l.angle=0}{v2}
\fmfv{decor.size=0, label=${\scs (2)}$, l.dist=1mm, l.angle=90}{v3}
\end{fmfgraph*}}}
\quad & = & \quad 
- \hspace*{0.3cm} 
\parbox{21mm}{\begin{center}
\begin{fmfgraph*}(17,7)
\setval
\fmfforce{0w,1/2h}{v1}
\fmfforce{5/17w,1/2h}{v2}
\fmfforce{12/17w,1/2h}{v3}
\fmfforce{1w,1/2h}{v4}
\fmfforce{1/2w,0h}{v5}
\fmfforce{1/2w,1h}{v6}
\fmf{boson}{v1,v2}
\fmf{boson}{v3,v4}
\fmf{boson}{v5,v6}
\fmf{fermion,right=0.4}{v2,v5}
\fmf{fermion,right=0.4}{v5,v3}
\fmf{fermion,right=0.4}{v3,v6}
\fmf{fermion,right=0.4}{v6,v2}
\fmfdot{v2,v3,v5,v6}
\fmfv{decor.size=0, label=${\scs 1}$, l.dist=1mm, l.angle=-180}{v1}
\fmfv{decor.size=0, label=${\scs 2}$, l.dist=1mm, l.angle=0}{v4}
\end{fmfgraph*}
\end{center} }
\hspace*{0.3cm} - \hspace*{0.3cm} 
\parbox{18mm}{\begin{center}
\begin{fmfgraph*}(15,5)
\setval
\fmfforce{0w,0h}{v1}
\fmfforce{1/3w,0h}{v2}
\fmfforce{2/3w,0h}{v3}
\fmfforce{1w,0h}{v4}
\fmfforce{1/3w,1h}{v5}
\fmfforce{2/3w,1h}{v6}
\fmf{boson}{v1,v2}
\fmf{boson}{v3,v4}
\fmf{boson,right=0.4}{v5,v6}
\fmf{fermion,right=0.4}{v5,v2}
\fmf{fermion,right=0.4}{v6,v5}
\fmf{fermion,right=0.4}{v3,v6}
\fmf{fermion,right=0.4}{v2,v3}
\fmfdot{v2,v3,v5,v6}
\fmfv{decor.size=0, label=${\scs 1}$, l.dist=1mm, l.angle=-180}{v1}
\fmfv{decor.size=0, label=${\scs 2}$, l.dist=1mm, l.angle=0}{v4}
\end{fmfgraph*}
\end{center} }
\hspace*{0.3cm} - \hspace*{0.3cm} 
\parbox{18mm}{\begin{center}
\begin{fmfgraph*}(15,5)
\setval
\fmfforce{0w,0h}{v1}
\fmfforce{1/3w,0h}{v2}
\fmfforce{2/3w,0h}{v3}
\fmfforce{1w,0h}{v4}
\fmfforce{1/3w,1h}{v5}
\fmfforce{2/3w,1h}{v6}
\fmf{boson}{v1,v2}
\fmf{boson}{v3,v4}
\fmf{boson,right=0.4}{v5,v6}
\fmf{fermion,left=0.4}{v2,v5}
\fmf{fermion,left=0.4}{v5,v6}
\fmf{fermion,left=0.4}{v6,v3}
\fmf{fermion,left=0.4}{v3,v2}
\fmfdot{v2,v3,v5,v6}
\fmfv{decor.size=0, label=${\scs 1}$, l.dist=1mm, l.angle=-180}{v1}
\fmfv{decor.size=0, label=${\scs 2}$, l.dist=1mm, l.angle=0}{v4}
\end{fmfgraph*}
\end{center} }
\no \\
\la{PHO}
&& \quad + \hspace*{0.3cm}
\parbox{28mm}{\begin{center}
\begin{fmfgraph*}(25,5)
\setval
\fmfforce{0w,1/2h}{v1}
\fmfforce{1/5w,1/2h}{v2}
\fmfforce{2/5w,1/2h}{v3}
\fmfforce{3/5w,1/2h}{v4}
\fmfforce{4/5w,1/2h}{v5}
\fmfforce{5/5w,1/2h}{v6}
\fmf{boson}{v1,v2}
\fmf{fermion,right=1}{v2,v3,v2}
\fmf{boson}{v3,v4}
\fmf{fermion,right=1}{v4,v5,v4}
\fmf{boson}{v5,v6}
\fmfdot{v2,v3,v4,v5}
\fmfv{decor.size=0, label=${\scs 1}$, l.dist=1mm, l.angle=-180}{v1}
\fmfv{decor.size=0, label=${\scs 2}$, l.dist=1mm, l.angle=0}{v6}
\end{fmfgraph*}
\end{center} }
\hspace*{0.3cm} + \hspace*{0.3cm} 
\parbox{18mm}{\begin{center}
\begin{fmfgraph*}(15,15)
\setval
\fmfforce{0w,1/6h}{v1}
\fmfforce{1/3w,1/6h}{v2}
\fmfforce{2/3w,1/6h}{v3}
\fmfforce{1w,1/6h}{v4}
\fmfforce{1/2w,1/3h}{v5}
\fmfforce{1/2w,2/3h}{v6}
\fmfi{fermion}{fullcircle rotated 270 
scaled 1/3w shifted (1/2w,5/6h)}
\fmf{boson}{v1,v2}
\fmf{boson}{v3,v4}
\fmf{boson}{v5,v6}
\fmf{fermion,right=0.4}{v5,v2}
\fmf{fermion,right=0.4}{v3,v5}
\fmf{fermion,right=1}{v2,v3}
\fmfdot{v2,v3,v5,v6}
\fmfv{decor.size=0, label=${\scs 1}$, l.dist=1mm, l.angle=-180}{v1}
\fmfv{decor.size=0, label=${\scs 2}$, l.dist=1mm, l.angle=0}{v4}
\end{fmfgraph*}
\end{center} }
\hspace*{0.3cm} + \hspace*{0.3cm} 
\parbox{18mm}{\begin{center}
\begin{fmfgraph*}(15,15)
\setval
\fmfforce{0w,1/6h}{v1}
\fmfforce{1/3w,1/6h}{v2}
\fmfforce{2/3w,1/6h}{v3}
\fmfforce{1w,1/6h}{v4}
\fmfforce{1/2w,1/3h}{v5}
\fmfforce{1/2w,2/3h}{v6}
\fmfi{fermion}{fullcircle rotated 270 
scaled 1/3w shifted (1/2w,5/6h)}
\fmf{boson}{v1,v2}
\fmf{boson}{v3,v4}
\fmf{boson}{v5,v6}
\fmf{fermion,left=0.4}{v2,v5}
\fmf{fermion,left=0.4}{v5,v3}
\fmf{fermion,left=1}{v3,v2}
\fmfdot{v2,v3,v5,v6}
\fmfv{decor.size=0, label=${\scs 1}$, l.dist=1mm, l.angle=-180}{v1}
\fmfv{decor.size=0, label=${\scs 2}$, l.dist=1mm, l.angle=0}{v4}
\end{fmfgraph*}
\end{center} }
\hspace*{0.4cm} .
\eeq
From the Feynman diagrams of the connected electron and photon two-point function as well as the connected
three-point functiongenerated so far, we read off a simple rule for their weights.
They are given by $(-1)^l$ with $l$ being the number of electron 
loops \cite{QED}. The same Feynman diagrams have been obtained in
Ref. \cite{QED} by amputating lines or vertices in the connected vacuum diagrams.
\subsection{Connected Vacuum Diagrams}
\la{CONNVAC}
The connected vacuum diagrams of QED can be generated together with
their weights in two different ways. First, they can be constructed from the above diagrams of the connected electron
and photon two-point function as well as the connected three-point function.
Second, we derive from the above functional identities a nonlinear functional differential equation
for the vacuum energy and convert it into a graphical recursion
relation which directly generates the connected vacuum diagrams as in Ref.~\cite{QED}.
\subsubsection{Relation to the Diagrams of the Connected $n$-Point Functions}
After having iteratively solved the closed set of graphical 
recursion relations (\r{CSD1})--(\r{CSD3}) for the diagrams of the
connected electron and photon two-point function as well as of the 
connected three-point function, the corresponding connected
vacuum diagrams can be constructed loopwise as follows. Going back to the defining
equations (\r{S2})--(\r{V2}) for $\fulls$, $\fulld$, $\fullg$, we obtain
with (\r{A2}), (\r{ID2}), and the functional chain rule (\r{FCR}) 
three functional differential equations for the vacuum energy:
\beq
\int_{12} S_{12} \frac{\delta W}{\delta S_{12}} & = &
- \int_{12} S^{-1}_{21} \fulls_{12} \, , \la{FDW1} \\
\int_{12} D_{12} \frac{\delta W}{\delta D_{12}} & = &
\frac{1}{2} \int_{12} D^{-1}_{12} \fulld_{12} + 
\frac{1}{2} \int_{123456} V_{123} V_{456} \fulls_{21} \fulls_{54}
D_{36} + \frac{1}{2} \int_{12} D_{12} J_1 J_2 + \int_{1234} V_{123} 
\fulls_{21} D_{34} J_4 \, , \la{FDW2} \\
\int_{123} V_{123} \frac{\delta W}{\delta V_{123}} & = & \int_{123} V_{123}
\fullg_{213} + \int_{123456} V_{123} V_{456} \fulls_{21} \fulls_{54}
D_{36} + \int_{1234} V_{123} \fulls_{21} D_{34} J_4 \, . \la{FDW3}
\eeq
Their graphical representations are
\beq
\parbox{8mm}{\begin{center}
\begin{fmfgraph*}(2.5,5)
\setval
\fmfstraight
\fmfforce{1w,0h}{v1}
\fmfforce{1w,1h}{v2}
\fmf{electron,left=1}{v1,v2}
\fmfv{decor.size=0, label=${\scs 2}$, l.dist=1mm, l.angle=0}{v1}
\fmfv{decor.size=0, label=${\scs 1}$, l.dist=1mm, l.angle=0}{v2}
\end{fmfgraph*}
\end{center}}
\hspace*{0.3cm} \dephi{W}{1}{2} \quad & = & \quad - \hspace*{0.2cm}
\parbox{8mm}{\centerline{
\begin{fmfgraph}(5,5)
\setval
\fmfforce{0w,0.5h}{v1}
\fmfforce{1w,0.5h}{v2}
\fmf{heavy,width=0.2mm,right=1}{v2,v1}
\fmf{fermion,right=1}{v1,v2}
\end{fmfgraph} }} 
\hspace*{0.4cm} , \la{VAC1} \\
\parbox{8mm}{\begin{center}
\begin{fmfgraph*}(2.5,5)
\setval
\fmfstraight
\fmfforce{1w,0h}{v1}
\fmfforce{1w,1h}{v2}
\fmf{photon,left=1}{v1,v2}
\fmfv{decor.size=0, label=${\scs 2}$, l.dist=1mm, l.angle=0}{v1}
\fmfv{decor.size=0, label=${\scs 1}$, l.dist=1mm, l.angle=0}{v2}
\end{fmfgraph*}
\end{center}}
\hspace*{0.3cm} \dbphi{W}{1}{2} \quad & = & \quad \frac{1}{2} \hspace*{0.2cm}
\parbox{8mm}{\centerline{
\begin{fmfgraph}(5,5)
\setval
\fmfforce{0w,0.5h}{v1}
\fmfforce{1w,0.5h}{v2}
\fmf{dbl_wiggly,width=0.2mm,right=1}{v2,v1}
\fmf{boson,right=1}{v1,v2}
\end{fmfgraph} }} 
\hspace*{0.2cm} + \hspace*{0.2cm} \frac{1}{2} \hspace*{0.2cm}
\parbox{19mm}{\centerline{
\begin{fmfgraph}(15,5)
\setval
\fmfforce{1/3w,0.5h}{v1}
\fmfforce{2/3w,0.5h}{v2}
\fmf{boson}{v1,v2}
\fmfi{heavy,width=0.2mm,right=1}{reverse fullcircle 
scaled 1/3w shifted (1/6w,0.5h)}
\fmfi{heavy,width=0.2mm,right=1}{fullcircle rotated 
180 scaled 1/3w shifted (5/6w,0.5h)}
\fmfdot{v1,v2}
\end{fmfgraph}
}}
\hspace*{0.2cm} + \hspace*{0.2cm} \frac{1}{2} \hspace*{0.2cm}
\parbox{14mm}{\centerline{
\begin{fmfgraph}(11,6)
\setval
\fmfforce{3/11w,0.5h}{v1}
\fmfforce{8/11w,0.5h}{v2}
\fmfforce{0w,0h}{v3}
\fmfforce{0w,1h}{v4}
\fmfforce{1w,0h}{v5}
\fmfforce{1w,1h}{v6}
\fmf{double,width=0.2mm}{v3,v1,v4}
\fmf{double,width=0.2mm}{v5,v2,v6}
\fmf{photon}{v1,v2}
\fmfdot{v1,v2}
\end{fmfgraph} }} 
\hspace*{0.2cm} - \hspace*{0.2cm}
\parbox{16mm}{\centerline{
\begin{fmfgraph}(13,6)
\setval
\fmfforce{3/13w,0.5h}{v1}
\fmfforce{8/13w,0.5h}{v2}
\fmfforce{1w,0.5h}{v3}
\fmfforce{0w,0h}{v4}
\fmfforce{0w,1h}{v5}
\fmfforce{10.5/13w,5.5/6h}{v6}
\fmfforce{10.5/13w,0.5/6h}{v7}
\fmf{double,width=0.2mm,right=1}{v6,v7}
\fmf{heavy,width=0.2mm,right=1}{v7,v6}
\fmf{double,width=0.2mm}{v4,v1,v5}
\fmf{photon}{v1,v2}
\fmfdot{v2,v1}
\end{fmfgraph} }} 
\hspace*{0.4cm} , \la{VAC2} \\
\parbox{5mm}{\begin{center}
\begin{fmfgraph*}(3,4)
\setval
\fmfstraight
\fmfforce{0w,1/2h}{v1}
\fmfforce{1w,1/2h}{v2}
\fmfforce{1w,1.25h}{v3}
\fmfforce{1w,-0.25h}{v4}
\fmf{photon}{v1,v2}
\fmf{electron}{v1,v3}
\fmf{electron}{v4,v1}
\fmfv{decor.size=0, label=${\scs 3}$, l.dist=1mm, l.angle=0}{v2}
\fmfv{decor.size=0, label=${\scs 1}$, l.dist=1mm, l.angle=0}{v3}
\fmfv{decor.size=0, label=${\scs 2}$, l.dist=1mm, l.angle=0}{v4}
\fmfdot{v1}
\end{fmfgraph*}
\end{center}}
\quad \dvertex{W}{3}{2}{1} \quad & = & \quad \hspace*{0.3cm} - \hspace*{0.3cm}
\parbox{10mm}{\begin{center}
\begin{fmfgraph}(7,6)
\setval
\fmfforce{0w,1/2h}{v1}
\fmfforce{5/7w,1/2h}{v2}
\fmfforce{6/7w,1/2h}{v3}
\fmfforce{1w,1/2h}{v4}
\fmfforce{0.8w,0.67h}{v5}
\fmfforce{0.8w,0.33h}{v6}
\fmf{boson}{v1,v2}
\fmf{plain,left=1}{v2,v4,v2}
\fmf{fermion,right=0.9}{v5,v1}
\fmf{fermion,right=0.9}{v1,v6}
\fmfdot{v1}
\end{fmfgraph}
\end{center} }
\quad + \quad
\parbox{19mm}{\centerline{
\begin{fmfgraph}(15,5)
\setval
\fmfforce{1/3w,0.5h}{v1}
\fmfforce{2/3w,0.5h}{v2}
\fmf{boson}{v1,v2}
\fmfi{heavy,width=0.2mm,right=1}{reverse fullcircle 
scaled 1/3w shifted (1/6w,0.5h)}
\fmfi{heavy,width=0.2mm,right=1}{fullcircle rotated 
180 scaled 1/3w shifted (5/6w,0.5h)}
\fmfdot{v1,v2}
\end{fmfgraph}
}}
\quad - \quad
\parbox{16mm}{\centerline{
\begin{fmfgraph}(13,6)
\setval
\fmfforce{3/13w,0.5h}{v1}
\fmfforce{8/13w,0.5h}{v2}
\fmfforce{1w,0.5h}{v3}
\fmfforce{0w,0h}{v4}
\fmfforce{0w,1h}{v5}
\fmfforce{10.5/13w,5.5/6h}{v6}
\fmfforce{10.5/13w,0.5/6h}{v7}
\fmf{double,width=0.2mm,right=1}{v6,v7}
\fmf{heavy,width=0.2mm,right=1}{v7,v6}
\fmf{double,width=0.2mm}{v4,v1,v5}
\fmf{photon}{v1,v2}
\fmfdot{v2,v1}
\end{fmfgraph} }} 
\hspace*{0.4cm} , \la{VAC3} \\ && \no 
\eeq
where the first term on the right-hand side of (\r{VAC1}) and (\r{VAC2}) pictures the closing of the external legs
of the connected electron and photon two-point function, respectively.
All three equations have in common that the terms on the left-hand side
count the number of a graphical element of each connected vacuum diagram.
Indeed, when performing the operation $\int d E \delta/\delta E$ with $E=S,D,V$, 
the functional derivative $\delta/\delta E$
removes successively an electron line, a photon line, or a three-vertex 
in all possible ways, which is subsequently reinserted by the integration $\int d E$.\\

If the interaction $V$ vanishes, the Eqs. (\r{VAC1})--(\r{VAC3}) are solved
by the free contribution to the vacuum energy (\r{W0}) with the graphical representation 
\beq
\la{ZEGA0}
W^{({\rm free})} \quad = \quad - \hspace*{0.3cm}
\parbox{8mm}{\centerline{
\begin{fmfgraph}(5,5)
\setval
\fmfforce{0w,0.5h}{v1}
\fmfforce{1w,0.5h}{v2}
\fmf{fermion,right=1}{v2,v1}
\fmf{plain,left=1}{v2,v1}
\end{fmfgraph} }} 
\hspace*{0.3cm} + \hspace*{0.3cm} \frac{1}{2}\hspace*{0.3cm}
\parbox{8mm}{\centerline{
\begin{fmfgraph}(5,5)
\setval
\fmfi{boson}{reverse fullcircle scaled 1w shifted (0.5w,0.5h)}
\end{fmfgraph}
}}
\hspace*{0.2cm} + \hspace*{0.2cm} \frac{1}{2} \hspace*{0.2cm}
\parbox{14mm}{\centerline{
\begin{fmfgraph}(11,6)
\setval
\fmfforce{3/11w,0.5h}{v1}
\fmfforce{8/11w,0.5h}{v2}
\fmfforce{0w,0h}{v3}
\fmfforce{0w,1h}{v4}
\fmfforce{1w,0h}{v5}
\fmfforce{1w,1h}{v6}
\fmf{double,width=0.2mm}{v3,v1,v4}
\fmf{double,width=0.2mm}{v5,v2,v6}
\fmf{photon}{v1,v2}
\fmfdot{v1,v2}
\end{fmfgraph} }} 
\hspace*{0.4cm} ,
\eeq
due to (\r{EL1}) and (\r{EL2}). For a non-vanishing interaction $V$, the
Eqs. (\r{VAC1})--(\r{VAC3}) produce corrections to (\r{ZEGA0}) which we
shall denote with $W^{({\rm int})}$. Thus the vacuum energy $W$ 
decomposes according to
\beq
\la{WDE}
W = W^{({\rm free})} + W^{({\rm int})} \, .
\eeq
In the following we recursively determine $W^{({\rm int})}$ in a
graphical way for a vanishing external current, so that we can neglect the
last two terms in (\r{VAC2}) and the last term in (\r{VAC3}). Performing
a loopwise expansion of the interaction part of the vacuum energy
\beq
\la{WLOOP}
W^{({\rm int})} = \sum_{l=2}^{\infty} W^{(l)} \, , 
\eeq
we use the following eigenvalue problems for $l \ge 2$:
\beq
\parbox{8mm}{\begin{center}
\begin{fmfgraph*}(2.5,5)
\setval
\fmfstraight
\fmfforce{1w,0h}{v1}
\fmfforce{1w,1h}{v2}
\fmf{electron,left=1}{v1,v2}
\fmfv{decor.size=0, label=${\scs 2}$, l.dist=1mm, l.angle=0}{v1}
\fmfv{decor.size=0, label=${\scs 1}$, l.dist=1mm, l.angle=0}{v2}
\end{fmfgraph*}
\end{center}}
\hspace*{0.3cm} \dephi{W^{(l)}}{1}{2} \quad & = & \quad 2 (l-1) \,\,W^{(l)} 
\hspace*{0.4cm} , \la{EW1} \\
\parbox{8mm}{\begin{center}
\begin{fmfgraph*}(2.5,5)
\setval
\fmfstraight
\fmfforce{1w,0h}{v1}
\fmfforce{1w,1h}{v2}
\fmf{photon,left=1}{v1,v2}
\fmfv{decor.size=0, label=${\scs 2}$, l.dist=1mm, l.angle=0}{v1}
\fmfv{decor.size=0, label=${\scs 1}$, l.dist=1mm, l.angle=0}{v2}
\end{fmfgraph*}
\end{center}}
\hspace*{0.3cm} \dbphi{W^{(l)}}{1}{2} \quad & = & \quad (l-1) \,\, 
W^{(l)}  \hspace*{0.4cm} ,  \la{EW2} \\
\parbox{5mm}{\begin{center}
\begin{fmfgraph*}(3,4)
\setval
\fmfstraight
\fmfforce{0w,1/2h}{v1}
\fmfforce{1w,1/2h}{v2}
\fmfforce{1w,1.25h}{v3}
\fmfforce{1w,-0.25h}{v4}
\fmf{photon}{v1,v2}
\fmf{electron}{v1,v3}
\fmf{electron}{v4,v1}
\fmfv{decor.size=0, label=${\scs 3}$, l.dist=1mm, l.angle=0}{v2}
\fmfv{decor.size=0, label=${\scs 1}$, l.dist=1mm, l.angle=0}{v3}
\fmfv{decor.size=0, label=${\scs 2}$, l.dist=1mm, l.angle=0}{v4}
\fmfdot{v1}
\end{fmfgraph*}
\end{center}}
\quad \dvertex{W^{(l)}}{3}{2}{1} \quad & = & \quad 2 (l-1) \,\,W^{(l)} 
\hspace*{0.4cm} . \la{EW3} \\ && \no
\eeq
With these we explicitly solve (\r{VAC1})--(\r{VAC3}) for the expansion 
coefficients $W^{(l)}$ and obtain for $l \ge 2$:
\beq
\la{VA1}
W^{(l)} \quad & = & \quad - \frac{1}{2(l-1)} \hspace*{0.2cm}
\parbox{8mm}{\centerline{
\begin{fmfgraph*}(5,5)
\setval
\fmfforce{0w,0.5h}{v1}
\fmfforce{1w,0.5h}{v2}
\fmfforce{0.5w,1.1h}{v3}
\fmf{heavy,width=0.2mm,right=1}{v2,v1}
\fmf{fermion,right=1}{v1,v2}
\fmfv{decor.size=0, label=${\scs (l-1)}$, l.dist=1mm, l.angle=90}{v3}
\end{fmfgraph*} }} 
\hspace*{0.4cm} , \\
\la{VA2}
W^{(l)} \quad & = & \quad \frac{1}{2(l-1)} \hspace*{0.2cm} 
\left\{ \hspace*{0.2cm}
\parbox{8mm}{\centerline{
\begin{fmfgraph*}(5,5)
\setval
\fmfforce{0w,0.5h}{v1}
\fmfforce{1w,0.5h}{v2}
\fmfforce{0.5w,1.1h}{v3}
\fmf{dbl_wiggly,width=0.2mm,right=1}{v2,v1}
\fmf{boson,right=1}{v1,v2}
\fmfv{decor.size=0, label=${\scs (l-1)}$, l.dist=1mm, l.angle=90}{v3}
\end{fmfgraph*} }} 
\hspace*{0.2cm} + \hspace*{0.2cm} \sum_{k=0}^{l-2} \hspace*{0.2cm}
\parbox{19mm}{\centerline{
\begin{fmfgraph*}(15,5)
\setval
\fmfforce{1/3w,0.5h}{v1}
\fmfforce{2/3w,0.5h}{v2}
\fmfforce{1/6w,1.1h}{v3}
\fmfforce{5/6w,1.1h}{v4}
\fmf{boson}{v1,v2}
\fmfi{heavy,width=0.2mm,right=1}{reverse fullcircle 
scaled 1/3w shifted (1/6w,0.5h)}
\fmfi{heavy,width=0.2mm,right=1}{fullcircle rotated 
180 scaled 1/3w shifted (5/6w,0.5h)}
\fmfv{decor.size=0, label=${\scs (k)}$, l.dist=1mm, l.angle=90}{v3}
\fmfv{decor.size=0, label=${\scs (l-k-2)}$, l.dist=1mm, l.angle=90}{v4}
\fmfdot{v1,v2}
\end{fmfgraph*}
}}
\hspace*{0.2cm} \right\} \hspace*{0.4cm} , \\
\la{VA3}
W^{(l)} \quad & = & \quad \frac{1}{2(l-1)} 
\hspace*{0.2cm} \left\{ \hspace*{0.2cm} - \hspace*{0.2cm}
\parbox{14mm}{\centerline{
\begin{fmfgraph*}(11,6)
\setval
\fmfforce{0w,0.5h}{v1}
\fmfforce{5/11w,0.5h}{v2}
\fmfforce{1w,0.5h}{v3}
\fmfforce{8/11w,0.5h}{v4}
\fmfforce{8/11w,0h}{v5}
\fmfforce{8/11w,1h}{v6}
\fmfforce{7/11w,0.97h}{v7}
\fmfforce{7/11w,0.03h}{v8}
\fmf{plain,left=1}{v2,v3,v2}
\fmf{boson}{v1,v2}
\fmf{fermion,right=0.7}{v1,v8}
\fmf{fermion,right=0.7}{v7,v1}
\fmfdot{v1}
\fmfv{decor.size=0, label=${\scs l-2}$, l.dist=0mm, l.angle=90}{v4}
\end{fmfgraph*} }} 
+ \hspace*{0.2cm} \sum_{k=0}^{l-2} \hspace*{0.2cm}
\parbox{19mm}{\centerline{
\begin{fmfgraph*}(15,5)
\setval
\fmfforce{1/3w,0.5h}{v1}
\fmfforce{2/3w,0.5h}{v2}
\fmfforce{1/6w,1.1h}{v3}
\fmfforce{5/6w,1.1h}{v4}
\fmf{boson}{v1,v2}
\fmfi{heavy,width=0.2mm,right=1}{reverse fullcircle 
scaled 1/3w shifted (1/6w,0.5h)}
\fmfi{heavy,width=0.2mm,right=1}{fullcircle rotated 
180 scaled 1/3w shifted (5/6w,0.5h)}
\fmfv{decor.size=0, label=${\scs (k)}$, l.dist=1mm, l.angle=90}{v3}
\fmfv{decor.size=0, label=${\scs (l-k-2)}$, l.dist=1mm, l.angle=90}{v4}
\fmfdot{v1,v2}
\end{fmfgraph*}
}}
\hspace*{0.2cm} \right\} \hspace*{0.4cm} . 
\eeq
Inserting (\r{EL1})--(\r{PHO}) for the lower loop contributions of the 
connected electron and photon two-point function as well 
as the connected  three-point function in one of the equations
(\r{VA1})--(\r{VA3}), we find the vacuum energy for two loops
\beq
\la{W2}
W^{(2)} \quad = \quad 
%
%
- \hspace*{0.1cm} \frac{1}{2} \hspace*{0.3cm}  
\parbox{10mm}{\centerline{
\begin{fmfgraph}(6,6)
\setval
\fmfforce{0w,1/2h}{v1}
\fmfforce{1w,1/2h}{v2}
\fmf{boson}{v1,v2}
\fmf{fermion,right=1}{v1,v2}
\fmf{fermion,right=1}{v2,v1}
\fmfdot{v1,v2}
\end{fmfgraph}
}} 
%
%
\hspace*{0.3cm} + \hspace*{0.1cm} \frac{1}{2} \hspace*{0.3cm}  
\parbox{19mm}{\centerline{
\begin{fmfgraph}(15,7)
\setval
\fmfforce{0.33w,0.5h}{v1}
\fmfforce{0.66w,0.5h}{v2}
\fmf{boson}{v1,v2}
\fmfi{fermion}{reverse fullcircle scaled 0.33w shifted (0.165w,0.5h)}
\fmfi{fermion}{fullcircle rotated 180 scaled 0.33w shifted (0.825w,0.5h)}
\fmfdot{v1,v2}
\end{fmfgraph}
}}
\eeq
and for three loops
\beq
\la{W3}
W^{(3)} \quad = \quad 
%
%
- \hspace*{0.1cm} \frac{1}{4}\hspace*{0.2cm}
\parbox{8mm}{\begin{center}
\begin{fmfgraph}(5,5)
\setval
\fmfforce{0w,0h}{v1}
\fmfforce{0w,1h}{v2}
\fmfforce{1w,0h}{v3}
\fmfforce{1w,1h}{v4}
\fmf{fermion,right=0.4}{v2,v1}
\fmf{fermion,right=0.4}{v4,v2}
\fmf{fermion,right=0.4}{v3,v4}
\fmf{fermion,right=0.4}{v1,v3}
\fmf{boson}{v1,v4}
\fmf{boson}{v3,v2}
\fmfdot{v1,v2,v3,v4}
\end{fmfgraph}
\end{center}}
%
%
\hspace*{0.2cm}+ \hspace*{0.1cm} \frac{1}{4} \hspace*{0.2cm}
\parbox{14mm}{\begin{center}
\begin{fmfgraph}(12,12)
\setval
\fmfforce{3.5/12w,1/2h}{v1}
\fmfforce{8.5/12w,1/2h}{v2}
\fmfforce{1/2w,0h}{v3}
\fmfforce{1/2w,3.5/12h}{v4}
\fmfforce{1/2w,8.5/12h}{v5}
\fmfforce{1/2w,1h}{v6}
\fmf{fermion,right=1}{v3,v6,v3}
\fmf{boson}{v3,v4}
\fmf{boson}{v5,v6}
\fmf{fermion,left=1}{v4,v5,v4}
\fmfdot{v3,v4,v5,v6}
\end{fmfgraph}
\end{center}}
%
%
\hspace*{0.2cm} - \hspace*{0.1cm} \frac{1}{2}\hspace*{0.2cm}
\parbox{8mm}{\begin{center}
\begin{fmfgraph}(5,5)
\setval
\fmfforce{0w,0h}{v1}
\fmfforce{0w,1h}{v2}
\fmfforce{1w,0h}{v3}
\fmfforce{1w,1h}{v4}
\fmf{fermion,right=0.4}{v2,v1}
\fmf{fermion,right=0.4}{v4,v2}
\fmf{fermion,right=0.4}{v3,v4}
\fmf{fermion,right=0.4}{v1,v3}
\fmf{boson,right=0.4}{v1,v2}
\fmf{boson,right=0.4}{v4,v3}
\fmfdot{v1,v2,v3,v4}
\end{fmfgraph}
\end{center}}
%
%
\hspace*{0.2cm} - \hspace*{0.1cm} \frac{1}{2} \hspace*{0.2cm} 
\parbox{28mm}{\begin{center}
\begin{fmfgraph}(25,5)
\setval
\fmfforce{1/5w,0.5h}{v1}
\fmfforce{2/5w,0.5h}{v2}
\fmfforce{3/5w,0.5h}{v3}
\fmfforce{4/5w,0.5h}{v4}
\fmf{fermion,left=1}{v2,v3,v2}
\fmf{boson}{v1,v2}
\fmf{boson}{v3,v4}
\fmfi{fermion}{reverse fullcircle scaled 1/5w shifted (1/10w,0.5h)}
\fmfi{fermion}{fullcircle rotated 180 scaled 1/5w shifted (9/10w,0.5h)}
\fmfdot{v1,v2,v3,v4}
\end{fmfgraph}
\end{center}}
%
%
\hspace*{0.2cm} + \hspace*{0.1cm}
\parbox{19mm}{\begin{center}
\begin{fmfgraph}(16,6)
\setval
\fmfforce{3/16w,0h}{v1}
\fmfforce{3/16w,1h}{v2}
\fmfforce{6/16w,1/2h}{v3}
\fmfforce{11/16w,1/2h}{v4}
\fmf{fermion,right=0.4}{v1,v3,v2}
\fmf{fermion,right=1}{v2,v1}
\fmf{boson}{v3,v4}
\fmf{boson}{v1,v2}
\fmfi{fermion}{fullcircle rotated 180 scaled 0.313w shifted (0.844w,0.5h)}
\fmfdot{v1,v2,v3,v4}
\end{fmfgraph}
\end{center}}
\hspace*{0.4cm} .
\eeq 
\end{fmffile}
\begin{fmffile}{sd12}
\subsubsection{Graphical Recursion Relation}
\la{GRR2}
Each of the three functional differential equations for the vacuum
energy (\r{FDW1})--(\r{FDW3}) can be used to derive a graphical recursion
relation which directly leads to the connected vacuum diagrams. Here we
restrict ourselves to the functional differential equation (\r{FDW1}) which is
based on counting the number of electron lines of the connected
vacuum diagrams. Inserting the Eqs. (\r{S2}), (\r{IES}), and (\r{FIDE})
for the connected electron two-point function and
the connected three-point function, we obtain from (\r{FDW1}) via the functional chain rule (\r{FCR}) 
\beq
\la{WW}
\delta_{11} \int_1 - \int_{12} S^{-1}_{12} \frac{\delta W}{\delta S^{-1}_{12}}
= \int_{123456} V_{123} V_{456} D_{36} \left\{ \frac{\delta^2 W}{\delta
S^{-1}_{12} \delta S_{45}^{-1}} + \frac{\delta W}{\delta S^{-1}_{12}}
\frac{\delta W}{\delta S^{-1}_{45}} \right\} + 
\int_{1234} V_{123} D_{34} J_4 \frac{\delta W}{\delta S^{-1}_{12}} \, .
\eeq
If the interaction $V$ vanishes, this equation is solved by the free vacuum energy (\r{W0}) which has the functional derivatives
\beq
\la{WD}
\frac{\delta W^{({\rm free})}}{\delta S^{-1}_{12}} = S_{21} \, , \hspace*{1cm}
\frac{\delta^2 W^{({\rm free})}}{\delta S^{-1}_{12} \delta S^{-1}_{45}} = - 
S_{24} S_{51} \, .
\eeq
For a non-vanishing interaction $V$, the right-hand side of (\r{WW})
corrects (\r{W0}) by the interaction part of the vacuum energy
$W^{({\rm int})}$. Inserting the decomposition (\r{WDE}) into (\r{WW}),
and using (\r{WD}), we obtain together with the functional chain rule (\r{FCR}) the following functional
differential equation for the interaction part of the vacuum energy:
\beq
&&
\int_{12} S_{12} \frac{\delta W^{({\rm int})}}{\delta S_{12}} =  
\int_{123456} V_{123} V_{456} S_{21} S_{54} D_{36} 
- \int_{123456} V_{123} V_{456} S_{24} S_{51} D_{36} 
-2 \int_{12345678} V_{123} V_{456} D_{36} S_{21} S_{74} S_{58}
\frac{\delta W^{({\rm int})}}{\delta S_{78}} 
\no \\ && \hspace*{0.5cm}
+2 \int_{12345678} V_{123} V_{456} D_{36} S_{51} S_{28} S_{74} 
\frac{\delta W^{({\rm int})}}{\delta S_{78}} 
+  \int_{123456789\bar{1}} V_{123} V_{456} D_{36}
S_{71} S_{28} S_{94} S_{5\bar{1}} 
\frac{\delta W^{({\rm int})}}{\delta S_{78}}
\frac{\delta W^{({\rm int})}}{\delta S_{9\bar{1}} } 
\no \\ && \hspace*{0.5cm}
+ \int_{123456789\bar{1}} V_{123} V_{456} D_{36}
S_{71} S_{28} S_{94} S_{5\bar{1}} 
\frac{\delta^2 W^{({\rm int})}}{\delta S_{78} S_{9\bar{1}} } 
+ \int_{1234} V_{123} D_{34} S_{21} J_4
- \int_{123456} V_{123} S_{51} S_{26} D_{34} J_4   
\frac{\delta W^{({\rm int})}}{\delta S_{56}} \, .
\eeq
Its graphical representation is
\beq
\parbox{8mm}{\begin{center}
\begin{fmfgraph*}(2.5,5)
\setval
\fmfstraight
\fmfforce{1w,0h}{v1}
\fmfforce{1w,1h}{v2}
\fmf{electron,left=1}{v1,v2}
\fmfv{decor.size=0, label=${\scs 2}$, l.dist=1mm, l.angle=0}{v1}
\fmfv{decor.size=0, label=${\scs 1}$, l.dist=1mm, l.angle=0}{v2}
\end{fmfgraph*}
\end{center}}
\hspace*{0.3cm} \dephi{W^{({\rm int})}}{1}{2} \quad & = & \quad
%
%
- \hspace*{0.3cm}  
\parbox{10mm}{\centerline{
\begin{fmfgraph}(6,6)
\setval
\fmfforce{0w,1/2h}{v1}
\fmfforce{1w,1/2h}{v2}
\fmf{boson}{v1,v2}
\fmf{fermion,right=1}{v1,v2}
\fmf{fermion,right=1}{v2,v1}
\fmfdot{v1,v2}
\end{fmfgraph}
}} 
%
%
\hspace*{0.3cm} + \hspace*{0.3cm}  
\parbox{19mm}{\centerline{
\begin{fmfgraph}(15,7)
\setval
\fmfforce{0.33w,0.5h}{v1}
\fmfforce{0.66w,0.5h}{v2}
\fmf{boson}{v1,v2}
\fmfi{fermion}{reverse fullcircle scaled 0.33w shifted (0.165w,0.5h)}
\fmfi{fermion}{fullcircle rotated 180 scaled 0.33w shifted (0.825w,0.5h)}
\fmfdot{v1,v2}
\end{fmfgraph}
}}
%
%
\hspace*{0.3cm} + \hspace*{0.1cm}  2 \hspace*{0.3cm}  
\parbox{10.5mm}{\begin{center}
\begin{fmfgraph*}(7.5,5)
\setval
\fmfstraight
\fmfforce{1/3w,0h}{v1}
\fmfforce{1/3w,1h}{v2}
\fmfforce{1w,0h}{v3}
\fmfforce{1w,1h}{v4}
\fmf{photon}{v1,v2}
\fmf{fermion}{v3,v1}
\fmf{fermion}{v2,v4}
\fmf{fermion,left=1}{v1,v2}
\fmfv{decor.size=0, label=${\scs 2}$, l.dist=1mm, l.angle=0}{v3}
\fmfv{decor.size=0, label=${\scs 1}$, l.dist=1mm, l.angle=0}{v4}
\fmfdot{v1,v2}
\end{fmfgraph*}
\end{center}}
\hspace*{0.3cm} \dephi{W^{({\rm int})}}{1}{2}
%
%
\hspace*{0.3cm} - \hspace*{0.1cm} 2 \hspace*{0.3cm}
\parbox{17mm}{\begin{center}
\begin{fmfgraph*}(14,8)
\setval
\fmfstraight
\fmfforce{2.5/14w,1.5/8h}{v1}
\fmfforce{2.5/14w,6.5/8h}{v2}
\fmfforce{5/14w,1/2h}{v3}
\fmfforce{10/14w,1/2h}{v4}
\fmfforce{1w,0h}{v5}
\fmfforce{1w,1h}{v6}
\fmf{photon}{v3,v4}
\fmf{plain,right=1}{v1,v2}
\fmf{fermion,right=1}{v2,v1}
\fmf{fermion}{v4,v6}
\fmf{fermion}{v5,v4}
\fmfv{decor.size=0, label=${\scs 2}$, l.dist=1mm, l.angle=0}{v5}
\fmfv{decor.size=0, label=${\scs 1}$, l.dist=1mm, l.angle=0}{v6}
\fmfdot{v3,v4}
\end{fmfgraph*}
\end{center}}
\hspace*{0.3cm} \dephi{W^{({\rm int})}}{1}{2}
\no \\
&& \quad + \hspace*{0.3cm}
%
%
\parbox{8mm}{\begin{center}
\begin{fmfgraph*}(5,9)
\setval
\fmfstraight
\fmfforce{0w,1/6h}{v1}
\fmfforce{0w,5/6h}{v2}
\fmfforce{1w,0h}{v3}
\fmfforce{1w,1/3h}{v4}
\fmfforce{1w,2/3h}{v5}
\fmfforce{1w,1h}{v6}
\fmf{photon}{v2,v1}
\fmf{fermion}{v3,v1}
\fmf{fermion}{v1,v4}
\fmf{fermion}{v5,v2}
\fmf{fermion}{v2,v6}
\fmfv{decor.size=0, label=${\scs 1}$, l.dist=1mm, l.angle=0}{v6}
\fmfv{decor.size=0, label=${\scs 2}$, l.dist=1mm, l.angle=0}{v5}
\fmfv{decor.size=0, label=${\scs 3}$, l.dist=1mm, l.angle=0}{v4}
\fmfv{decor.size=0, label=${\scs 4}$, l.dist=1mm, l.angle=0}{v3}
\fmfdot{v1,v2}
\end{fmfgraph*}
\end{center}}
\hspace*{0.3cm} \ddfermi{W^{({\rm int})}}{1}{2}{3}{4}
%
%
\hspace*{0.3cm}+ \hspace*{0.3cm} \dephi{W^{({\rm int})}}{1}{2} \hspace*{0.3cm}
\parbox{16mm}{\begin{center}
\begin{fmfgraph*}(13,8)
\setval
\fmfstraight
\fmfforce{0w,0h}{v1}
\fmfforce{0w,1h}{v2}
\fmfforce{4/13w,1/2h}{v3}
\fmfforce{9/13w,1/2h}{v4}
\fmfforce{1w,0h}{v5}
\fmfforce{1w,1h}{v6}
\fmf{photon}{v3,v4}
\fmf{fermion}{v4,v6}
\fmf{fermion}{v5,v4}
\fmf{fermion}{v1,v3}
\fmf{fermion}{v3,v2}
\fmfv{decor.size=0, label=${\scs 1}$, l.dist=1mm, l.angle=-180}{v2}
\fmfv{decor.size=0, label=${\scs 2}$, l.dist=1mm, l.angle=-180}{v1}
\fmfv{decor.size=0, label=${\scs 3}$, l.dist=1mm, l.angle=0}{v6}
\fmfv{decor.size=0, label=${\scs 4}$, l.dist=1mm, l.angle=0}{v5}
\fmfdot{v3,v4}
\end{fmfgraph*}
\end{center}}
\hspace*{0.3cm} \dephi{W^{({\rm int})}}{3}{4} \no \\
&& \quad 
%
%
- \hspace*{0.3cm}
\parbox{16mm}{\centerline{
\begin{fmfgraph}(13,6)
\setval
\fmfforce{3/13w,0.5h}{v1}
\fmfforce{8/13w,0.5h}{v2}
\fmfforce{0w,0h}{v3}
\fmfforce{0w,1h}{v4}
\fmfforce{10.5/13w,0.5/6h}{v5}
\fmfforce{10.5/13w,5.5/6h}{v6}
\fmf{plain,right=1}{v6,v5}
\fmf{fermion,right=1}{v5,v6}
\fmf{double,width=0.2mm}{v3,v1,v4}
\fmf{photon}{v1,v2}
\fmfdot{v1,v2}
\end{fmfgraph} }} 
%
%
\hspace*{0.3cm} + \hspace*{0.3cm}
\parbox{15mm}{\centerline{
\begin{fmfgraph*}(12,8)
\setval
\fmfforce{3/12w,0.5h}{v1}
\fmfforce{8/12w,0.5h}{v2}
\fmfforce{0w,0h}{v3}
\fmfforce{0w,1h}{v4}
\fmfforce{1w,0h}{v5}
\fmfforce{1w,1h}{v6}
\fmf{double,width=0.2mm}{v3,v1,v4}
\fmf{fermion}{v5,v2,v6}
\fmf{photon}{v1,v2}
\fmfv{decor.size=0, label=${\scs 1}$, l.dist=1mm, l.angle=0}{v5}
\fmfv{decor.size=0, label=${\scs 2}$, l.dist=1mm, l.angle=0}{v6}
\fmfdot{v1,v2}
\end{fmfgraph*} }} 
\hspace*{0.3cm} \dephi{W^{({\rm int})}}{1}{2}
\hspace*{0.4cm} . \la{VACGR} 
\eeq
As before, we illustrate the graphical recursive solution only for a
vanishing external current so that we can drop the last two terms in
(\r{VACGR}). Thus inserting the loop expansion (\r{WLOOP})
and using the eigenvalue problem
(\r{EW1}), we obtain a graphical recursion relation for the expansion
coefficients $W^{(l)}$ of the vacuum energy for $l \ge 3$ \cite{QED}:
\beq
W^{(l)} \quad &=& \quad \frac{1}{l-1} \hspace*{0.2cm} \left\{ \hspace*{0.2cm}
\parbox{10.5mm}{\begin{center}
\begin{fmfgraph*}(7.5,5)
\setval
\fmfstraight
\fmfforce{1/3w,0h}{v1}
\fmfforce{1/3w,1h}{v2}
\fmfforce{1w,0h}{v3}
\fmfforce{1w,1h}{v4}
\fmf{photon}{v1,v2}
\fmf{fermion}{v3,v1}
\fmf{fermion}{v2,v4}
\fmf{fermion,left=1}{v1,v2}
\fmfv{decor.size=0, label=${\scs 2}$, l.dist=1mm, l.angle=0}{v3}
\fmfv{decor.size=0, label=${\scs 1}$, l.dist=1mm, l.angle=0}{v4}
\fmfdot{v1,v2}
\end{fmfgraph*}
\end{center}}
\hspace*{0.3cm} \dephi{W^{(l-1)}}{1}{2}
\hspace*{0.3cm} - \hspace*{0.3cm}
\parbox{17mm}{\begin{center}
\begin{fmfgraph*}(14,8)
\setval
\fmfstraight
\fmfforce{2.5/14w,1.5/8h}{v1}
\fmfforce{2.5/14w,6.5/8h}{v2}
\fmfforce{5/14w,1/2h}{v3}
\fmfforce{10/14w,1/2h}{v4}
\fmfforce{1w,0h}{v5}
\fmfforce{1w,1h}{v6}
\fmf{photon}{v3,v4}
\fmf{plain,right=1}{v1,v2}
\fmf{fermion,right=1}{v2,v1}
\fmf{fermion}{v4,v6}
\fmf{fermion}{v5,v4}
\fmfv{decor.size=0, label=${\scs 2}$, l.dist=1mm, l.angle=0}{v5}
\fmfv{decor.size=0, label=${\scs 1}$, l.dist=1mm, l.angle=0}{v6}
\fmfdot{v3,v4}
\end{fmfgraph*}
\end{center}}
\hspace*{0.3cm} \dephi{W^{(l-1)}}{1}{2}
\right. \no \\
&& \left. \quad + \hspace*{0.1cm}\frac{1}{2} \hspace*{0.3cm}
\parbox{8mm}{\begin{center}
\begin{fmfgraph*}(5,9)
\setval
\fmfstraight
\fmfforce{0w,1/6h}{v1}
\fmfforce{0w,5/6h}{v2}
\fmfforce{1w,0h}{v3}
\fmfforce{1w,1/3h}{v4}
\fmfforce{1w,2/3h}{v5}
\fmfforce{1w,1h}{v6}
\fmf{photon}{v2,v1}
\fmf{fermion}{v3,v1}
\fmf{fermion}{v1,v4}
\fmf{fermion}{v5,v2}
\fmf{fermion}{v2,v6}
\fmfv{decor.size=0, label=${\scs 1}$, l.dist=1mm, l.angle=0}{v6}
\fmfv{decor.size=0, label=${\scs 2}$, l.dist=1mm, l.angle=0}{v5}
\fmfv{decor.size=0, label=${\scs 3}$, l.dist=1mm, l.angle=0}{v4}
\fmfv{decor.size=0, label=${\scs 4}$, l.dist=1mm, l.angle=0}{v3}
\fmfdot{v1,v2}
\end{fmfgraph*}
\end{center}}
\hspace*{0.3cm} \ddfermi{W^{(l-1)}}{1}{2}{3}{4}
\hspace*{0.3cm}+ \hspace*{0.1cm}
\frac{1}{2} \hspace*{0.1cm}\sum_{k=2}^{l-2}
\hspace*{0.3cm} \dephi{W^{(k)}}{1}{2} \hspace*{0.3cm}
\parbox{16mm}{\begin{center}
\begin{fmfgraph*}(13,8)
\setval
\fmfstraight
\fmfforce{0w,0h}{v1}
\fmfforce{0w,1h}{v2}
\fmfforce{4/13w,1/2h}{v3}
\fmfforce{9/13w,1/2h}{v4}
\fmfforce{1w,0h}{v5}
\fmfforce{1w,1h}{v6}
\fmf{photon}{v3,v4}
\fmf{fermion}{v4,v6}
\fmf{fermion}{v5,v4}
\fmf{fermion}{v1,v3}
\fmf{fermion}{v3,v2}
\fmfv{decor.size=0, label=${\scs 1}$, l.dist=1mm, l.angle=-180}{v2}
\fmfv{decor.size=0, label=${\scs 2}$, l.dist=1mm, l.angle=-180}{v1}
\fmfv{decor.size=0, label=${\scs 3}$, l.dist=1mm, l.angle=0}{v6}
\fmfv{decor.size=0, label=${\scs 4}$, l.dist=1mm, l.angle=0}{v5}
\fmfdot{v3,v4}
\end{fmfgraph*}
\end{center}}
\hspace*{0.3cm} \dephi{W^{(l-k)}}{3}{4} \hspace*{0.2cm} \right\} 
\hspace*{0.4cm} . \la{VACL}
\eeq
They are solved starting from $W^{(2)}$ in (\r{W2}). We start with the
amputation of one or two electron lines in (\r{W2}):
\beq
\dephi{W^{(2)}}{1}{2} \hspace*{0.3cm} =  \hspace*{0.3cm}
%
%
\parbox{17mm}{\begin{center}
\begin{fmfgraph*}(12.12,8)
\setval
\fmfstraight
\fmfforce{0/12.12w,2/8h}{v1}
\fmfforce{0/12.12w,6/8h}{v2}
\fmfforce{2.12/12.12w,1/2h}{v3}
\fmfforce{7.12/12.12w,1/2h}{v4}
\fmfforce{9.62/12.12w,1.5/8h}{v5}
\fmfforce{9.62/12.12w,6.5/8h}{v6}
\fmf{photon}{v3,v4}
\fmf{plain,left=1}{v5,v6}
\fmf{fermion,right=1}{v5,v6}
\fmf{fermion}{v3,v1}
\fmf{fermion}{v2,v3}
\fmfv{decor.size=0, label=${\scs 2}$, l.dist=1mm, l.angle=-180}{v1}
\fmfv{decor.size=0, label=${\scs 1}$, l.dist=1mm, l.angle=-180}{v2}
\fmfdot{v3,v4}
\end{fmfgraph*}
\end{center}}
%
%
\hspace*{0.1cm} - \hspace*{0.1cm}
\parbox{10.5mm}{\begin{center}
\begin{fmfgraph*}(5.5,5)
\setval
\fmfstraight
\fmfforce{3/5.5w,0h}{v1}
\fmfforce{3/5.5w,1h}{v2}
\fmfforce{0w,0h}{v3}
\fmfforce{0w,1h}{v4}
\fmf{photon}{v1,v2}
\fmf{fermion}{v1,v3}
\fmf{fermion}{v4,v2}
\fmf{fermion,left=1}{v2,v1}
\fmfv{decor.size=0, label=${\scs 2}$, l.dist=1mm, l.angle=-180}{v3}
\fmfv{decor.size=0, label=${\scs 1}$, l.dist=1mm, l.angle=-180}{v4}
\fmfdot{v1,v2}
\end{fmfgraph*}
\end{center}}
\hspace*{0.5cm}, \hspace*{0.5cm}
\ddfermi{W^{(2)}}{1}{2}{3}{4} \hspace*{0.3cm} = \hspace*{0.3cm}
%
%
\parbox{8mm}{\begin{center}
\begin{fmfgraph*}(3,9)
\setval
\fmfstraight
\fmfforce{1w,1/6h}{v1}
\fmfforce{1w,5/6h}{v2}
\fmfforce{0w,0h}{v3}
\fmfforce{0w,1/3h}{v4}
\fmfforce{0w,2/3h}{v5}
\fmfforce{0w,1h}{v6}
\fmf{photon}{v2,v1}
\fmf{fermion}{v1,v3}
\fmf{fermion}{v4,v1}
\fmf{fermion}{v2,v5}
\fmf{fermion}{v6,v2}
\fmfv{decor.size=0, label=${\scs 1}$, l.dist=1mm, l.angle=-180}{v6}
\fmfv{decor.size=0, label=${\scs 2}$, l.dist=1mm, l.angle=-180}{v5}
\fmfv{decor.size=0, label=${\scs 3}$, l.dist=1mm, l.angle=-180}{v4}
\fmfv{decor.size=0, label=${\scs 4}$, l.dist=1mm, l.angle=-180}{v3}
\fmfdot{v1,v2}
\end{fmfgraph*}
\end{center}}
%
%
\hspace*{0.1cm} - \hspace*{0.1cm}
\parbox{8mm}{\begin{center}
\begin{fmfgraph*}(3,9)
\setval
\fmfstraight
\fmfforce{1w,1/6h}{v1}
\fmfforce{1w,5/6h}{v2}
\fmfforce{0w,0h}{v3}
\fmfforce{0w,1/3h}{v4}
\fmfforce{0w,2/3h}{v5}
\fmfforce{0w,1h}{v6}
\fmf{photon}{v2,v1}
\fmf{fermion}{v1,v3}
\fmf{fermion}{v4,v1}
\fmf{fermion}{v2,v5}
\fmf{fermion}{v6,v2}
\fmfv{decor.size=0, label=${\scs 1}$, l.dist=1mm, l.angle=-180}{v6}
\fmfv{decor.size=0, label=${\scs 4}$, l.dist=1mm, l.angle=-180}{v5}
\fmfv{decor.size=0, label=${\scs 3}$, l.dist=1mm, l.angle=-180}{v4}
\fmfv{decor.size=0, label=${\scs 2}$, l.dist=1mm, l.angle=-180}{v3}
\fmfdot{v1,v2}
\end{fmfgraph*}
\end{center}}
\hspace*{0.4cm} . \la{AM}
\eeq
Inserting (\r{AM}) into (\r{VACL}), we reobtain the three-loop contribution
of the vacuum energy from (\r{W3}). The corresponding calculation of 
the four-loop correction $W^{(4)}$ leads altogether to $20$ connected vacuum
diagrams:
\beq
W^{(4)} \hspace*{0.2cm} & = & \hspace*{0.2cm}
%
%
\frac{1}{6}\hspace*{0.1cm}
\parbox{16mm}{\begin{center}
\begin{fmfgraph}(12,12)
\setval
\fmfforce{3.5/12w,1/2h}{v1}
\fmfforce{8.5/12w,1/2h}{v2}
\fmfforce{1/2w,0h}{v3}
\fmfforce{1/2w,3.5/12h}{v4}
\fmfforce{1/2w,8.5/12h}{v5}
\fmfforce{1/2w,1h}{v6}
\fmfforce{8.165/12w,7.25/12h}{v7}
\fmfforce{3.835/12w,7.25/12h}{v8}
\fmfforce{11.196/12w,9/12h}{v9}
\fmfforce{0.834/12w,9/12h}{v10}
\fmf{fermion,right=0.6}{v3,v9,v10,v3}
\fmf{boson}{v7,v9}
\fmf{boson}{v3,v4}
\fmf{boson}{v8,v10}
\fmf{fermion,right=0.55}{v4,v7,v8,v4}
\fmfdot{v3,v4,v7,v8,v9,v10}
\end{fmfgraph}
\end{center}}
%
%
\hspace*{0.1cm} + \hspace*{0.05cm} \frac{1}{6}\hspace*{0.1cm}
\parbox{16mm}{\begin{center}
\begin{fmfgraph}(12,12)
\setval
\fmfforce{3.5/12w,1/2h}{v1}
\fmfforce{8.5/12w,1/2h}{v2}
\fmfforce{1/2w,0h}{v3}
\fmfforce{1/2w,3.5/12h}{v4}
\fmfforce{1/2w,8.5/12h}{v5}
\fmfforce{1/2w,1h}{v6}
\fmfforce{8.165/12w,7.25/12h}{v7}
\fmfforce{3.835/12w,7.25/12h}{v8}
\fmfforce{11.196/12w,9/12h}{v9}
\fmfforce{0.834/12w,9/12h}{v10}
\fmf{fermion,right=0.6}{v3,v9,v10,v3}
\fmf{boson}{v7,v9}
\fmf{boson}{v3,v4}
\fmf{boson}{v8,v10}
\fmf{fermion,left=0.55}{v4,v8,v7,v4}
\fmfdot{v3,v4,v7,v8,v9,v10}
\end{fmfgraph}
\end{center}}
%
%
\hspace*{0.1cm}- \hspace*{0.05cm} \frac{1}{6}\hspace*{0.1cm}
\parbox{11mm}{\begin{center}
\begin{fmfgraph}(8,8)
\setval
\fmfforce{0w,0.5h}{v1}
\fmfforce{0.25w,0.933h}{v2}
\fmfforce{0.75w,0.933h}{v3}
\fmfforce{1w,0.5h}{v4}
\fmfforce{0.75w,0.067h}{v5}
\fmfforce{0.25w,0.067h}{v6}
\fmf{fermion,right=0.3}{v1,v6,v5,v4,v3,v2,v1}
\fmf{boson}{v1,v4}
\fmf{boson}{v2,v5}
\fmf{boson}{v3,v6}
\fmfdot{v1,v2,v3,v4,v5,v6}
\end{fmfgraph}
\end{center}}
%
%
\hspace*{0.1cm}- \hspace*{0.05cm}\frac{1}{2}\hspace*{0.1cm}
\parbox{11mm}{\begin{center}
\begin{fmfgraph}(8,8)
\setval
\fmfforce{0w,0.5h}{v1}
\fmfforce{0.25w,0.933h}{v2}
\fmfforce{0.75w,0.933h}{v3}
\fmfforce{1w,0.5h}{v4}
\fmfforce{0.75w,0.067h}{v5}
\fmfforce{0.25w,0.067h}{v6}
\fmf{fermion,right=0.3}{v1,v6,v5,v4,v3,v2,v1}
\fmf{boson}{v1,v4}
\fmf{boson}{v2,v6}
\fmf{boson}{v3,v5}
\fmfdot{v1,v2,v3,v4,v5,v6}
\end{fmfgraph}
\end{center}}
%
%
\hspace*{0.1cm}+ \hspace*{0.05cm}\frac{1}{2}\hspace*{0.1cm}
\parbox{14mm}{\begin{center}
\begin{fmfgraph}(12,12)
\setval
\fmfforce{3.5/12w,1/2h}{v1}
\fmfforce{8.5/12w,1/2h}{v2}
\fmfforce{1/2w,0h}{v3}
\fmfforce{1/2w,3.5/12h}{v4}
\fmfforce{1/2w,8.5/12h}{v5}
\fmfforce{1/2w,1h}{v6}
\fmf{fermion,right=1}{v3,v6,v3}
\fmf{boson}{v1,v2}
\fmf{boson}{v3,v4}
\fmf{boson}{v5,v6}
\fmf{fermion,left=0.4}{v4,v1,v5,v2,v4}
\fmfdot{v1,v2,v3,v4,v5,v6}
\end{fmfgraph}
\end{center}}
%
%
\hspace*{0.1cm}- \hspace*{0.05cm}\frac{1}{6}\hspace*{0.3cm}
\parbox{12mm}{\begin{center}
\begin{fmfgraph}(10,10)
\setval
\fmfforce{0w,1/2h}{v1}
\fmfforce{1/4w,0h}{v2}
\fmfforce{3/4w,0h}{v3}
\fmfforce{1w,1/2h}{v4}
\fmfforce{3/4w,1h}{v5}
\fmfforce{1/4w,1h}{v6}
\fmfforce{1/2w,1/2h}{v7}
\fmf{fermion,right=1}{v6,v5}
\fmf{fermion,right=1}{v5,v6}
\fmf{fermion,right=1}{v1,v2}
\fmf{fermion,right=1}{v2,v1}
\fmf{fermion,right=1}{v3,v4}
\fmf{fermion,right=1}{v4,v3}
\fmf{boson}{v2,v3}
\fmf{boson}{v4,v5}
\fmf{boson}{v6,v1}
\fmfdot{v1,v2,v3,v4,v5,v6}
\end{fmfgraph}
\end{center}}
%
%
\hspace*{0.1cm}- \hspace*{0.05cm}\frac{1}{3}\hspace*{0.1cm}
\parbox{11mm}{\begin{center}
\begin{fmfgraph}(8,8)
\setval
\fmfforce{0w,0.5h}{v1}
\fmfforce{0.25w,0.933h}{v2}
\fmfforce{0.75w,0.933h}{v3}
\fmfforce{1w,0.5h}{v4}
\fmfforce{0.75w,0.067h}{v5}
\fmfforce{0.25w,0.067h}{v6}
\fmf{fermion,right=0.3}{v1,v6,v5,v4,v3,v2,v1}
\fmf{boson,right=0.7}{v2,v3}
\fmf{boson,right=0.7}{v4,v5}
\fmf{boson,right=0.7}{v6,v1}
\fmfdot{v1,v2,v3,v4,v5,v6}
\end{fmfgraph}
\end{center}} 
\no \\
&& \hspace*{0.2cm}
%
%
- \hspace*{0.1cm}\frac{1}{2}\hspace*{0.15cm}
\parbox{11mm}{\begin{center}
\begin{fmfgraph}(8,8)
\setval
\fmfforce{0w,0.5h}{v1}
\fmfforce{0.25w,0.933h}{v2}
\fmfforce{0.75w,0.933h}{v3}
\fmfforce{1w,0.5h}{v4}
\fmfforce{0.75w,0.067h}{v5}
\fmfforce{0.25w,0.067h}{v6}
\fmf{fermion,right=0.3}{v1,v6,v5,v4,v3,v2,v1}
\fmf{boson,right=0.7}{v2,v3}
\fmf{boson}{v1,v4}
\fmf{boson,right=0.7}{v5,v6}
\fmfdot{v1,v2,v3,v4,v5,v6}
\end{fmfgraph}
\end{center}}
%
%
\hspace*{0.15cm}- \hspace*{0.2cm}
\parbox{11mm}{\begin{center}
\begin{fmfgraph}(8,8)
\setval
\fmfforce{0w,0.5h}{v1}
\fmfforce{0.25w,0.933h}{v2}
\fmfforce{0.75w,0.933h}{v3}
\fmfforce{1w,0.5h}{v4}
\fmfforce{0.75w,0.067h}{v5}
\fmfforce{0.25w,0.067h}{v6}
\fmf{fermion,right=0.3}{v1,v6,v5,v4,v3,v2,v1}
\fmf{boson,right=0.7}{v2,v3}
\fmf{boson,right=0.2}{v4,v6}
\fmf{boson,right=0.2}{v5,v1}
\fmfdot{v1,v2,v3,v4,v5,v6}
\end{fmfgraph}
\end{center}}
%
%
\hspace*{0.15cm}+ \hspace*{0.2cm}
\parbox{14mm}{\begin{center}
\begin{fmfgraph}(12,12)
\setval
\fmfforce{3.5/12w,1/2h}{v1}
\fmfforce{8.5/12w,1/2h}{v2}
\fmfforce{1/2w,0h}{v3}
\fmfforce{1/2w,3.5/12h}{v4}
\fmfforce{1/2w,8.5/12h}{v5}
\fmfforce{1/2w,1h}{v6}
\fmfforce{0.7/12w,3.5/12h}{v7}
\fmfforce{0.7/12w,8.5/12h}{v8}
\fmf{fermion,right=1}{v3,v6}
\fmf{boson}{v3,v4}
\fmf{boson}{v5,v6}
\fmf{fermion,right=0.3}{v6,v8}
\fmf{fermion,right=0.25}{v8,v7}
\fmf{boson,left=0.25}{v8,v7}
\fmf{fermion,right=0.3}{v7,v3}
\fmf{fermion,right=1}{v5,v4,v5}
\fmfdot{v3,v4,v5,v6,v7,v8}
\end{fmfgraph}
\end{center}}
%
%
\hspace*{0.15cm}+\hspace*{0.2cm}
\parbox{20mm}{\begin{center}
\begin{fmfgraph}(18,8)
\setval
\fmfforce{0.055w,0.825h}{v1}
\fmfforce{0.055w,0.175h}{v2}
\fmfforce{0.33w,0.925h}{v3}
\fmfforce{0.33w,0.075h}{v4}
\fmfforce{8/18w,1/2h}{v5}
\fmfforce{13/18w,1/2h}{v6}
\fmfforce{15.5/18w,6.5/8h}{v7}
\fmfforce{15.5/18w,1.5/8h}{v8}
\fmf{fermion,right=1}{v8,v7}
\fmf{plain,right=1}{v7,v8}
\fmf{fermion,right=0.35}{v5,v3}
\fmf{fermion,right=0.35}{v3,v1}
\fmf{fermion,right=0.35}{v1,v2}
\fmf{fermion,right=0.35}{v2,v4}
\fmf{fermion,right=0.35}{v4,v5}
\fmf{boson}{v6,v5}
\fmf{boson}{v1,v4}
\fmf{boson}{v2,v3}
\fmfdot{v1,v2,v3,v4,v5,v6}
\end{fmfgraph}
\end{center}}
%
%
\hspace*{0.15cm}+ \hspace*{0.2cm}
\parbox{20mm}{\begin{center}
\begin{fmfgraph}(18,8)
\setval
\fmfforce{0.055w,0.825h}{v1}
\fmfforce{0.055w,0.175h}{v2}
\fmfforce{0.33w,0.925h}{v3}
\fmfforce{0.33w,0.075h}{v4}
\fmfforce{8/18w,1/2h}{v5}
\fmfforce{13/18w,1/2h}{v6}
\fmfforce{15.5/18w,6.5/8h}{v7}
\fmfforce{15.5/18w,1.5/8h}{v8}
\fmf{fermion,right=1}{v8,v7}
\fmf{plain,right=1}{v7,v8}
\fmf{fermion,right=0.35}{v5,v3}
\fmf{fermion,right=0.35}{v3,v1}
\fmf{fermion,right=0.35}{v1,v2}
\fmf{fermion,right=0.35}{v2,v4}
\fmf{fermion,right=0.35}{v4,v5}
\fmf{boson}{v6,v5}
\fmf{boson}{v3,v4}
\fmf{boson,left=0.35}{v1,v2}
\fmfdot{v1,v2,v3,v4,v5,v6}
\end{fmfgraph}
\end{center}}
%
%
\hspace*{0.15cm}+ \hspace*{0.2cm}
\parbox{20mm}{\begin{center}
\begin{fmfgraph}(18,8)
\setval
\fmfforce{0.055w,0.825h}{v1}
\fmfforce{0.055w,0.175h}{v2}
\fmfforce{0.33w,0.925h}{v3}
\fmfforce{0.33w,0.075h}{v4}
\fmfforce{8/18w,1/2h}{v5}
\fmfforce{13/18w,1/2h}{v6}
\fmfforce{15.5/18w,6.5/8h}{v7}
\fmfforce{15.5/18w,1.5/8h}{v8}
\fmf{fermion,right=1}{v8,v7}
\fmf{plain,right=1}{v7,v8}
\fmf{fermion,right=0.35}{v5,v3}
\fmf{fermion,right=0.35}{v3,v1}
\fmf{fermion,right=0.35}{v1,v2}
\fmf{fermion,right=0.35}{v2,v4}
\fmf{fermion,right=0.35}{v4,v5}
\fmf{boson}{v6,v5}
\fmf{boson,left=0.35}{v3,v1}
\fmf{boson,left=0.35}{v2,v4}
\fmfdot{v1,v2,v3,v4,v5,v6}
\end{fmfgraph}
\end{center}}
\no \\ && \hspace*{0.2cm}
%
%
+ \hspace*{0.1cm}\frac{1}{2}\hspace*{0.2cm}
\parbox{20mm}{\begin{center}
\begin{fmfgraph}(17,6)
\setval
\fmfforce{3/17w,1h}{v1}
\fmfforce{3/17w,0h}{v2}
\fmfforce{6/17w,1/2h}{v3}
\fmfforce{11/17w,1/2h}{v4}
\fmfforce{14/17w,1h}{v5}
\fmfforce{14/17w,0h}{v6}
\fmf{boson}{v1,v2}
\fmf{boson}{v3,v4}
\fmf{boson}{v5,v6}
\fmf{fermion,right=1}{v1,v2}
\fmf{fermion,right=0.4}{v2,v3}
\fmf{fermion,right=0.4}{v3,v1}
\fmf{fermion,right=1}{v6,v5}
\fmf{fermion,right=0.4}{v5,v4}
\fmf{fermion,right=0.4}{v4,v6}
\fmfdot{v1,v2,v3,v4,v5,v6}
\end{fmfgraph}
\end{center}}
%
%
\hspace*{0.2cm} - \hspace*{0.2cm}
\parbox{28mm}{\begin{center}
\begin{fmfgraph}(25,5)
\setval
\fmfforce{1/10w,1h}{v1}
\fmfforce{1/10w,0h}{v2}
\fmfforce{1/5w,1/2h}{v3}
\fmfforce{2/5w,1/2h}{v6}
\fmfforce{5/10w,1h}{v4}
\fmfforce{5/10w,0h}{v5}
\fmfforce{9/10w,1h}{v7}
\fmfforce{9/10w,0h}{v8}
\fmf{boson}{v3,v6}
\fmf{boson}{v4,v7}
\fmf{boson}{v5,v8}
\fmf{fermion,right=1}{v1,v2}
\fmf{plain,right=1}{v2,v1}
\fmf{fermion,right=1}{v5,v4}
\fmf{fermion,right=0.4}{v4,v6}
\fmf{fermion,right=0.4}{v6,v5}
\fmf{fermion,right=1}{v8,v7}
\fmf{fermion,right=1}{v7,v8}
\fmfdot{v3,v4,v5,v6,v7,v8}
\end{fmfgraph}
\end{center}}
%
%
\hspace*{0.2cm}- \hspace*{0.1cm}\frac{1}{2}\hspace*{0.2cm}
\parbox{29mm}{\begin{center}
\begin{fmfgraph}(26,6)
\setval
\fmfforce{2.5/26w,5.5/6h}{v1}
\fmfforce{2.5/26w,0.5/6h}{v2}
\fmfforce{5/26w,1/2h}{v3}
\fmfforce{10/26w,1/2h}{v4}
\fmfforce{13/26w,1h}{v5}
\fmfforce{13/26w,0h}{v6}
\fmfforce{16/26w,1/2h}{v7}
\fmfforce{21/26w,1/2h}{v8}
\fmfforce{23.5/26w,5.5/6h}{v9}
\fmfforce{23.5/26w,0.5/6h}{v10}
\fmf{fermion,right=1}{v1,v2}
\fmf{plain,right=1}{v2,v1}
\fmf{fermion,right=0.4}{v5,v4,v6,v7,v5}
\fmf{fermion,right=1}{v10,v9}
\fmf{plain,right=1}{v9,v10}
\fmf{boson}{v3,v4}
\fmf{boson}{v5,v6}
\fmf{boson}{v7,v8}
\fmfdot{v3,v4,v5,v6,v7,v8}
\end{fmfgraph}
\end{center}}
%
%
\hspace*{0.2cm}- \hspace*{0.2cm}
\parbox{29mm}{\begin{center}
\begin{fmfgraph}(26,6)
\setval
\fmfforce{2.5/26w,5.5/6h}{v1}
\fmfforce{2.5/26w,0.5/6h}{v2}
\fmfforce{5/26w,1/2h}{v3}
\fmfforce{10/26w,1/2h}{v4}
\fmfforce{11.5/26w,0.9h}{v5}
\fmfforce{14.5/26w,0.9h}{v6}
\fmfforce{16/26w,1/2h}{v7}
\fmfforce{21/26w,1/2h}{v8}
\fmfforce{23.5/26w,5.5/6h}{v9}
\fmfforce{23.5/26w,0.5/6h}{v10}
\fmf{fermion,right=1}{v4,v7}
\fmf{plain,left=1}{v7,v4}
\fmf{fermion,right=0.3}{v7,v6}
\fmf{fermion,right=0.3}{v6,v5}
\fmf{fermion,right=0.3}{v5,v4}
\fmf{fermion,right=1}{v1,v2}
\fmf{plain,right=1}{v2,v1}
\fmf{fermion,right=1}{v10,v9}
\fmf{plain,right=1}{v9,v10}
\fmf{boson}{v3,v4}
\fmf{boson}{v7,v8}
\fmf{boson,right=0.7}{v5,v6}
\fmfdot{v3,v4,v5,v6,v7,v8}
\end{fmfgraph}
\end{center}}
\no \\ && \hspace*{0.2cm}
%
%
- \hspace*{0.2cm}
\parbox{29mm}{\begin{center}
\begin{fmfgraph}(26,6)
\setval
\fmfforce{3/26w,1h}{v1}
\fmfforce{3/26w,0h}{v2}
\fmfforce{6/26w,1/2h}{v3}
\fmfforce{11/26w,1/2h}{v4}
\fmfforce{16/26w,1/2h}{v5}
\fmfforce{21/26w,1/2h}{v6}
\fmfforce{23.5/26w,5.5/6h}{v7}
\fmfforce{23.5/26w,0.5/6h}{v8}
\fmf{fermion,right=1}{v1,v2}
\fmf{fermion,right=0.4}{v2,v3}
\fmf{fermion,right=0.4}{v3,v1}
\fmf{fermion,right=1}{v4,v5}
\fmf{fermion,right=1}{v5,v4}
\fmf{fermion,right=1}{v8,v7}
\fmf{plain,right=1}{v7,v8}
\fmf{boson}{v1,v2}
\fmf{boson}{v3,v4}
\fmf{boson}{v5,v6}
\fmfdot{v1,v2,v3,v4,v5,v6}
\end{fmfgraph}
\end{center}}
%
%
\hspace*{0.2cm}+ \hspace*{0.1cm}\frac{1}{2}\hspace*{0.2cm}
\parbox{38mm}{\begin{center}
\begin{fmfgraph}(35,5)
\setval
\fmfforce{1/14w,1h}{v1}
\fmfforce{1/14w,0h}{v2}
\fmfforce{1/7w,1/2h}{v3}
\fmfforce{2/7w,1/2h}{v4}
\fmfforce{3/7w,1/2h}{v5}
\fmfforce{4/7w,1/2h}{v6}
\fmfforce{5/7w,1/2h}{v7}
\fmfforce{6/7w,1/2h}{v8}
\fmfforce{13/14w,1h}{v9}
\fmfforce{13/14w,0h}{v10}
\fmf{boson}{v3,v4}
\fmf{boson}{v5,v6}
\fmf{boson}{v7,v8}
\fmf{fermion,right=1}{v1,v2}
\fmf{plain,right=1}{v2,v1}
\fmf{fermion,right=1}{v4,v5}
\fmf{fermion,right=1}{v5,v4}
\fmf{fermion,right=1}{v6,v7}
\fmf{fermion,right=1}{v7,v6}
\fmf{fermion,right=1}{v10,v9}
\fmf{plain,right=1}{v9,v10}
\fmfdot{v3,v4,v5,v6,v7,v8}
\end{fmfgraph}
\end{center}}
%
%
\hspace*{0.2cm}+ \hspace*{0.1cm}\frac{1}{3}\hspace*{0.2cm}
\parbox{28mm}{\begin{center}
\begin{fmfgraph}(25,15)
\setval
\fmfforce{1/10w,1/3h}{v1}
\fmfforce{1/10w,0h}{v2}
\fmfforce{1/5w,1/6h}{v3}
\fmfforce{2/5w,1/6h}{v4}
\fmfforce{3/5w,1/6h}{v5}
\fmfforce{4/5w,1/6h}{v6}
\fmfforce{9/10w,1/3h}{v7}
\fmfforce{9/10w,0h}{v8}
\fmfforce{1/2w,1/3h}{v9}
\fmfforce{1/2w,2/3h}{v10}
\fmfforce{2/5w,5/6h}{v11}
\fmfforce{3/5w,5/6h}{v12}
\fmf{boson}{v3,v4}
\fmf{boson}{v5,v6}
\fmf{boson}{v9,v10}
\fmf{fermion,right=1}{v1,v2}
\fmf{plain,right=1}{v2,v1}
\fmf{fermion,right=1}{v4,v5}
\fmf{fermion,right=0.4}{v5,v9}
\fmf{fermion,right=0.4}{v9,v4}
\fmf{fermion,right=1}{v8,v7}
\fmf{plain,right=1}{v7,v8}
\fmf{fermion,right=1}{v12,v11}
\fmf{plain,right=1}{v11,v12}
\fmfdot{v3,v4,v5,v6,v9,v10}
\end{fmfgraph}
\end{center}}
\hspace*{0.4cm}.
\la{W4}
\eeq
From the vacuum diagrams (\r{W2}), (\r{W3}), and (\ref{W4}),
we observe a simple mnemonic rule 
for the weights of the connected vacuum diagrams in QED \cite{QED}. 
At least up to four loops, each weight is
equal to the reciprocal number of electron lines, which
are generated by cutting the same electron two-point diagrams.
The sign is given by $(-1)^l$, where $l$ denotes the number of electron
loops. Let us also note that the total weight, which is the sum of all weights of the connected 
vacuum diagrams in the loop order under considertation, vanishes in QED. 
These simple weights are a consequence of the Fermi statistics and the three-vertex 
of the interaction in (\r{AC}). The weights of the vacuum diagrams in other  
theories, like $\phi^4$-theory~\cite{PHI4,Boris,asym,Verena}, 
follow more complicated rules.
\section{One-Particle Irreducible Feynman Diagrams}
\la{IRRFD}
So far, we have generated all connected Feynman diagrams of QED. We now eliminate the
one-particle reducible Feynman diagrams. This is done as usual with the help of
a functional Legendre transform with respect to the current which we introduce in Subsection \ref{E2}.
With this we derive in Subsection
\r{CSCD2} a closed set of Schwinger-Dyson equations for the one-particle $n$-point
functions. In Subsection \r{GRR22} they are converted
into graphical recursion relations for the
corresponding one-particle irreducible Feynman diagrams needed for renormalizing QED.
Finally, the one-particle irreducible vacuum diagrams are constructed graphically 
in Subsection \r{CONNVAC2}.
\subsection{Functional Legendre Transform With Respect to the Current}
\la{E2}
We set up the functional Legendre transform with respect to the current which converts
the vacuum energy $W$ to the effective energy of the first kind $\Gamma_1$. In particular, we investigate the
respective functional derivatives of $W$ and $\Gamma_1$ and the field-theoretic relations between them.
\subsubsection{Effective Energy of the First Kind}
\label{FLTFK}
Starting from the vacuum energy $W[J,S^{-1},D^{-1},V]$ we introduce the new field
\beq
\la{LT1}
\fulla_1 [J,S^{-1},D^{-1},V] = \left. \frac{\delta 
W[J,S^{-1},D^{-1},V]}{\delta J_1}
\right|_{S^{-1},D^{-1},V} \, ,
\eeq
which implicitly defines $J$ as a functional of $\fulla$:
\beq
\la{LT2}
J_1 = J_1 [\fulla,S^{-1},D^{-1},V] \, .
\eeq
From Eq.~(\r{A2}) we read off that $\fulla$ coincides with the field expectation value of the photon in the presence
of the current $J$. The functional 
Legendre transform of the vacuum energy  $W[J,S^{-1},D^{-1},V]$
with respect to the current $J$
results in the effective energy of the first kind
\beq
\Gamma_1 [ \fulla , S^{-1}, D^{-1},V] & = & \int_1 J_1[ \fulla , S^{-1}, 
D^{-1},V] \,
\left. \frac{\delta  W\left[J[\fulla,S^{-1},D^{-1},V],
S^{-1},D^{-1},V\right]}{\delta
J_1[ \fulla , S^{-1}, D^{-1},V]} \right|_{S^{-1},D^{-1},V} \no \\ 
& & - W\left[J[\fulla,S^{-1},D^{-1},V],S^{-1},D^{-1},V\right] \, ,
\eeq
which simplifies due to (\r{LT1}):
\beq
\la{EE}
\Gamma_1 [ \fulla , S^{-1}, D^{-1},V] = \int_1 J_1[ \fulla , S^{-1}, D^{-1},V] 
\fulla_1
- W\left[J[\fulla,S^{-1},D^{-1},V],S^{-1},D^{-1},V\right] \, .
\eeq
Taking into account the functional chain rule, it leads to  
the equation of state 
\beq
\la{SE}
\left. \frac{\delta \Gamma_1 [ \fulla , S^{-1}, D^{-1},V]}{\delta \fulla_1} 
\right|_{S^{-1}, D^{-1},V} = J_1 [\fulla,S^{-1},D^{-1},V] \, . 
\eeq
Performing a loop expansion, the contributions to the effective
energy of the first kind (\r{EE}) may be displayed as one-particle 
irreducible vacuum diagrams which are 
constructed according to the Feynman rules (\r{vac07}), (\r{vac08}), and
(\r{vac09}). In addition, a dot with a wiggly line represents 
the field expectation value of the photon
\beq
\parbox{10mm}{\begin{center}
\begin{fmfgraph*}(5,5)
\setval
\fmfforce{0w,1/2h}{v1}
\fmfforce{1w,1/2h}{v2}
\fmf{boson}{v1,v2}
\fmfdot{v1}
\fmfv{decor.size=0, label=${\scs 1}$, l.dist=1mm, l.angle=0}{v2}
\end{fmfgraph*}
\end{center}}
\quad \equiv \quad  \fulla_{1} \,\, . \la{FEV}
\eeq
If the interaction $V$ vanishes, the vacuum energy (\r{W0}) leads  with (\r{LT1}) to the field expectation
value
\beq
\mbox{\boldmath $A$}^{{\rm c}}_1 [J,S^{-1},D^{-1},0] = \int_2 D_{12} J_2 \, ,
\eeq
which is inverted to give
\beq
J_1 [\fulla,S^{-1},D^{-1},0] = \int_2 D_{12}^{-1} \fulla_2 \, ,
\eeq
leading to the free effective energy of the first kind:
\beq
\la{FEF}
\Gamma_1^{({\rm free})} = \Gamma_1 [ \fulla , S^{-1}, D^{-1} ,0] =  - \,\,\mbox{Tr} 
\ln S^{-1} + \frac{1}{2} \,\,\mbox{Tr} \ln D^{-1} 
+ \frac{1}{2} \int_{12} D_{12}^{-1} \fulla_1 \fulla_2 \, .
\eeq
Its graphical representation is 
\beq
\la{ZEGA}
- \Gamma^{({\rm free})}_1 \quad = \quad 
- \hspace*{0.3cm}
\parbox{8mm}{\centerline{
\begin{fmfgraph}(5,5)
\setval
\fmfforce{0w,0.5h}{v1}
\fmfforce{1w,0.5h}{v2}
\fmf{fermion,right=1}{v2,v1}
\fmf{plain,left=1}{v2,v1}
\end{fmfgraph} }} 
\hspace*{0.3cm} + \hspace*{0.3cm} \frac{1}{2}\hspace*{0.3cm}
\parbox{8mm}{\centerline{
\begin{fmfgraph}(5,5)
\setval
\fmfi{boson}{reverse fullcircle scaled 1w shifted (0.5w,0.5h)}
\end{fmfgraph}
}}
\hspace*{0.3cm} - \hspace*{0.1cm} \frac{1}{2} \hspace*{0.3cm}
\parbox{8mm}{\begin{center}
\begin{fmfgraph}(5,5)
\setval
\fmfstraight
\fmfforce{0w,1/2h}{v1}
\fmfforce{1w,1/2h}{v2}
\fmf{boson}{v1,v2}
\fmfdot{v1,v2}
\end{fmfgraph}
\end{center}} 
\hspace*{0.4cm} .
\eeq
In order to investigate in detail the field-theoretic consequences of the functional Legendre transform of the first kind,
it is advantageous to start with the effective energy of the first kind
$\Gamma_1 [ \fulla , S^{-1}, D^{-1},V]$
and to introduce the current $J$ via the equation of state (\r{SE}).
This implicitly defines the field expectation value of the photon as a functional of the current, i.e.
\beq
\fulla_1 = \fulla_1 [J,S^{-1},D^{-1},V] \, .
\eeq
Thus the vacuum energy is recovered by the inverse functional Legendre transform  
\beq
\la{FEE}
W [ J , S^{-1}, D^{-1},V] = \int_1 J_1 \fulla_1[ J , S^{-1}, D^{-1},V] 
- \Gamma_1 \left[\fulla[J,S^{-1},D^{-1},V],S^{-1},D^{-1},V\right] \, .
\eeq
With this we derive useful relations between the functional
derivatives of the vacuum energy $W$ and the effective energy of the
first kind $\Gamma_1$, respectively.
\subsubsection{Functional Derivatives}
Taking into account the functional chain rule, the first
functional derivatives of the vacuum energy $W$ read
(\r{LT1}) and
\beq
\left.\frac{\delta W[J,S^{-1},D^{-1},V]}{\delta 
S^{-1}_{12}}\right|_{J,D^{-1},V} &= & -
\left.
\frac{\delta \Gamma_1 \left[\fulla[J,S^{-1},
D^{-1},V],S^{-1},D^{-1},V\right]}{\delta
S^{-1}_{12}} \right|_{\fullas,D^{-1},V} \, , \la{DD1}\\
\left.\frac{\delta W[J,S^{-1},D^{-1},V]}{\delta 
D^{-1}_{12}}\right|_{J,S^{-1},V} &= & -
\left.
\frac{\delta \Gamma_1 \left[\fulla[J,S^{-1},D^{-1},V],
S^{-1},D^{-1},V\right]}{\delta D^{-1}_{12}} 
\right|_{\fullas,S^{-1},V} \, , \la{DD2}\\
\left.\frac{\delta W[J,S^{-1},D^{-1},V]}{\delta 
V_{123}}\right|_{J,S^{-1},D^{-1}} 
&= & -
\left.\frac{\delta \Gamma_1 \left[\fulla[J,S^{-1},D^{-1},V],
S^{-1},D^{-1},V\right]}{\delta
V_{123}} \right|_{\fullas,S^{-1},D^{-1}} \, . \la{D2b}
\eeq
To evaluate second functional derivatives of the 
vacuum energy $W$ is 
more involved. First, we observe
\beq
\left.\frac{\delta^2 W[J,S^{-1},D^{-1},V]}{\delta J_2 \delta J_1} 
\right|_{S^{-1},D^{-1},V} & = &
\left.\frac{\delta \fulla_1[J,S^{-1},D^{-1},V]}{\delta J_2} 
\right|_{S^{-1},D^{-1},V} \no \\ 
& = & \left( \left.
\frac{\delta J_2[\fulla[J,S^{-1},D^{-1},V], S^{-1},D^{-1},V]}{\delta 
\fulla_1[J,S^{-1},D^{-1},V]}
\right|_{S^{-1},D^{-1},V} \right)^{-1} \no \\
&= &\left( \left.\frac{\delta^2 \Gamma_1 
\left[\fulla[J,S^{-1},D^{-1},V],S^{-1},
D^{-1},V\right]}{\delta \fulla_1 [J,S^{-1},D^{-1},V]
\delta \fulla_2 [J,S^{-1},D^{-1},V]}
\right|_{S^{-1},D^{-1},V} \right)^{-1} \, ,
\la{SCH}
\eeq
where we have used (\r{LT1}), (\r{SE}), and the fact that the derivative of
a functional equals the inverse of the derivative of the inverse functional.
To precise the meaning of relation (\r{SCH}), we rederive it from
another point of view. Considering the functional identity
\beq
\left. \frac{\delta J_1 \left[ \fulla [ J , S^{-1}, 
D^{-1}, V],S^{-1}, D^{-1}, V\right]}{\delta J_2} 
\right|_{S^{-1}, D^{-1}, V} = \delta_{12} \, ,
\eeq
we apply the functional chain rule  together with (\r{LT1}) and (\r{SE}). Thus we result in
\beq
\int_3 
\left.\frac{\delta^2 \Gamma_1 
\left[\fulla[J,S^{-1},D^{-1},V],S^{-1},D^{-1},V\right]}{\delta 
\fulla_1 [J,S^{-1},D^{-1},V]
\delta \fulla_3[J,S^{-1},D^{-1},V]}
\right|_{S^{-1},D^{-1},V} \, 
\left.\frac{\delta^2 W[J,S^{-1},D^{-1},V]}{\delta J_3 \delta J_2} 
\right|_{S^{-1},D^{-1},V} = \delta_{12} \, , \la{FID}
\eeq
which coincides with (\r{SCH}).
Furthermore we obtain from (\r{DD1}) by applying again the functional
chain rule and relation (\r{SCH}):
\beq
&& \frac{\delta}{\delta J_3} \left( \left.
\frac{\delta W[J,S^{-1},D^{-1},V]}{\delta S^{-1}_{12}}\right|_{J,D^{-1},V} 
\right)_{S^{-1},D^{-1},V} 
= - \int_4 \left( \left.
\frac{\delta^2 \Gamma_1 \left[\fulla[J,S^{-1},D^{-1},V],S^{-1},
D^{-1},V\right]}{\delta \fulla_3 [J,S^{-1},D^{-1},V]
\delta \fulla_4[J,S^{-1},D^{-1},V]}
\right|_{S^{-1},D^{-1},V} \right)^{-1}
\no\\ && 
\hspace*{2cm}\times\,\, \frac{\delta}{\delta\fulla_4[J,S^{-1},D^{-1},V]}
\left( \left.
\frac{\delta \Gamma_1 \left[\fulla[J,S^{-1},D^{-1},V],
S^{-1},D^{-1},V\right]]}{\delta S^{-1}_{12}} 
\right|_{\fullas,D^{-1},V} \right)_{S^{-1},D^{-1},V} \, .
\la{D3}
\eeq
\subsubsection{Field-Theoretic Identities}
Performing the functional Legendre transform with respect to the current, the
compatibility relation (\r{CP1}) between functional derivatives with
respect to the current $J$ and the photon kernel $D^{-1}$ yields
\beq
\la{D5}
\left( \left.
\frac{\delta^2 \Gamma_1 [\fulla,S^{-1},D^{-1},V]}{\delta \fulla_2 
\delta \fulla_1}
\right|_{S^{-1},D^{-1},V} \right)^{-1} = 2 \, \left. \frac{\delta 
\Gamma_1 [\fulla,S^{-1},D^{-1},V]}{\delta D^{-1}_{12}} 
\right|_{\fullas,D^{-1},V} - \fulla_1 \fulla_2
\eeq 
due to (\r{LT1}), (\r{DD2}), and (\r{SCH}). In a similar way, 
the compatibility relation (\r{CP2}) between functional derivatives with
respect to the current $J$, the electron kernel $S^{-1}$,
and the interaction $V$ is converted using (\r{LT1}), (\r{DD1}), (\r{D2b}), and (\r{D3}) to 
\beq
&& \int_4 \frac{\delta}{\delta \fulla_4} \left( \left.
\frac{\delta\Gamma_1 [\fulla,S^{-1},D^{-1},V]}{\delta 
S^{-1}_{12}} \right|_{\fullas,D^{-1},V}  
\right)_{S^{-1},D^{-1},V} 
\left( \left.
\frac{\delta^2 \Gamma_1 [\fulla,S^{-1},D^{-1},V]}{\delta \fulla_3 
\delta \fulla_4} \right|_{S^{-1},D^{-1},V} \right)^{-1}
\no \\
&& \hspace*{2cm} = \left.
\frac{\delta \Gamma_1 [\fulla,S^{-1},D^{-1},V]}{\delta 
V_{123}}\right|_{\fullas,S^{-1},D^{-1}} 
- \left.
\frac{\delta \Gamma_1 [\fulla,S^{-1},D^{-1},V]}{\delta S^{-1}_{12}} 
\right|_{\fullas,D^{-1},V}\,\fulla_3 \, .
\la{D6}
\eeq
The functional Legendre transform has also consequences for the 
connected $n$-point functions (\r{S2})--(\r{V2}).
Taking into account (\r{LT1}) and (\r{DD1})--(\r{D2b}), we obtain
\beq
\fulls_{12} &=& 
-\, \frac{\delta \Gamma_1}{\delta S^{-1}_{21}} 
 \,, \la{S3} \\
\fulld_{12} &=& 
2 \, \frac{\delta \Gamma_1}{\delta D^{-1}_{12}} - \fulla_1 \fulla_2
 \,, \la{DD3} \\
\fullg_{123} &=& - \, \frac{\delta \Gamma_1}{\delta V_{213}} 
- \fulls_{12} \fulla_3 \,. \la{V3}
\eeq
From (\r{DD3}) we read off that, for instance, cutting a photon line of the
one-particle irreducible vacuum diagrams in all possible ways leads to the
diagrams contributing to the connected two-point function
$\fulld_{12}$. \\

The connected $n$-point functions are related to the one-particle
irreducible $n$-point functions. The electron and photon self-energy, 
which we shall denote by $\Sigma^e$ and  $\Sigma^\gamma$, is defined according to
\beq
\Sigma^e_{12} &\equiv& S^{-1}_{12} - \mbox{\boldmath $S$}^{{\rm c}\,-1}_{12} 
\,, \la{SE1} \\ 
\Sigma^\gamma_{12} &\equiv& D^{-1}_{12} - \mbox{\boldmath $D$}^{{\rm c}\,-1}_{12} 
\,, \la{SP1} 
\eeq
where $\mbox{\boldmath $S$}^{{\rm c}\,-1}_{12}$ and $\mbox{\boldmath $D$}^{{\rm c}\,-1}_{12}$ represent the inverse of the
propagators $\fulls_{12}$ and $\fulld_{12}$, respectively:
\beq
\int_{2} \fulls_{12} \mbox{\boldmath $S$}^{{\rm c}\,-1}_{23} &= &\delta_{13} \, , \la{FIA1} \\
\int_{2} \fulld_{12} \mbox{\boldmath $D$}^{{\rm c}\,-1}_{23} &= & \delta_{13} \, .\la{FIB2}
\eeq
Due to (\r{SE1}) and (\r{SP1})
the connected electron and photon two-point functions follow from the Dyson equations
\beq
\fulls_{12} &= &S_{12} + \int_{34} S_{13} \Sigma^e_{34} \fulls_{42} \, , \la{SE2}\\
\fulld_{12} &= &D_{12} + \int_{34} D_{13} \Sigma^\gamma_{34} \fulld_{42} \, . \la{SP2}
\eeq
Representing the self-energies $\Sigma^e$ and $\Sigma^\gamma$ by a two-vertex with a big open dot
\beq
\parbox{11mm}{\centerline{
\begin{fmfgraph*}(8,5)
\setval
\fmfforce{0w,0.5h}{v1}
\fmfforce{3/8w,0.5h}{v2}
\fmfforce{5/8w,0.5h}{v3}
\fmfforce{5.2/8w,0.5h}{v3b}
\fmfforce{1w,0.5h}{v4}
\fmf{fermion}{v2,v1}
\fmf{fermion}{v4,v3b}
\fmf{double,width=0.2mm,left=1}{v2,v3,v2}
\fmfv{decor.size=0, label=${\scs 1}$, l.dist=1mm, l.angle=-180}{v1}
\fmfv{decor.size=0, label=${\scs 2}$, l.dist=1mm, l.angle=0}{v4}
\end{fmfgraph*} }}
\quad & \equiv & \quad \Sigma^e_{12} \, , \\
\parbox{11mm}{\centerline{
\begin{fmfgraph*}(8,5)
\setval
\fmfforce{0w,0.5h}{v1}
\fmfforce{3/8w,0.5h}{v2}
\fmfforce{5/8w,0.5h}{v3}
\fmfforce{5.2/8w,0.5h}{v3b}
\fmfforce{1w,0.5h}{v4}
\fmf{boson}{v2,v1}
\fmf{boson}{v4,v3b}
\fmf{double,width=0.2mm,left=1}{v2,v3,v2}
\fmfv{decor.size=0, label=${\scs 1}$, l.dist=1mm, l.angle=-180}{v1}
\fmfv{decor.size=0, label=${\scs 2}$, l.dist=1mm, l.angle=0}{v4}
\end{fmfgraph*} }}
\quad & \equiv & \quad \Sigma^\gamma_{12} \, ,
\eeq
the Dyson equations (\r{SE2}) and (\r{SP2}) read graphically
\beq
\parbox{10mm}{\centerline{
\begin{fmfgraph*}(7,3)
\setval
\fmfleft{v1}
\fmfright{v2}
\fmf{heavy,width=0.2mm}{v2,v1}
\fmfv{decor.size=0, label=${\scs 1}$, l.dist=1mm, l.angle=-180}{v1}
\fmfv{decor.size=0, label=${\scs 2}$, l.dist=1mm, l.angle=0}{v2}
\end{fmfgraph*}}}  
\quad & = &\quad 
\parbox{10mm}{\centerline{
\begin{fmfgraph*}(7,3)
\setval
\fmfleft{v1}
\fmfright{v2}
\fmf{fermion}{v2,v1}
\fmfv{decor.size=0, label=${\scs 1}$, l.dist=1mm, l.angle=-180}{v1}
\fmfv{decor.size=0, label=${\scs 2}$, l.dist=1mm, l.angle=0}{v2}
\end{fmfgraph*}}} 
\hspace*{0.2cm} + \hspace*{0.2cm}
\parbox{15mm}{\centerline{
\begin{fmfgraph*}(12,5)
\setval
\fmfforce{0w,0.5h}{v1}
\fmfforce{5/12w,0.5h}{v2}
\fmfforce{7/12w,0.5h}{v3}
\fmfforce{7.2/12w,0.5h}{v3b}
\fmfforce{1w,0.5h}{v4}
\fmf{fermion}{v2,v1}
\fmf{heavy,width=0.2mm}{v4,v3b}
\fmf{double,width=0.2mm,left=1}{v2,v3,v2}
\fmfv{decor.size=0, label=${\scs 1}$, l.dist=1mm, l.angle=-180}{v1}
\fmfv{decor.size=0, label=${\scs 2}$, l.dist=1mm, l.angle=0}{v4}
\end{fmfgraph*} }}
\hspace*{0.4cm} , \la{GE23}
\\
\parbox{10mm}{\centerline{
\begin{fmfgraph*}(7,3)
\setval
\fmfleft{v1}
\fmfright{v2}
\fmf{dbl_wiggly,width=0.2mm}{v2,v1}
\fmfv{decor.size=0, label=${\scs 1}$, l.dist=1mm, l.angle=-180}{v1}
\fmfv{decor.size=0, label=${\scs 2}$, l.dist=1mm, l.angle=0}{v2}
\end{fmfgraph*}}}  
\quad & = &\quad  
\parbox{10mm}{\centerline{
\begin{fmfgraph*}(7,3)
\setval
\fmfleft{v1}
\fmfright{v2}
\fmf{boson}{v1,v2}
\fmfv{decor.size=0, label=${\scs 1}$, l.dist=1mm, l.angle=-180}{v1}
\fmfv{decor.size=0, label=${\scs 2}$, l.dist=1mm, l.angle=0}{v2}
\end{fmfgraph*}}}
\hspace*{0.2cm} + \hspace*{0.2cm}
\parbox{15mm}{\centerline{
\begin{fmfgraph*}(12,5)
\setval
\fmfforce{0w,0.5h}{v1}
\fmfforce{5/12w,0.5h}{v2}
\fmfforce{7/12w,0.5h}{v3}
\fmfforce{7.2/12w,0.5h}{v3b}
\fmfforce{1w,0.5h}{v4}
\fmf{boson}{v2,v1}
\fmf{dbl_wiggly,width=0.2mm}{v4,v3b}
\fmf{double,width=0.2mm,left=1}{v2,v3,v2}
\fmfv{decor.size=0, label=${\scs 1}$, l.dist=1mm, l.angle=-180}{v1}
\fmfv{decor.size=0, label=${\scs 2}$, l.dist=1mm, l.angle=0}{v4}
\end{fmfgraph*} }}
\hspace*{0.4cm} .\la{GE24}
\eeq
The one-particle irreducible three-point function $\tau$ is defined by
\beq
\la{DIV}
\fullg_{123} = - \int_{456} \fulls_{14} \fulls_{52} \fulld_{36}
\tau_{456} \, .
\eeq
Representing the one-particle irreducible three-point function $\tau$ by a three-vertex with a big open dot
\beq
\la{TAU}
\parbox{10mm}{\centerline{
\begin{fmfgraph*}(6.93,9)
\setval
\fmfforce{1w,0h}{v1}
\fmfforce{0w,0h}{v2}
\fmfforce{0.5w,2/3h}{v3}
\fmfforce{0.5w,3/9h}{v4}
\fmfforce{0.5w,2.5/9h}{v5}
\fmfforce{0.5w,1.5/9h}{v6}
\fmfforce{0.5w,1/9h}{v7}
\fmfforce{0.333w,1.5/9h}{v8}
\fmfforce{0.667w,1.5/9h}{v9}
\fmf{photon}{v3,v4}
\fmf{fermion}{v8,v2}
\fmf{fermion}{v1,v9}
\fmf{double,left=1,width=0.2mm}{v4,v7,v4}
\fmfv{decor.size=0, label=${\scs 1}$, l.dist=1mm, l.angle=-150}{v2}
\fmfv{decor.size=0, label=${\scs 2}$, l.dist=1mm, l.angle=-30}{v1}
\fmfv{decor.size=0, label=${\scs 3}$, l.dist=1mm, l.angle=90}{v3}
\end{fmfgraph*} }}
\quad \equiv \quad - \,\,\tau_{123} \, ,
\eeq
relation (\r{DIV}) is pictured by
\beq
\parbox{13.39mm}{\centerline{
\begin{fmfgraph*}(10.39,9)
\setval
\fmfforce{1/2w,2/9h}{v1}
\fmfforce{1/2w,4/9h}{v2}
\fmfforce{1/2w,1h}{v3}
\fmfforce{0w,0h}{v4}
\fmfforce{1w,0h}{v5}
\fmfforce{0.5834w,2.5/9h}{v6}
\fmfforce{0.4166w,2.5/9h}{v7}
\fmf{plain,left=1}{v1,v2,v1}
\fmf{boson}{v2,v3}
\fmf{fermion}{v7,v4}
\fmf{fermion}{v5,v6}
\fmfv{decor.size=0, label=${\scs 1}$, l.dist=1mm, l.angle=-150}{v4}
\fmfv{decor.size=0, label=${\scs 2}$, l.dist=1mm, l.angle=-30}{v5}
\fmfv{decor.size=0, label=${\scs 3}$, l.dist=1mm, l.angle=90}{v3}
\end{fmfgraph*} }} 
\quad = \quad 
\parbox{10mm}{\centerline{
\begin{fmfgraph*}(10.39,9)
\setval
\fmfforce{1w,0h}{v1}
\fmfforce{0w,0h}{v2}
\fmfforce{0.5w,1h}{v3}
\fmfforce{0.5w,4/9h}{v4}
\fmfforce{0.5w,2/9h}{v7}
\fmfforce{0.4038w,2.5/9h}{v8}
\fmfforce{0.596w,2.5/9h}{v9}
\fmf{dbl_wiggly,width=0.2mm}{v3,v4}
\fmf{heavy,width=0.2mm}{v8,v2}
\fmf{heavy,width=0.2mm}{v1,v9}
\fmf{double,left=1,width=0.2mm}{v4,v7,v4}
\fmfv{decor.size=0, label=${\scs 1}$, l.dist=1mm, l.angle=-150}{v2}
\fmfv{decor.size=0, label=${\scs 2}$, l.dist=1mm, l.angle=-30}{v1}
\fmfv{decor.size=0, label=${\scs 3}$, l.dist=1mm, l.angle=90}{v3}
\end{fmfgraph*} }}
\hspace*{0.4cm} \, . 
\\ \no
\eeq
For later purposes we note that the one-particle three-point function
$\tau$ may be also defined by
\beq
\la{LP}
\frac{\delta^2 \Gamma_1}{\delta S^{-1}_{12} \delta \fulla_3} = 
\int_{45} \fulls_{24} \fulls_{51} \tau_{453} \, ,
\eeq
as follows from (\r{D5})--(\r{S3}), (\r{V3}), and (\r{DIV}).
\subsection{Closed Set of Schwinger-Dyson Equations for One-Particle Irreducible 
$n$-Point Functions}
\label{CSCD2}
Now we calculate the effect of the  functional Legendre transform with respect to the
current upon the trivial functional identities (\r{ID1}), (\r{ID2}), and (\r{ID3})
which immediately followed from the definition of the functional integral.
This leads to a closed set of equations determining the electron and photon self-energy as well as the one-particle irreducible 
three-point function.
\subsubsection{Electron Self-Energy}
In order to determine the electron self-energy $\Sigma^e$, we start
with the first functional differential equation (\r{ID1}) for the vacuum energy $W$
and perform the functional Legendre transform of the first kind defined in Subsection \r{E2}.
Inserting (\r{LT1}), (\r{DD1}), and (\r{D3})
by taking into account the compatibility relation (\r{D5}),
we thus obtain
\beq
\delta_{12} + \int_3 S^{-1}_{13} \frac{\delta \Gamma_1}{\delta S^{-1}_{23}}
+ \int_{345} V_{134} \frac{\delta^2 \Gamma_1}{\delta S^{-1}_{23}
\delta \fulla_5} \left\{ 2 \frac{\delta \Gamma_1}{\delta D^{-1}_{45}}
- \fulla_4 \fulla_5 \right\} + \int_{34} V_{134} \fulla_4
\frac{\delta \Gamma_1}{\delta S^{-1}_{23}} = 0 \, .
\eeq
With the definition of the connected electron and photon two-point function (\r{S3}) and (\r{DD3}), as well as the one-particle
irreducible three-point function (\r{LP}), we get
\beq
\delta_{12} - \int_3 S^{-1}_{13} \fulls_{32} + \int_{34567} V_{134}
\tau_{567} \fulls_{62} \fulls_{35} \fulld_{47} - \int_{34}
V_{134} \fulls_{32} \fulla_{4} = 0 \, .
\eeq
This reduces to the Schwinger-Dyson equation for the electron self-energy (\r{SE2})
\beq
\la{CSD4}
\Sigma^e_{12} = \int_{3456} V_{134} \tau_{526} \fulls_{35} \fulld_{46}
- \int_3 V_{123} \fulla_{3} \, , 
\eeq
which graphically reads
\beq
\la{SDI1}
\parbox{11mm}{\centerline{
\begin{fmfgraph*}(8,5)
\setval
\fmfforce{0w,0.5h}{v1}
\fmfforce{3/8w,0.5h}{v2}
\fmfforce{5/8w,0.5h}{v3}
\fmfforce{5.2/8w,0.5h}{v3b}
\fmfforce{1w,0.5h}{v4}
\fmf{fermion}{v2,v1}
\fmf{fermion}{v4,v3b}
\fmf{double,width=0.2mm,left=1}{v2,v3,v2}
\fmfv{decor.size=0, label=${\scs 1}$, l.dist=1mm, l.angle=-180}{v1}
\fmfv{decor.size=0, label=${\scs 2}$, l.dist=1mm, l.angle=0}{v4}
\end{fmfgraph*} }}
\quad = \quad 
\parbox{16mm}{\centerline{
\begin{fmfgraph*}(13,3)
\setval
\fmfforce{0w,0h}{v1}
\fmfforce{3/13w,0h}{v2}
\fmfforce{8/13w,0h}{v3}
\fmfforce{7.8/13w,0h}{v3b}
\fmfforce{10/13w,0h}{v4}
\fmfforce{10.2/13w,0h}{v4b}
\fmfforce{1w,0h}{v5}
\fmfforce{12/13w,0h}{v6}
\fmfforce{8.8/13w,0.33h}{v7}
\fmf{double,width=0.2mm,left=1}{v3,v4,v3}
\fmf{dbl_wiggly,width=0.2mm,left=0.85}{v2,v7}
\fmf{fermion}{v5,v4b}
\fmf{heavy,width=0.2mm}{v3b,v2}
\fmf{fermion}{v2,v1}
\fmfv{decor.size=0, label=${\scs 1}$, l.dist=1mm, l.angle=-180}{v1}
\fmfv{decor.size=0, label=${\scs 2}$, l.dist=1mm, l.angle=0}{v5}
\fmfdot{v2}
\end{fmfgraph*} }} 
\hspace*{0.2cm} + \hspace*{0.2cm}
\parbox{9mm}{\begin{center}
\begin{fmfgraph*}(6,5)
\setval
\fmfforce{0w,0h}{v1}
\fmfforce{1/2w,0h}{v2}
\fmfforce{1/2w,1h}{v2b}
\fmfforce{1w,0h}{v3}
\fmf{fermion}{v3,v2}
\fmf{fermion}{v2,v1}
\fmf{boson}{v2,v2b}
\fmfdot{v2,v2b}
\fmfv{decor.size=0, label=${\scs 1}$, l.dist=1mm, l.angle=-180}{v1}
\fmfv{decor.size=0, label=${\scs 2}$, l.dist=1mm, l.angle=0}{v3}
\end{fmfgraph*}
\end{center}} 
\hspace*{0.4cm} .
\eeq
\subsubsection{Photon Self-Energy}
The photon self-energy $\Sigma^{\gamma}$ follows in the same way from
the third functional identity (\r{ID3}). We perform the
functional Legendre transform of the first kind by using (\r{LT1}),
(\r{DD1}), (\r{DD2}), (\r{D3}), and the compatibility relation (\r{D5}):
\beq
\la{ZW1B}
\delta_{12} + \frac{\delta \Gamma_1}{\delta \fulla_{1}} \,\fulla_{2}
- 2 \int_{3} D^{-1}_{13}\, \frac{\delta \Gamma_1}{\delta D^{-1}_{23}}
- \int_{345} V_{341} \,\frac{\delta^2 \Gamma_1}{\delta S^{-1}_{34} \delta
\fulla_5} \, \left\{ 2 \frac{\delta \Gamma_1}{\delta D^{-1}_{25}}
- \fulla_2 \fulla_5 \right\} - \int_{34} V_{341} \,
\frac{\delta \Gamma_1}{\delta S^{-1}_{34}} \,\fulla_2 = 0 \, .
\eeq
The functional derivative with respect to the field expectation value $\fulla$ in the second term can be eliminated
by performing the functional Legendre transform of the first kind in the second functional identity (\r{ID2}). 
by performing the functional Legendre transform of the first kind.
With (\r{LT1}), (\r{SE}), and (\r{DD1}) we obtain
\beq
\la{ZW1}
\frac{\delta \Gamma_1}{\delta \fulla_1} = \int_2 D^{-1}_{12} \fulla_2
+ \int_{23} V_{231} \, \frac{\delta \Gamma_1}{\delta S^{-1}_{23}} \, .
\eeq
Inserting this into (\r{ZW1B}) leads to
\beq
\delta_{12} - \int_3 D^{-1}_{13} \left\{ 2 \frac{\delta \Gamma_1}{\delta D^{-1}_{23}} - \fulla_2 \fulla_3 \right\}
- \int_{345} V_{341} \frac{\delta^2 \Gamma_1}{\delta S^{-1}_{34} \delta \fulla_5} \left\{ 2 \frac{\delta \Gamma_1}{\delta D^{-1}_{25}} 
- \fulla_2 \fulla_5 \right\} = 0 \, .
\eeq
Using once more the relations (\r{DD3}) and (\r{LP}), we find
\beq
\delta_{12} - \int_3 D^{-1}_{13} \fulld_{23} - \int_{34567} V_{341}
\tau_{675} \fulls_{46} \fulls_{73} \fulld_{25} = 0 \, .
\eeq
Thus the Schwinger-Dyson equation for the photon self-energy (\r{SP1}) reads
\beq
\la{CSD5}
\Sigma^{\gamma}_{12} = - \int_{3456} V_{341} \tau_{562} \fulls_{63}
\fulls_{45} \, ,
\eeq
with the graphical representation 
\beq
\la{SDI2}
\parbox{11mm}{\centerline{
\begin{fmfgraph*}(8,5)
\setval
\fmfforce{0w,0.5h}{v1}
\fmfforce{3/8w,0.5h}{v2}
\fmfforce{5/8w,0.5h}{v3}
\fmfforce{5.2/8w,0.5h}{v3b}
\fmfforce{1w,0.5h}{v4}
\fmf{boson}{v2,v1}
\fmf{boson}{v4,v3b}
\fmf{double,width=0.2mm,left=1}{v2,v3,v2}
\fmfv{decor.size=0, label=${\scs 1}$, l.dist=1mm, l.angle=-180}{v1}
\fmfv{decor.size=0, label=${\scs 2}$, l.dist=1mm, l.angle=0}{v4}
\end{fmfgraph*} }}
\quad = \quad  - \hspace*{0.2cm}
\parbox{16mm}{\centerline{
\begin{fmfgraph*}(13,6)
\setval
\fmfforce{0w,1/2h}{v1}
\fmfforce{3/13w,1/2h}{v2}
\fmfforce{8/13w,1/2h}{v3}
\fmfforce{10/13w,1/2h}{v4}
\fmfforce{10.2/13w,1/2h}{v4b}
\fmfforce{1w,1/2h}{v5}
\fmfforce{9/13w,1/2h}{v6}
\fmfforce{8.8/13w,2/3h}{v7}
\fmfforce{8.8/13w,1/3h}{v8}
\fmf{double,width=0.2mm,left=1}{v3,v4,v3}
\fmf{heavy,width=0.2mm,right=0.85}{v7,v2}
\fmf{heavy,width=0.2mm,right=0.85}{v2,v8}
\fmf{boson}{v5,v4b}
\fmf{boson}{v2,v1}
\fmfv{decor.size=0, label=${\scs 1}$, l.dist=1mm, l.angle=-180}{v1}
\fmfv{decor.size=0, label=${\scs 2}$, l.dist=1mm, l.angle=0}{v5}
\fmfdot{v2}
\end{fmfgraph*} }} 
\hspace*{0.4cm} .
\eeq
\subsubsection{One-Particle Irreducible Three-Point Function}
The iteration of the integral equations (\r{SDI1}) and (\r{SDI2})
for the electron and photon self-energy $\Sigma^{e}$ and
$\Sigma^{\gamma}$ requires the knowledge of the one-particle
irreducible three-point function $\tau$. To obtain this, we further evaluate
(\r{LP}) by inserting (\r{ZW1}). Thus the one-particle irreducible 
three-point function $\tau$ follows from
\beq
\la{ZW2}
\tau_{123} = \int_{4567} \mbox{\boldmath $S$}^{{\rm c}\,-1}_{14} 
\mbox{\boldmath $S$}^{{\rm c}\,-1}_{52} \, 
\frac{\delta^2 \Gamma_1}{\delta S^{-1}_{54} \delta S^{-1}_{67}} \, V_{673}\, .
\eeq
To express the right-hand side in terms of the electron self-energy,
we apply a functional derivative with respect to the electron kernel
$S^{-1}_{45}$ to the identity
\beq
\int_2 \fulls_{12} \mbox{\boldmath $S$}^{{\rm c}\,-1}_{23} = \delta_{13} \, , 
\eeq
yielding
\beq
\la{NNNN}
\frac{\delta \mbox{\boldmath $S$}^{{\rm c}\,-1}_{12}}{\delta S^{-1}_{45}} = - \int_{67} 
\mbox{\boldmath $S$}^{{\rm c}\,-1}_{17} \mbox{\boldmath $S$}^{{\rm c}\,-1}_{62} 
\frac{\delta \fulls_{76}}{\delta S^{-1}_{45}} \, ,
\eeq
so that we obtain together with (\r{S3})
\beq
\la{ZW3}
\frac{\delta \mbox{\boldmath $S$}^{{\rm c}\,-1}_{12}}{\delta S^{-1}_{45}}
= \int_{67} \mbox{\boldmath $S$}^{{\rm c}\,-1}_{17}
\mbox{\boldmath $S$}^{{\rm c}\,-1}_{62} 
\frac{\delta^2 \Gamma_1}{\delta S^{-1}_{45} \delta S^{-1}_{67}} \, .
\eeq
From (\r{ZW2}) and (\r{ZW3}) we conclude
\beq
\la{NNNNN}
\tau_{123} = \int_{45} 
\frac{\delta \mbox{\boldmath $S$}^{{\rm c}\,-1}_{12}}{\delta S^{-1}_{45}}
V_{453} \, .
\eeq
Inserting (\r{SE1}) and using the functional chain rule (\r{FCR}),
we finally arrive at a functional integrodifferential equation for the
one-particle irreducible three-point function
\beq
\la{TRHEEG}
\tau_{123} = V_{123} + \int_{4567} V_{453} S_{64} S_{57} 
\frac{\delta \Sigma^{e}_{12}}{\delta S_{67}} \, ,
\eeq
whose graphical representation is
\beq
\la{SDI3}
\parbox{10mm}{\centerline{
\begin{fmfgraph*}(6.93,6)
\setval
\fmfforce{1w,0h}{v1}
\fmfforce{0w,0h}{v2}
\fmfforce{0.5w,1h}{v3}
\fmfforce{0.5w,3/6h}{v4}
\fmfforce{0.5w,2.5/6h}{v5}
\fmfforce{0.5w,1.5/6h}{v6}
\fmfforce{0.5w,1/6h}{v7}
\fmfforce{0.333w,1.5/6h}{v8}
\fmfforce{0.667w,1.5/6h}{v9}
\fmf{photon}{v3,v4}
\fmf{fermion}{v8,v2}
\fmf{fermion}{v1,v9}
\fmf{double,left=1,width=0.2mm}{v4,v7,v4}
\fmfv{decor.size=0, label=${\scs 1}$, l.dist=1mm, l.angle=-150}{v2}
\fmfv{decor.size=0, label=${\scs 2}$, l.dist=1mm, l.angle=-30}{v1}
\fmfv{decor.size=0, label=${\scs 3}$, l.dist=1mm, l.angle=90}{v3}
\end{fmfgraph*} }}
\quad = \quad 
\parbox{8mm}{\centerline{
\begin{fmfgraph*}(5,4.33)
\setval
\fmfforce{1w,0h}{v1}
\fmfforce{0w,0h}{v2}
\fmfforce{0.5w,1h}{v3}
\fmfforce{0.5w,0.2886h}{vm}
\fmf{fermion}{v1,vm}
\fmf{fermion}{vm,v2}
\fmf{photon}{v3,vm}
\fmfv{decor.size=0,label=${\scs 2}$,l.dist=1mm}{v1}
\fmfv{decor.size=0,label=${\scs 1}$,l.dist=1mm}{v2}
\fmfv{decor.size=0,label=${\scs 3}$,l.dist=1mm}{v3}
\fmfdot{vm}
\end{fmfgraph*} }}
\hspace*{0.3cm} + \hspace*{0.3cm}
\dephi{
\parbox{14mm}{\centerline{
\begin{fmfgraph*}(8,5)
\setval
\fmfforce{0w,0.5h}{v1}
\fmfforce{3/8w,0.5h}{v2}
\fmfforce{5/8w,0.5h}{v3}
\fmfforce{5.2/8w,0.5h}{v3b}
\fmfforce{1w,0.5h}{v4}
\fmf{fermion}{v2,v1}
\fmf{fermion}{v4,v3b}
\fmf{double,width=0.2mm,left=1}{v2,v3,v2}
\fmfv{decor.size=0, label=${\scs 1}$, l.dist=1mm, l.angle=-180}{v1}
\fmfv{decor.size=0, label=${\scs 2}$, l.dist=1mm, l.angle=0}{v4}
\end{fmfgraph*} }}
}{4}{5}
\hspace*{0.5cm}
\parbox{10.5mm}{\begin{center}
\begin{fmfgraph*}(5.5,8.66)
\setval
\fmfforce{0w,0h}{v1}
\fmfforce{0w,1h}{v2}
\fmfforce{2.5/5.5w,1/2h}{v3}
\fmfforce{1w,1/2h}{v4}
\fmf{fermion}{v1,v3}
\fmf{fermion}{v3,v2}
\fmf{boson}{v3,v4}
\fmfdot{v3}
\fmfv{decor.size=0, label=${\scs 4}$, l.dist=1mm, l.angle=180}{v2}
\fmfv{decor.size=0, label=${\scs 5}$, l.dist=1mm, l.angle=-180}{v1}
\fmfv{decor.size=0, label=${\scs 3}$, l.dist=1mm, l.angle=0}{v4}
\end{fmfgraph*}
\end{center}}
\hspace*{0.4cm}.
\eeq
Thus the diagrams of the one-particle irreducible three-point function follow from those of the electron self-energy
by inserting a three-vertex in an electron line in all possible ways.
Note that the graphical content (\r{SDI3}) of the functional integrodifferential equation
(\r{TRHEEG}) can be heuristically deduced from the local current conservation law of QED and
its corresponding Ward identity (see, for example, the detailed discussion in Ref.~\ci{Drell}).\\

Another way to determine the one-particle irreducible three-point function follows from inserting the Schwinger-Dyson
equation for the electron self-energy (\r{CSD4}) into (\r{TRHEEG}). By doing so, we take into account
\beq
\int_{45} \fulls_{14} \fulls_{52} \tau_{453} = \int_{4567} S_{64} S_{57} \frac{\delta \fulls_{12}}{\delta S_{67}} V_{453} \, ,
\eeq
which is derived from (\r{NNNN}), (\r{NNNNN}), and the functional chain rule (\r{FCR}). Thus we otain as an alternative
functional integrodifferential equation for the one-particle irreducible three-point function:
\beq
\tau_{123} &= &V_{123} + \int_{456789} V_{145} \tau_{627} V_{893} \fulls_{48} \fulls_{69} \fulld_{57}
+ \int_{456789\bar{1}\bar{2}} V_{145} \tau_{627} V_{893} \fulls_{46} S_{\bar{1}8} S_{9\bar{2}} 
\frac{\delta \fulld_{57}}{\delta S_{\bar{1}\bar{2}}} \no \\
&& + \int_{456789\bar{1}\bar{2}} V_{145} \frac{\delta \tau_{627}}{\delta S_{\bar{1}\bar{2}}} V_{893} \fulld_{57}
\fulls_{46} S_{\bar{1}8} S_{9\bar{2}} \, .
\eeq
Its graphical representation reads
\beq
\parbox{10mm}{\centerline{
\begin{fmfgraph*}(6.93,6)
\setval
\fmfforce{1w,0h}{v1}
\fmfforce{0w,0h}{v2}
\fmfforce{0.5w,1h}{v3}
\fmfforce{0.5w,3/6h}{v4}
\fmfforce{0.5w,2.5/6h}{v5}
\fmfforce{0.5w,1.5/6h}{v6}
\fmfforce{0.5w,1/6h}{v7}
\fmfforce{0.333w,1.5/6h}{v8}
\fmfforce{0.667w,1.5/6h}{v9}
\fmf{photon}{v3,v4}
\fmf{fermion}{v8,v2}
\fmf{fermion}{v1,v9}
\fmf{double,left=1,width=0.2mm}{v4,v7,v4}
\fmfv{decor.size=0, label=${\scs 1}$, l.dist=1mm, l.angle=-150}{v2}
\fmfv{decor.size=0, label=${\scs 2}$, l.dist=1mm, l.angle=-30}{v1}
\fmfv{decor.size=0, label=${\scs 3}$, l.dist=1mm, l.angle=90}{v3}
\end{fmfgraph*} }}
\quad = \quad 
%
%
\parbox{8mm}{\centerline{
\begin{fmfgraph*}(5,4.33)
\setval
\fmfforce{1w,0h}{v1}
\fmfforce{0w,0h}{v2}
\fmfforce{0.5w,1h}{v3}
\fmfforce{0.5w,0.2886h}{vm}
\fmf{fermion}{v1,vm}
\fmf{fermion}{vm,v2}
\fmf{photon}{v3,vm}
\fmfv{decor.size=0,label=${\scs 2}$,l.dist=1mm}{v1}
\fmfv{decor.size=0,label=${\scs 1}$,l.dist=1mm}{v2}
\fmfv{decor.size=0,label=${\scs 3}$,l.dist=1mm}{v3}
\fmfdot{vm}
\end{fmfgraph*} }}
%
%
\hspace*{0.3cm} + \hspace*{0.1cm}
\parbox{23mm}{\centerline{
\begin{fmfgraph*}(20,8)
\setval
\fmfforce{0/20w,0.5h}{v1}
\fmfforce{3/20w,0.5h}{v2}
\fmfforce{8/20w,0.5h}{v3a}
\fmfforce{10/20w,0.5h}{v3b}
\fmfforce{9/20w,5/8h}{v3c}
\fmfforce{15/20w,0.5h}{v4a}
\fmfforce{17/20w,0.5h}{v4b}
\fmfforce{16/20w,3/8h}{v4c}
\fmfforce{1w,0.5086h}{v5}
\fmfforce{9/20w,8/8h}{v6}
\fmf{double,width=0.2mm,left=1}{v3a,v3b,v3a}
\fmf{double,width=0.2mm,left=1}{v4a,v4b,v4a}
\fmf{fermion}{v2,v1}
\fmf{heavy,width=0.2mm}{v3a,v2}
\fmf{heavy,width=0.2mm}{v4a,v3b}
\fmf{fermion}{v5,v4b}
\fmf{boson}{v3c,v6}
\fmf{dbl_wiggly,width=0.2mm,right=0.4}{v2,v4c}
\fmfdot{v2}
\fmfv{decor.size=0,label=${\scs 2}$,l.dist=0.5mm}{v5}
\fmfv{decor.size=0,label=${\scs 1}$,l.dist=0.5mm}{v1}
\fmfv{decor.size=0,label=${\scs 3}$,l.dist=0.5mm}{v6}
\end{fmfgraph*}
}}
%
%
\hspace*{0.3cm} + \hspace*{0.1cm}
\parbox{11mm}{\centerline{
\begin{fmfgraph*}(8,7)
\setval
\fmfforce{0w,1/7h}{v1}
\fmfforce{3/8w,1/7h}{v2a}
\fmfforce{5/8w,1/7h}{v2b}
\fmfforce{4/8w,2/7h}{v2c}
\fmfforce{1w,1/7h}{v3}
\fmfforce{1/8w,1h}{v4}
\fmfforce{4/8w,1h}{v5}
\fmfforce{1w,1h}{v6}
\fmf{double,width=0.2mm,left=1}{v2a,v2b,v2a}
\fmf{fermion}{v1,v2a}
\fmf{heavy,width=0.2mm}{v2c,v5}
\fmf{fermion}{v4,v5}
\fmf{boson}{v2b,v3}
\fmf{boson}{v5,v6}
\fmfdot{v5}
\fmfv{decor.size=0,label=${\scs 2}$,l.dist=0.5mm, l.angle=-180}{v1}
\fmfv{decor.size=0,label=${\scs 1}$,l.dist=0.5mm, l.angle=-180}{v4}
\fmfv{decor.size=0,label=${\scs 4}$,l.dist=0.5mm, l.angle=0}{v6}
\fmfv{decor.size=0,label=${\scs 5}$,l.dist=0.5mm, l.angle=0}{v3}
\end{fmfgraph*}
}}
\hspace*{0.2cm}
%
%
\dephi{
\,\parbox{12mm}{\centerline{
\begin{fmfgraph*}(7,3)
\setval
\fmfleft{v1}
\fmfright{v2}
\fmf{dbl_wiggly,width=0.2mm}{v2,v1}
\fmfv{decor.size=0, label=${\scs 4}$, l.dist=1mm, l.angle=-180}{v1}
\fmfv{decor.size=0, label=${\scs 5}$, l.dist=1mm, l.angle=0}{v2}
\end{fmfgraph*}}}  
}{6}{7}
\hspace*{0.2cm}
%
%
\parbox{10.5mm}{\begin{center}
\begin{fmfgraph*}(5.5,8.66)
\setval
\fmfforce{0w,0h}{v1}
\fmfforce{0w,1h}{v2}
\fmfforce{2.5/5.5w,1/2h}{v3}
\fmfforce{1w,1/2h}{v4}
\fmf{fermion}{v1,v3}
\fmf{fermion}{v3,v2}
\fmf{boson}{v3,v4}
\fmfdot{v3}
\fmfv{decor.size=0, label=${\scs 4}$, l.dist=1mm, l.angle=180}{v2}
\fmfv{decor.size=0, label=${\scs 5}$, l.dist=1mm, l.angle=-180}{v1}
\fmfv{decor.size=0, label=${\scs 3}$, l.dist=1mm, l.angle=0}{v4}
\end{fmfgraph*}
\end{center}}
%
%
\hspace*{0.3cm} + \hspace*{0.1cm}
\parbox{10.5mm}{\begin{center}
\begin{fmfgraph*}(5.5,8.66)
\setval
\fmfforce{0w,1/2h}{v1}
\fmfforce{3/5.5w,1/2h}{v2}
\fmfforce{1w,0h}{v3}
\fmfforce{1w,1h}{v4}
\fmf{fermion}{v2,v1}
\fmf{heavy,width=0.2mm}{v4,v2}
\fmf{dbl_wiggly,width=0.2mm}{v3,v2}
\fmfdot{v2}
\fmfv{decor.size=0, label=${\scs 1}$, l.dist=1mm, l.angle=-180}{v1}
\fmfv{decor.size=0, label=${\scs 4}$, l.dist=1mm, l.angle=0}{v4}
\fmfv{decor.size=0, label=${\scs 5}$, l.dist=1mm, l.angle=0}{v3}
\end{fmfgraph*}
\end{center}}
\hspace*{0.2cm}
%
%
\dephi{
\,\parbox{12mm}{\centerline{
\begin{fmfgraph*}(6.93,6)
\setval
\fmfforce{1w,2.5/6h}{v1}
\fmfforce{0w,2.5/6h}{v2}
\fmfforce{0.5w,8.5/6h}{v3}
\fmfforce{0.5w,5.5/6h}{v4}
\fmfforce{0.5w,5/6h}{v5}
\fmfforce{0.5w,4/6h}{v6}
\fmfforce{0.5w,3.5/6h}{v7}
\fmfforce{0.333w,4/6h}{v8}
\fmfforce{0.667w,4/6h}{v9}
\fmf{photon}{v3,v4}
\fmf{fermion}{v8,v2}
\fmf{fermion}{v1,v9}
\fmf{double,left=1,width=0.2mm}{v4,v7,v4}
\fmfv{decor.size=0, label=${\scs 4}$, l.dist=1mm, l.angle=-150}{v2}
\fmfv{decor.size=0, label=${\scs 6}$, l.dist=1mm, l.angle=-30}{v1}
\fmfv{decor.size=0, label=${\scs 5}$, l.dist=1mm, l.angle=90}{v3}
\end{fmfgraph*} }}
}{7}{8}
\hspace*{0.2cm}
%
%
\parbox{10.5mm}{\begin{center}
\begin{fmfgraph*}(5.5,8.66)
\setval
\fmfforce{0w,0h}{v1}
\fmfforce{0w,1h}{v2}
\fmfforce{2.5/5.5w,1/2h}{v3}
\fmfforce{1w,1/2h}{v4}
\fmf{fermion}{v1,v3}
\fmf{fermion}{v3,v2}
\fmf{boson}{v3,v4}
\fmfdot{v3}
\fmfv{decor.size=0, label=${\scs 7}$, l.dist=1mm, l.angle=180}{v2}
\fmfv{decor.size=0, label=${\scs 8}$, l.dist=1mm, l.angle=-180}{v1}
\fmfv{decor.size=0, label=${\scs 3}$, l.dist=1mm, l.angle=0}{v4}
\end{fmfgraph*}
\end{center}}
\,\, . \la{WAA}
\eeq
Thus the closed set of Schwinger-Dyson equations is given by (\r{SDI1}), (\r{SDI2}), and (\r{SDI3}) or (\r{WAA})
which have to be supplemented by the Dyson equations (\r{GE23}), (\ref{GE24}).
\end{fmffile}
\begin{fmffile}{sd13}
\subsection{Graphical Recursion Relations}
\label{GRR22}
We now demonstrate how the diagrams of the connected electron and 
photon two-point function, the electron and photon self-energy as well as the one-particle irreducible 
three-point function are recursively determined in a graphical way, generating all one-particle 
irreducible Feynman diagrams which are needed for the renormalization of QED. To simplify
the discussion, we restrict ourselves to a vanishing field
expectation value, so that we can neglect the last term in Eq. (\r{SDI1}).
Performing a loop expansion of the connected electron and photon 
two-point function
\beq
\label{L3}
\parbox{20mm}{\centerline{
\begin{fmfgraph*}(7,3)
\setval
\fmfleft{v1}
\fmfright{v2}
\fmf{heavy,width=0.2mm}{v2,v1}
\fmfv{decor.size=0, label=${\scs 1}$, l.dist=1mm, l.angle=-180}{v1}
\fmfv{decor.size=0, label=${\scs 2}$, l.dist=1mm, l.angle=0}{v2}
\end{fmfgraph*}}}  
& = & \quad \sum_{l=0}^{\infty}
\parbox{15mm}{\centerline{
\begin{fmfgraph*}(7,3)
\setval
\fmfleft{v1}
\fmfright{v2}
\fmfforce{0.5w,2/3h}{v3}
\fmf{heavy,width=0.2mm}{v2,v1}
\fmfv{decor.size=0, label=${\scs 1}$, l.dist=1mm, l.angle=-180}{v1}
\fmfv{decor.size=0, label=${\scs 2}$, l.dist=1mm, l.angle=0}{v2}
\fmfv{decor.size=0, label=${\scs (l)}$, l.dist=1mm, l.angle=90}{v3}
\end{fmfgraph*}}}  \hspace*{0.4cm} , \\
\la{L4}
\parbox{20mm}{\centerline{
\begin{fmfgraph*}(7,3)
\setval
\fmfleft{v1}
\fmfright{v2}
\fmf{dbl_wiggly,width=0.2mm}{v2,v1}
\fmfv{decor.size=0, label=${\scs 1}$, l.dist=1mm, l.angle=-180}{v1}
\fmfv{decor.size=0, label=${\scs 2}$, l.dist=1mm, l.angle=0}{v2}
\end{fmfgraph*}}}  
& = &\quad
\sum_{l=0}^{\infty}
\parbox{15mm}{\centerline{
\begin{fmfgraph*}(7,3)
\setval
\fmfleft{v1}
\fmfright{v2}
\fmfforce{0.5w,2/3h}{v3}
\fmf{dbl_wiggly,width=0.2mm}{v2,v1}
\fmfv{decor.size=0, label=${\scs 1}$, l.dist=1mm, l.angle=-180}{v1}
\fmfv{decor.size=0, label=${\scs 2}$, l.dist=1mm, l.angle=0}{v2}
\fmfv{decor.size=0, label=${\scs (l)}$, l.dist=1mm, l.angle=90}{v3}
\end{fmfgraph*}}}  
\hspace*{0.4cm} ,
\eeq
of their corresponding self-energies
\beq
\parbox{11mm}{\centerline{
\begin{fmfgraph*}(8,5)
\setval
\fmfforce{0w,0.5h}{v1}
\fmfforce{3/8w,0.5h}{v2}
\fmfforce{5/8w,0.5h}{v3}
\fmfforce{5.2/8w,0.5h}{v3b}
\fmfforce{1w,0.5h}{v4}
\fmf{fermion}{v2,v1}
\fmf{fermion}{v4,v3b}
\fmf{double,width=0.2mm,left=1}{v2,v3,v2}
\fmfv{decor.size=0, label=${\scs 1}$, l.dist=1mm, l.angle=-180}{v1}
\fmfv{decor.size=0, label=${\scs 2}$, l.dist=1mm, l.angle=0}{v4}
\end{fmfgraph*} }}
\quad & = & \quad \sum_{l=1}^{\infty} 
\parbox{18mm}{\centerline{
\begin{fmfgraph*}(11,5)
\setval
\fmfforce{0w,1/2h}{v1}
\fmfforce{3/11w,1/2h}{v2}
\fmfforce{8/11w,1/2h}{v3}
\fmfforce{8.2/11w,1/2h}{v3b}
\fmfforce{1w,1/2h}{v4}
\fmfforce{1/2w,1/2h}{v5}
\fmf{electron}{v2,v1}
\fmf{electron}{v4,v3b}
\fmf{double,width=0.2mm,left=1}{v2,v3,v2}
\fmfv{decor.size=0, label=${\scs l}$, l.dist=0mm, l.angle=0}{v5}
\fmfv{decor.size=0, label=${\scs 1}$, l.dist=1mm, l.angle=-180}{v1}
\fmfv{decor.size=0, label=${\scs 2}$, l.dist=1mm, l.angle=0}{v4}
\end{fmfgraph*} }}
\hspace*{0.4cm} , \\
\parbox{11mm}{\centerline{
\begin{fmfgraph*}(8,5)
\setval
\fmfforce{0w,0.5h}{v1}
\fmfforce{3/8w,0.5h}{v2}
\fmfforce{5/8w,0.5h}{v3}
\fmfforce{5.2/8w,0.5h}{v3b}
\fmfforce{1w,0.5h}{v4}
\fmf{boson}{v2,v1}
\fmf{boson}{v4,v3b}
\fmf{double,width=0.2mm,left=1}{v2,v3,v2}
\fmfv{decor.size=0, label=${\scs 1}$, l.dist=1mm, l.angle=-180}{v1}
\fmfv{decor.size=0, label=${\scs 2}$, l.dist=1mm, l.angle=0}{v4}
\end{fmfgraph*} }}
\quad & = & \quad \sum_{l=1}^{\infty} 
\parbox{18mm}{\centerline{
\begin{fmfgraph*}(11,5)
\setval
\fmfforce{0w,1/2h}{v1}
\fmfforce{3/11w,1/2h}{v2}
\fmfforce{8/11w,1/2h}{v3}
\fmfforce{8.2/11w,1/2h}{v3b}
\fmfforce{1w,1/2h}{v4}
\fmfforce{1/2w,1/2h}{v5}
\fmf{boson}{v2,v1}
\fmf{boson}{v4,v3b}
\fmf{double,width=0.2mm,left=1}{v2,v3,v2}
\fmfv{decor.size=0, label=${\scs l}$, l.dist=0mm, l.angle=0}{v5}
\fmfv{decor.size=0, label=${\scs 1}$, l.dist=1mm, l.angle=-180}{v1}
\fmfv{decor.size=0, label=${\scs 2}$, l.dist=1mm, l.angle=0}{v4}
\end{fmfgraph*} }}
\hspace*{0.4cm}, 
\eeq
as well as of the one-particle irreducible three-point function
\beq
\parbox{10mm}{\centerline{
\begin{fmfgraph*}(6.93,6)
\setval
\fmfforce{1w,0h}{v1}
\fmfforce{0w,0h}{v2}
\fmfforce{0.5w,1h}{v3}
\fmfforce{0.5w,3/6h}{v4}
\fmfforce{0.5w,2.5/6h}{v5}
\fmfforce{0.5w,1.5/6h}{v6}
\fmfforce{0.5w,1/6h}{v7}
\fmfforce{0.333w,1.5/6h}{v8}
\fmfforce{0.667w,1.5/6h}{v9}
\fmf{photon}{v3,v4}
\fmf{fermion}{v8,v2}
\fmf{fermion}{v1,v9}
\fmf{double,left=1,width=0.2mm}{v4,v7,v4}
\fmfv{decor.size=0, label=${\scs 1}$, l.dist=1mm, l.angle=-150}{v2}
\fmfv{decor.size=0, label=${\scs 2}$, l.dist=1mm, l.angle=-30}{v1}
\fmfv{decor.size=0, label=${\scs 3}$, l.dist=1mm, l.angle=90}{v3}
\end{fmfgraph*} }}
\quad = \quad \sum_{l=0}^{\infty}
\parbox{17mm}{\centerline{
\begin{fmfgraph*}(10,8.66)
\setval
\fmfforce{1w,0h}{v1}
\fmfforce{0w,0h}{v2}
\fmfforce{0.5w,0.95h}{v3}
\fmfforce{0.25w,0.18h}{v4}
\fmfforce{0.75w,0.18h}{v5}
\fmfforce{0.5w,0.6h}{v6}
\fmfforce{0.5w,-0.0114h}{vm1}
\fmfforce{0.5w,0.2886h}{vm}
\fmfforce{0.5w,0.5886h}{vm2}
\fmf{fermion}{v4,v2}
\fmf{fermion}{v1,v5}
\fmf{photon}{v3,v6}
\fmf{double,width=0.2mm,left=1}{vm1,vm2,vm1}
\fmfv{decor.size=0,label=${\scs l}$, l.dist=0mm, l.angle=0}{vm}
\fmfv{decor.size=0,label=${\scs 2}$,l.dist=0.5mm}{v1}
\fmfv{decor.size=0,label=${\scs 1}$,l.dist=0.5mm}{v2}
\fmfv{decor.size=0,label=${\scs 3}$,l.dist=0.5mm}{v3}
\end{fmfgraph*}
}} \hspace*{0.4cm} ,
\eeq
we obtain from (\r{SE2}), (\ref{SP2})
(\r{SDI1}), (\r{SDI2}), and (\r{SDI3}) or (\r{WAA}) 
the following closed set of graphical recursion relations:
\beq
\parbox{15mm}{\centerline{
\begin{fmfgraph*}(7,3)
\setval
\fmfleft{v1}
\fmfright{v2}
\fmfforce{0.5w,2/3h}{v3}
\fmf{heavy,width=0.2mm}{v2,v1}
\fmfv{decor.size=0, label=${\scs 1}$, l.dist=1mm, l.angle=-180}{v1}
\fmfv{decor.size=0, label=${\scs 2}$, l.dist=1mm, l.angle=0}{v2}
\fmfv{decor.size=0, label=${\scs (l)}$, l.dist=1mm, l.angle=90}{v3}
\end{fmfgraph*}}}  
\quad & = & \quad 
\sum_{k=1}^l \hspace*{0.2cm}
\parbox{18mm}{\centerline{
\begin{fmfgraph*}(15,5)
\setval
\fmfforce{0w,1/2h}{v1}
\fmfforce{5/15w,1/2h}{v2}
\fmfforce{10/15w,1/2h}{v3}
\fmfforce{10.2/15w,1/2h}{v3b}
\fmfforce{1w,1/2h}{v4}
\fmfforce{7.5/15w,1/2h}{v5}
\fmfforce{13/15w,0.9h}{v6}
\fmf{fermion}{v2,v1}
\fmf{heavy,width=0.2mm}{v4,v3b}
\fmf{double,width=0.2mm,left=1}{v2,v3,v2}
\fmfv{decor.size=0, label=${\scs (l-k)}$, l.dist=0.5mm, l.angle=90}{v6}
\fmfv{decor.size=0, label=${\scs k}$, l.dist=0mm, l.angle=0}{v5}
\fmfv{decor.size=0, label=${\scs 1}$, l.dist=1mm, l.angle=-180}{v1}
\fmfv{decor.size=0, label=${\scs 2}$, l.dist=1mm, l.angle=0}{v4}
\end{fmfgraph*} }}
\hspace*{0.4cm} , \la{R1}
\\
\parbox{15mm}{\centerline{
\begin{fmfgraph*}(7,3)
\setval
\fmfleft{v1}
\fmfright{v2}
\fmfforce{0.5w,2/3h}{v3}
\fmf{dbl_wiggly,width=0.2mm}{v2,v1}
\fmfv{decor.size=0, label=${\scs 1}$, l.dist=1mm, l.angle=-180}{v1}
\fmfv{decor.size=0, label=${\scs 2}$, l.dist=1mm, l.angle=0}{v2}
\fmfv{decor.size=0, label=${\scs (l)}$, l.dist=1mm, l.angle=90}{v3}
\end{fmfgraph*}}} 
\quad & = &\quad  
\sum_{k=1}^l \hspace*{0.2cm}
\parbox{18mm}{\centerline{
\begin{fmfgraph*}(15,5)
\setval
\fmfforce{0w,1/2h}{v1}
\fmfforce{5/15w,1/2h}{v2}
\fmfforce{10/15w,1/2h}{v3}
\fmfforce{10.2/15w,1/2h}{v3b}
\fmfforce{1w,1/2h}{v4}
\fmfforce{7.5/15w,1/2h}{v5}
\fmfforce{13/15w,0.9h}{v6}
\fmf{boson}{v2,v1}
\fmf{dbl_wiggly,width=0.2mm}{v4,v3b}
\fmf{double,width=0.2mm,left=1}{v2,v3,v2}
\fmfv{decor.size=0, label=${\scs (l-k)}$, l.dist=0.5mm, l.angle=90}{v6}
\fmfv{decor.size=0, label=${\scs k}$, l.dist=0mm, l.angle=0}{v5}
\fmfv{decor.size=0, label=${\scs 1}$, l.dist=1mm, l.angle=-180}{v1}
\fmfv{decor.size=0, label=${\scs 2}$, l.dist=1mm, l.angle=0}{v4}
\end{fmfgraph*} }}
\hspace*{0.4cm} , \la{R2} \\
\la{R3}
\parbox{18mm}{\centerline{
\begin{fmfgraph*}(11,5)
\setval
\fmfforce{0w,1/2h}{v1}
\fmfforce{3/11w,1/2h}{v2}
\fmfforce{8/11w,1/2h}{v3}
\fmfforce{8.2/11w,1/2h}{v3b}
\fmfforce{1w,1/2h}{v4}
\fmfforce{1/2w,1/2h}{v5}
\fmf{electron}{v2,v1}
\fmf{electron}{v4,v3b}
\fmf{double,width=0.2mm,left=1}{v2,v3,v2}
\fmfv{decor.size=0, label=${\scs l}$, l.dist=0mm, l.angle=0}{v5}
\fmfv{decor.size=0, label=${\scs 1}$, l.dist=1mm, l.angle=-180}{v1}
\fmfv{decor.size=0, label=${\scs 2}$, l.dist=1mm, l.angle=0}{v4}
\end{fmfgraph*} }}
\quad & = & \quad \sum_{k_1=0}^{l} \sum_{k_2=0}^{l-k_1-1} 
\parbox{20mm}{\begin{center}
\begin{fmfgraph*}(17,6)
\setval
\fmfforce{0w,0.3h}{v1}
\fmfforce{3/17w,0.3h}{v2}
\fmfforce{8/17w,0.3h}{v3}
\fmfforce{11/17w,-0.2h}{v4}
\fmfforce{11/17w,0.3h}{v5}
\fmfforce{11/17w,0.8h}{v6}
\fmfforce{1w,0.3h}{v7}
\fmfforce{10/17w,0.77h}{v8}
\fmfforce{14/17w,0.3h}{v9}
\fmfforce{14.2/17w,0.3h}{v9b}
\fmfforce{5.5/17w,0.3h}{v10}
\fmfforce{1/2w,1.2h}{v11}
\fmf{heavy,width=0.2mm}{v3,v2}
\fmf{fermion}{v2,v1}
\fmf{fermion}{v7,v9b}
\fmf{dbl_wiggly,width=0.2mm,left=0.7}{v2,v8}
\fmf{double,width=0.2mm,left=1}{v4,v6,v4}
\fmfdot{v2}
\fmfv{decor.size=0, label=${\scs k_2}$, l.dist=0mm, l.angle=0}{v5}
\fmfv{decor.size=0, label=${\scs 1}$, l.dist=1mm, l.angle=-180}{v1}
\fmfv{decor.size=0, label=${\scs 2}$, l.dist=1mm, l.angle=0}{v7}
\fmfv{decor.size=0, label=${\scs (k_1)}$, l.dist=1mm, l.angle=-90}{v10}
\fmfv{decor.size=0, label=${\scs (l-k_1-k_2-1)}$,l.dist=1mm,l.angle=90}{v11}
\end{fmfgraph*}
\end{center} }
\hspace*{0.4cm} ,\\
\la{R4}
\parbox{18mm}{\centerline{
\begin{fmfgraph*}(11,5)
\setval
\fmfforce{0w,1/2h}{v1}
\fmfforce{3/11w,1/2h}{v2}
\fmfforce{8/11w,1/2h}{v3}
\fmfforce{8.2/11w,1/2h}{v3b}
\fmfforce{1w,1/2h}{v4}
\fmfforce{1/2w,1/2h}{v5}
\fmf{boson}{v2,v1}
\fmf{boson}{v4,v3b}
\fmf{double,width=0.2mm,left=1}{v2,v3,v2}
\fmfv{decor.size=0, label=${\scs l}$, l.dist=0mm, l.angle=0}{v5}
\fmfv{decor.size=0, label=${\scs 1}$, l.dist=1mm, l.angle=-180}{v1}
\fmfv{decor.size=0, label=${\scs 2}$, l.dist=1mm, l.angle=0}{v4}
\end{fmfgraph*} }}
\quad & = & \quad - \,\, \sum_{k_1=0}^{l} \sum_{k_2=0}^{l-k_1-1} 
\parbox{20mm}{\begin{center}
\begin{fmfgraph*}(17,6)
\setval
\fmfforce{0w,0.5h}{v1}
\fmfforce{3/17w,0.5h}{v2}
\fmfforce{8/17w,0.5h}{v3}
\fmfforce{11/17w,0h}{v4}
\fmfforce{11/17w,1/2h}{v5}
\fmfforce{11/17w,1h}{v6}
\fmfforce{1w,0.5h}{v7}
\fmfforce{10/17w,0.97h}{v8}
\fmfforce{14/17w,1/2h}{v9}
\fmfforce{14.2/17w,1/2h}{v9b}
\fmfforce{10/17w,0.03h}{v10}
\fmfforce{7/17w,-0.3h}{v11}
\fmfforce{7/17w,1.3h}{v12}
\fmf{boson}{v2,v1}
\fmf{boson}{v7,v9b}
\fmf{heavy,width=0.2mm,right=0.7}{v8,v2}
\fmf{heavy,width=0.2mm,right=0.7}{v2,v10}
\fmf{double,width=0.2mm,left=1}{v4,v6,v4}
\fmfdot{v2}
\fmfv{decor.size=0, label=${\scs k_2}$, l.dist=0mm, l.angle=0}{v5}
\fmfv{decor.size=0, label=${\scs 1}$, l.dist=1mm, l.angle=-180}{v1}
\fmfv{decor.size=0, label=${\scs 2}$, l.dist=1mm, l.angle=0}{v7}
\fmfv{decor.size=0, label=${\scs (k_1)}$, l.dist=1mm, l.angle=-90}{v11}
\fmfv{decor.size=0, label=${\scs (l-k_1-k_2-1)}$, l.dist=1mm, l.angle=90}{v12}
\end{fmfgraph*}
\end{center} }
\hspace*{0.4cm},
\eeq
and
\beq
\la{R5}
\parbox{17mm}{\centerline{
\begin{fmfgraph*}(10,8.66)
\setval
\fmfforce{1w,0h}{v1}
\fmfforce{0w,0h}{v2}
\fmfforce{0.5w,0.95h}{v3}
\fmfforce{0.25w,0.18h}{v4}
\fmfforce{0.75w,0.18h}{v5}
\fmfforce{0.5w,0.6h}{v6}
\fmfforce{0.5w,-0.0114h}{vm1}
\fmfforce{0.5w,0.2886h}{vm}
\fmfforce{0.5w,0.5886h}{vm2}
\fmf{fermion}{v4,v2}
\fmf{fermion}{v1,v5}
\fmf{photon}{v3,v6}
\fmf{double,width=0.2mm,left=1}{vm1,vm2,vm1}
\fmfv{decor.size=0,label=${\scs l}$, l.dist=0mm, l.angle=0}{vm}
\fmfv{decor.size=0,label=${\scs 2}$,l.dist=0.5mm}{v1}
\fmfv{decor.size=0,label=${\scs 1}$,l.dist=0.5mm}{v2}
\fmfv{decor.size=0,label=${\scs 3}$,l.dist=0.5mm}{v3}
\end{fmfgraph*}
}}
\quad & = & \quad
\dephi{
\parbox{17mm}{\centerline{
\begin{fmfgraph*}(11,7)
\setval
\fmfforce{0w,1/2h}{v1}
\fmfforce{3/11w,1/2h}{v2}
\fmfforce{8/11w,1/2h}{v3}
\fmfforce{8.2/11w,1/2h}{v3b}
\fmfforce{1w,1/2h}{v4}
\fmfforce{1/2w,1/2h}{v5}
\fmf{electron}{v2,v1}
\fmf{electron}{v4,v3b}
\fmf{double,width=0.2mm,left=1}{v2,v3,v2}
\fmfv{decor.size=0, label=${\scs l}$, l.dist=0mm, l.angle=0}{v5}
\fmfv{decor.size=0, label=${\scs 1}$, l.dist=1mm, l.angle=-180}{v1}
\fmfv{decor.size=0, label=${\scs 2}$, l.dist=1mm, l.angle=0}{v4}
\end{fmfgraph*} }}
}{4}{5}
\hspace*{0.3cm}
\parbox{10.5mm}{\begin{center}
\begin{fmfgraph*}(5.5,8.66)
\setval
\fmfforce{0w,0h}{v1}
\fmfforce{0w,1h}{v2}
\fmfforce{2.5/5.5w,1/2h}{v3}
\fmfforce{1w,1/2h}{v4}
\fmf{fermion}{v1,v3}
\fmf{fermion}{v3,v2}
\fmf{boson}{v3,v4}
\fmfdot{v3}
\fmfv{decor.size=0, label=${\scs 4}$, l.dist=1mm, l.angle=180}{v2}
\fmfv{decor.size=0, label=${\scs 5}$, l.dist=1mm, l.angle=-180}{v1}
\fmfv{decor.size=0, label=${\scs 3}$, l.dist=1mm, l.angle=0}{v4}
\end{fmfgraph*}
\end{center}}
\eeq
or
\beq
\eeq
These are solved starting from the zeroth-loop contribution to the connected electron
and photon two-point function
\beq
\la{SST1}
\parbox{15mm}{\centerline{
\begin{fmfgraph*}(7,3)
\setval
\fmfleft{v1}
\fmfright{v2}
\fmfforce{0.5w,2/3h}{v3}
\fmf{heavy,width=0.2mm}{v2,v1}
\fmfv{decor.size=0, label=${\scs 1}$, l.dist=1mm, l.angle=-180}{v1}
\fmfv{decor.size=0, label=${\scs 2}$, l.dist=1mm, l.angle=0}{v2}
\fmfv{decor.size=0, label=${\scs (0)}$, l.dist=1mm, l.angle=90}{v3}
\end{fmfgraph*}}}
\quad & = &  
\parbox{20mm}{\centerline{
\begin{fmfgraph*}(7,3)
\setval
\fmfleft{v1}
\fmfright{v2}
\fmf{fermion}{v2,v1}
\fmfv{decor.size=0, label=${\scs 1}$, l.dist=1mm, l.angle=-180}{v1}
\fmfv{decor.size=0, label=${\scs 2}$, l.dist=1mm, l.angle=0}{v2}
\end{fmfgraph*}}}
\hspace*{0.4cm} , \\
\la{SST2}
\parbox{15mm}{\centerline{
\begin{fmfgraph*}(7,3)
\setval
\fmfleft{v1}
\fmfright{v2}
\fmfforce{0.5w,2/3h}{v3}
\fmf{dbl_wiggly,width=0.2mm}{v2,v1}
\fmfv{decor.size=0, label=${\scs 1}$, l.dist=1mm, l.angle=-180}{v1}
\fmfv{decor.size=0, label=${\scs 2}$, l.dist=1mm, l.angle=0}{v2}
\fmfv{decor.size=0, label=${\scs (0)}$, l.dist=1mm, l.angle=90}{v3}
\end{fmfgraph*}}}
\quad & = &
\parbox{20mm}{\centerline{
\begin{fmfgraph*}(7,3)
\setval
\fmfleft{v1}
\fmfright{v2}
\fmf{boson}{v1,v2}
\fmfv{decor.size=0, label=${\scs 1}$, l.dist=1mm, l.angle=-180}{v1}
\fmfv{decor.size=0, label=${\scs 2}$, l.dist=1mm, l.angle=0}{v2}
\end{fmfgraph*}}}
\hspace*{0.4cm} , 
\eeq
and the one-particle irreducible three-point function
\beq
\la{SST3}
\parbox{17mm}{\centerline{
\begin{fmfgraph*}(10,8.66)
\setval
\fmfforce{1w,0h}{v1}
\fmfforce{0w,0h}{v2}
\fmfforce{0.5w,0.95h}{v3}
\fmfforce{0.25w,0.18h}{v4}
\fmfforce{0.75w,0.18h}{v5}
\fmfforce{0.5w,0.6h}{v6}
\fmfforce{0.5w,-0.0114h}{vm1}
\fmfforce{0.5w,0.2886h}{vm}
\fmfforce{0.5w,0.5886h}{vm2}
\fmf{fermion}{v4,v2}
\fmf{fermion}{v1,v5}
\fmf{photon}{v3,v6}
\fmf{double,width=0.2mm,left=1}{vm1,vm2,vm1}
\fmfv{decor.size=0,label=${\scs 0}$, l.dist=0mm, l.angle=0}{vm}
\fmfv{decor.size=0,label=${\scs 2}$,l.dist=0.5mm}{v1}
\fmfv{decor.size=0,label=${\scs 1}$,l.dist=0.5mm}{v2}
\fmfv{decor.size=0,label=${\scs 3}$,l.dist=0.5mm}{v3}
\end{fmfgraph*}
}}
\quad &= &
\parbox{8mm}{\centerline{
\begin{fmfgraph*}(5,8.33)
\setval
\fmfforce{1w,2/8.33h}{v1}
\fmfforce{0w,2/8.33h}{v2}
\fmfforce{0.5w,6.33/8.33h}{v3}
\fmfforce{0.5w,3.24/8.33h}{vm}
\fmf{fermion}{v1,vm}
\fmf{fermion}{vm,v2}
\fmf{photon}{v3,vm}
\fmfv{decor.size=0,label=${\scs 2}$,l.dist=1mm}{v1}
\fmfv{decor.size=0,label=${\scs 1}$,l.dist=1mm}{v2}
\fmfv{decor.size=0,label=${\scs 3}$,l.dist=1mm}{v3}
\fmfdot{vm}
\end{fmfgraph*} }}
\hspace*{0.4cm}.
\\ && \no 
\eeq
\end{fmffile}
\begin{fmffile}{sd14}
\hspace*{-0.4cm}
First, we insert (\r{SST1})--(\r{SST3}) into (\r{R3}) and (\r{R4})
to obtain the one-loop contribution to the electron and the photon self-energy:
\beq
\la{SELF1}
\parbox{18mm}{\centerline{
\begin{fmfgraph*}(11,5)
\setval
\fmfforce{0w,1/2h}{v1}
\fmfforce{3/11w,1/2h}{v2}
\fmfforce{8/11w,1/2h}{v3}
\fmfforce{8.2/11w,1/2h}{v3b}
\fmfforce{1w,1/2h}{v4}
\fmfforce{1/2w,1/2h}{v5}
\fmf{electron}{v2,v1}
\fmf{electron}{v4,v3b}
\fmf{double,width=0.2mm,left=1}{v2,v3,v2}
\fmfv{decor.size=0, label=${\scs 1}$, l.dist=0mm, l.angle=0}{v5}
\fmfv{decor.size=0, label=${\scs 1}$, l.dist=1mm, l.angle=-180}{v1}
\fmfv{decor.size=0, label=${\scs 2}$, l.dist=1mm, l.angle=0}{v4}
\end{fmfgraph*} }}
\quad & = & \quad 
\parbox{14mm}{\begin{center}
\begin{fmfgraph*}(11,5)
\setval
\fmfforce{0w,0.5h}{v1}
\fmfforce{3/11w,0.5h}{v2}
\fmfforce{8/11w,0.5h}{v3}
\fmfforce{1w,0.5h}{v4}
\fmf{fermion}{v4,v3,v2,v1}
\fmf{boson,left=1}{v2,v3}
\fmfdot{v2,v3}
\fmfv{decor.size=0, label=${\scs 1}$, l.dist=1mm, l.angle=-180}{v1}
\fmfv{decor.size=0, label=${\scs 2}$, l.dist=1mm, l.angle=0}{v4}
\end{fmfgraph*}
\end{center} }
\hspace*{0.4cm} ,  \\
\parbox{18mm}{\centerline{
\begin{fmfgraph*}(11,5)
\setval
\fmfforce{0w,1/2h}{v1}
\fmfforce{3/11w,1/2h}{v2}
\fmfforce{8/11w,1/2h}{v3}
\fmfforce{8.2/11w,1/2h}{v3b}
\fmfforce{1w,1/2h}{v4}
\fmfforce{1/2w,1/2h}{v5}
\fmf{boson}{v2,v1}
\fmf{boson}{v4,v3b}
\fmf{double,width=0.2mm,left=1}{v2,v3,v2}
\fmfv{decor.size=0, label=${\scs 1}$, l.dist=0mm, l.angle=0}{v5}
\fmfv{decor.size=0, label=${\scs 1}$, l.dist=1mm, l.angle=-180}{v1}
\fmfv{decor.size=0, label=${\scs 2}$, l.dist=1mm, l.angle=0}{v4}
\end{fmfgraph*} }}
\quad & = & \quad - \hspace*{0.3cm}
\parbox{14mm}{\begin{center}
\begin{fmfgraph*}(11,5)
\setval
\fmfforce{0w,0.5h}{v1}
\fmfforce{3/11w,0.5h}{v2}
\fmfforce{8/11w,0.5h}{v3}
\fmfforce{1w,0.5h}{v4}
\fmf{boson}{v1,v2}
\fmf{fermion,right}{v2,v3,v2}
\fmf{boson}{v3,v4}
\fmfdot{v2,v3}
\fmfv{decor.size=0, label=${\scs 1}$, l.dist=1mm, l.angle=-180}{v1}
\fmfv{decor.size=0, label=${\scs 2}$, l.dist=1mm, l.angle=0}{v4}
\end{fmfgraph*}
\end{center}} \hspace*{0.4cm} . \la{PPEL1}
\eeq
With this we find from (\r{R1}) and (\r{R2}) the corresponding connected
two-point functions in the one-loop order:
\beq
\parbox{15mm}{\centerline{
\begin{fmfgraph*}(7,3)
\setval
\fmfleft{v1}
\fmfright{v2}
\fmfforce{0.5w,2/3h}{v3}
\fmf{heavy,width=0.2mm}{v2,v1}
\fmfv{decor.size=0, label=${\scs 1}$, l.dist=1mm, l.angle=-180}{v1}
\fmfv{decor.size=0, label=${\scs 2}$, l.dist=1mm, l.angle=0}{v2}
\fmfv{decor.size=0, label=${\scs (1)}$, l.dist=1mm, l.angle=90}{v3}
\end{fmfgraph*}}}
\quad & = & \quad 
\parbox{18mm}{\begin{center}
\begin{fmfgraph*}(15,5)
\setval
\fmfforce{0w,0.5h}{v1}
\fmfforce{1/3w,0.5h}{v2}
\fmfforce{2/3w,0.5h}{v3}
\fmfforce{1w,0.5h}{v4}
\fmf{fermion}{v4,v3,v2,v1}
\fmf{boson,left=1}{v2,v3}
\fmfdot{v2,v3}
\fmfv{decor.size=0, label=${\scs 1}$, l.dist=1mm, l.angle=-180}{v1}
\fmfv{decor.size=0, label=${\scs 2}$, l.dist=1mm, l.angle=0}{v4}
\end{fmfgraph*}
\end{center} }
\hspace*{0.4cm}, \la{SES1} \\
\parbox{15mm}{\centerline{
\begin{fmfgraph*}(7,3)
\setval
\fmfleft{v1}
\fmfright{v2}
\fmfforce{0.5w,2/3h}{v3}
\fmf{dbl_wiggly,width=0.2mm}{v2,v1}
\fmfv{decor.size=0, label=${\scs 1}$, l.dist=1mm, l.angle=-180}{v1}
\fmfv{decor.size=0, label=${\scs 2}$, l.dist=1mm, l.angle=0}{v2}
\fmfv{decor.size=0, label=${\scs (1)}$, l.dist=1mm, l.angle=90}{v3}
\end{fmfgraph*}}}
\quad & = & \quad - \hspace*{0.3cm}
\parbox{18mm}{\begin{center}
\begin{fmfgraph*}(15,5)
\setval
\fmfforce{0w,0.5h}{v1}
\fmfforce{1/3w,0.5h}{v2}
\fmfforce{2/3w,0.5h}{v3}
\fmfforce{1w,0.5h}{v4}
\fmf{boson}{v1,v2}
\fmf{fermion,right}{v2,v3,v2}
\fmf{boson}{v3,v4}
\fmfdot{v2,v3}
\fmfv{decor.size=0, label=${\scs 1}$, l.dist=1mm, l.angle=-180}{v1}
\fmfv{decor.size=0, label=${\scs 2}$, l.dist=1mm, l.angle=0}{v4}
\end{fmfgraph*}
\end{center}} \hspace*{0.4cm} .
\eeq
Amputating one electron line from (\r{SELF1}), 
\beq
\dephi{
\parbox{17mm}{\centerline{
\begin{fmfgraph*}(11,7)
\setval
\fmfforce{0w,1/2h}{v1}
\fmfforce{3/11w,1/2h}{v2}
\fmfforce{8/11w,1/2h}{v3}
\fmfforce{8.2/11w,1/2h}{v3b}
\fmfforce{1w,1/2h}{v4}
\fmfforce{1/2w,1/2h}{v5}
\fmf{electron}{v2,v1}
\fmf{electron}{v4,v3b}
\fmf{double,width=0.2mm,left=1}{v2,v3,v2}
\fmfv{decor.size=0, label=${\scs l}$, l.dist=0mm, l.angle=0}{v5}
\fmfv{decor.size=0, label=${\scs 1}$, l.dist=1mm, l.angle=-180}{v1}
\fmfv{decor.size=0, label=${\scs 2}$, l.dist=1mm, l.angle=0}{v4}
\end{fmfgraph*} }}
}{4}{5}
\quad & = & \quad 
\parbox{12mm}{\begin{center}
\begin{fmfgraph*}(9.243,4.243)
\setval
\fmfforce{0w,0h}{v1}
\fmfforce{0w,1h}{v2}
\fmfforce{2.12/9.243w,1/2h}{v3}
\fmfforce{7.12/9.243w,1/2h}{v4}
\fmfforce{1w,0h}{v5}
\fmfforce{1w,1h}{v6}
\fmf{boson}{v3,v4}
\fmf{fermion}{v3,v1}
\fmf{fermion}{v2,v3}
\fmf{fermion}{v5,v4}
\fmf{fermion}{v4,v6}
\fmfv{decor.size=0, label=${\scs 1}$, l.dist=1mm, l.angle=-180}{v1}
\fmfv{decor.size=0, label=${\scs 2}$, l.dist=1mm, l.angle=0}{v5}
\fmfv{decor.size=0, label=${\scs 4}$, l.dist=1mm, l.angle=-180}{v2}
\fmfv{decor.size=0, label=${\scs 5}$, l.dist=1mm, l.angle=0}{v6}
\fmfdot{v3,v4}
\end{fmfgraph*}
\end{center} }
\hspace*{0.4cm} ,
\eeq
we determine from (\r{R5}) the one-loop contribution to the one-particle irreducible three-point function
\beq
\mbox{} && \no \\
\parbox{17mm}{\centerline{
\begin{fmfgraph*}(10,8.66)
\setval
\fmfforce{1w,0h}{v1}
\fmfforce{0w,0h}{v2}
\fmfforce{0.5w,0.95h}{v3}
\fmfforce{0.25w,0.18h}{v4}
\fmfforce{0.75w,0.18h}{v5}
\fmfforce{0.5w,0.6h}{v6}
\fmfforce{0.5w,-0.0114h}{vm1}
\fmfforce{0.5w,0.2886h}{vm}
\fmfforce{0.5w,0.5886h}{vm2}
\fmf{fermion}{v4,v2}
\fmf{fermion}{v1,v5}
\fmf{photon}{v3,v6}
\fmf{double,width=0.2mm,left=1}{vm1,vm2,vm1}
\fmfv{decor.size=0,label=${\scs 1}$, l.dist=0mm, l.angle=0}{vm}
\fmfv{decor.size=0,label=${\scs 2}$,l.dist=0.5mm}{v1}
\fmfv{decor.size=0,label=${\scs 1}$,l.dist=0.5mm}{v2}
\fmfv{decor.size=0,label=${\scs 3}$,l.dist=0.5mm}{v3}
\end{fmfgraph*}
}}
\quad & = & \quad
\parbox{19mm}{\begin{center}
\begin{fmfgraph*}(16,3)
\setval
\fmfforce{0w,0h}{v1}
\fmfforce{3/16w,0h}{v2}
\fmfforce{8/16w,0h}{v3}
\fmfforce{13/16w,0h}{v4}
\fmfforce{1w,0h}{v5}
\fmfforce{1/2w,1h}{v6}
\fmf{fermion}{v5,v4,v3,v2,v1}
\fmf{boson,right=0.75}{v2,v4}
\fmf{boson}{v3,v6}
\fmfdot{v2,v3,v4,v6}
\fmfv{decor.size=0, label=${\scs 1}$, l.dist=1mm, l.angle=-180}{v1}
\fmfv{decor.size=0, label=${\scs 2}$, l.dist=1mm, l.angle=0}{v5}
\fmfv{decor.size=0, label=${\scs 3}$, l.dist=1mm, l.angle=90}{v6}
\end{fmfgraph*}
\end{center} }
\hspace*{0.4cm}. \la{VVV1} \\ \mbox{} && \no \\
\eeq
Using the Eqs. (\r{R3}) and (\r{R4}), we then find the electron and photon self-energy with two loops:
\beq
\la{SELF2}
\parbox{18mm}{\centerline{
\begin{fmfgraph*}(11,5)
\setval
\fmfforce{0w,1/2h}{v1}
\fmfforce{3/11w,1/2h}{v2}
\fmfforce{8/11w,1/2h}{v3}
\fmfforce{8.2/11w,1/2h}{v3b}
\fmfforce{1w,1/2h}{v4}
\fmfforce{1/2w,1/2h}{v5}
\fmf{electron}{v2,v1}
\fmf{electron}{v4,v3b}
\fmf{double,width=0.2mm,left=1}{v2,v3,v2}
\fmfv{decor.size=0, label=${\scs 2}$, l.dist=0mm, l.angle=0}{v5}
\fmfv{decor.size=0, label=${\scs 1}$, l.dist=1mm, l.angle=-180}{v1}
\fmfv{decor.size=0, label=${\scs 2}$, l.dist=1mm, l.angle=0}{v4}
\end{fmfgraph*} }}
\quad & = & \quad
%
%
\hspace*{0.3cm} 
\parbox{24mm}{\begin{center}
\begin{fmfgraph*}(21,10)
\setval
\fmfforce{0w,1/2h}{v1}
\fmfforce{3/21w,1/2h}{v2}
\fmfforce{8/21w,1/2h}{v3}
\fmfforce{13/21w,1/2h}{v4}
\fmfforce{18/21w,1/2h}{v5}
\fmfforce{1w,1/2h}{v6}
\fmf{fermion}{v6,v5,v4,v3,v2,v1}
\fmf{boson,left=0.75}{v2,v4}
\fmf{boson,right=0.75}{v3,v5}
\fmfdot{v2,v3,v4,v5}
\fmfv{decor.size=0, label=${\scs 1}$, l.dist=1mm, l.angle=-180}{v1}
\fmfv{decor.size=0, label=${\scs 2}$, l.dist=1mm, l.angle=0}{v6}
\end{fmfgraph*}
\end{center} }
%
%
\hspace*{0.3cm} - \hspace*{0.3cm}  
\parbox{14mm}{\begin{center}
\begin{fmfgraph*}(11,7.5)
\setval
\fmfforce{0w,0h}{v1}
\fmfforce{3/11w,0h}{v2}
\fmfforce{8/11w,0h}{v3}
\fmfforce{3/3w,0h}{v4}
\fmfforce{3/11w,2/3h}{v5}
\fmfforce{8/11w,2/3h}{v6}
\fmf{fermion}{v4,v3,v2,v1}
\fmf{fermion,right=1}{v5,v6}
\fmf{fermion,right=1}{v6,v5}
\fmf{boson}{v2,v5}
\fmf{boson}{v3,v6}
\fmfdot{v2,v3,v5,v6}
\fmfv{decor.size=0, label=${\scs 1}$, l.dist=1mm, l.angle=-180}{v1}
\fmfv{decor.size=0, label=${\scs 2}$, l.dist=1mm, l.angle=0}{v4}
\end{fmfgraph*}
\end{center} }
\hspace*{0.3cm} + \hspace*{0.3cm} 
%
%
\parbox{24mm}{\begin{center}
\begin{fmfgraph*}(21,5)
\setval
\fmfforce{0w,0h}{v1}
\fmfforce{3/21w,0h}{v2}
\fmfforce{8/21w,0h}{v3}
\fmfforce{13/21w,0h}{v4}
\fmfforce{18/21w,0h}{v5}
\fmfforce{1w,0h}{v6}
\fmf{fermion}{v6,v5,v4,v3,v2,v1}
\fmf{boson,left=0.75}{v2,v5}
\fmf{boson,left=1}{v3,v4}
\fmfdot{v2,v3,v4,v5}
\fmfv{decor.size=0, label=${\scs 1}$, l.dist=1mm, l.angle=-180}{v1}
\fmfv{decor.size=0, label=${\scs 2}$, l.dist=1mm, l.angle=0}{v6}
\end{fmfgraph*}
\end{center} }
\hspace*{0.4cm} , \\
\parbox{18mm}{\centerline{
\begin{fmfgraph*}(11,5)
\setval
\fmfforce{0w,1/2h}{v1}
\fmfforce{3/11w,1/2h}{v2}
\fmfforce{8/11w,1/2h}{v3}
\fmfforce{8.2/11w,1/2h}{v3b}
\fmfforce{1w,1/2h}{v4}
\fmfforce{1/2w,1/2h}{v5}
\fmf{boson}{v2,v1}
\fmf{boson}{v4,v3b}
\fmf{double,width=0.2mm,left=1}{v2,v3,v2}
\fmfv{decor.size=0, label=${\scs 2}$, l.dist=0mm, l.angle=0}{v5}
\fmfv{decor.size=0, label=${\scs 1}$, l.dist=1mm, l.angle=-180}{v1}
\fmfv{decor.size=0, label=${\scs 2}$, l.dist=1mm, l.angle=0}{v4}
\end{fmfgraph*} }}
\quad & = & \quad
%
%
- \hspace*{0.3cm} 
\parbox{16mm}{\begin{center}
\begin{fmfgraph*}(13,7)
\setval
\fmfforce{0w,1/2h}{v1}
\fmfforce{3/13w,1/2h}{v2}
\fmfforce{10/13w,1/2h}{v3}
\fmfforce{1w,1/2h}{v4}
\fmfforce{1/2w,0h}{v5}
\fmfforce{1/2w,1h}{v6}
\fmf{boson}{v1,v2}
\fmf{boson}{v3,v4}
\fmf{boson}{v5,v6}
\fmf{fermion,right=0.4}{v2,v5}
\fmf{fermion,right=0.4}{v5,v3}
\fmf{fermion,right=0.4}{v3,v6}
\fmf{fermion,right=0.4}{v6,v2}
\fmfdot{v2,v3,v5,v6}
\fmfv{decor.size=0, label=${\scs 1}$, l.dist=1mm, l.angle=-180}{v1}
\fmfv{decor.size=0, label=${\scs 2}$, l.dist=1mm, l.angle=0}{v4}
\end{fmfgraph*}
\end{center} }
%
%
\hspace*{0.3cm} - \hspace*{0.3cm} 
\parbox{14mm}{\begin{center}
\begin{fmfgraph*}(11,5)
\setval
\fmfforce{0w,0h}{v1}
\fmfforce{3/11w,0h}{v2}
\fmfforce{8/11w,0h}{v3}
\fmfforce{1w,0h}{v4}
\fmfforce{3/11w,1h}{v5}
\fmfforce{8/11w,1h}{v6}
\fmf{boson}{v1,v2}
\fmf{boson}{v3,v4}
\fmf{boson,right=0.4}{v5,v6}
\fmf{fermion,right=0.4}{v5,v2}
\fmf{fermion,right=0.4}{v6,v5}
\fmf{fermion,right=0.4}{v3,v6}
\fmf{fermion,right=0.4}{v2,v3}
\fmfdot{v2,v3,v5,v6}
\fmfv{decor.size=0, label=${\scs 1}$, l.dist=1mm, l.angle=-180}{v1}
\fmfv{decor.size=0, label=${\scs 2}$, l.dist=1mm, l.angle=0}{v4}
\end{fmfgraph*}
\end{center} }
%
%
\hspace*{0.3cm} - \hspace*{0.3cm} 
\parbox{14mm}{\begin{center}
\begin{fmfgraph*}(11,5)
\setval
\fmfforce{0w,0h}{v1}
\fmfforce{3/11w,0h}{v2}
\fmfforce{8/11w,0h}{v3}
\fmfforce{1w,0h}{v4}
\fmfforce{3/11w,1h}{v5}
\fmfforce{8/11w,1h}{v6}
\fmf{boson}{v1,v2}
\fmf{boson}{v3,v4}
\fmf{boson,right=0.4}{v5,v6}
\fmf{fermion,left=0.4}{v2,v5}
\fmf{fermion,left=0.4}{v5,v6}
\fmf{fermion,left=0.4}{v6,v3}
\fmf{fermion,left=0.4}{v3,v2}
\fmfdot{v2,v3,v5,v6}
\fmfv{decor.size=0, label=${\scs 1}$, l.dist=1mm, l.angle=-180}{v1}
\fmfv{decor.size=0, label=${\scs 2}$, l.dist=1mm, l.angle=0}{v4}
\end{fmfgraph*}
\end{center} }
\hspace*{0.4cm}  . \la{PPEL2}
\eeq
The corresponding contributions to the connected two-point functions
are according to (\r{R1}) and (\r{R2}):
\beq
\parbox{15mm}{\centerline{
\begin{fmfgraph*}(7,3)
\setval
\fmfleft{v1}
\fmfright{v2}
\fmfforce{0.5w,2/3h}{v3}
\fmf{heavy,width=0.2mm}{v2,v1}
\fmfv{decor.size=0, label=${\scs 1}$, l.dist=1mm, l.angle=-180}{v1}
\fmfv{decor.size=0, label=${\scs 2}$, l.dist=1mm, l.angle=0}{v2}
\fmfv{decor.size=0, label=${\scs (2)}$, l.dist=1mm, l.angle=90}{v3}
\end{fmfgraph*}}}
\quad & = & \quad 
%
%
\parbox{28mm}{\begin{center}
\begin{fmfgraph*}(25,10)
\setval
\fmfforce{0w,1/2h}{v1}
\fmfforce{1/5w,1/2h}{v2}
\fmfforce{2/5w,1/2h}{v3}
\fmfforce{3/5w,1/2h}{v4}
\fmfforce{4/5w,1/2h}{v5}
\fmfforce{1w,1/2h}{v6}
\fmf{fermion}{v6,v5,v4,v3,v2,v1}
\fmf{boson,left=0.75}{v2,v4}
\fmf{boson,right=0.75}{v3,v5}
\fmfdot{v2,v3,v4,v5}
\fmfv{decor.size=0, label=${\scs 1}$, l.dist=1mm, l.angle=-180}{v1}
\fmfv{decor.size=0, label=${\scs 2}$, l.dist=1mm, l.angle=0}{v6}
\end{fmfgraph*}
\end{center} }
%
%
\hspace*{0.3cm} - \hspace*{0.3cm} 
\parbox{18mm}{\begin{center}
\begin{fmfgraph*}(15,7.5)
\setval
\fmfforce{0w,0h}{v1}
\fmfforce{1/3w,0h}{v2}
\fmfforce{2/3w,0h}{v3}
\fmfforce{3/3w,0h}{v4}
\fmfforce{1/3w,2/3h}{v5}
\fmfforce{2/3w,2/3h}{v6}
\fmf{fermion}{v4,v3,v2,v1}
\fmf{fermion,right=1}{v5,v6}
\fmf{fermion,right=1}{v6,v5}
\fmf{boson}{v2,v5}
\fmf{boson}{v3,v6}
\fmfdot{v2,v3,v5,v6}
\fmfv{decor.size=0, label=${\scs 1}$, l.dist=1mm, l.angle=-180}{v1}
\fmfv{decor.size=0, label=${\scs 2}$, l.dist=1mm, l.angle=0}{v4}
\end{fmfgraph*}
\end{center} }
%
%
\hspace*{0.3cm} + \hspace*{0.3cm} 
\parbox{28mm}{\begin{center}
\begin{fmfgraph*}(25,5)
\setval
\fmfforce{0w,0h}{v1}
\fmfforce{1/5w,0h}{v2}
\fmfforce{2/5w,0h}{v3}
\fmfforce{3/5w,0h}{v4}
\fmfforce{4/5w,0h}{v5}
\fmfforce{1w,0h}{v6}
\fmf{fermion}{v6,v5,v4,v3,v2,v1}
\fmf{boson,left=0.75}{v2,v5}
\fmf{boson,left=1}{v3,v4}
\fmfdot{v2,v3,v4,v5}
\fmfv{decor.size=0, label=${\scs 1}$, l.dist=1mm, l.angle=-180}{v1}
\fmfv{decor.size=0, label=${\scs 2}$, l.dist=1mm, l.angle=0}{v6}
\end{fmfgraph*}
\end{center} }
%
%
\hspace*{0.3cm} + \hspace*{0.3cm} 
\parbox{28mm}{\begin{center}
\begin{fmfgraph*}(25,2.5)
\setval
\fmfforce{0w,0h}{v1}
\fmfforce{1/5w,0h}{v2}
\fmfforce{2/5w,0h}{v3}
\fmfforce{3/5w,0h}{v4}
\fmfforce{4/5w,0h}{v5}
\fmfforce{1w,0h}{v6}
\fmf{fermion}{v6,v5,v4,v3,v2,v1}
\fmf{boson,left=1}{v2,v3}
\fmf{boson,left=1}{v4,v5}
\fmfdot{v2,v3,v4,v5}
\fmfv{decor.size=0, label=${\scs 1}$, l.dist=1mm, l.angle=-180}{v1}
\fmfv{decor.size=0, label=${\scs 2}$, l.dist=1mm, l.angle=0}{v6}
\end{fmfgraph*}
\end{center} }
\hspace*{0.4cm} , \la{SES2} \\
\parbox{15mm}{\centerline{
\begin{fmfgraph*}(7,3)
\setval
\fmfleft{v1}
\fmfright{v2}
\fmfforce{0.5w,2/3h}{v3}
\fmf{dbl_wiggly,width=0.2mm}{v2,v1}
\fmfv{decor.size=0, label=${\scs 1}$, l.dist=1mm, l.angle=-180}{v1}
\fmfv{decor.size=0, label=${\scs 2}$, l.dist=1mm, l.angle=0}{v2}
\fmfv{decor.size=0, label=${\scs (2)}$, l.dist=1mm, l.angle=90}{v3}
\end{fmfgraph*}}}
\quad & = & \quad 
%
%
- \hspace*{0.3cm} 
\parbox{20mm}{\begin{center}
\begin{fmfgraph*}(17,7)
\setval
\fmfforce{0w,1/2h}{v1}
\fmfforce{5/17w,1/2h}{v2}
\fmfforce{12/17w,1/2h}{v3}
\fmfforce{1w,1/2h}{v4}
\fmfforce{1/2w,0h}{v5}
\fmfforce{1/2w,1h}{v6}
\fmf{boson}{v1,v2}
\fmf{boson}{v3,v4}
\fmf{boson}{v5,v6}
\fmf{fermion,right=0.4}{v2,v5}
\fmf{fermion,right=0.4}{v5,v3}
\fmf{fermion,right=0.4}{v3,v6}
\fmf{fermion,right=0.4}{v6,v2}
\fmfdot{v2,v3,v5,v6}
\fmfv{decor.size=0, label=${\scs 1}$, l.dist=1mm, l.angle=-180}{v1}
\fmfv{decor.size=0, label=${\scs 2}$, l.dist=1mm, l.angle=0}{v4}
\end{fmfgraph*}
\end{center} }
%
%
\hspace*{0.3cm} - \hspace*{0.3cm} 
\parbox{18mm}{\begin{center}
\begin{fmfgraph*}(15,5)
\setval
\fmfforce{0w,0h}{v1}
\fmfforce{1/3w,0h}{v2}
\fmfforce{2/3w,0h}{v3}
\fmfforce{1w,0h}{v4}
\fmfforce{1/3w,1h}{v5}
\fmfforce{2/3w,1h}{v6}
\fmf{boson}{v1,v2}
\fmf{boson}{v3,v4}
\fmf{boson,right=0.4}{v5,v6}
\fmf{fermion,right=0.4}{v5,v2}
\fmf{fermion,right=0.4}{v6,v5}
\fmf{fermion,right=0.4}{v3,v6}
\fmf{fermion,right=0.4}{v2,v3}
\fmfdot{v2,v3,v5,v6}
\fmfv{decor.size=0, label=${\scs 1}$, l.dist=1mm, l.angle=-180}{v1}
\fmfv{decor.size=0, label=${\scs 2}$, l.dist=1mm, l.angle=0}{v4}
\end{fmfgraph*}
\end{center} }
%
%
\hspace*{0.3cm} - \hspace*{0.3cm} 
\parbox{18mm}{\begin{center}
\begin{fmfgraph*}(15,5)
\setval
\fmfforce{0w,0h}{v1}
\fmfforce{1/3w,0h}{v2}
\fmfforce{2/3w,0h}{v3}
\fmfforce{1w,0h}{v4}
\fmfforce{1/3w,1h}{v5}
\fmfforce{2/3w,1h}{v6}
\fmf{boson}{v1,v2}
\fmf{boson}{v3,v4}
\fmf{boson,right=0.4}{v5,v6}
\fmf{fermion,left=0.4}{v2,v5}
\fmf{fermion,left=0.4}{v5,v6}
\fmf{fermion,left=0.4}{v6,v3}
\fmf{fermion,left=0.4}{v3,v2}
\fmfdot{v2,v3,v5,v6}
\fmfv{decor.size=0, label=${\scs 1}$, l.dist=1mm, l.angle=-180}{v1}
\fmfv{decor.size=0, label=${\scs 2}$, l.dist=1mm, l.angle=0}{v4}
\end{fmfgraph*}
\end{center} }
%
%
\hspace*{0.3cm} + \hspace*{0.3cm}
\parbox{28mm}{\begin{center}
\begin{fmfgraph*}(25,5)
\setval
\fmfforce{0w,1/2h}{v1}
\fmfforce{1/5w,1/2h}{v2}
\fmfforce{2/5w,1/2h}{v3}
\fmfforce{3/5w,1/2h}{v4}
\fmfforce{4/5w,1/2h}{v5}
\fmfforce{5/5w,1/2h}{v6}
\fmf{boson}{v1,v2}
\fmf{fermion,right=1}{v2,v3,v2}
\fmf{boson}{v3,v4}
\fmf{fermion,right=1}{v4,v5,v4}
\fmf{boson}{v5,v6}
\fmfdot{v2,v3,v4,v5}
\fmfv{decor.size=0, label=${\scs 1}$, l.dist=1mm, l.angle=-180}{v1}
\fmfv{decor.size=0, label=${\scs 2}$, l.dist=1mm, l.angle=0}{v6}
\end{fmfgraph*}
\end{center} }
\hspace*{0.4cm} . \la{SES2B}
\eeq
Comparing (\r{PHOA}) and (\r{PHO}) with (\r{SES2}) and (\r{SES2B}) shows the graphical consequence of the
functional Legendre transform with respect to the current by the example of the connected two-point function. It systematically
eliminates the one-particle reducible diagrams carrying any kind of tadpol correction.
The subsequent amputation of one electron line in (\r{SELF2}) leads to
\beq
\dephi{
\parbox{18mm}{\centerline{
\begin{fmfgraph*}(11,7)
\setval
\fmfforce{0w,1/2h}{v1}
\fmfforce{3/11w,1/2h}{v2}
\fmfforce{8/11w,1/2h}{v3}
\fmfforce{1w,1/2h}{v4}
\fmfforce{1/2w,1/2h}{v5}
\fmf{electron}{v2,v1}
\fmf{electron}{v4,v3}
\fmf{double,width=0.2mm,left=1}{v2,v3,v2}
\fmfv{decor.size=0, label=${\scs 2}$, l.dist=0mm, l.angle=0}{v5}
\fmfv{decor.size=0, label=${\scs 1}$, l.dist=1mm, l.angle=-180}{v1}
\fmfv{decor.size=0, label=${\scs 2}$, l.dist=1mm, l.angle=0}{v4}
\end{fmfgraph*} }}
}{4}{5}
\quad & = & \quad  
%
%
- \hspace*{0.3cm}
\parbox{26.39mm}{\centerline{
\begin{fmfgraph*}(19.24,5)
\setval
\fmfforce{0w,0.38/5h}{v1}
\fmfforce{2.12/19.24w,1/2h}{v2}
\fmfforce{7.12/19.24w,1/2h}{v3}
\fmfforce{12.12/19.24w,1/2h}{v4}
\fmfforce{17.12/19.24w,1/2h}{v5}
\fmfforce{1w,0.38/5h}{v6}
\fmfforce{0w,4.62/5h}{v7}
\fmfforce{1w,4.62/5h}{v8}
\fmf{electron}{v2,v1}
\fmf{electron}{v6,v5}
\fmf{electron}{v7,v2}
\fmf{electron}{v5,v8}
\fmf{boson}{v2,v3}
\fmf{boson}{v4,v5}
\fmf{electron,right=1}{v3,v4,v3}
\fmfv{decor.size=0, label=${\scs 1}$, l.dist=1mm, l.angle=-180}{v1}
\fmfv{decor.size=0, label=${\scs 2}$, l.dist=1mm, l.angle=0}{v6}
\fmfv{decor.size=0, label=${\scs 4}$, l.dist=1mm, l.angle=-180}{v7}
\fmfv{decor.size=0, label=${\scs 5}$, l.dist=1mm, l.angle=0}{v8}
\fmfdot{v2,v3,v4,v5}
\end{fmfgraph*} 
}}
%
%
\hspace*{0.3cm}- \hspace*{0.3cm}
\parbox{14mm}{\centerline{
\begin{fmfgraph*}(11,5)
\setval
\fmfforce{0w,0h}{v1}
\fmfforce{3/11w,0h}{v2}
\fmfforce{8/11w,0h}{v3}
\fmfforce{1w,0h}{v4}
\fmfforce{0w,1h}{v5}
\fmfforce{3/11w,1h}{v6}
\fmfforce{8/11w,1h}{v7}
\fmfforce{1w,1h}{v8}
\fmf{electron}{v4,v3}
\fmf{electron}{v2,v1}
\fmf{electron,left=0.4}{v3,v2}
\fmf{electron}{v7,v8}
\fmf{electron,left=0.4}{v6,v7}
\fmf{electron}{v5,v6}
\fmf{boson}{v2,v7}
\fmf{boson}{v3,v6}
\fmfv{decor.size=0, label=${\scs 1}$, l.dist=1mm, l.angle=-180}{v1}
\fmfv{decor.size=0, label=${\scs 2}$, l.dist=1mm, l.angle=0}{v4}
\fmfv{decor.size=0, label=${\scs 5}$, l.dist=1mm, l.angle=0}{v8}
\fmfv{decor.size=0, label=${\scs 4}$, l.dist=1mm, l.angle=-180}{v5}
\fmfdot{v2,v3,v6,v7}
\end{fmfgraph*} 
}}
%
%
\hspace*{0.3cm}- \hspace*{0.3cm}
\parbox{14mm}{\centerline{
\begin{fmfgraph*}(11,5)
\setval
\fmfforce{0w,0h}{v1}
\fmfforce{3/11w,0h}{v2}
\fmfforce{8/11w,0h}{v3}
\fmfforce{1w,0h}{v4}
\fmfforce{0w,1h}{v5}
\fmfforce{3/11w,1h}{v6}
\fmfforce{8/11w,1h}{v7}
\fmfforce{1w,1h}{v8}
\fmf{electron}{v4,v3}
\fmf{electron}{v2,v1}
\fmf{electron,left=0.4}{v3,v2}
\fmf{electron}{v7,v8}
\fmf{electron,left=0.4}{v6,v7}
\fmf{electron}{v5,v6}
\fmf{boson,left=0.4}{v2,v6}
\fmf{boson,right=0.4}{v3,v7}
\fmfv{decor.size=0, label=${\scs 1}$, l.dist=1mm, l.angle=-180}{v1}
\fmfv{decor.size=0, label=${\scs 2}$, l.dist=1mm, l.angle=0}{v4}
\fmfv{decor.size=0, label=${\scs 4}$, l.dist=1mm, l.angle=-180}{v5}
\fmfv{decor.size=0, label=${\scs 5}$, l.dist=1mm, l.angle=0}{v8}
\fmfdot{v2,v3,v6,v7}
\end{fmfgraph*} 
}}
%
%
\hspace*{0.3cm}+ \hspace*{0.3cm}
\parbox{14mm}{\centerline{
\begin{fmfgraph*}(11,5)
\setval
\fmfforce{0w,0h}{v1}
\fmfforce{3/11w,0h}{v2}
\fmfforce{8/11w,0h}{v3}
\fmfforce{1w,0h}{v4}
\fmfforce{0w,1h}{v5}
\fmfforce{3/11w,1h}{v6}
\fmfforce{8/11w,1h}{v7}
\fmfforce{1w,1h}{v8}
\fmf{electron}{v4,v3}
\fmf{electron}{v2,v1}
\fmf{boson}{v7,v2}
\fmf{electron}{v7,v8}
\fmf{boson}{v6,v3}
\fmf{electron}{v5,v6}
\fmf{fermion,right=0.4}{v6,v2}
\fmf{fermion,right=0.4}{v3,v7}
\fmfv{decor.size=0, label=${\scs 1}$, l.dist=1mm, l.angle=-180}{v1}
\fmfv{decor.size=0, label=${\scs 2}$, l.dist=1mm, l.angle=0}{v4}
\fmfv{decor.size=0, label=${\scs 4}$, l.dist=1mm, l.angle=-180}{v5}
\fmfv{decor.size=0, label=${\scs 5}$, l.dist=1mm, l.angle=0}{v8}
\fmfdot{v2,v3,v6,v7}
\end{fmfgraph*} 
}}
\no \\
&& \quad 
\no \\
&& \quad 
\no \\
&& \quad
%
%
+ \hspace*{0.3cm}
\parbox{13.12mm}{\centerline{
\begin{fmfgraph*}(10.12,10)
\setval
\fmfforce{0w,0h}{v1}
\fmfforce{3/10.12w,0h}{v2}
\fmfforce{3/10.12w,5/10h}{v3}
\fmfforce{3/11w,10/10h}{v4}
\fmfforce{0w,1h}{v5}
\fmfforce{8/10.12w,5/10h}{v6}
\fmfforce{1w,2.88/10h}{v7}
\fmfforce{1w,7.12/10h}{v8}
\fmf{electron}{v2,v1}
\fmf{electron}{v3,v2}
\fmf{electron}{v4,v3}
\fmf{electron}{v5,v4}
\fmf{electron}{v7,v6}
\fmf{electron}{v6,v8}
\fmf{boson}{v3,v6}
\fmf{boson,left=0.6}{v2,v4}
\fmfv{decor.size=0, label=${\scs 1}$, l.dist=1mm, l.angle=-180}{v1}
\fmfv{decor.size=0, label=${\scs 2}$, l.dist=1mm, l.angle=0}{v7}
\fmfv{decor.size=0, label=${\scs 4}$, l.dist=1mm, l.angle=-180}{v5}
\fmfv{decor.size=0, label=${\scs 5}$, l.dist=1mm, l.angle=0}{v8}
\fmfdot{v2,v3,v6,v4}
\end{fmfgraph*} 
}}
%
%
\hspace*{0.3cm}+ \hspace*{0.3cm}
\parbox{13.12mm}{\centerline{
\begin{fmfgraph*}(10.12,10)
\setval
\fmfforce{0w,2.88/10h}{v1}
\fmfforce{2.12/10.12w,1/2h}{v2}
\fmfforce{0w,7.12/10h}{v3}
\fmfforce{7.12/10.12w,0h}{v4}
\fmfforce{7.12/10.12w,5/10h}{v5}
\fmfforce{7.12/10.12w,1h}{v6}
\fmfforce{1w,1h}{v7}
\fmfforce{1w,0h}{v8}
\fmf{electron}{v2,v1}
\fmf{electron}{v3,v2}
\fmf{electron}{v4,v5}
\fmf{electron}{v5,v6}
\fmf{electron}{v6,v7}
\fmf{electron}{v8,v4}
\fmf{boson}{v2,v5}
\fmf{boson,left=0.6}{v6,v4}
\fmfv{decor.size=0, label=${\scs 1}$, l.dist=1mm, l.angle=-180}{v1}
\fmfv{decor.size=0, label=${\scs 2}$, l.dist=1mm, l.angle=0}{v8}
\fmfv{decor.size=0, label=${\scs 4}$, l.dist=1mm, l.angle=-180}{v3}
\fmfv{decor.size=0, label=${\scs 5}$, l.dist=1mm, l.angle=0}{v7}
\fmfdot{v2,v5,v6,v4}
\end{fmfgraph*} 
}}
%
%
\hspace*{0.3cm}+ \hspace*{0.3cm}
\parbox{14mm}{\centerline{
\begin{fmfgraph*}(11,5)
\setval
\fmfforce{0w,0h}{v1}
\fmfforce{3/11w,0h}{v2}
\fmfforce{8/11w,0h}{v3}
\fmfforce{1w,0h}{v4}
\fmfforce{0w,1h}{v5}
\fmfforce{3/11w,1h}{v6}
\fmfforce{8/11w,1h}{v7}
\fmfforce{1w,1h}{v8}
\fmf{electron}{v4,v3}
\fmf{electron}{v2,v1}
\fmf{boson,left=0.4}{v3,v2}
\fmf{electron}{v7,v8}
\fmf{boson,left=0.4}{v6,v7}
\fmf{electron}{v5,v6}
\fmf{fermion,right=0.4}{v6,v2}
\fmf{fermion,right=0.4}{v3,v7}
\fmfv{decor.size=0, label=${\scs 1}$, l.dist=1mm, l.angle=-180}{v1}
\fmfv{decor.size=0, label=${\scs 2}$, l.dist=1mm, l.angle=0}{v4}
\fmfv{decor.size=0, label=${\scs 4}$, l.dist=1mm, l.angle=-180}{v5}
\fmfv{decor.size=0, label=${\scs 5}$, l.dist=1mm, l.angle=0}{v8}
\fmfdot{v2,v3,v6,v7}
\end{fmfgraph*} 
}}
%
%
\hspace*{0.3cm}+ \hspace*{0.3cm}
\parbox{14mm}{\centerline{
\begin{fmfgraph*}(10.12,12.12)
\setval
\fmfforce{0w,2.12/12.12h}{v1}
\fmfforce{3/10.12w,2.12/12.12h}{v2}
\fmfforce{8/10.12w,2.12/12.12h}{v3}
\fmfforce{1w,0h}{v4}
\fmfforce{0w,1h}{v5}
\fmfforce{3/10.12w,7.12/12.12h}{v6}
\fmfforce{3/10.12w,1h}{v7}
\fmfforce{1w,4.24/12.12h}{v8}
\fmf{electron}{v2,v1}
\fmf{electron}{v6,v2}
\fmf{electron}{v7,v6}
\fmf{electron}{v5,v7}
\fmf{electron}{v3,v8}
\fmf{electron}{v4,v3}
\fmf{boson}{v3,v2}
\fmf{boson,right=1}{v6,v7}
\fmfv{decor.size=0, label=${\scs 1}$, l.dist=1mm, l.angle=-180}{v1}
\fmfv{decor.size=0, label=${\scs 2}$, l.dist=1mm, l.angle=0}{v4}
\fmfv{decor.size=0, label=${\scs 4}$, l.dist=1mm, l.angle=-180}{v5}
\fmfv{decor.size=0, label=${\scs 5}$, l.dist=1mm, l.angle=0}{v8}
\fmfdot{v2,v3,v6,v7}
\end{fmfgraph*} 
}}
%
%
\hspace*{0.3cm}+ \hspace*{0.3cm}
\parbox{14mm}{\centerline{
\begin{fmfgraph*}(10.12,12.12)
\setval
\fmfforce{0w,0h}{v1}
\fmfforce{2.12/10.12w,2.12/12.12h}{v2}
\fmfforce{7.12/10.12w,2.12/12.12h}{v3}
\fmfforce{1w,2.12/12.12h}{v4}
\fmfforce{0w,4.24/12.12h}{v5}
\fmfforce{7.12/10.12w,7.12/12.12h}{v6}
\fmfforce{7.12/10.12w,1h}{v7}
\fmfforce{1w,1h}{v8}
\fmf{electron}{v2,v1}
\fmf{electron}{v5,v2}
\fmf{electron}{v4,v3}
\fmf{electron}{v3,v6}
\fmf{electron}{v6,v7}
\fmf{electron}{v7,v8}
\fmf{boson}{v3,v2}
\fmf{boson,left=1}{v6,v7}
\fmfv{decor.size=0, label=${\scs 1}$, l.dist=1mm, l.angle=-180}{v1}
\fmfv{decor.size=0, label=${\scs 2}$, l.dist=1mm, l.angle=0}{v4}
\fmfv{decor.size=0, label=${\scs 4}$, l.dist=1mm, l.angle=-180}{v5}
\fmfv{decor.size=0, label=${\scs 5}$, l.dist=1mm, l.angle=0}{v8}
\fmfdot{v2,v3,v6,v7}
\end{fmfgraph*} 
}}
\hspace*{0.4cm} ,
\eeq
so we find from (\r{R5}) the one-particle irreducible three-point function with two loops:
\beq
\parbox{17mm}{\centerline{
\begin{fmfgraph*}(10,8.66)
\setval
\fmfforce{1w,0h}{v1}
\fmfforce{0w,0h}{v2}
\fmfforce{0.5w,0.95h}{v3}
\fmfforce{0.25w,0.18h}{v4}
\fmfforce{0.75w,0.18h}{v5}
\fmfforce{0.5w,0.6h}{v6}
\fmfforce{0.5w,-0.0114h}{vm1}
\fmfforce{0.5w,0.2886h}{vm}
\fmfforce{0.5w,0.5886h}{vm2}
\fmf{fermion}{v4,v2}
\fmf{fermion}{v1,v5}
\fmf{photon}{v3,v6}
\fmf{double,width=0.2mm,left=1}{vm1,vm2,vm1}
\fmfv{decor.size=0,label=${\scs 2}$, l.dist=0mm, l.angle=0}{vm}
\fmfv{decor.size=0,label=${\scs 2}$,l.dist=0.5mm}{v1}
\fmfv{decor.size=0,label=${\scs 1}$,l.dist=0.5mm}{v2}
\fmfv{decor.size=0,label=${\scs 3}$,l.dist=0.5mm}{v3}
\end{fmfgraph*}
}}
\quad & = & \quad  
%
%
\parbox{14mm}{\centerline{
\begin{fmfgraph*}(11,12)
\setval
\fmfforce{0w,0/12h}{v1}
\fmfforce{3/11w,0/12h}{v2}
\fmfforce{8/11w,0/12h}{v3}
\fmfforce{1w,0/12h}{v4}
\fmfforce{3/11w,5/12h}{v5}
\fmfforce{8/11w,5/12h}{v6}
\fmfforce{1/2w,6/12h}{v7}
\fmfforce{1/2w,9/12h}{v8}
\fmf{electron}{v2,v1}
\fmf{electron}{v4,v3}
\fmf{boson}{v8,v7}
\fmf{boson}{v2,v6}
\fmf{fermion,right=0.4}{v5,v2}
\fmf{fermion,right=0.2}{v7,v5}
\fmf{fermion,right=0.2}{v6,v7}
\fmf{fermion,right=0.4}{v3,v6}
\fmf{boson}{v3,v5}
\fmfdot{v2,v3,v5,v6,v7}
\fmfv{decor.size=0, label=${\scs 3}$, l.dist=1mm, l.angle=90}{v8}
\fmfv{decor.size=0, label=${\scs 1}$, l.dist=1mm, l.angle=-180}{v1}
\fmfv{decor.size=0, label=${\scs 2}$, l.dist=1mm, l.angle=0}{v4}
\end{fmfgraph*}
}}
%
%
\hspace*{0.3cm} - \hspace*{0.3cm}
\parbox{14mm}{\centerline{
\begin{fmfgraph*}(11,10.5)
\setval
\fmfforce{0w,0h}{v1}
\fmfforce{3/11w,0h}{v2}
\fmfforce{8/11w,0h}{v3}
\fmfforce{1w,0h}{v4}
\fmfforce{3/11w,5/10.5h}{v5}
\fmfforce{8/11w,5/10.5h}{v6}
\fmfforce{1/2w,7.5/10.5h}{v7}
\fmfforce{1/2w,1h}{v8}
\fmf{electron}{v4,v3,v2,v1}
\fmf{boson}{v8,v7}
\fmf{boson}{v2,v5}
\fmf{boson}{v3,v6}
\fmf{electron,right=1}{v5,v6}
\fmf{electron,right=0.4}{v6,v7}
\fmf{electron,right=0.4}{v7,v5}
\fmfdot{v2,v3,v5,v6,v7}
\fmfv{decor.size=0, label=${\scs 3}$, l.dist=1mm, l.angle=90}{v8}
\fmfv{decor.size=0, label=${\scs 1}$, l.dist=1mm, l.angle=-180}{v1}
\fmfv{decor.size=0, label=${\scs 2}$, l.dist=1mm, l.angle=0}{v4}
\end{fmfgraph*}
}}
%
%
\hspace*{0.3cm} - \hspace*{0.3cm}
\parbox{14mm}{\centerline{
\begin{fmfgraph*}(11,10.5)
\setval
\fmfforce{0w,0h}{v1}
\fmfforce{3/11w,0h}{v2}
\fmfforce{8/11w,0h}{v3}
\fmfforce{1w,0h}{v4}
\fmfforce{3/11w,5/10.5h}{v5}
\fmfforce{8/11w,5/10.5h}{v6}
\fmfforce{1/2w,7.5/10.5h}{v7}
\fmfforce{1/2w,1h}{v8}
\fmf{electron}{v4,v3,v2,v1}
\fmf{boson}{v8,v7}
\fmf{boson}{v2,v5}
\fmf{boson}{v3,v6}
\fmf{electron,left=1}{v6,v5}
\fmf{electron,left=0.4}{v7,v6}
\fmf{electron,left=0.4}{v5,v7}
\fmfdot{v2,v3,v5,v6,v7}
\fmfv{decor.size=0, label=${\scs 3}$, l.dist=1mm, l.angle=90}{v8}
\fmfv{decor.size=0, label=${\scs 1}$, l.dist=1mm, l.angle=-180}{v1}
\fmfv{decor.size=0, label=${\scs 2}$, l.dist=1mm, l.angle=0}{v4}
\end{fmfgraph*}
}}
%
%
\hspace*{0.3cm}+ \hspace*{0.3cm}
\parbox{14mm}{\centerline{
\begin{fmfgraph*}(11,12)
\setval
\fmfforce{0w,0/12h}{v1}
\fmfforce{3/11w,0/12h}{v2}
\fmfforce{8/11w,0/12h}{v3}
\fmfforce{1w,0/12h}{v4}
\fmfforce{3/11w,5/12h}{v5}
\fmfforce{8/11w,5/12h}{v6}
\fmfforce{1/2w,6/12h}{v7}
\fmfforce{1/2w,9/12h}{v8}
\fmf{electron}{v2,v1}
\fmf{electron}{v4,v3}
\fmf{boson}{v8,v7}
\fmf{boson,right=0.4}{v2,v3}
\fmf{fermion,right=0.4}{v5,v2}
\fmf{fermion,right=0.2}{v7,v5}
\fmf{fermion,right=0.2}{v6,v7}
\fmf{fermion,right=0.4}{v3,v6}
\fmf{boson,right=0.4}{v5,v6}
\fmfdot{v2,v3,v5,v6,v7}
\fmfv{decor.size=0, label=${\scs 3}$, l.dist=1mm, l.angle=90}{v8}
\fmfv{decor.size=0, label=${\scs 1}$, l.dist=1mm, l.angle=-180}{v1}
\fmfv{decor.size=0, label=${\scs 2}$, l.dist=1mm, l.angle=0}{v4}
\end{fmfgraph*}
}}
%
%
\hspace*{0.3cm} + \hspace*{0.3cm}
\parbox{14mm}{\centerline{
\begin{fmfgraph*}(11,7)
\setval
\fmfforce{0w,0/7h}{v1}
\fmfforce{3/11w,0/7h}{v2}
\fmfforce{8/11w,0/7h}{v3}
\fmfforce{1w,0/7h}{v4}
\fmfforce{3/11w,5/7h}{v5}
\fmfforce{8/11w,5/7h}{v6}
\fmfforce{9/11w,2.5/7h}{v7}
\fmfforce{1w,5/7h}{v8}
\fmf{electron}{v2,v1}
\fmf{electron}{v4,v3}
\fmf{boson}{v8,v7}
\fmf{boson}{v2,v6}
\fmf{fermion,right=0.4}{v5,v2}
\fmf{fermion,right=0.4}{v6,v5}
\fmf{boson}{v3,v5}
\fmf{fermion,right=0.2}{v3,v7}
\fmf{fermion,right=0.2}{v7,v6}
\fmfdot{v2,v3,v5,v6,v7}
\fmfv{decor.size=0, label=${\scs 3}$, l.dist=1mm, l.angle=45}{v8}
\fmfv{decor.size=0, label=${\scs 1}$, l.dist=1mm, l.angle=-180}{v1}
\fmfv{decor.size=0, label=${\scs 2}$, l.dist=1mm, l.angle=0}{v4}
\end{fmfgraph*}
}}
\no \\ && \no \\ && \no \\ 
&& \quad 
%
%
+ \hspace*{0.3cm}
\parbox{14mm}{\centerline{
\begin{fmfgraph*}(11,7)
\setval
\fmfforce{0w,0/7h}{v1}
\fmfforce{3/11w,0/7h}{v2}
\fmfforce{8/11w,0/7h}{v3}
\fmfforce{1w,0/7h}{v4}
\fmfforce{3/11w,5/7h}{v5}
\fmfforce{8/11w,5/7h}{v6}
\fmfforce{2/11w,2.5/7h}{v7}
\fmfforce{0w,5/7h}{v8}
\fmf{electron}{v2,v1}
\fmf{electron}{v4,v3}
\fmf{boson}{v8,v7}
\fmf{boson}{v2,v6}
\fmf{fermion,right=0.4}{v6,v5}
\fmf{fermion,right=0.4}{v3,v6}
\fmf{boson}{v5,v3}
\fmf{fermion,right=0.2}{v7,v2}
\fmf{fermion,right=0.2}{v5,v7}
\fmfdot{v2,v3,v5,v6,v7}
\fmfv{decor.size=0, label=${\scs 3}$, l.dist=1mm, l.angle=135}{v8}
\fmfv{decor.size=0, label=${\scs 1}$, l.dist=1mm, l.angle=-180}{v1}
\fmfv{decor.size=0, label=${\scs 2}$, l.dist=1mm, l.angle=0}{v4}
\end{fmfgraph*}
}}
%
%
\hspace*{0.3cm}- \hspace*{0.3cm}
\parbox{14mm}{\centerline{
\begin{fmfgraph*}(11,13)
\setval
\fmfforce{0w,7.5/13h}{v1}
\fmfforce{3/11w,7.5/13h}{v2}
\fmfforce{8/11w,7.5/13h}{v3}
\fmfforce{1w,7.5/13h}{v4}
\fmfforce{3/11w,2.5/13h}{v5}
\fmfforce{8/11w,2.5/13h}{v6}
\fmfforce{1/2w,10/13h}{v7}
\fmfforce{1/2w,1h}{v8}
\fmf{electron}{v2,v1}
\fmf{electron}{v4,v3}
\fmf{boson}{v8,v7}
\fmf{boson}{v2,v5}
\fmf{boson}{v3,v6}
\fmf{electron,right=1}{v6,v5,v6}
\fmf{electron,right=0.4}{v3,v7}
\fmf{electron,right=0.4}{v7,v2}
\fmfdot{v2,v3,v5,v6,v7}
\fmfv{decor.size=0, label=${\scs 3}$, l.dist=1mm, l.angle=90}{v8}
\fmfv{decor.size=0, label=${\scs 1}$, l.dist=1mm, l.angle=-180}{v1}
\fmfv{decor.size=0, label=${\scs 2}$, l.dist=1mm, l.angle=0}{v4}
\end{fmfgraph*}
}}
%
%
\hspace*{0.3cm} + \hspace*{0.3cm}
\parbox{14mm}{\centerline{
\begin{fmfgraph*}(11,7)
\setval
\fmfforce{0w,2/7h}{v1}
\fmfforce{3/11w,2/7h}{v2}
\fmfforce{8/11w,2/7h}{v3}
\fmfforce{1w,2/7h}{v4}
\fmfforce{3/11w,1h}{v5}
\fmfforce{8/11w,1h}{v6}
\fmfforce{9/11w,4.5/7h}{v7}
\fmfforce{1w,1h}{v8}
\fmf{electron}{v2,v1}
\fmf{electron}{v4,v3}
\fmf{boson}{v8,v7}
\fmf{boson,right=0.4}{v2,v3}
\fmf{fermion,right=0.4}{v5,v2}
\fmf{fermion,right=0.4}{v6,v5}
\fmf{boson,right=0.4}{v5,v6}
\fmf{fermion,right=0.2}{v3,v7}
\fmf{fermion,right=0.2}{v7,v6}
\fmfdot{v2,v3,v5,v6,v7}
\fmfv{decor.size=0, label=${\scs 3}$, l.dist=1mm, l.angle=45}{v8}
\fmfv{decor.size=0, label=${\scs 1}$, l.dist=1mm, l.angle=-180}{v1}
\fmfv{decor.size=0, label=${\scs 2}$, l.dist=1mm, l.angle=0}{v4}
\end{fmfgraph*}
}}
%
%
\hspace*{0.3cm} + \hspace*{0.3cm}
\parbox{14mm}{\centerline{
\begin{fmfgraph*}(11,7)
\setval
\fmfforce{0w,2/7h}{v1}
\fmfforce{3/11w,2/7h}{v2}
\fmfforce{8/11w,2/7h}{v3}
\fmfforce{1w,2/7h}{v4}
\fmfforce{3/11w,1h}{v5}
\fmfforce{8/11w,1h}{v6}
\fmfforce{2/11w,4.5/7h}{v7}
\fmfforce{0w,1h}{v8}
\fmf{electron}{v2,v1}
\fmf{electron}{v4,v3}
\fmf{boson}{v8,v7}
\fmf{boson,right=0.4}{v2,v3}
\fmf{fermion,right=0.4}{v5,v2}
\fmf{fermion,right=0.4}{v6,v5}
\fmf{fermion,right=0.4}{v3,v6}
\fmf{boson,right=0.4}{v5,v6}
\fmf{fermion,right=0.2}{v7,v2}
\fmf{fermion,right=0.2}{v5,v7}
\fmfdot{v2,v3,v5,v6,v7}
\fmfv{decor.size=0, label=${\scs 3}$, l.dist=1mm, l.angle=135}{v8}
\fmfv{decor.size=0, label=${\scs 1}$, l.dist=1mm, l.angle=-180}{v1}
\fmfv{decor.size=0, label=${\scs 2}$, l.dist=1mm, l.angle=0}{v4}
\end{fmfgraph*}
}}
\hspace*{0.4cm} . \la{VVV2}
\eeq
As a consequence, we obtain from (\r{R3}) and (\r{R4}) the three-loop contribution of the electron self-energy
\beq
\parbox{18mm}{\centerline{
\begin{fmfgraph*}(11,5)
\setval
\fmfforce{0w,1/2h}{v1}
\fmfforce{3/11w,1/2h}{v2}
\fmfforce{8/11w,1/2h}{v3}
\fmfforce{1w,1/2h}{v4}
\fmfforce{1/2w,1/2h}{v5}
\fmf{fermion}{v2,v1}
\fmf{fermion}{v4,v3}
\fmf{double,width=0.2mm,left=1}{v2,v3,v2}
\fmfv{decor.size=0, label=${\scs 3}$, l.dist=0mm, l.angle=0}{v5}
\fmfv{decor.size=0, label=${\scs 1}$, l.dist=1mm, l.angle=-180}{v1}
\fmfv{decor.size=0, label=${\scs 2}$, l.dist=1mm, l.angle=0}{v4}
\end{fmfgraph*} }}
\quad & = & \quad
%
%
\parbox{15mm}{\centerline{
\begin{fmfgraph*}(8,8)
\setval
\fmfforce{0w,1/2h}{v1}
\fmfforce{1w,1/2h}{v2}
\fmfforce{3/4w,0.933h}{v3}
\fmfforce{1/4w,0.933h}{v4}
\fmfforce{3/4w,0.067h}{v5}
\fmfforce{1/4w,0.067h}{v6}
\fmfforce{-0.1w,0.067h}{v7}
\fmfforce{1.1w,0.067h}{v8}
\fmf{fermion,right=0.3}{v5,v2}
\fmf{fermion,right=0.3}{v2,v3}
\fmf{fermion,right=0.3}{v3,v4}
\fmf{fermion,right=0.3}{v4,v1}
\fmf{fermion,right=0.3}{v1,v6}
\fmf{boson}{v4,v5}
\fmf{boson}{v3,v6}
\fmf{boson}{v2,v1}
\fmf{electron}{v6,v7}
\fmf{electron}{v8,v5}
\fmfv{decor.size=0, label=${\scs 1}$, l.dist=1mm, l.angle=-180}{v7}
\fmfv{decor.size=0, label=${\scs 2}$, l.dist=1mm, l.angle=0}{v8}
\fmfdot{v1,v2,v3,v4,v5,v6}
\end{fmfgraph*}
}}
%
%
\hspace*{0.3cm}- \hspace*{0.3cm}
\parbox{19mm}{\centerline{
\begin{fmfgraph*}(16,10)
\setval
\fmfforce{0w,0h}{v1}
\fmfforce{3/16w,0h}{v2}
\fmfforce{8/16w,0h}{v3}
\fmfforce{13/16w,0h}{v4}
\fmfforce{1w,0h}{v5}
\fmfforce{3/16w,1/2h}{v6}
\fmfforce{8/16w,1/2h}{v7}
\fmfforce{13/16w,1/2h}{v8}
\fmf{fermion}{v5,v4,v3,v2,v1}
\fmf{fermion}{v8,v7,v6}
\fmf{boson}{v2,v6}
\fmf{boson}{v3,v7}
\fmf{boson}{v4,v8}
\fmf{fermion,left=0.5}{v6,v8}
\fmfv{decor.size=0, label=${\scs 1}$, l.dist=1mm, l.angle=-180}{v1}
\fmfv{decor.size=0, label=${\scs 2}$, l.dist=1mm, l.angle=0}{v5}
\fmfdot{v2,v3,v4,v6,v7,v8}
\end{fmfgraph*}
}}
%
%
\hspace*{0.3cm} - \hspace*{0.3cm}
\parbox{19mm}{\centerline{
\begin{fmfgraph*}(16,10)
\setval
\fmfforce{0w,0h}{v1}
\fmfforce{3/16w,0h}{v2}
\fmfforce{8/16w,0h}{v3}
\fmfforce{13/16w,0h}{v4}
\fmfforce{1w,0h}{v5}
\fmfforce{3/16w,1/2h}{v6}
\fmfforce{8/16w,1/2h}{v7}
\fmfforce{13/16w,1/2h}{v8}
\fmf{fermion}{v5,v4,v3,v2,v1}
\fmf{fermion}{v6,v7,v8}
\fmf{boson}{v2,v6}
\fmf{boson}{v3,v7}
\fmf{boson}{v4,v8}
\fmf{fermion,right=0.5}{v8,v6}
\fmfv{decor.size=0, label=${\scs 1}$, l.dist=1mm, l.angle=-180}{v1}
\fmfv{decor.size=0, label=${\scs 2}$, l.dist=1mm, l.angle=0}{v5}
\fmfdot{v2,v3,v4,v6,v7,v8}
\end{fmfgraph*}
}}
%
%
\hspace*{0.3cm} + \hspace*{0.3cm}
\parbox{15mm}{\centerline{
\begin{fmfgraph*}(8,8)
\setval
\fmfforce{0w,1/2h}{v1}
\fmfforce{1w,1/2h}{v2}
\fmfforce{3/4w,0.933h}{v3}
\fmfforce{1/4w,0.933h}{v4}
\fmfforce{3/4w,0.067h}{v5}
\fmfforce{1/4w,0.067h}{v6}
\fmfforce{-0.1w,0.067h}{v7}
\fmfforce{1.1w,0.067h}{v8}
\fmf{fermion,right=0.3}{v5,v2}
\fmf{fermion,right=0.3}{v2,v3}
\fmf{fermion,right=0.3}{v3,v4}
\fmf{fermion,right=0.3}{v4,v1}
\fmf{fermion,right=0.3}{v1,v6}
\fmf{boson}{v6,v3}
\fmf{boson}{v1,v5}
\fmf{boson}{v2,v4}
\fmf{electron}{v6,v7}
\fmf{electron}{v8,v5}
\fmfv{decor.size=0, label=${\scs 1}$, l.dist=1mm, l.angle=-180}{v7}
\fmfv{decor.size=0, label=${\scs 2}$, l.dist=1mm, l.angle=0}{v8}
\fmfdot{v1,v2,v3,v4,v5,v6}
\end{fmfgraph*}
}}
%
%
\hspace*{0.3cm} + \hspace*{0.3cm}
\parbox{15mm}{\centerline{
\begin{fmfgraph*}(8,8)
\setval
\fmfforce{0w,1/2h}{v1}
\fmfforce{1w,1/2h}{v2}
\fmfforce{3/4w,0.933h}{v3}
\fmfforce{1/4w,0.933h}{v4}
\fmfforce{3/4w,0.067h}{v5}
\fmfforce{1/4w,0.067h}{v6}
\fmfforce{-0.1w,0.067h}{v7}
\fmfforce{1.1w,0.067h}{v8}
\fmf{fermion,right=0.3}{v5,v2}
\fmf{fermion,right=0.3}{v2,v3}
\fmf{fermion,right=0.3}{v3,v4}
\fmf{fermion,right=0.3}{v4,v1}
\fmf{fermion,right=0.3}{v1,v6}
\fmf{boson}{v4,v5}
\fmf{boson}{v1,v3}
\fmf{boson}{v2,v6}
\fmf{electron}{v6,v7}
\fmf{electron}{v8,v5}
\fmfv{decor.size=0, label=${\scs 1}$, l.dist=1mm, l.angle=-180}{v7}
\fmfv{decor.size=0, label=${\scs 2}$, l.dist=1mm, l.angle=0}{v8}
\fmfdot{v1,v2,v3,v4,v5,v6}
\end{fmfgraph*}
}}
\no \\
&& \quad \no \\
&& \quad 
\no \\
&& \quad 
%
%
+ \hspace*{0.3cm}
\parbox{15mm}{\centerline{
\begin{fmfgraph*}(8,8)
\setval
\fmfforce{0w,1/2h}{v1}
\fmfforce{1w,1/2h}{v2}
\fmfforce{3/4w,0.933h}{v3}
\fmfforce{1/4w,0.933h}{v4}
\fmfforce{3/4w,0.067h}{v5}
\fmfforce{1/4w,0.067h}{v6}
\fmfforce{-0.1w,0.067h}{v7}
\fmfforce{1.1w,0.067h}{v8}
\fmf{fermion,right=0.3}{v5,v2}
\fmf{fermion,right=0.3}{v2,v3}
\fmf{fermion,right=0.3}{v3,v4}
\fmf{fermion,right=0.3}{v4,v1}
\fmf{fermion,right=0.3}{v1,v6}
\fmf{boson}{v4,v6}
\fmf{boson}{v3,v5}
\fmf{boson}{v2,v1}
\fmf{electron}{v6,v7}
\fmf{electron}{v8,v5}
\fmfv{decor.size=0, label=${\scs 1}$, l.dist=1mm, l.angle=-180}{v7}
\fmfv{decor.size=0, label=${\scs 2}$, l.dist=1mm, l.angle=0}{v8}
\fmfdot{v1,v2,v3,v4,v5,v6}
\end{fmfgraph*}
}}
%
%
\hspace*{0.3cm}+ \hspace*{0.3cm}
\parbox{15mm}{\centerline{
\begin{fmfgraph*}(8,8)
\setval
\fmfforce{0w,1/2h}{v1}
\fmfforce{1w,1/2h}{v2}
\fmfforce{3/4w,0.933h}{v3}
\fmfforce{1/4w,0.933h}{v4}
\fmfforce{3/4w,0.067h}{v5}
\fmfforce{1/4w,0.067h}{v6}
\fmfforce{-0.1w,0.067h}{v7}
\fmfforce{1.1w,0.067h}{v8}
\fmf{boson,right=0.3}{v3,v4}
\fmf{fermion,left=1}{v3,v2,v3}
\fmf{fermion,right=1}{v4,v1,v4}
\fmf{boson,right=0.3}{v1,v6}
\fmf{boson,right=0.3}{v5,v2}
\fmf{fermion,left=0.3}{v5,v6}
\fmf{electron}{v6,v7}
\fmf{electron}{v8,v5}
\fmfv{decor.size=0, label=${\scs 1}$, l.dist=1mm, l.angle=-180}{v7}
\fmfv{decor.size=0, label=${\scs 2}$, l.dist=1mm, l.angle=0}{v8}
\fmfdot{v1,v2,v3,v4,v5,v6}
\end{fmfgraph*}
}}
%
%
\hspace*{0.3cm}- \hspace*{0.3cm} 
\parbox{15mm}{\centerline{
\begin{fmfgraph*}(11,7.5)
\setval
\fmfforce{0w,0h}{v1}
\fmfforce{3/11w,0h}{v2}
\fmfforce{8/11w,0h}{v3}
\fmfforce{1w,0h}{v4}
\fmfforce{3/11w,2/3h}{v5}
\fmfforce{8/11w,2/3h}{v6}
\fmfforce{1/2w,1/3h}{v7}
\fmfforce{1/2w,1h}{v8}
\fmf{fermion}{v4,v3,v2,v1}
\fmf{boson}{v2,v5}
\fmf{boson}{v3,v6}
\fmf{fermion,right=0.4}{v5,v7}
\fmf{fermion,right=0.4}{v7,v6}
\fmf{fermion,right=0.4}{v6,v8}
\fmf{fermion,right=0.4}{v8,v5}
\fmf{boson}{v7,v8}
\fmfv{decor.size=0, label=${\scs 1}$, l.dist=1mm, l.angle=-180}{v1}
\fmfv{decor.size=0, label=${\scs 2}$, l.dist=1mm, l.angle=0}{v4}
\fmfdot{v2,v3,v5,v6,v7,v8}
\end{fmfgraph*}
}}
%
%
\hspace*{0.3cm} - \hspace*{0.3cm} 
\parbox{19mm}{\centerline{
\begin{fmfgraph*}(16,10)
\setval
\fmfforce{0w,1/4h}{v1}
\fmfforce{3/16w,1/4h}{v2}
\fmfforce{8/16w,1/4h}{v3}
\fmfforce{13/16w,1/4h}{v4}
\fmfforce{1w,1/4h}{v5}
\fmfforce{8/16w,3/4h}{v6}
\fmfforce{13/16w,3/4h}{v7}
\fmfforce{10.5/16w,0h}{v8}
\fmf{fermion}{v3,v2,v1}
\fmf{fermion}{v5,v4}
\fmf{boson}{v3,v6}
\fmf{boson}{v4,v7}
\fmf{boson,right=0.4}{v2,v8}
\fmf{fermion,left=0.4}{v8,v3}
\fmf{fermion,left=0.4}{v4,v8}
\fmf{fermion,right=1}{v6,v7,v6}
\fmfv{decor.size=0, label=${\scs 1}$, l.dist=1mm, l.angle=-180}{v1}
\fmfv{decor.size=0, label=${\scs 2}$, l.dist=1mm, l.angle=0}{v5}
\fmfdot{v2,v3,v4,v6,v7,v8}
\end{fmfgraph*}
}}
%
%
\hspace*{0.3cm}- \hspace*{0.3cm} 
\parbox{19mm}{\centerline{
\begin{fmfgraph*}(16,10)
\setval
\fmfforce{0w,1/4h}{v1}
\fmfforce{3/16w,1/4h}{v2}
\fmfforce{8/16w,1/4h}{v3}
\fmfforce{13/16w,1/4h}{v4}
\fmfforce{1w,1/4h}{v5}
\fmfforce{3/16w,3/4h}{v6}
\fmfforce{8/16w,3/4h}{v7}
\fmfforce{5.5/16w,0h}{v8}
\fmf{fermion}{v2,v1}
\fmf{fermion}{v5,v4,v3}
\fmf{boson}{v2,v6}
\fmf{boson}{v3,v7}
\fmf{boson,right=0.4}{v8,v4}
\fmf{fermion,left=0.4}{v3,v8}
\fmf{fermion,left=0.4}{v8,v2}
\fmf{fermion,right=1}{v6,v7,v6}
\fmfv{decor.size=0, label=${\scs 1}$, l.dist=1mm, l.angle=-180}{v1}
\fmfv{decor.size=0, label=${\scs 2}$, l.dist=1mm, l.angle=0}{v5}
\fmfdot{v2,v3,v4,v6,v7,v8}
\end{fmfgraph*}
}}
\no \\
&& \quad 
\no \\
&& \quad \no \\
&& \quad 
%
%
+ \hspace*{0.3cm}
\parbox{15mm}{\centerline{
\begin{fmfgraph*}(8,8)
\setval
\fmfforce{0w,1/2h}{v1}
\fmfforce{1w,1/2h}{v2}
\fmfforce{3/4w,0.933h}{v3}
\fmfforce{1/4w,0.933h}{v4}
\fmfforce{3/4w,0.067h}{v5}
\fmfforce{1/4w,0.067h}{v6}
\fmfforce{-0.1w,0.067h}{v7}
\fmfforce{1.1w,0.067h}{v8}
\fmf{fermion,right=0.3}{v2,v3}
\fmf{boson,left=0.3}{v2,v3}
\fmf{fermion,right=0.3}{v3,v4}
\fmf{fermion,right=0.3}{v4,v1}
\fmf{boson,left=0.3}{v4,v1}
\fmf{fermion,right=0.3}{v1,v6}
\fmf{boson,right=0.3}{v6,v5}
\fmf{fermion,right=0.3}{v5,v2}
\fmf{electron}{v6,v7}
\fmf{electron}{v8,v5}
\fmfv{decor.size=0, label=${\scs 1}$, l.dist=1mm, l.angle=-180}{v7}
\fmfv{decor.size=0, label=${\scs 2}$, l.dist=1mm, l.angle=0}{v8}
\fmfdot{v1,v2,v3,v4,v5,v6}
\end{fmfgraph*}
}}
%
%
\hspace*{0.3cm}+ \hspace*{0.3cm}
\parbox{15mm}{\centerline{
\begin{fmfgraph*}(8,8)
\setval
\fmfforce{0w,1/2h}{v1}
\fmfforce{1w,1/2h}{v2}
\fmfforce{3/4w,0.933h}{v3}
\fmfforce{1/4w,0.933h}{v4}
\fmfforce{3/4w,0.067h}{v5}
\fmfforce{1/4w,0.067h}{v6}
\fmfforce{-0.1w,0.067h}{v7}
\fmfforce{1.1w,0.067h}{v8}
\fmf{fermion,right=0.3}{v3,v4}
\fmf{boson,left=0.3}{v3,v4}
\fmf{fermion,right=0.3}{v2,v3}
\fmf{fermion,right=0.3}{v4,v1}
\fmf{boson}{v2,v1}
\fmf{fermion,right=0.3}{v1,v6}
\fmf{boson,right=0.3}{v6,v5}
\fmf{fermion,right=0.3}{v5,v2}
\fmf{electron}{v6,v7}
\fmf{electron}{v8,v5}
\fmfv{decor.size=0, label=${\scs 1}$, l.dist=1mm, l.angle=-180}{v7}
\fmfv{decor.size=0, label=${\scs 2}$, l.dist=1mm, l.angle=0}{v8}
\fmfdot{v1,v2,v3,v4,v5,v6}
\end{fmfgraph*}
}}
%
%
\hspace*{0.3cm} + \hspace*{0.3cm}
\parbox{15mm}{\centerline{
\begin{fmfgraph*}(8,8)
\setval
\fmfforce{0w,1/2h}{v1}
\fmfforce{1w,1/2h}{v2}
\fmfforce{3/4w,0.933h}{v3}
\fmfforce{1/4w,0.933h}{v4}
\fmfforce{3/4w,0.067h}{v5}
\fmfforce{1/4w,0.067h}{v6}
\fmfforce{-0.1w,0.067h}{v7}
\fmfforce{1.1w,0.067h}{v8}
\fmf{fermion,right=0.3}{v3,v4}
\fmf{boson}{v3,v1}
\fmf{fermion,right=0.3}{v2,v3}
\fmf{fermion,right=0.3}{v4,v1}
\fmf{boson}{v4,v2}
\fmf{fermion,right=0.3}{v1,v6}
\fmf{boson,right=0.3}{v6,v5}
\fmf{fermion,right=0.3}{v5,v2}
\fmf{electron}{v6,v7}
\fmf{electron}{v8,v5}
\fmfv{decor.size=0, label=${\scs 1}$, l.dist=1mm, l.angle=-180}{v7}
\fmfv{decor.size=0, label=${\scs 2}$, l.dist=1mm, l.angle=0}{v8}
\fmfdot{v1,v2,v3,v4,v5,v6}
\end{fmfgraph*}
}}
%
%
\hspace*{0.3cm}- \hspace*{0.3cm}
\parbox{15mm}{\centerline{
\begin{fmfgraph*}(8,8)
\setval
\fmfforce{0w,1/2h}{v1}
\fmfforce{1w,1/2h}{v2}
\fmfforce{3/4w,0.933h}{v3}
\fmfforce{1/4w,0.933h}{v4}
\fmfforce{3/4w,0.067h}{v5}
\fmfforce{1/4w,0.067h}{v6}
\fmfforce{-0.1w,0.067h}{v7}
\fmfforce{1.1w,0.067h}{v8}
\fmf{fermion,right=0.3}{v3,v4}
\fmf{boson,left=0.3}{v3,v4}
\fmf{fermion,right=0.3}{v2,v3}
\fmf{fermion,right=0.3}{v4,v1}
\fmf{boson,right=0.3}{v1,v6}
\fmf{boson,right=0.3}{v5,v2}
\fmf{fermion,left=0.3}{v5,v6}
\fmf{electron}{v1,v2}
\fmf{electron}{v6,v7}
\fmf{electron}{v8,v5}
\fmfv{decor.size=0, label=${\scs 1}$, l.dist=1mm, l.angle=-180}{v7}
\fmfv{decor.size=0, label=${\scs 2}$, l.dist=1mm, l.angle=0}{v8}
\fmfdot{v1,v2,v3,v4,v5,v6}
\end{fmfgraph*}
}}
%
%
\hspace*{0.3cm}- \hspace*{0.3cm} 
\parbox{15mm}{\centerline{
\begin{fmfgraph*}(8,8)
\setval
\fmfforce{0w,1/2h}{v1}
\fmfforce{1w,1/2h}{v2}
\fmfforce{3/4w,0.933h}{v3}
\fmfforce{1/4w,0.933h}{v4}
\fmfforce{3/4w,0.067h}{v5}
\fmfforce{1/4w,0.067h}{v6}
\fmfforce{-0.1w,0.067h}{v7}
\fmfforce{1.1w,0.067h}{v8}
\fmf{fermion,left=0.3}{v4,v3}
\fmf{boson,left=0.3}{v3,v4}
\fmf{fermion,left=0.3}{v3,v2}
\fmf{fermion,left=0.3}{v1,v4}
\fmf{boson,right=0.3}{v1,v6}
\fmf{boson,right=0.3}{v5,v2}
\fmf{fermion,left=0.3}{v5,v6}
\fmf{electron}{v2,v1}
\fmf{electron}{v6,v7}
\fmf{electron}{v8,v5}
\fmfv{decor.size=0, label=${\scs 1}$, l.dist=1mm, l.angle=-180}{v7}
\fmfv{decor.size=0, label=${\scs 2}$, l.dist=1mm, l.angle=0}{v8}
\fmfdot{v1,v2,v3,v4,v5,v6}
\end{fmfgraph*}
}}
\no \\
&& \quad \vspace*{0.2cm}
\no \\&& \quad \vspace*{0.2cm}
\no \\
&& \quad 
%
%
+ \hspace*{0.3cm}
\parbox{15mm}{\centerline{
\begin{fmfgraph*}(8,8)
\setval
\fmfforce{0w,1/2h}{v1}
\fmfforce{1w,1/2h}{v2}
\fmfforce{3/4w,0.933h}{v3}
\fmfforce{1/4w,0.933h}{v4}
\fmfforce{3/4w,0.067h}{v5}
\fmfforce{1/4w,0.067h}{v6}
\fmfforce{-0.1w,0.067h}{v7}
\fmfforce{1.1w,0.067h}{v8}
\fmf{fermion,right=0.3}{v5,v2}
\fmf{fermion,right=0.3}{v2,v3}
\fmf{fermion,right=0.3}{v3,v4}
\fmf{fermion,right=0.3}{v4,v1}
\fmf{fermion,right=0.3}{v1,v6}
\fmf{boson,left=0.3}{v2,v3}
\fmf{boson}{v4,v6}
\fmf{boson}{v1,v5}
\fmf{electron}{v6,v7}
\fmf{electron}{v8,v5}
\fmfv{decor.size=0, label=${\scs 1}$, l.dist=1mm, l.angle=-180}{v7}
\fmfv{decor.size=0, label=${\scs 2}$, l.dist=1mm, l.angle=0}{v8}
\fmfdot{v1,v2,v3,v4,v5,v6}
\end{fmfgraph*}
}}
%
%
\hspace*{0.3cm}+ \hspace*{0.3cm}
\parbox{15mm}{\centerline{
\begin{fmfgraph*}(8,8)
\setval
\fmfforce{0w,1/2h}{v1}
\fmfforce{1w,1/2h}{v2}
\fmfforce{3/4w,0.933h}{v3}
\fmfforce{1/4w,0.933h}{v4}
\fmfforce{3/4w,0.067h}{v5}
\fmfforce{1/4w,0.067h}{v6}
\fmfforce{-0.1w,0.067h}{v7}
\fmfforce{1.1w,0.067h}{v8}
\fmf{fermion,right=0.3}{v5,v2}
\fmf{fermion,right=0.3}{v2,v3}
\fmf{fermion,right=0.3}{v3,v4}
\fmf{fermion,right=0.3}{v4,v1}
\fmf{fermion,right=0.3}{v1,v6}
\fmf{boson,right=0.3}{v4,v3}
\fmf{boson}{v1,v5}
\fmf{boson}{v2,v6}
\fmf{electron}{v6,v7}
\fmf{electron}{v8,v5}
\fmfv{decor.size=0, label=${\scs 1}$, l.dist=1mm, l.angle=-180}{v7}
\fmfv{decor.size=0, label=${\scs 2}$, l.dist=1mm, l.angle=0}{v8}
\fmfdot{v1,v2,v3,v4,v5,v6}
\end{fmfgraph*}
}}
%
%
\hspace*{0.3cm} + \hspace*{0.3cm}
\parbox{15mm}{\centerline{
\begin{fmfgraph*}(8,8)
\setval
\fmfforce{0w,1/2h}{v1}
\fmfforce{1w,1/2h}{v2}
\fmfforce{3/4w,0.933h}{v3}
\fmfforce{1/4w,0.933h}{v4}
\fmfforce{3/4w,0.067h}{v5}
\fmfforce{1/4w,0.067h}{v6}
\fmfforce{-0.1w,0.067h}{v7}
\fmfforce{1.1w,0.067h}{v8}
\fmf{fermion,right=0.3}{v5,v2}
\fmf{fermion,right=0.3}{v2,v3}
\fmf{fermion,right=0.3}{v3,v4}
\fmf{fermion,right=0.3}{v4,v1}
\fmf{fermion,right=0.3}{v1,v6}
\fmf{boson,left=0.3}{v4,v1}
\fmf{boson}{v3,v5}
\fmf{boson}{v2,v6}
\fmf{electron}{v6,v7}
\fmf{electron}{v8,v5}
\fmfv{decor.size=0, label=${\scs 1}$, l.dist=1mm, l.angle=-180}{v7}
\fmfv{decor.size=0, label=${\scs 2}$, l.dist=1mm, l.angle=0}{v8}
\fmfdot{v1,v2,v3,v4,v5,v6}
\end{fmfgraph*}
}}
%
%
\hspace*{0.3cm}- \hspace*{0.3cm}
\parbox{15mm}{\centerline{
\begin{fmfgraph*}(8,8)
\setval
\fmfforce{0w,1/2h}{v1}
\fmfforce{1w,1/2h}{v2}
\fmfforce{3/4w,0.933h}{v3}
\fmfforce{1/4w,0.933h}{v4}
\fmfforce{3/4w,0.067h}{v5}
\fmfforce{1/4w,0.067h}{v6}
\fmfforce{-0.1w,0.067h}{v7}
\fmfforce{1.1w,0.067h}{v8}
\fmf{fermion,right=1}{v4,v3,v4}
\fmf{boson,right=0.3}{v2,v3}
\fmf{boson,right=0.3}{v4,v1}
\fmf{fermion}{v2,v1}
\fmf{fermion,right=0.3}{v1,v6}
\fmf{boson,right=0.3}{v6,v5}
\fmf{fermion,right=0.3}{v5,v2}
\fmf{electron}{v6,v7}
\fmf{electron}{v8,v5}
\fmfv{decor.size=0, label=${\scs 1}$, l.dist=1mm, l.angle=-180}{v7}
\fmfv{decor.size=0, label=${\scs 2}$, l.dist=1mm, l.angle=0}{v8}
\fmfdot{v1,v2,v3,v4,v5,v6}
\end{fmfgraph*}
}}
%
%
\hspace*{0.3cm} - \hspace*{0.3cm}
\parbox{18mm}{\centerline{
\begin{fmfgraph*}(8,8)
\setval
\fmfforce{0w,1/2h}{v1}
\fmfforce{1w,1/2h}{v2}
\fmfforce{3/4w,0.933h}{v3}
\fmfforce{1/4w,0.933h}{v4}
\fmfforce{3/4w,0.067h}{v5}
\fmfforce{1/4w,0.067h}{v6}
\fmfforce{-3/8w,1/2h}{v7}
\fmfforce{11/8w,1/2h}{v8}
\fmf{fermion,right=1}{v4,v3,v4}
\fmf{fermion,left=0.3}{v6,v1}
\fmf{fermion,left=0.3}{v5,v6}
\fmf{fermion,left=0.3}{v2,v5}
\fmf{boson,right=0.3}{v4,v1}
\fmf{boson,right=0.3}{v2,v3}
\fmf{boson,left=0.3}{v6,v5}
\fmf{electron}{v1,v7}
\fmf{electron}{v8,v2}
\fmfv{decor.size=0, label=${\scs 1}$, l.dist=1mm, l.angle=-180}{v7}
\fmfv{decor.size=0, label=${\scs 2}$, l.dist=1mm, l.angle=0}{v8}
\fmfdot{v1,v2,v3,v4,v5,v6}
\end{fmfgraph*}
}}
\la{SELF3}
\eeq
\end{fmffile}
\begin{fmffile}{sd15}
\hspace*{-0.3cm} and of the photon self-energy
\beq
\parbox{18mm}{\centerline{
\begin{fmfgraph*}(11,5)
\setval
\fmfforce{0w,1/2h}{v1}
\fmfforce{3/11w,1/2h}{v2}
\fmfforce{8/11w,1/2h}{v3}
\fmfforce{1w,1/2h}{v4}
\fmfforce{1/2w,1/2h}{v5}
\fmf{boson}{v2,v1}
\fmf{boson}{v4,v3}
\fmf{double,width=0.2mm,left=1}{v2,v3,v2}
\fmfv{decor.size=0, label=${\scs 3}$, l.dist=0mm, l.angle=0}{v5}
\fmfv{decor.size=0, label=${\scs 1}$, l.dist=1mm, l.angle=-180}{v1}
\fmfv{decor.size=0, label=${\scs 2}$, l.dist=1mm, l.angle=0}{v4}
\end{fmfgraph*} }}
\quad & = & \quad
%
%
\parbox{21.5mm}{\centerline{
\begin{fmfgraph*}(18.5,5)
\setval
\fmfforce{0w,1/2h}{v1}
\fmfforce{3/18.5w,1/2h}{v2}
\fmfforce{5.5/18.5w,0h}{v3}
\fmfforce{5.5/18.5w,1h}{v4}
\fmfforce{13/18.5w,0h}{v5}
\fmfforce{13/18.5w,1h}{v6}
\fmfforce{15.5/18.5w,1/2h}{v7}
\fmfforce{1w,1/2h}{v8}
\fmf{boson}{v1,v2}
\fmf{fermion,right=0.4}{v4,v2}
\fmf{fermion,right=0.4}{v2,v3}
\fmf{fermion,right=1}{v3,v4}
\fmf{boson}{v4,v6}
\fmf{boson}{v3,v5}
\fmf{fermion,right=0.4}{v7,v6}
\fmf{fermion,right=0.4}{v5,v7}
\fmf{fermion,right=1}{v6,v5}
\fmf{boson}{v8,v7}
\fmfv{decor.size=0, label=${\scs 1}$, l.dist=1mm, l.angle=-180}{v1}
\fmfv{decor.size=0, label=${\scs 2}$, l.dist=1mm, l.angle=0}{v8}
\fmfdot{v2,v3,v4,v5,v6,v7}
\end{fmfgraph*}
}}
%
%
\hspace*{0.3cm} + \hspace*{0.3cm}
\parbox{21.5mm}{\centerline{
\begin{fmfgraph*}(18.5,5)
\setval
\fmfforce{0w,1/2h}{v1}
\fmfforce{3/18.5w,1/2h}{v2}
\fmfforce{5.5/18.5w,0h}{v3}
\fmfforce{5.5/18.5w,1h}{v4}
\fmfforce{13/18.5w,0h}{v5}
\fmfforce{13/18.5w,1h}{v6}
\fmfforce{15.5/18.5w,1/2h}{v7}
\fmfforce{1w,1/2h}{v8}
\fmf{boson}{v1,v2}
\fmf{fermion,right=0.4}{v4,v2}
\fmf{fermion,right=0.4}{v2,v3}
\fmf{fermion,right=1}{v3,v4}
\fmf{boson}{v4,v6}
\fmf{boson}{v3,v5}
\fmf{fermion,left=0.4}{v6,v7}
\fmf{fermion,left=0.4}{v7,v5}
\fmf{fermion,left=1}{v5,v6}
\fmf{boson}{v8,v7}
\fmfv{decor.size=0, label=${\scs 1}$, l.dist=1mm, l.angle=-180}{v1}
\fmfv{decor.size=0, label=${\scs 2}$, l.dist=1mm, l.angle=0}{v8}
\fmfdot{v2,v3,v4,v5,v6,v7}
\end{fmfgraph*}
}}
%
%
\hspace*{0.3cm} - \hspace*{0.3cm}
\parbox{18mm}{\centerline{
\begin{fmfgraph*}(8,8)
\setval
\fmfforce{0w,1/2h}{v1}
\fmfforce{1w,1/2h}{v2}
\fmfforce{3/4w,0.933h}{v3}
\fmfforce{1/4w,0.933h}{v4}
\fmfforce{3/4w,0.067h}{v5}
\fmfforce{1/4w,0.067h}{v6}
\fmfforce{-3/8w,1/2h}{v7}
\fmfforce{11/8w,1/2h}{v8}
\fmf{fermion,right=0.3}{v3,v4}
\fmf{fermion,right=0.3}{v1,v6}
\fmf{fermion,right=0.3}{v6,v5}
\fmf{fermion,right=0.3}{v5,v2}
\fmf{fermion,right=0.3}{v4,v1}
\fmf{fermion,right=0.3}{v2,v3}
\fmf{boson}{v4,v5}
\fmf{boson}{v3,v6}
\fmf{boson}{v8,v2}
\fmf{boson}{v7,v1}
\fmfv{decor.size=0, label=${\scs 1}$, l.dist=1mm, l.angle=-180}{v7}
\fmfv{decor.size=0, label=${\scs 2}$, l.dist=1mm, l.angle=0}{v8}
\fmfdot{v1,v2,v3,v4,v5,v6}
\end{fmfgraph*}
}}
%
%
\hspace*{0.3cm}- \hspace*{0.3cm}
\parbox{18mm}{\centerline{
\begin{fmfgraph*}(8,8)
\setval
\fmfforce{0w,1/2h}{v1}
\fmfforce{1w,1/2h}{v2}
\fmfforce{3/4w,0.933h}{v3}
\fmfforce{1/4w,0.933h}{v4}
\fmfforce{3/4w,0.067h}{v5}
\fmfforce{1/4w,0.067h}{v6}
\fmfforce{-3/8w,1/2h}{v7}
\fmfforce{11/8w,1/2h}{v8}
\fmf{fermion,right=0.3}{v3,v4}
\fmf{fermion,right=0.3}{v1,v6}
\fmf{fermion,right=0.3}{v6,v5}
\fmf{fermion,right=0.3}{v5,v2}
\fmf{fermion,right=0.3}{v4,v1}
\fmf{fermion,right=0.3}{v2,v3}
\fmf{boson}{v4,v6}
\fmf{boson}{v3,v5}
\fmf{boson}{v8,v2}
\fmf{boson}{v7,v1}
\fmfv{decor.size=0, label=${\scs 1}$, l.dist=1mm, l.angle=-180}{v7}
\fmfv{decor.size=0, label=${\scs 2}$, l.dist=1mm, l.angle=0}{v8}
\fmfdot{v1,v2,v3,v4,v5,v6}
\end{fmfgraph*}
}}
%
%
\hspace*{0.3cm}- \hspace*{0.3cm}
\parbox{18mm}{\centerline{
\begin{fmfgraph*}(8,8)
\setval
\fmfforce{1/2w,0h}{v1}
\fmfforce{1/2w,1h}{v2}
\fmfforce{0.933w,3/4h}{v3}
\fmfforce{0.933w,1/4h}{v4}
\fmfforce{0.067w,3/4h}{v5}
\fmfforce{0.067w,1/4h}{v6}
\fmfforce{1.233w,1/4h}{v7}
\fmfforce{-0.233w,1/4h}{v8}
\fmf{fermion,right=0.3}{v2,v5}
\fmf{fermion,right=0.3}{v3,v2}
\fmf{fermion,right=0.3}{v4,v3}
\fmf{fermion,right=0.3}{v1,v4}
\fmf{fermion,right=0.3}{v6,v1}
\fmf{fermion,right=0.3}{v5,v6}
\fmf{boson}{v1,v2}
\fmf{boson,left=0.6}{v3,v5}
\fmf{boson}{v6,v8}
\fmf{boson}{v4,v7}
\fmfv{decor.size=0, label=${\scs 1}$, l.dist=1mm, l.angle=-180}{v8}
\fmfv{decor.size=0, label=${\scs 2}$, l.dist=1mm, l.angle=0}{v7}
\fmfdot{v1,v2,v3,v4,v5,v6}
\end{fmfgraph*}
}}
\no \\
&& \quad 
\no \\
&& \quad 
\no \\
&& \quad 
%
%
- \hspace*{0.3cm}
\parbox{18mm}{\centerline{
\begin{fmfgraph*}(8,8)
\setval
\fmfforce{1/2w,0h}{v1}
\fmfforce{1/2w,1h}{v2}
\fmfforce{0.933w,3/4h}{v3}
\fmfforce{0.933w,1/4h}{v4}
\fmfforce{0.067w,3/4h}{v5}
\fmfforce{0.067w,1/4h}{v6}
\fmfforce{1.233w,1/4h}{v7}
\fmfforce{-0.233w,1/4h}{v8}
\fmf{fermion,left=0.3}{v5,v2}
\fmf{fermion,left=0.3}{v2,v3}
\fmf{fermion,left=0.3}{v3,v4}
\fmf{fermion,left=0.3}{v4,v1}
\fmf{fermion,left=0.3}{v1,v6}
\fmf{fermion,left=0.3}{v6,v5}
\fmf{boson}{v1,v2}
\fmf{boson,left=0.6}{v3,v5}
\fmf{boson}{v6,v8}
\fmf{boson}{v4,v7}
\fmfv{decor.size=0, label=${\scs 1}$, l.dist=1mm, l.angle=-180}{v8}
\fmfv{decor.size=0, label=${\scs 2}$, l.dist=1mm, l.angle=0}{v7}
\fmfdot{v1,v2,v3,v4,v5,v6}
\end{fmfgraph*}
}}
%
%
\hspace*{0.3cm}+ \hspace*{0.3cm}
\parbox{20mm}{\centerline{
\begin{fmfgraph*}(10,10)
\setval
\fmfforce{0w,1/2h}{v1}
\fmfforce{1w,1/2h}{v2}
\fmfforce{1/2w,0h}{v3}
\fmfforce{1/2w,1/4h}{v4}
\fmfforce{1/2w,3/4h}{v5}
\fmfforce{1/2w,1h}{v6}
\fmfforce{-0.3w,1/2h}{v7}
\fmfforce{1.3w,1/2h}{v8}
\fmf{fermion,right=0.4}{v1,v3}
\fmf{fermion,right=0.4}{v3,v2}
\fmf{fermion,right=0.4}{v2,v6}
\fmf{fermion,right=0.4}{v6,v1}
\fmf{fermion,right=1}{v4,v5,v4}
\fmf{boson}{v1,v7}
\fmf{boson}{v8,v2}
\fmf{boson}{v3,v4}
\fmf{boson}{v5,v6}
\fmfv{decor.size=0, label=${\scs 1}$, l.dist=1mm, l.angle=-180}{v7}
\fmfv{decor.size=0, label=${\scs 2}$, l.dist=1mm, l.angle=0}{v8}
\fmfdot{v1,v2,v3,v4,v5,v6}
\end{fmfgraph*}
}}
%
%
\hspace*{0.3cm}- \hspace*{0.3cm}
\parbox{15mm}{\centerline{
\begin{fmfgraph*}(8,8)
\setval
\fmfforce{0w,1/2h}{v1}
\fmfforce{1w,1/2h}{v2}
\fmfforce{3/4w,0.933h}{v3}
\fmfforce{1/4w,0.933h}{v4}
\fmfforce{3/4w,0.067h}{v5}
\fmfforce{1/4w,0.067h}{v6}
\fmfforce{-0.1w,0.067h}{v7}
\fmfforce{1.1w,0.067h}{v8}
\fmf{fermion,right=0.3}{v2,v3}
\fmf{boson,left=0.3}{v2,v3}
\fmf{fermion,right=0.3}{v3,v4}
\fmf{fermion,right=0.3}{v4,v1}
\fmf{boson,left=0.3}{v4,v1}
\fmf{fermion,right=0.3}{v1,v6}
\fmf{fermion,right=0.3}{v6,v5}
\fmf{fermion,right=0.3}{v5,v2}
\fmf{boson}{v6,v7}
\fmf{boson}{v8,v5}
\fmfv{decor.size=0, label=${\scs 1}$, l.dist=1mm, l.angle=-180}{v7}
\fmfv{decor.size=0, label=${\scs 2}$, l.dist=1mm, l.angle=0}{v8}
\fmfdot{v1,v2,v3,v4,v5,v6}
\end{fmfgraph*}
}}
%
%
\hspace*{0.3cm} - \hspace*{0.3cm}
\parbox{15mm}{\centerline{
\begin{fmfgraph*}(8,8)
\setval
\fmfforce{0w,1/2h}{v1}
\fmfforce{1w,1/2h}{v2}
\fmfforce{3/4w,0.933h}{v3}
\fmfforce{1/4w,0.933h}{v4}
\fmfforce{3/4w,0.067h}{v5}
\fmfforce{1/4w,0.067h}{v6}
\fmfforce{-0.1w,0.067h}{v7}
\fmfforce{1.1w,0.067h}{v8}
\fmf{fermion,left=0.3}{v3,v2}
\fmf{boson,left=0.3}{v2,v3}
\fmf{fermion,left=0.3}{v4,v3}
\fmf{fermion,left=0.3}{v1,v4}
\fmf{boson,left=0.3}{v4,v1}
\fmf{fermion,left=0.3}{v6,v1}
\fmf{fermion,left=0.3}{v5,v6}
\fmf{fermion,left=0.3}{v2,v5}
\fmf{boson}{v6,v7}
\fmf{boson}{v8,v5}
\fmfv{decor.size=0, label=${\scs 1}$, l.dist=1mm, l.angle=-180}{v7}
\fmfv{decor.size=0, label=${\scs 2}$, l.dist=1mm, l.angle=0}{v8}
\fmfdot{v1,v2,v3,v4,v5,v6}
\end{fmfgraph*}
}}
%
%
\hspace*{0.3cm} - \hspace*{0.3cm}
\parbox{15mm}{\centerline{
\begin{fmfgraph*}(8,8)
\setval
\fmfforce{0w,1/2h}{v1}
\fmfforce{1w,1/2h}{v2}
\fmfforce{3/4w,0.933h}{v3}
\fmfforce{1/4w,0.933h}{v4}
\fmfforce{3/4w,0.067h}{v5}
\fmfforce{1/4w,0.067h}{v6}
\fmfforce{-0.1w,0.067h}{v7}
\fmfforce{1.1w,0.067h}{v8}
\fmf{fermion,right=0.3}{v3,v4}
\fmf{boson,left=0.3}{v3,v4}
\fmf{fermion,right=0.3}{v2,v3}
\fmf{fermion,right=0.3}{v4,v1}
\fmf{boson}{v2,v1}
\fmf{fermion,right=0.3}{v1,v6}
\fmf{fermion,right=0.3}{v6,v5}
\fmf{fermion,right=0.3}{v5,v2}
\fmf{boson}{v6,v7}
\fmf{boson}{v8,v5}
\fmfv{decor.size=0, label=${\scs 1}$, l.dist=1mm, l.angle=-180}{v7}
\fmfv{decor.size=0, label=${\scs 2}$, l.dist=1mm, l.angle=0}{v8}
\fmfdot{v1,v2,v3,v4,v5,v6}
\end{fmfgraph*}
}}
\no \\
&& \quad 
\no \\
&& \quad 
\no \\
&& \quad 
%
%
- \hspace*{0.3cm}
\parbox{15mm}{\centerline{
\begin{fmfgraph*}(8,8)
\setval
\fmfforce{0w,1/2h}{v1}
\fmfforce{1w,1/2h}{v2}
\fmfforce{3/4w,0.933h}{v3}
\fmfforce{1/4w,0.933h}{v4}
\fmfforce{3/4w,0.067h}{v5}
\fmfforce{1/4w,0.067h}{v6}
\fmfforce{-0.1w,0.067h}{v7}
\fmfforce{1.1w,0.067h}{v8}
\fmf{fermion,left=0.3}{v4,v3}
\fmf{boson,left=0.3}{v3,v4}
\fmf{fermion,left=0.3}{v3,v2}
\fmf{fermion,left=0.3}{v1,v4}
\fmf{boson}{v2,v1}
\fmf{fermion,left=0.3}{v6,v1}
\fmf{fermion,left=0.3}{v5,v6}
\fmf{fermion,left=0.3}{v2,v5}
\fmf{boson}{v6,v7}
\fmf{boson}{v8,v5}
\fmfv{decor.size=0, label=${\scs 1}$, l.dist=1mm, l.angle=-180}{v7}
\fmfv{decor.size=0, label=${\scs 2}$, l.dist=1mm, l.angle=0}{v8}
\fmfdot{v1,v2,v3,v4,v5,v6}
\end{fmfgraph*}
}}
%
%
\hspace*{0.3cm} - \hspace*{0.3cm}
\parbox{15mm}{\centerline{
\begin{fmfgraph*}(8,8)
\setval
\fmfforce{0w,1/2h}{v1}
\fmfforce{1w,1/2h}{v2}
\fmfforce{3/4w,0.933h}{v3}
\fmfforce{1/4w,0.933h}{v4}
\fmfforce{3/4w,0.067h}{v5}
\fmfforce{1/4w,0.067h}{v6}
\fmfforce{-0.1w,0.067h}{v7}
\fmfforce{1.1w,0.067h}{v8}
\fmf{fermion,right=0.3}{v3,v4}
\fmf{boson}{v3,v1}
\fmf{fermion,right=0.3}{v2,v3}
\fmf{fermion,right=0.3}{v4,v1}
\fmf{boson}{v4,v2}
\fmf{fermion,right=0.3}{v1,v6}
\fmf{fermion,right=0.3}{v6,v5}
\fmf{fermion,right=0.3}{v5,v2}
\fmf{boson}{v6,v7}
\fmf{boson}{v8,v5}
\fmfv{decor.size=0, label=${\scs 1}$, l.dist=1mm, l.angle=-180}{v7}
\fmfv{decor.size=0, label=${\scs 2}$, l.dist=1mm, l.angle=0}{v8}
\fmfdot{v1,v2,v3,v4,v5,v6}
\end{fmfgraph*}
}}
%
%
\hspace*{0.3cm}- \hspace*{0.3cm}
\parbox{15mm}{\centerline{
\begin{fmfgraph*}(8,8)
\setval
\fmfforce{0w,1/2h}{v1}
\fmfforce{1w,1/2h}{v2}
\fmfforce{3/4w,0.933h}{v3}
\fmfforce{1/4w,0.933h}{v4}
\fmfforce{3/4w,0.067h}{v5}
\fmfforce{1/4w,0.067h}{v6}
\fmfforce{-0.1w,0.067h}{v7}
\fmfforce{1.1w,0.067h}{v8}
\fmf{fermion,left=0.3}{v4,v3}
\fmf{boson}{v3,v1}
\fmf{fermion,left=0.3}{v3,v2}
\fmf{fermion,left=0.3}{v1,v4}
\fmf{boson}{v4,v2}
\fmf{fermion,left=0.3}{v6,v1}
\fmf{fermion,left=0.3}{v5,v6}
\fmf{fermion,left=0.3}{v2,v5}
\fmf{boson}{v6,v7}
\fmf{boson}{v8,v5}
\fmfv{decor.size=0, label=${\scs 1}$, l.dist=1mm, l.angle=-180}{v7}
\fmfv{decor.size=0, label=${\scs 2}$, l.dist=1mm, l.angle=0}{v8}
\fmfdot{v1,v2,v3,v4,v5,v6}
\end{fmfgraph*}
}}
%
%
\hspace*{0.3cm} - \hspace*{0.3cm}
\parbox{18mm}{\centerline{
\begin{fmfgraph*}(8,8)
\setval
\fmfforce{0w,1/2h}{v1}
\fmfforce{1w,1/2h}{v2}
\fmfforce{3/4w,0.933h}{v3}
\fmfforce{1/4w,0.933h}{v4}
\fmfforce{3/4w,0.067h}{v5}
\fmfforce{1/4w,0.067h}{v6}
\fmfforce{-3/8w,1/2h}{v7}
\fmfforce{11/8w,1/2h}{v8}
\fmf{fermion,right=0.3}{v3,v4}
\fmf{fermion,right=0.3}{v1,v6}
\fmf{fermion,right=0.3}{v6,v5}
\fmf{fermion,right=0.3}{v5,v2}
\fmf{fermion,right=0.3}{v4,v1}
\fmf{fermion,right=0.3}{v2,v3}
\fmf{boson,right=0.3}{v4,v3}
\fmf{boson,right=0.3}{v5,v6}
\fmf{boson}{v8,v2}
\fmf{boson}{v7,v1}
\fmfv{decor.size=0, label=${\scs 1}$, l.dist=1mm, l.angle=-180}{v7}
\fmfv{decor.size=0, label=${\scs 2}$, l.dist=1mm, l.angle=0}{v8}
\fmfdot{v1,v2,v3,v4,v5,v6}
\end{fmfgraph*}
}}
%
%
\hspace*{0.3cm} - \hspace*{0.3cm}
\parbox{18mm}{\centerline{
\begin{fmfgraph*}(8,8)
\setval
\fmfforce{1/2w,0h}{v1}
\fmfforce{1/2w,1h}{v2}
\fmfforce{0.933w,3/4h}{v3}
\fmfforce{0.933w,1/4h}{v4}
\fmfforce{0.067w,3/4h}{v5}
\fmfforce{0.067w,1/4h}{v6}
\fmfforce{1.233w,1/4h}{v7}
\fmfforce{-0.233w,1/4h}{v8}
\fmf{fermion,right=0.3}{v2,v5}
\fmf{fermion,right=0.3}{v3,v2}
\fmf{fermion,right=0.3}{v4,v3}
\fmf{fermion,right=0.3}{v1,v4}
\fmf{fermion,right=0.3}{v6,v1}
\fmf{fermion,right=0.3}{v5,v6}
\fmf{boson}{v1,v5}
\fmf{boson,right=0.3}{v2,v3}
\fmf{boson}{v6,v8}
\fmf{boson}{v4,v7}
\fmfv{decor.size=0, label=${\scs 1}$, l.dist=1mm, l.angle=-180}{v8}
\fmfv{decor.size=0, label=${\scs 2}$, l.dist=1mm, l.angle=0}{v7}
\fmfdot{v1,v2,v3,v4,v5,v6}
\end{fmfgraph*}
}}
\no \\
&& \quad 
\no \\
&& \quad 
\no \\
&& \quad 
%
%
- \hspace*{0.3cm}
\parbox{18mm}{\centerline{
\begin{fmfgraph*}(8,8)
\setval
\fmfforce{1/2w,0h}{v1}
\fmfforce{1/2w,1h}{v2}
\fmfforce{0.933w,3/4h}{v3}
\fmfforce{0.933w,1/4h}{v4}
\fmfforce{0.067w,3/4h}{v5}
\fmfforce{0.067w,1/4h}{v6}
\fmfforce{1.233w,1/4h}{v7}
\fmfforce{-0.233w,1/4h}{v8}
\fmf{fermion,right=0.3}{v2,v5}
\fmf{fermion,right=0.3}{v3,v2}
\fmf{fermion,right=0.3}{v4,v3}
\fmf{fermion,right=0.3}{v1,v4}
\fmf{fermion,right=0.3}{v6,v1}
\fmf{fermion,right=0.3}{v5,v6}
\fmf{boson}{v1,v3}
\fmf{boson,left=0.3}{v2,v5}
\fmf{boson}{v6,v8}
\fmf{boson}{v4,v7}
\fmfv{decor.size=0, label=${\scs 1}$, l.dist=1mm, l.angle=-180}{v8}
\fmfv{decor.size=0, label=${\scs 2}$, l.dist=1mm, l.angle=0}{v7}
\fmfdot{v1,v2,v3,v4,v5,v6}
\end{fmfgraph*}
}}
%
%
\hspace*{0.3cm}- \hspace*{0.3cm}
\parbox{18mm}{\centerline{
\begin{fmfgraph*}(8,8)
\setval
\fmfforce{1/2w,0h}{v1}
\fmfforce{1/2w,1h}{v2}
\fmfforce{0.933w,3/4h}{v3}
\fmfforce{0.933w,1/4h}{v4}
\fmfforce{0.067w,3/4h}{v5}
\fmfforce{0.067w,1/4h}{v6}
\fmfforce{1.233w,1/4h}{v7}
\fmfforce{-0.233w,1/4h}{v8}
\fmf{fermion,right=0.3}{v2,v5}
\fmf{fermion,right=0.3}{v3,v2}
\fmf{fermion,right=0.3}{v4,v3}
\fmf{fermion,right=0.3}{v1,v4}
\fmf{fermion,right=0.3}{v6,v1}
\fmf{fermion,right=0.3}{v5,v6}
\fmf{boson}{v1,v3}
\fmf{boson,left=0.3}{v2,v5}
\fmf{boson}{v6,v8}
\fmf{boson}{v4,v7}
\fmfv{decor.size=0, label=${\scs 1}$, l.dist=1mm, l.angle=-180}{v8}
\fmfv{decor.size=0, label=${\scs 2}$, l.dist=1mm, l.angle=0}{v7}
\fmfdot{v1,v2,v3,v4,v5,v6}
\end{fmfgraph*}
}}
%
%
\hspace*{0.3cm}- \hspace*{0.3cm}
\parbox{18mm}{\centerline{
\begin{fmfgraph*}(8,8)
\setval
\fmfforce{1/2w,0h}{v1}
\fmfforce{1/2w,1h}{v2}
\fmfforce{0.933w,3/4h}{v3}
\fmfforce{0.933w,1/4h}{v4}
\fmfforce{0.067w,3/4h}{v5}
\fmfforce{0.067w,1/4h}{v6}
\fmfforce{1.233w,1/4h}{v7}
\fmfforce{-0.233w,1/4h}{v8}
\fmf{fermion,right=0.3}{v2,v5}
\fmf{fermion,right=0.3}{v3,v2}
\fmf{fermion,right=0.3}{v4,v3}
\fmf{fermion,right=0.3}{v1,v4}
\fmf{fermion,right=0.3}{v6,v1}
\fmf{fermion,right=0.3}{v5,v6}
\fmf{boson}{v1,v5}
\fmf{boson,right=0.3}{v2,v3}
\fmf{boson}{v6,v8}
\fmf{boson}{v4,v7}
\fmfv{decor.size=0, label=${\scs 1}$, l.dist=1mm, l.angle=-180}{v8}
\fmfv{decor.size=0, label=${\scs 2}$, l.dist=1mm, l.angle=0}{v7}
\fmfdot{v1,v2,v3,v4,v5,v6}
\end{fmfgraph*}
}}
%
%
\hspace*{0.3cm}+ \hspace*{0.3cm}
\parbox{15mm}{\centerline{
\begin{fmfgraph*}(8,8)
\setval
\fmfforce{0w,1/2h}{v1}
\fmfforce{1w,1/2h}{v2}
\fmfforce{3/4w,0.933h}{v3}
\fmfforce{1/4w,0.933h}{v4}
\fmfforce{3/4w,0.067h}{v5}
\fmfforce{1/4w,0.067h}{v6}
\fmfforce{-0.1w,0.067h}{v7}
\fmfforce{1.1w,0.067h}{v8}
\fmf{fermion,right=1}{v4,v3,v4}
\fmf{boson,right=0.3}{v2,v3}
\fmf{boson,right=0.3}{v4,v1}
\fmf{fermion}{v2,v1}
\fmf{fermion,right=0.3}{v1,v6}
\fmf{fermion,right=0.3}{v6,v5}
\fmf{fermion,right=0.3}{v5,v2}
\fmf{boson}{v6,v7}
\fmf{boson}{v8,v5}
\fmfv{decor.size=0, label=${\scs 1}$, l.dist=1mm, l.angle=-180}{v7}
\fmfv{decor.size=0, label=${\scs 2}$, l.dist=1mm, l.angle=0}{v8}
\fmfdot{v1,v2,v3,v4,v5,v6}
\end{fmfgraph*}
}}
%
%
\hspace*{0.3cm}+ \hspace*{0.3cm}
\parbox{15mm}{\centerline{
\begin{fmfgraph*}(8,8)
\setval
\fmfforce{0w,1/2h}{v1}
\fmfforce{1w,1/2h}{v2}
\fmfforce{3/4w,0.933h}{v3}
\fmfforce{1/4w,0.933h}{v4}
\fmfforce{3/4w,0.067h}{v5}
\fmfforce{1/4w,0.067h}{v6}
\fmfforce{-0.1w,0.067h}{v7}
\fmfforce{1.1w,0.067h}{v8}
\fmf{fermion,right=1}{v4,v3,v4}
\fmf{boson,right=0.3}{v2,v3}
\fmf{boson,right=0.3}{v4,v1}
\fmf{fermion}{v1,v2}
\fmf{fermion,left=0.3}{v6,v1}
\fmf{fermion,left=0.3}{v5,v6}
\fmf{fermion,left=0.3}{v2,v5}
\fmf{boson}{v6,v7}
\fmf{boson}{v8,v5}
\fmfv{decor.size=0, label=${\scs 1}$, l.dist=1mm, l.angle=-180}{v7}
\fmfv{decor.size=0, label=${\scs 2}$, l.dist=1mm, l.angle=0}{v8}
\fmfdot{v1,v2,v3,v4,v5,v6}
\end{fmfgraph*}
}}
\hspace*{0.4cm}. \la{PPEL3} \\ && \no 
\eeq
With this, the corresponding connected two-point functions for three loops follow from (\r{R1}) and (\r{R2}):
\beq
\parbox{15mm}{\centerline{
\begin{fmfgraph*}(7,3)
\setval
\fmfleft{v1}
\fmfright{v2}
\fmfforce{0.5w,2/3h}{v3}
\fmf{heavy,width=0.2mm}{v2,v1}
\fmfv{decor.size=0, label=${\scs 1}$, l.dist=1mm, l.angle=-180}{v1}
\fmfv{decor.size=0, label=${\scs 2}$, l.dist=1mm, l.angle=0}{v2}
\fmfv{decor.size=0, label=${\scs (3)}$, l.dist=1mm, l.angle=90}{v3}
\end{fmfgraph*}}}
\hspace*{0.2cm} & = & \hspace*{0.2cm} 
%
%
\parbox{19mm}{\centerline{
\begin{fmfgraph*}(8,8)
\setval
\fmfforce{0w,1/2h}{v1}
\fmfforce{1w,1/2h}{v2}
\fmfforce{3/4w,0.933h}{v3}
\fmfforce{1/4w,0.933h}{v4}
\fmfforce{3/4w,0.067h}{v5}
\fmfforce{1/4w,0.067h}{v6}
\fmfforce{-0.375w,0.067h}{v7}
\fmfforce{1.375w,0.067h}{v8}
\fmf{fermion,right=0.3}{v5,v2}
\fmf{fermion,right=0.3}{v2,v3}
\fmf{fermion,right=0.3}{v3,v4}
\fmf{fermion,right=0.3}{v4,v1}
\fmf{fermion,right=0.3}{v1,v6}
\fmf{boson}{v4,v5}
\fmf{boson}{v3,v6}
\fmf{boson}{v2,v1}
\fmf{electron}{v6,v7}
\fmf{electron}{v8,v5}
\fmfv{decor.size=0, label=${\scs 1}$, l.dist=1mm, l.angle=-180}{v7}
\fmfv{decor.size=0, label=${\scs 2}$, l.dist=1mm, l.angle=0}{v8}
\fmfdot{v1,v2,v3,v4,v5,v6}
\end{fmfgraph*}
}}
%
%
\hspace*{0.3cm}- \hspace*{0.3cm}
\parbox{23mm}{\centerline{
\begin{fmfgraph*}(16,10)
\setval
\fmfforce{-2/16w,0h}{v1}
\fmfforce{3/16w,0h}{v2}
\fmfforce{8/16w,0h}{v3}
\fmfforce{13/16w,0h}{v4}
\fmfforce{18/16w,0h}{v5}
\fmfforce{3/16w,1/2h}{v6}
\fmfforce{8/16w,1/2h}{v7}
\fmfforce{13/16w,1/2h}{v8}
\fmf{fermion}{v5,v4,v3,v2,v1}
\fmf{fermion}{v8,v7,v6}
\fmf{boson}{v2,v6}
\fmf{boson}{v3,v7}
\fmf{boson}{v4,v8}
\fmf{fermion,left=0.5}{v6,v8}
\fmfv{decor.size=0, label=${\scs 1}$, l.dist=1mm, l.angle=-180}{v1}
\fmfv{decor.size=0, label=${\scs 2}$, l.dist=1mm, l.angle=0}{v5}
\fmfdot{v2,v3,v4,v6,v7,v8}
\end{fmfgraph*}
}}
%
%
\hspace*{0.3cm} - \hspace*{0.3cm}
\parbox{23mm}{\centerline{
\begin{fmfgraph*}(16,10)
\setval
\fmfforce{-2/16w,0h}{v1}
\fmfforce{3/16w,0h}{v2}
\fmfforce{8/16w,0h}{v3}
\fmfforce{13/16w,0h}{v4}
\fmfforce{18/16w,0h}{v5}
\fmfforce{3/16w,1/2h}{v6}
\fmfforce{8/16w,1/2h}{v7}
\fmfforce{13/16w,1/2h}{v8}
\fmf{fermion}{v5,v4,v3,v2,v1}
\fmf{fermion}{v6,v7,v8}
\fmf{boson}{v2,v6}
\fmf{boson}{v3,v7}
\fmf{boson}{v4,v8}
\fmf{fermion,right=0.5}{v8,v6}
\fmfv{decor.size=0, label=${\scs 1}$, l.dist=1mm, l.angle=-180}{v1}
\fmfv{decor.size=0, label=${\scs 2}$, l.dist=1mm, l.angle=0}{v5}
\fmfdot{v2,v3,v4,v6,v7,v8}
\end{fmfgraph*}
}}
%
%
\hspace*{0.2cm}+ \hspace*{0.2cm}
\parbox{18mm}{\centerline{
\begin{fmfgraph*}(8,8)
\setval
\fmfforce{0w,1/2h}{v1}
\fmfforce{1w,1/2h}{v2}
\fmfforce{3/4w,0.933h}{v3}
\fmfforce{1/4w,0.933h}{v4}
\fmfforce{3/4w,0.067h}{v5}
\fmfforce{1/4w,0.067h}{v6}
\fmfforce{-0.375w,0.067h}{v7}
\fmfforce{1.375w,0.067h}{v8}
\fmf{fermion,right=0.3}{v5,v2}
\fmf{fermion,right=0.3}{v2,v3}
\fmf{fermion,right=0.3}{v3,v4}
\fmf{fermion,right=0.3}{v4,v1}
\fmf{fermion,right=0.3}{v1,v6}
\fmf{boson}{v6,v3}
\fmf{boson}{v1,v5}
\fmf{boson}{v2,v4}
\fmf{electron}{v6,v7}
\fmf{electron}{v8,v5}
\fmfv{decor.size=0, label=${\scs 1}$, l.dist=1mm, l.angle=-180}{v7}
\fmfv{decor.size=0, label=${\scs 2}$, l.dist=1mm, l.angle=0}{v8}
\fmfdot{v1,v2,v3,v4,v5,v6}
\end{fmfgraph*}
}}
%
%
\hspace*{0.3cm}+ \hspace*{0.3cm}
\parbox{19mm}{\centerline{
\begin{fmfgraph*}(8,8)
\setval
\fmfforce{0w,1/2h}{v1}
\fmfforce{1w,1/2h}{v2}
\fmfforce{3/4w,0.933h}{v3}
\fmfforce{1/4w,0.933h}{v4}
\fmfforce{3/4w,0.067h}{v5}
\fmfforce{1/4w,0.067h}{v6}
\fmfforce{-0.375w,0.067h}{v7}
\fmfforce{1.375w,0.067h}{v8}
\fmf{fermion,right=0.3}{v5,v2}
\fmf{fermion,right=0.3}{v2,v3}
\fmf{fermion,right=0.3}{v3,v4}
\fmf{fermion,right=0.3}{v4,v1}
\fmf{fermion,right=0.3}{v1,v6}
\fmf{boson}{v4,v5}
\fmf{boson}{v1,v3}
\fmf{boson}{v2,v6}
\fmf{electron}{v6,v7}
\fmf{electron}{v8,v5}
\fmfv{decor.size=0, label=${\scs 1}$, l.dist=1mm, l.angle=-180}{v7}
\fmfv{decor.size=0, label=${\scs 2}$, l.dist=1mm, l.angle=0}{v8}
\fmfdot{v1,v2,v3,v4,v5,v6}
\end{fmfgraph*}
}}
\no \\
&& \quad 
\no \\
&& \quad 
\no \\
&& \quad 
%
%
+ \hspace*{0.3cm}
\parbox{19mm}{\centerline{
\begin{fmfgraph*}(8,8)
\setval
\fmfforce{0w,1/2h}{v1}
\fmfforce{1w,1/2h}{v2}
\fmfforce{3/4w,0.933h}{v3}
\fmfforce{1/4w,0.933h}{v4}
\fmfforce{3/4w,0.067h}{v5}
\fmfforce{1/4w,0.067h}{v6}
\fmfforce{-0.375w,0.067h}{v7}
\fmfforce{1.375w,0.067h}{v8}
\fmf{fermion,right=0.3}{v5,v2}
\fmf{fermion,right=0.3}{v2,v3}
\fmf{fermion,right=0.3}{v3,v4}
\fmf{fermion,right=0.3}{v4,v1}
\fmf{fermion,right=0.3}{v1,v6}
\fmf{boson}{v4,v6}
\fmf{boson}{v3,v5}
\fmf{boson}{v2,v1}
\fmf{electron}{v6,v7}
\fmf{electron}{v8,v5}
\fmfv{decor.size=0, label=${\scs 1}$, l.dist=1mm, l.angle=-180}{v7}
\fmfv{decor.size=0, label=${\scs 2}$, l.dist=1mm, l.angle=0}{v8}
\fmfdot{v1,v2,v3,v4,v5,v6}
\end{fmfgraph*}
}}
%
%
\hspace*{0.3cm} + \hspace*{0.3cm}
\parbox{19mm}{\centerline{
\begin{fmfgraph*}(8,8)
\setval
\fmfforce{0w,1/2h}{v1}
\fmfforce{1w,1/2h}{v2}
\fmfforce{3/4w,0.933h}{v3}
\fmfforce{1/4w,0.933h}{v4}
\fmfforce{3/4w,0.067h}{v5}
\fmfforce{1/4w,0.067h}{v6}
\fmfforce{-0.375w,0.067h}{v7}
\fmfforce{1.375w,0.067h}{v8}
\fmf{boson,right=0.3}{v3,v4}
\fmf{fermion,left=1}{v3,v2,v3}
\fmf{fermion,right=1}{v4,v1,v4}
\fmf{boson,right=0.3}{v1,v6}
\fmf{boson,right=0.3}{v5,v2}
\fmf{fermion,left=0.3}{v5,v6}
\fmf{electron}{v6,v7}
\fmf{electron}{v8,v5}
\fmfv{decor.size=0, label=${\scs 1}$, l.dist=1mm, l.angle=-180}{v7}
\fmfv{decor.size=0, label=${\scs 2}$, l.dist=1mm, l.angle=0}{v8}
\fmfdot{v1,v2,v3,v4,v5,v6}
\end{fmfgraph*}
}}
%
%
\hspace*{0.3cm}- \hspace*{0.3cm} 
\parbox{19mm}{\centerline{
\begin{fmfgraph*}(11,7.5)
\setval
\fmfforce{-2/11w,0h}{v1}
\fmfforce{3/11w,0h}{v2}
\fmfforce{8/11w,0h}{v3}
\fmfforce{13/11w,0h}{v4}
\fmfforce{3/11w,2/3h}{v5}
\fmfforce{8/11w,2/3h}{v6}
\fmfforce{1/2w,1/3h}{v7}
\fmfforce{1/2w,1h}{v8}
\fmf{fermion}{v4,v3,v2,v1}
\fmf{boson}{v2,v5}
\fmf{boson}{v3,v6}
\fmf{fermion,right=0.4}{v5,v7}
\fmf{fermion,right=0.4}{v7,v6}
\fmf{fermion,right=0.4}{v6,v8}
\fmf{fermion,right=0.4}{v8,v5}
\fmf{boson}{v7,v8}
\fmfv{decor.size=0, label=${\scs 1}$, l.dist=1mm, l.angle=-180}{v1}
\fmfv{decor.size=0, label=${\scs 2}$, l.dist=1mm, l.angle=0}{v4}
\fmfdot{v2,v3,v5,v6,v7,v8}
\end{fmfgraph*}
}}
%
%
\hspace*{0.3cm} - \hspace*{0.3cm} 
\parbox{23mm}{\centerline{
\begin{fmfgraph*}(16,10)
\setval
\fmfforce{-2/16w,1/4h}{v1}
\fmfforce{3/16w,1/4h}{v2}
\fmfforce{8/16w,1/4h}{v3}
\fmfforce{13/16w,1/4h}{v4}
\fmfforce{18/16w,1/4h}{v5}
\fmfforce{8/16w,3/4h}{v6}
\fmfforce{13/16w,3/4h}{v7}
\fmfforce{10.5/16w,0h}{v8}
\fmf{fermion}{v3,v2,v1}
\fmf{fermion}{v5,v4}
\fmf{boson}{v3,v6}
\fmf{boson}{v4,v7}
\fmf{boson,right=0.4}{v2,v8}
\fmf{fermion,left=0.4}{v8,v3}
\fmf{fermion,left=0.4}{v4,v8}
\fmf{fermion,right=1}{v6,v7,v6}
\fmfv{decor.size=0, label=${\scs 1}$, l.dist=1mm, l.angle=-180}{v1}
\fmfv{decor.size=0, label=${\scs 2}$, l.dist=1mm, l.angle=0}{v5}
\fmfdot{v2,v3,v4,v6,v7,v8}
\end{fmfgraph*}
}}
%
%
\hspace*{0.3cm}- \hspace*{0.3cm} 
\parbox{23mm}{\centerline{
\begin{fmfgraph*}(16,10)
\setval
\fmfforce{-2/16w,1/4h}{v1}
\fmfforce{3/16w,1/4h}{v2}
\fmfforce{8/16w,1/4h}{v3}
\fmfforce{13/16w,1/4h}{v4}
\fmfforce{18/16w,1/4h}{v5}
\fmfforce{3/16w,3/4h}{v6}
\fmfforce{8/16w,3/4h}{v7}
\fmfforce{5.5/16w,0h}{v8}
\fmf{fermion}{v2,v1}
\fmf{fermion}{v5,v4,v3}
\fmf{boson}{v2,v6}
\fmf{boson}{v3,v7}
\fmf{boson,right=0.4}{v8,v4}
\fmf{fermion,left=0.4}{v3,v8}
\fmf{fermion,left=0.4}{v8,v2}
\fmf{fermion,right=1}{v6,v7,v6}
\fmfv{decor.size=0, label=${\scs 1}$, l.dist=1mm, l.angle=-180}{v1}
\fmfv{decor.size=0, label=${\scs 2}$, l.dist=1mm, l.angle=0}{v5}
\fmfdot{v2,v3,v4,v6,v7,v8}
\end{fmfgraph*}
}}
\no \\
&& \quad 
\no \\
&& \quad 
\no \\
&& \quad 
+ \hspace*{0.2cm}
%
%
\parbox{18mm}{\centerline{
\begin{fmfgraph*}(8,8)
\setval
\fmfforce{0w,1/2h}{v1}
\fmfforce{1w,1/2h}{v2}
\fmfforce{3/4w,0.933h}{v3}
\fmfforce{1/4w,0.933h}{v4}
\fmfforce{3/4w,0.067h}{v5}
\fmfforce{1/4w,0.067h}{v6}
\fmfforce{-0.375w,0.067h}{v7}
\fmfforce{1.375w,0.067h}{v8}
\fmf{fermion,right=0.3}{v2,v3}
\fmf{boson,left=0.3}{v2,v3}
\fmf{fermion,right=0.3}{v3,v4}
\fmf{fermion,right=0.3}{v4,v1}
\fmf{boson,left=0.3}{v4,v1}
\fmf{fermion,right=0.3}{v1,v6}
\fmf{boson,right=0.3}{v6,v5}
\fmf{fermion,right=0.3}{v5,v2}
\fmf{electron}{v6,v7}
\fmf{electron}{v8,v5}
\fmfv{decor.size=0, label=${\scs 1}$, l.dist=1mm, l.angle=-180}{v7}
\fmfv{decor.size=0, label=${\scs 2}$, l.dist=1mm, l.angle=0}{v8}
\fmfdot{v1,v2,v3,v4,v5,v6}
\end{fmfgraph*}
}}
%
%
\hspace*{0.2cm} + \hspace*{0.2cm}
\parbox{18mm}{\centerline{
\begin{fmfgraph*}(8,8)
\setval
\fmfforce{0w,1/2h}{v1}
\fmfforce{1w,1/2h}{v2}
\fmfforce{3/4w,0.933h}{v3}
\fmfforce{1/4w,0.933h}{v4}
\fmfforce{3/4w,0.067h}{v5}
\fmfforce{1/4w,0.067h}{v6}
\fmfforce{-0.375w,0.067h}{v7}
\fmfforce{1.375w,0.067h}{v8}
\fmf{fermion,right=0.3}{v3,v4}
\fmf{boson,left=0.3}{v3,v4}
\fmf{fermion,right=0.3}{v2,v3}
\fmf{fermion,right=0.3}{v4,v1}
\fmf{boson}{v2,v1}
\fmf{fermion,right=0.3}{v1,v6}
\fmf{boson,right=0.3}{v6,v5}
\fmf{fermion,right=0.3}{v5,v2}
\fmf{electron}{v6,v7}
\fmf{electron}{v8,v5}
\fmfv{decor.size=0, label=${\scs 1}$, l.dist=1mm, l.angle=-180}{v7}
\fmfv{decor.size=0, label=${\scs 2}$, l.dist=1mm, l.angle=0}{v8}
\fmfdot{v1,v2,v3,v4,v5,v6}
\end{fmfgraph*}
}}
%
%
\hspace*{0.2cm} + \hspace*{0.2cm}
\parbox{18mm}{\centerline{
\begin{fmfgraph*}(8,8)
\setval
\fmfforce{0w,1/2h}{v1}
\fmfforce{1w,1/2h}{v2}
\fmfforce{3/4w,0.933h}{v3}
\fmfforce{1/4w,0.933h}{v4}
\fmfforce{3/4w,0.067h}{v5}
\fmfforce{1/4w,0.067h}{v6}
\fmfforce{-0.375w,0.067h}{v7}
\fmfforce{1.375w,0.067h}{v8}
\fmf{fermion,right=0.3}{v3,v4}
\fmf{boson}{v3,v1}
\fmf{fermion,right=0.3}{v2,v3}
\fmf{fermion,right=0.3}{v4,v1}
\fmf{boson}{v4,v2}
\fmf{fermion,right=0.3}{v1,v6}
\fmf{boson,right=0.3}{v6,v5}
\fmf{fermion,right=0.3}{v5,v2}
\fmf{electron}{v6,v7}
\fmf{electron}{v8,v5}
\fmfv{decor.size=0, label=${\scs 1}$, l.dist=1mm, l.angle=-180}{v7}
\fmfv{decor.size=0, label=${\scs 2}$, l.dist=1mm, l.angle=0}{v8}
\fmfdot{v1,v2,v3,v4,v5,v6}
\end{fmfgraph*}
}}
%
%
- \hspace*{0.2cm}
\parbox{18mm}{\centerline{
\begin{fmfgraph*}(8,8)
\setval
\fmfforce{0w,1/2h}{v1}
\fmfforce{1w,1/2h}{v2}
\fmfforce{3/4w,0.933h}{v3}
\fmfforce{1/4w,0.933h}{v4}
\fmfforce{3/4w,0.067h}{v5}
\fmfforce{1/4w,0.067h}{v6}
\fmfforce{-0.375w,0.067h}{v7}
\fmfforce{1.375w,0.067h}{v8}
\fmf{fermion,right=0.3}{v3,v4}
\fmf{boson,left=0.3}{v3,v4}
\fmf{fermion,right=0.3}{v2,v3}
\fmf{fermion,right=0.3}{v4,v1}
\fmf{boson,right=0.3}{v1,v6}
\fmf{boson,right=0.3}{v5,v2}
\fmf{fermion,left=0.3}{v5,v6}
\fmf{electron}{v1,v2}
\fmf{electron}{v6,v7}
\fmf{electron}{v8,v5}
\fmfv{decor.size=0, label=${\scs 1}$, l.dist=1mm, l.angle=-180}{v7}
\fmfv{decor.size=0, label=${\scs 2}$, l.dist=1mm, l.angle=0}{v8}
\fmfdot{v1,v2,v3,v4,v5,v6}
\end{fmfgraph*}
}}
%
%
\hspace*{0.2cm}- \hspace*{0.2cm} 
\parbox{18mm}{\centerline{
\begin{fmfgraph*}(8,8)
\setval
\fmfforce{0w,1/2h}{v1}
\fmfforce{1w,1/2h}{v2}
\fmfforce{3/4w,0.933h}{v3}
\fmfforce{1/4w,0.933h}{v4}
\fmfforce{3/4w,0.067h}{v5}
\fmfforce{1/4w,0.067h}{v6}
\fmfforce{-0.375w,0.067h}{v7}
\fmfforce{1.375w,0.067h}{v8}
\fmf{fermion,left=0.3}{v4,v3}
\fmf{boson,left=0.3}{v3,v4}
\fmf{fermion,left=0.3}{v3,v2}
\fmf{fermion,left=0.3}{v1,v4}
\fmf{boson,right=0.3}{v1,v6}
\fmf{boson,right=0.3}{v5,v2}
\fmf{fermion,left=0.3}{v5,v6}
\fmf{electron}{v2,v1}
\fmf{electron}{v6,v7}
\fmf{electron}{v8,v5}
\fmfv{decor.size=0, label=${\scs 1}$, l.dist=1mm, l.angle=-180}{v7}
\fmfv{decor.size=0, label=${\scs 2}$, l.dist=1mm, l.angle=0}{v8}
\fmfdot{v1,v2,v3,v4,v5,v6}
\end{fmfgraph*}
}}
\no \\
&& \quad 
\no \\
&& \quad 
\no \\
&& \quad 
%
%
+ \hspace*{0.3cm}
\parbox{19mm}{\centerline{
\begin{fmfgraph*}(8,8)
\setval
\fmfforce{0w,1/2h}{v1}
\fmfforce{1w,1/2h}{v2}
\fmfforce{3/4w,0.933h}{v3}
\fmfforce{1/4w,0.933h}{v4}
\fmfforce{3/4w,0.067h}{v5}
\fmfforce{1/4w,0.067h}{v6}
\fmfforce{-0.375w,0.067h}{v7}
\fmfforce{1.375w,0.067h}{v8}
\fmf{fermion,right=0.3}{v5,v2}
\fmf{fermion,right=0.3}{v2,v3}
\fmf{fermion,right=0.3}{v3,v4}
\fmf{fermion,right=0.3}{v4,v1}
\fmf{fermion,right=0.3}{v1,v6}
\fmf{boson,left=0.3}{v2,v3}
\fmf{boson}{v4,v6}
\fmf{boson}{v1,v5}
\fmf{electron}{v6,v7}
\fmf{electron}{v8,v5}
\fmfv{decor.size=0, label=${\scs 1}$, l.dist=1mm, l.angle=-180}{v7}
\fmfv{decor.size=0, label=${\scs 2}$, l.dist=1mm, l.angle=0}{v8}
\fmfdot{v1,v2,v3,v4,v5,v6}
\end{fmfgraph*}
}}
%
%
\hspace*{0.3cm}+ \hspace*{0.3cm}
\parbox{19mm}{\centerline{
\begin{fmfgraph*}(8,8)
\setval
\fmfforce{0w,1/2h}{v1}
\fmfforce{1w,1/2h}{v2}
\fmfforce{3/4w,0.933h}{v3}
\fmfforce{1/4w,0.933h}{v4}
\fmfforce{3/4w,0.067h}{v5}
\fmfforce{1/4w,0.067h}{v6}
\fmfforce{-0.375w,0.067h}{v7}
\fmfforce{1.375w,0.067h}{v8}
\fmf{fermion,right=0.3}{v5,v2}
\fmf{fermion,right=0.3}{v2,v3}
\fmf{fermion,right=0.3}{v3,v4}
\fmf{fermion,right=0.3}{v4,v1}
\fmf{fermion,right=0.3}{v1,v6}
\fmf{boson,right=0.3}{v4,v3}
\fmf{boson}{v1,v5}
\fmf{boson}{v2,v6}
\fmf{electron}{v6,v7}
\fmf{electron}{v8,v5}
\fmfv{decor.size=0, label=${\scs 1}$, l.dist=1mm, l.angle=-180}{v7}
\fmfv{decor.size=0, label=${\scs 2}$, l.dist=1mm, l.angle=0}{v8}
\fmfdot{v1,v2,v3,v4,v5,v6}
\end{fmfgraph*}
}}
%
%
\hspace*{0.3cm} + \hspace*{0.3cm}
\parbox{19mm}{\centerline{
\begin{fmfgraph*}(8,8)
\setval
\fmfforce{0w,1/2h}{v1}
\fmfforce{1w,1/2h}{v2}
\fmfforce{3/4w,0.933h}{v3}
\fmfforce{1/4w,0.933h}{v4}
\fmfforce{3/4w,0.067h}{v5}
\fmfforce{1/4w,0.067h}{v6}
\fmfforce{-0.375w,0.067h}{v7}
\fmfforce{1.375w,0.067h}{v8}
\fmf{fermion,right=0.3}{v5,v2}
\fmf{fermion,right=0.3}{v2,v3}
\fmf{fermion,right=0.3}{v3,v4}
\fmf{fermion,right=0.3}{v4,v1}
\fmf{fermion,right=0.3}{v1,v6}
\fmf{boson,left=0.3}{v4,v1}
\fmf{boson}{v3,v5}
\fmf{boson}{v2,v6}
\fmf{electron}{v6,v7}
\fmf{electron}{v8,v5}
\fmfv{decor.size=0, label=${\scs 1}$, l.dist=1mm, l.angle=-180}{v7}
\fmfv{decor.size=0, label=${\scs 2}$, l.dist=1mm, l.angle=0}{v8}
\fmfdot{v1,v2,v3,v4,v5,v6}
\end{fmfgraph*}
}}
%
%
\hspace*{0.3cm}- \hspace*{0.3cm}
\parbox{19mm}{\centerline{
\begin{fmfgraph*}(8,8)
\setval
\fmfforce{0w,1/2h}{v1}
\fmfforce{1w,1/2h}{v2}
\fmfforce{3/4w,0.933h}{v3}
\fmfforce{1/4w,0.933h}{v4}
\fmfforce{3/4w,0.067h}{v5}
\fmfforce{1/4w,0.067h}{v6}
\fmfforce{-0.375w,0.067h}{v7}
\fmfforce{1.375w,0.067h}{v8}
\fmf{fermion,right=1}{v4,v3,v4}
\fmf{boson,right=0.3}{v2,v3}
\fmf{boson,right=0.3}{v4,v1}
\fmf{fermion}{v2,v1}
\fmf{fermion,right=0.3}{v1,v6}
\fmf{boson,right=0.3}{v6,v5}
\fmf{fermion,right=0.3}{v5,v2}
\fmf{electron}{v6,v7}
\fmf{electron}{v8,v5}
\fmfv{decor.size=0, label=${\scs 1}$, l.dist=1mm, l.angle=-180}{v7}
\fmfv{decor.size=0, label=${\scs 2}$, l.dist=1mm, l.angle=0}{v8}
\fmfdot{v1,v2,v3,v4,v5,v6}
\end{fmfgraph*}
}}
%
%
\hspace*{0.3cm} - \hspace*{0.3cm}
\parbox{19mm}{\centerline{
\begin{fmfgraph*}(8,8)
\setval
\fmfforce{0w,1/2h}{v1}
\fmfforce{1w,1/2h}{v2}
\fmfforce{3/4w,0.933h}{v3}
\fmfforce{1/4w,0.933h}{v4}
\fmfforce{3/4w,0.067h}{v5}
\fmfforce{1/4w,0.067h}{v6}
\fmfforce{-5/8w,1/2h}{v7}
\fmfforce{13/8w,1/2h}{v8}
\fmf{fermion,right=1}{v4,v3,v4}
\fmf{fermion,left=0.3}{v6,v1}
\fmf{fermion,left=0.3}{v5,v6}
\fmf{fermion,left=0.3}{v2,v5}
\fmf{boson,right=0.3}{v4,v1}
\fmf{boson,right=0.3}{v2,v3}
\fmf{boson,left=0.3}{v6,v5}
\fmf{electron}{v1,v7}
\fmf{electron}{v8,v2}
\fmfv{decor.size=0, label=${\scs 1}$, l.dist=1mm, l.angle=-180}{v7}
\fmfv{decor.size=0, label=${\scs 2}$, l.dist=1mm, l.angle=0}{v8}
\fmfdot{v1,v2,v3,v4,v5,v6}
\end{fmfgraph*}
}}
\no \\
&& \no \\
&& \no \\
&& 
%
%
\quad - \hspace*{0.3cm}
\parbox{28mm}{\centerline{
\begin{fmfgraph*}(25,7.5)
\setval
\fmfforce{0w,0h}{v1}
\fmfforce{1/5w,0h}{v2}
\fmfforce{2/5w,0h}{v3}
\fmfforce{3/5w,0h}{v4}
\fmfforce{4/5w,0h}{v5}
\fmfforce{5/5w,0h}{v6}
\fmfforce{3/5w,2/3h}{v7}
\fmfforce{4/5w,2/3h}{v8}
\fmf{boson}{v4,v7}
\fmf{boson}{v5,v8}
\fmf{boson,right=1}{v3,v2}
\fmf{fermion}{v6,v5,v4,v3,v2,v1}
\fmf{fermion,right=1}{v7,v8,v7}
\fmfdot{v2,v3,v4,v5,v7,v8}
\fmfv{decor.size=0, label=${\scs 1}$, l.dist=1mm, l.angle=-180}{v1}
\fmfv{decor.size=0, label=${\scs 2}$, l.dist=1mm, l.angle=0}{v6}
\end{fmfgraph*}}}
%
%
\hspace*{0.3cm} - \hspace*{0.3cm}
\parbox{28mm}{\centerline{
\begin{fmfgraph*}(25,7.5)
\setval
\fmfforce{0w,0h}{v1}
\fmfforce{1/5w,0h}{v2}
\fmfforce{2/5w,0h}{v3}
\fmfforce{3/5w,0h}{v4}
\fmfforce{4/5w,0h}{v5}
\fmfforce{5/5w,0h}{v6}
\fmfforce{1/5w,2/3h}{v7}
\fmfforce{2/5w,2/3h}{v8}
\fmf{boson}{v2,v7}
\fmf{boson}{v3,v8}
\fmf{boson,right=1}{v5,v4}
\fmf{fermion}{v6,v5,v4,v3,v2,v1}
\fmf{fermion,right=1}{v7,v8,v7}
\fmfdot{v2,v3,v4,v5,v7,v8}
\fmfv{decor.size=0, label=${\scs 1}$, l.dist=1mm, l.angle=-180}{v1}
\fmfv{decor.size=0, label=${\scs 2}$, l.dist=1mm, l.angle=0}{v6}
\end{fmfgraph*}}}
%
%
\hspace*{0.3cm}  + \hspace*{0.3cm}
\parbox{28mm}{\centerline{
\begin{fmfgraph*}(25,7.5)
\setval
\fmfforce{0w,0h}{v1}
\fmfforce{1/5w,0h}{v2}
\fmfforce{2/5w,0h}{v3}
\fmfforce{3/5w,0h}{v4}
\fmfforce{4/5w,0h}{v5}
\fmfforce{5/5w,0h}{v6}
\fmfforce{3/5w,2/3h}{v7}
\fmfforce{4/5w,2/3h}{v8}
\fmf{fermion,right=0.4}{v7,v4}
\fmf{fermion,right=0.4}{v5,v8}
\fmf{boson,right=1}{v3,v2}
\fmf{fermion}{v6,v5}
\fmf{fermion}{v4,v3,v2,v1}
\fmf{boson,left=0.4}{v5,v4}
\fmf{boson,left=0.4}{v8,v7}
\fmf{fermion,right=0.4}{v8,v7}
\fmfdot{v2,v3,v4,v5,v7,v8}
\fmfv{decor.size=0, label=${\scs 1}$, l.dist=1mm, l.angle=-180}{v1}
\fmfv{decor.size=0, label=${\scs 2}$, l.dist=1mm, l.angle=0}{v6}
\end{fmfgraph*}}}
%
%
\hspace*{0.3cm} + \hspace*{0.3cm}
\parbox{28mm}{\centerline{
\begin{fmfgraph*}(25,7.5)
\setval
\fmfforce{0w,0h}{v1}
\fmfforce{1/5w,0h}{v2}
\fmfforce{2/5w,0h}{v3}
\fmfforce{3/5w,0h}{v4}
\fmfforce{4/5w,0h}{v5}
\fmfforce{5/5w,0h}{v6}
\fmfforce{1/5w,2/3h}{v7}
\fmfforce{2/5w,2/3h}{v8}
\fmf{fermion,right=0.4}{v7,v2}
\fmf{fermion,right=0.4}{v3,v8}
\fmf{boson,right=1}{v5,v4}
\fmf{fermion}{v6,v5,v4,v3}
\fmf{boson,left=0.4}{v3,v2}
\fmf{fermion}{v2,v1}
\fmf{fermion,right=0.4}{v8,v7}
\fmf{boson,right=0.4}{v7,v8}
\fmfdot{v2,v3,v4,v5,v7,v8}
\fmfv{decor.size=0, label=${\scs 1}$, l.dist=1mm, l.angle=-180}{v1}
\fmfv{decor.size=0, label=${\scs 2}$, l.dist=1mm, l.angle=0}{v6}
\end{fmfgraph*}}}
\no \\ 
&&\no \\ &&\no \\ 
&&
%
%
\quad + \hspace*{0.3cm}
\parbox{28mm}{\centerline{
\begin{fmfgraph*}(25,7.5)
\setval
\fmfforce{0w,0h}{v1}
\fmfforce{1/5w,0h}{v2}
\fmfforce{2/5w,0h}{v3}
\fmfforce{3/5w,0h}{v4}
\fmfforce{4/5w,0h}{v5}
\fmfforce{5/5w,0h}{v6}
\fmfforce{3/5w,2/3h}{v7}
\fmfforce{4/5w,2/3h}{v8}
\fmf{fermion,right=0.4}{v7,v4}
\fmf{fermion,right=0.4}{v5,v8}
\fmf{boson,right=1}{v3,v2}
\fmf{fermion}{v6,v5}
\fmf{fermion}{v4,v3,v2,v1}
\fmf{boson}{v5,v7}
\fmf{boson}{v4,v8}
\fmf{fermion,right=0.4}{v8,v7}
\fmfdot{v2,v3,v4,v5,v7,v8}
\fmfv{decor.size=0, label=${\scs 1}$, l.dist=1mm, l.angle=-180}{v1}
\fmfv{decor.size=0, label=${\scs 2}$, l.dist=1mm, l.angle=0}{v6}
\end{fmfgraph*}}}
%
%
\hspace*{0.3cm} + \hspace*{0.3cm}
\parbox{28mm}{\centerline{
\begin{fmfgraph*}(25,7.5)
\setval
\fmfforce{0w,0h}{v1}
\fmfforce{1/5w,0h}{v2}
\fmfforce{2/5w,0h}{v3}
\fmfforce{3/5w,0h}{v4}
\fmfforce{4/5w,0h}{v5}
\fmfforce{5/5w,0h}{v6}
\fmfforce{1/5w,2/3h}{v7}
\fmfforce{2/5w,2/3h}{v8}
\fmf{fermion,right=0.4}{v7,v2}
\fmf{fermion,right=0.4}{v3,v8}
\fmf{boson,right=1}{v5,v4}
\fmf{fermion}{v6,v5,v4,v3}
\fmf{fermion}{v2,v1}
\fmf{fermion,right=0.4}{v8,v7}
\fmf{boson}{v2,v8}
\fmf{boson}{v3,v7}
\fmfdot{v2,v3,v4,v5,v7,v8}
\fmfv{decor.size=0, label=${\scs 1}$, l.dist=1mm, l.angle=-180}{v1}
\fmfv{decor.size=0, label=${\scs 2}$, l.dist=1mm, l.angle=0}{v6}
\end{fmfgraph*} }}
%
%
\hspace*{0.3cm} + \hspace*{0.3cm}
\parbox{38mm}{\centerline{
\begin{fmfgraph*}(35,2.5)
\setval
\fmfforce{0w,0h}{v1}
\fmfforce{1/7w,0h}{v2}
\fmfforce{2/7w,0h}{v3}
\fmfforce{3/7w,0h}{v4}
\fmfforce{4/7w,0h}{v5}
\fmfforce{5/7w,0h}{v6}
\fmfforce{6/7w,0h}{v7}
\fmfforce{7/7w,0h}{v8}
\fmf{boson,right=1}{v5,v4}
\fmf{fermion}{v8,v7,v6,v5,v4,v3,v2,v1}
\fmf{boson,right=1}{v3,v2}
\fmf{boson,right=1}{v5,v4}
\fmf{boson,right=1}{v7,v6}
\fmfdot{v2,v3,v4,v5,v6,v7}
\fmfv{decor.size=0, label=${\scs 1}$, l.dist=1mm, l.angle=-180}{v1}
\fmfv{decor.size=0, label=${\scs 2}$, l.dist=1mm, l.angle=0}{v8}
\end{fmfgraph*}}}
\hspace*{0.4cm} , \la{SES3} \\ && \no \\  && \no \\
\parbox{15mm}{\centerline{
\begin{fmfgraph*}(7,3)
\setval
\fmfleft{v1}
\fmfright{v2}
\fmfforce{0.5w,2/3h}{v3}
\fmf{dbl_wiggly,width=0.2mm}{v2,v1}
\fmfv{decor.size=0, label=${\scs 1}$, l.dist=1mm, l.angle=-180}{v1}
\fmfv{decor.size=0, label=${\scs 2}$, l.dist=1mm, l.angle=0}{v2}
\fmfv{decor.size=0, label=${\scs (3)}$, l.dist=1mm, l.angle=90}{v3}
\end{fmfgraph*}}}
\hspace*{0.2cm} & = & \hspace*{0.2cm} 
%
%
\parbox{25.5mm}{\centerline{
\begin{fmfgraph*}(18.5,5)
\setval
\fmfforce{-2/18.5w,1/2h}{v1}
\fmfforce{3/18.5w,1/2h}{v2}
\fmfforce{5.5/18.5w,0h}{v3}
\fmfforce{5.5/18.5w,1h}{v4}
\fmfforce{13/18.5w,0h}{v5}
\fmfforce{13/18.5w,1h}{v6}
\fmfforce{15.5/18.5w,1/2h}{v7}
\fmfforce{20.5/18.5w,1/2h}{v8}
\fmf{boson}{v1,v2}
\fmf{fermion,right=0.4}{v4,v2}
\fmf{fermion,right=0.4}{v2,v3}
\fmf{fermion,right=1}{v3,v4}
\fmf{boson}{v4,v6}
\fmf{boson}{v3,v5}
\fmf{fermion,right=0.4}{v7,v6}
\fmf{fermion,right=0.4}{v5,v7}
\fmf{fermion,right=1}{v6,v5}
\fmf{boson}{v8,v7}
\fmfv{decor.size=0, label=${\scs 1}$, l.dist=1mm, l.angle=-180}{v1}
\fmfv{decor.size=0, label=${\scs 2}$, l.dist=1mm, l.angle=0}{v8}
\fmfdot{v2,v3,v4,v5,v6,v7}
\end{fmfgraph*}
}}
%
%
\hspace*{0.2cm} + \hspace*{0.2cm}
\parbox{25.5mm}{\centerline{
\begin{fmfgraph*}(18.5,5)
\setval
\fmfforce{-2/18.5w,1/2h}{v1}
\fmfforce{3/18.5w,1/2h}{v2}
\fmfforce{5.5/18.5w,0h}{v3}
\fmfforce{5.5/18.5w,1h}{v4}
\fmfforce{13/18.5w,0h}{v5}
\fmfforce{13/18.5w,1h}{v6}
\fmfforce{15.5/18.5w,1/2h}{v7}
\fmfforce{20.5/18.5w,1/2h}{v8}
\fmf{boson}{v1,v2}
\fmf{fermion,right=0.4}{v4,v2}
\fmf{fermion,right=0.4}{v2,v3}
\fmf{fermion,right=1}{v3,v4}
\fmf{boson}{v4,v6}
\fmf{boson}{v3,v5}
\fmf{fermion,left=0.4}{v6,v7}
\fmf{fermion,left=0.4}{v7,v5}
\fmf{fermion,left=1}{v5,v6}
\fmf{boson}{v8,v7}
\fmfv{decor.size=0, label=${\scs 1}$, l.dist=1mm, l.angle=-180}{v1}
\fmfv{decor.size=0, label=${\scs 2}$, l.dist=1mm, l.angle=0}{v8}
\fmfdot{v2,v3,v4,v5,v6,v7}
\end{fmfgraph*}
}}
%
%
\hspace*{0.2cm} - \hspace*{0.2cm}
\parbox{22mm}{\centerline{
\begin{fmfgraph*}(8,8)
\setval
\fmfforce{0w,1/2h}{v1}
\fmfforce{1w,1/2h}{v2}
\fmfforce{3/4w,0.933h}{v3}
\fmfforce{1/4w,0.933h}{v4}
\fmfforce{3/4w,0.067h}{v5}
\fmfforce{1/4w,0.067h}{v6}
\fmfforce{-5/8w,1/2h}{v7}
\fmfforce{13/8w,1/2h}{v8}
\fmf{fermion,right=0.3}{v3,v4}
\fmf{fermion,right=0.3}{v1,v6}
\fmf{fermion,right=0.3}{v6,v5}
\fmf{fermion,right=0.3}{v5,v2}
\fmf{fermion,right=0.3}{v4,v1}
\fmf{fermion,right=0.3}{v2,v3}
\fmf{boson}{v4,v5}
\fmf{boson}{v3,v6}
\fmf{boson}{v8,v2}
\fmf{boson}{v7,v1}
\fmfv{decor.size=0, label=${\scs 1}$, l.dist=1mm, l.angle=-180}{v7}
\fmfv{decor.size=0, label=${\scs 2}$, l.dist=1mm, l.angle=0}{v8}
\fmfdot{v1,v2,v3,v4,v5,v6}
\end{fmfgraph*}
}}
%
%
\hspace*{0.2cm}- \hspace*{0.2cm}
\parbox{22mm}{\centerline{
\begin{fmfgraph*}(8,8)
\setval
\fmfforce{0w,1/2h}{v1}
\fmfforce{1w,1/2h}{v2}
\fmfforce{3/4w,0.933h}{v3}
\fmfforce{1/4w,0.933h}{v4}
\fmfforce{3/4w,0.067h}{v5}
\fmfforce{1/4w,0.067h}{v6}
\fmfforce{-5/8w,1/2h}{v7}
\fmfforce{13/8w,1/2h}{v8}
\fmf{fermion,right=0.3}{v3,v4}
\fmf{fermion,right=0.3}{v1,v6}
\fmf{fermion,right=0.3}{v6,v5}
\fmf{fermion,right=0.3}{v5,v2}
\fmf{fermion,right=0.3}{v4,v1}
\fmf{fermion,right=0.3}{v2,v3}
\fmf{boson}{v4,v6}
\fmf{boson}{v3,v5}
\fmf{boson}{v8,v2}
\fmf{boson}{v7,v1}
\fmfv{decor.size=0, label=${\scs 1}$, l.dist=1mm, l.angle=-180}{v7}
\fmfv{decor.size=0, label=${\scs 2}$, l.dist=1mm, l.angle=0}{v8}
\fmfdot{v1,v2,v3,v4,v5,v6}
\end{fmfgraph*}
}}
%
%
\hspace*{0.2cm}- \hspace*{0.2cm}
\parbox{22mm}{\centerline{
\begin{fmfgraph*}(8,8)
\setval
\fmfforce{1/2w,0h}{v1}
\fmfforce{1/2w,1h}{v2}
\fmfforce{0.933w,3/4h}{v3}
\fmfforce{0.933w,1/4h}{v4}
\fmfforce{0.067w,3/4h}{v5}
\fmfforce{0.067w,1/4h}{v6}
\fmfforce{11.864/8w,1/4h}{v7}
\fmfforce{-3.864/8w,1/4h}{v8}
\fmf{fermion,right=0.3}{v2,v5}
\fmf{fermion,right=0.3}{v3,v2}
\fmf{fermion,right=0.3}{v4,v3}
\fmf{fermion,right=0.3}{v1,v4}
\fmf{fermion,right=0.3}{v6,v1}
\fmf{fermion,right=0.3}{v5,v6}
\fmf{boson}{v1,v2}
\fmf{boson,left=0.6}{v3,v5}
\fmf{boson}{v6,v8}
\fmf{boson}{v4,v7}
\fmfv{decor.size=0, label=${\scs 1}$, l.dist=1mm, l.angle=-180}{v8}
\fmfv{decor.size=0, label=${\scs 2}$, l.dist=1mm, l.angle=0}{v7}
\fmfdot{v1,v2,v3,v4,v5,v6}
\end{fmfgraph*}
}}
\no \\
&& \quad 
\no \\
&& \quad 
\no \\
&& \hspace*{0.2cm}
%
%
- \hspace*{0.2cm}
\parbox{22mm}{\centerline{
\begin{fmfgraph*}(8,8)
\setval
\fmfforce{1/2w,0h}{v1}
\fmfforce{1/2w,1h}{v2}
\fmfforce{0.933w,3/4h}{v3}
\fmfforce{0.933w,1/4h}{v4}
\fmfforce{0.067w,3/4h}{v5}
\fmfforce{0.067w,1/4h}{v6}
\fmfforce{11.864/8w,1/4h}{v7}
\fmfforce{-3.864/8w,1/4h}{v8}
\fmf{fermion,left=0.3}{v5,v2}
\fmf{fermion,left=0.3}{v2,v3}
\fmf{fermion,left=0.3}{v3,v4}
\fmf{fermion,left=0.3}{v4,v1}
\fmf{fermion,left=0.3}{v1,v6}
\fmf{fermion,left=0.3}{v6,v5}
\fmf{boson}{v1,v2}
\fmf{boson,left=0.6}{v3,v5}
\fmf{boson}{v6,v8}
\fmf{boson}{v4,v7}
\fmfv{decor.size=0, label=${\scs 1}$, l.dist=1mm, l.angle=-180}{v8}
\fmfv{decor.size=0, label=${\scs 2}$, l.dist=1mm, l.angle=0}{v7}
\fmfdot{v1,v2,v3,v4,v5,v6}
\end{fmfgraph*}
}}
%
%
\hspace*{0.2cm}+ \hspace*{0.2cm}
\parbox{24mm}{\centerline{
\begin{fmfgraph*}(10,10)
\setval
\fmfforce{0w,1/2h}{v1}
\fmfforce{1w,1/2h}{v2}
\fmfforce{1/2w,0h}{v3}
\fmfforce{1/2w,1/4h}{v4}
\fmfforce{1/2w,3/4h}{v5}
\fmfforce{1/2w,1h}{v6}
\fmfforce{-5/10w,1/2h}{v7}
\fmfforce{15/10w,1/2h}{v8}
\fmf{fermion,right=0.4}{v1,v3}
\fmf{fermion,right=0.4}{v3,v2}
\fmf{fermion,right=0.4}{v2,v6}
\fmf{fermion,right=0.4}{v6,v1}
\fmf{fermion,right=1}{v4,v5,v4}
\fmf{boson}{v1,v7}
\fmf{boson}{v8,v2}
\fmf{boson}{v3,v4}
\fmf{boson}{v5,v6}
\fmfv{decor.size=0, label=${\scs 1}$, l.dist=1mm, l.angle=-180}{v7}
\fmfv{decor.size=0, label=${\scs 2}$, l.dist=1mm, l.angle=0}{v8}
\fmfdot{v1,v2,v3,v4,v5,v6}
\end{fmfgraph*}
}}
%
%
\hspace*{0.2cm}- \hspace*{0.2cm}
\parbox{18mm}{\centerline{
\begin{fmfgraph*}(8,8)
\setval
\fmfforce{0w,1/2h}{v1}
\fmfforce{1w,1/2h}{v2}
\fmfforce{3/4w,0.933h}{v3}
\fmfforce{1/4w,0.933h}{v4}
\fmfforce{3/4w,0.067h}{v5}
\fmfforce{1/4w,0.067h}{v6}
\fmfforce{-0.375w,0.067h}{v7}
\fmfforce{1.375w,0.067h}{v8}
\fmf{fermion,right=0.3}{v2,v3}
\fmf{boson,left=0.3}{v2,v3}
\fmf{fermion,right=0.3}{v3,v4}
\fmf{fermion,right=0.3}{v4,v1}
\fmf{boson,left=0.3}{v4,v1}
\fmf{fermion,right=0.3}{v1,v6}
\fmf{fermion,right=0.3}{v6,v5}
\fmf{fermion,right=0.3}{v5,v2}
\fmf{boson}{v6,v7}
\fmf{boson}{v8,v5}
\fmfv{decor.size=0, label=${\scs 1}$, l.dist=1mm, l.angle=-180}{v7}
\fmfv{decor.size=0, label=${\scs 2}$, l.dist=1mm, l.angle=0}{v8}
\fmfdot{v1,v2,v3,v4,v5,v6}
\end{fmfgraph*}
}}
%
%
\hspace*{0.2cm}- \hspace*{0.2cm}
\parbox{18mm}{\centerline{
\begin{fmfgraph*}(8,8)
\setval
\fmfforce{0w,1/2h}{v1}
\fmfforce{1w,1/2h}{v2}
\fmfforce{3/4w,0.933h}{v3}
\fmfforce{1/4w,0.933h}{v4}
\fmfforce{3/4w,0.067h}{v5}
\fmfforce{1/4w,0.067h}{v6}
\fmfforce{-0.375w,0.067h}{v7}
\fmfforce{1.375w,0.067h}{v8}
\fmf{fermion,left=0.3}{v3,v2}
\fmf{boson,left=0.3}{v2,v3}
\fmf{fermion,left=0.3}{v4,v3}
\fmf{fermion,left=0.3}{v1,v4}
\fmf{boson,left=0.3}{v4,v1}
\fmf{fermion,left=0.3}{v6,v1}
\fmf{fermion,left=0.3}{v5,v6}
\fmf{fermion,left=0.3}{v2,v5}
\fmf{boson}{v6,v7}
\fmf{boson}{v8,v5}
\fmfv{decor.size=0, label=${\scs 1}$, l.dist=1mm, l.angle=-180}{v7}
\fmfv{decor.size=0, label=${\scs 2}$, l.dist=1mm, l.angle=0}{v8}
\fmfdot{v1,v2,v3,v4,v5,v6}
\end{fmfgraph*}
}}
%
%
\hspace*{0.3cm}- \hspace*{0.3cm}
\parbox{18mm}{\centerline{
\begin{fmfgraph*}(8,8)
\setval
\fmfforce{0w,1/2h}{v1}
\fmfforce{1w,1/2h}{v2}
\fmfforce{3/4w,0.933h}{v3}
\fmfforce{1/4w,0.933h}{v4}
\fmfforce{3/4w,0.067h}{v5}
\fmfforce{1/4w,0.067h}{v6}
\fmfforce{-0.375w,0.067h}{v7}
\fmfforce{1.375w,0.067h}{v8}
\fmf{fermion,right=0.3}{v3,v4}
\fmf{boson,left=0.3}{v3,v4}
\fmf{fermion,right=0.3}{v2,v3}
\fmf{fermion,right=0.3}{v4,v1}
\fmf{boson}{v2,v1}
\fmf{fermion,right=0.3}{v1,v6}
\fmf{fermion,right=0.3}{v6,v5}
\fmf{fermion,right=0.3}{v5,v2}
\fmf{boson}{v6,v7}
\fmf{boson}{v8,v5}
\fmfv{decor.size=0, label=${\scs 1}$, l.dist=1mm, l.angle=-180}{v7}
\fmfv{decor.size=0, label=${\scs 2}$, l.dist=1mm, l.angle=0}{v8}
\fmfdot{v1,v2,v3,v4,v5,v6}
\end{fmfgraph*}
}}
\no \\
&& \quad 
\no \\
&& \quad 
\no \\
&& \hspace*{0.2cm}
%
%
- \hspace*{0.3cm}
\parbox{18mm}{\centerline{
\begin{fmfgraph*}(8,8)
\setval
\fmfforce{0w,1/2h}{v1}
\fmfforce{1w,1/2h}{v2}
\fmfforce{3/4w,0.933h}{v3}
\fmfforce{1/4w,0.933h}{v4}
\fmfforce{3/4w,0.067h}{v5}
\fmfforce{1/4w,0.067h}{v6}
\fmfforce{-0.375w,0.067h}{v7}
\fmfforce{1.375w,0.067h}{v8}
\fmf{fermion,left=0.3}{v4,v3}
\fmf{boson,left=0.3}{v3,v4}
\fmf{fermion,left=0.3}{v3,v2}
\fmf{fermion,left=0.3}{v1,v4}
\fmf{boson}{v2,v1}
\fmf{fermion,left=0.3}{v6,v1}
\fmf{fermion,left=0.3}{v5,v6}
\fmf{fermion,left=0.3}{v2,v5}
\fmf{boson}{v6,v7}
\fmf{boson}{v8,v5}
\fmfv{decor.size=0, label=${\scs 1}$, l.dist=1mm, l.angle=-180}{v7}
\fmfv{decor.size=0, label=${\scs 2}$, l.dist=1mm, l.angle=0}{v8}
\fmfdot{v1,v2,v3,v4,v5,v6}
\end{fmfgraph*}
}}
%
%
\hspace*{0.3cm} - \hspace*{0.3cm}
\parbox{18mm}{\centerline{
\begin{fmfgraph*}(8,8)
\setval
\fmfforce{0w,1/2h}{v1}
\fmfforce{1w,1/2h}{v2}
\fmfforce{3/4w,0.933h}{v3}
\fmfforce{1/4w,0.933h}{v4}
\fmfforce{3/4w,0.067h}{v5}
\fmfforce{1/4w,0.067h}{v6}
\fmfforce{-0.375w,0.067h}{v7}
\fmfforce{1.375w,0.067h}{v8}
\fmf{fermion,right=0.3}{v3,v4}
\fmf{boson}{v3,v1}
\fmf{fermion,right=0.3}{v2,v3}
\fmf{fermion,right=0.3}{v4,v1}
\fmf{boson}{v4,v2}
\fmf{fermion,right=0.3}{v1,v6}
\fmf{fermion,right=0.3}{v6,v5}
\fmf{fermion,right=0.3}{v5,v2}
\fmf{boson}{v6,v7}
\fmf{boson}{v8,v5}
\fmfv{decor.size=0, label=${\scs 1}$, l.dist=1mm, l.angle=-180}{v7}
\fmfv{decor.size=0, label=${\scs 2}$, l.dist=1mm, l.angle=0}{v8}
\fmfdot{v1,v2,v3,v4,v5,v6}
\end{fmfgraph*}
}}
%
%
\hspace*{0.3cm} - \hspace*{0.3cm}
\parbox{18mm}{\centerline{
\begin{fmfgraph*}(8,8)
\setval
\fmfforce{0w,1/2h}{v1}
\fmfforce{1w,1/2h}{v2}
\fmfforce{3/4w,0.933h}{v3}
\fmfforce{1/4w,0.933h}{v4}
\fmfforce{3/4w,0.067h}{v5}
\fmfforce{1/4w,0.067h}{v6}
\fmfforce{-0.375w,0.067h}{v7}
\fmfforce{1.375w,0.067h}{v8}
\fmf{fermion,left=0.3}{v4,v3}
\fmf{boson}{v3,v1}
\fmf{fermion,left=0.3}{v3,v2}
\fmf{fermion,left=0.3}{v1,v4}
\fmf{boson}{v4,v2}
\fmf{fermion,left=0.3}{v6,v1}
\fmf{fermion,left=0.3}{v5,v6}
\fmf{fermion,left=0.3}{v2,v5}
\fmf{boson}{v6,v7}
\fmf{boson}{v8,v5}
\fmfv{decor.size=0, label=${\scs 1}$, l.dist=1mm, l.angle=-180}{v7}
\fmfv{decor.size=0, label=${\scs 2}$, l.dist=1mm, l.angle=0}{v8}
\fmfdot{v1,v2,v3,v4,v5,v6}
\end{fmfgraph*}
}}
%
%
\hspace*{0.3cm}- \hspace*{0.3cm}
\parbox{22mm}{\centerline{
\begin{fmfgraph*}(8,8)
\setval
\fmfforce{0w,1/2h}{v1}
\fmfforce{1w,1/2h}{v2}
\fmfforce{3/4w,0.933h}{v3}
\fmfforce{1/4w,0.933h}{v4}
\fmfforce{3/4w,0.067h}{v5}
\fmfforce{1/4w,0.067h}{v6}
\fmfforce{-5/8w,1/2h}{v7}
\fmfforce{13/8w,1/2h}{v8}
\fmf{fermion,right=0.3}{v3,v4}
\fmf{fermion,right=0.3}{v1,v6}
\fmf{fermion,right=0.3}{v6,v5}
\fmf{fermion,right=0.3}{v5,v2}
\fmf{fermion,right=0.3}{v4,v1}
\fmf{fermion,right=0.3}{v2,v3}
\fmf{boson,right=0.3}{v4,v3}
\fmf{boson,right=0.3}{v5,v6}
\fmf{boson}{v8,v2}
\fmf{boson}{v7,v1}
\fmfv{decor.size=0, label=${\scs 1}$, l.dist=1mm, l.angle=-180}{v7}
\fmfv{decor.size=0, label=${\scs 2}$, l.dist=1mm, l.angle=0}{v8}
\fmfdot{v1,v2,v3,v4,v5,v6}
\end{fmfgraph*}
}}
%
%
%
\hspace*{0.3cm}- \hspace*{0.3cm}
\parbox{22mm}{\centerline{
\begin{fmfgraph*}(8,8)
\setval
\fmfforce{1/2w,0h}{v1}
\fmfforce{1/2w,1h}{v2}
\fmfforce{0.933w,3/4h}{v3}
\fmfforce{0.933w,1/4h}{v4}
\fmfforce{0.067w,3/4h}{v5}
\fmfforce{0.067w,1/4h}{v6}
\fmfforce{11.864/8w,1/4h}{v7}
\fmfforce{-3.864/8w,1/4h}{v8}
\fmf{fermion,right=0.3}{v2,v5}
\fmf{fermion,right=0.3}{v3,v2}
\fmf{fermion,right=0.3}{v4,v3}
\fmf{fermion,right=0.3}{v1,v4}
\fmf{fermion,right=0.3}{v6,v1}
\fmf{fermion,right=0.3}{v5,v6}
\fmf{boson}{v1,v5}
\fmf{boson,right=0.3}{v2,v3}
\fmf{boson}{v6,v8}
\fmf{boson}{v4,v7}
\fmfv{decor.size=0, label=${\scs 1}$, l.dist=1mm, l.angle=-180}{v8}
\fmfv{decor.size=0, label=${\scs 2}$, l.dist=1mm, l.angle=0}{v7}
\fmfdot{v1,v2,v3,v4,v5,v6}
\end{fmfgraph*}
}}
\no \\%
&& \quad 
\no \\ && \quad 
\no \\&& \hspace*{0.2cm}
%
%
- \hspace*{0.3cm}
\parbox{22mm}{\centerline{
\begin{fmfgraph*}(8,8)
\setval
\fmfforce{1/2w,0h}{v1}
\fmfforce{1/2w,1h}{v2}
\fmfforce{0.933w,3/4h}{v3}
\fmfforce{0.933w,1/4h}{v4}
\fmfforce{0.067w,3/4h}{v5}
\fmfforce{0.067w,1/4h}{v6}
\fmfforce{11.864/8w,1/4h}{v7}
\fmfforce{-3.864/8w,1/4h}{v8}
\fmf{fermion,right=0.3}{v2,v5}
\fmf{fermion,right=0.3}{v3,v2}
\fmf{fermion,right=0.3}{v4,v3}
\fmf{fermion,right=0.3}{v1,v4}
\fmf{fermion,right=0.3}{v6,v1}
\fmf{fermion,right=0.3}{v5,v6}
\fmf{boson}{v1,v3}
\fmf{boson,left=0.3}{v2,v5}
\fmf{boson}{v6,v8}
\fmf{boson}{v4,v7}
\fmfv{decor.size=0, label=${\scs 1}$, l.dist=1mm, l.angle=-180}{v8}
\fmfv{decor.size=0, label=${\scs 2}$, l.dist=1mm, l.angle=0}{v7}
\fmfdot{v1,v2,v3,v4,v5,v6}
\end{fmfgraph*}
}}
%
%
\hspace*{0.3cm}- \hspace*{0.3cm}
\parbox{22mm}{\centerline{
\begin{fmfgraph*}(8,8)
\setval
\fmfforce{1/2w,0h}{v1}
\fmfforce{1/2w,1h}{v2}
\fmfforce{0.933w,3/4h}{v3}
\fmfforce{0.933w,1/4h}{v4}
\fmfforce{0.067w,3/4h}{v5}
\fmfforce{0.067w,1/4h}{v6}
\fmfforce{11.864/8w,1/4h}{v7}
\fmfforce{-3.864/8w,1/4h}{v8}
\fmf{fermion,right=0.3}{v2,v5}
\fmf{fermion,right=0.3}{v3,v2}
\fmf{fermion,right=0.3}{v4,v3}
\fmf{fermion,right=0.3}{v1,v4}
\fmf{fermion,right=0.3}{v6,v1}
\fmf{fermion,right=0.3}{v5,v6}
\fmf{boson}{v1,v3}
\fmf{boson,left=0.3}{v2,v5}
\fmf{boson}{v6,v8}
\fmf{boson}{v4,v7}
\fmfv{decor.size=0, label=${\scs 1}$, l.dist=1mm, l.angle=-180}{v8}
\fmfv{decor.size=0, label=${\scs 2}$, l.dist=1mm, l.angle=0}{v7}
\fmfdot{v1,v2,v3,v4,v5,v6}
\end{fmfgraph*}
}}
%
%
\hspace*{0.3cm}- \hspace*{0.3cm}
\parbox{22mm}{\centerline{
\begin{fmfgraph*}(8,8)
\setval
\fmfforce{1/2w,0h}{v1}
\fmfforce{1/2w,1h}{v2}
\fmfforce{0.933w,3/4h}{v3}
\fmfforce{0.933w,1/4h}{v4}
\fmfforce{0.067w,3/4h}{v5}
\fmfforce{0.067w,1/4h}{v6}
\fmfforce{11.864/8w,1/4h}{v7}
\fmfforce{-3.864/8w,1/4h}{v8}
\fmf{fermion,right=0.3}{v2,v5}
\fmf{fermion,right=0.3}{v3,v2}
\fmf{fermion,right=0.3}{v4,v3}
\fmf{fermion,right=0.3}{v1,v4}
\fmf{fermion,right=0.3}{v6,v1}
\fmf{fermion,right=0.3}{v5,v6}
\fmf{boson}{v1,v5}
\fmf{boson,right=0.3}{v2,v3}
\fmf{boson}{v6,v8}
\fmf{boson}{v4,v7}
\fmfv{decor.size=0, label=${\scs 1}$, l.dist=1mm, l.angle=-180}{v8}
\fmfv{decor.size=0, label=${\scs 2}$, l.dist=1mm, l.angle=0}{v7}
\fmfdot{v1,v2,v3,v4,v5,v6}
\end{fmfgraph*}
}}
%
%
\hspace*{0.3cm}+ \hspace*{0.3cm}
\parbox{19mm}{\centerline{
\begin{fmfgraph*}(8,8)
\setval
\fmfforce{0w,1/2h}{v1}
\fmfforce{1w,1/2h}{v2}
\fmfforce{3/4w,0.933h}{v3}
\fmfforce{1/4w,0.933h}{v4}
\fmfforce{3/4w,0.067h}{v5}
\fmfforce{1/4w,0.067h}{v6}
\fmfforce{-2.8/8w,0.067h}{v7}
\fmfforce{10.8/8w,0.067h}{v8}
\fmf{fermion,right=1}{v4,v3,v4}
\fmf{boson,right=0.3}{v2,v3}
\fmf{boson,right=0.3}{v4,v1}
\fmf{fermion}{v2,v1}
\fmf{fermion,right=0.3}{v1,v6}
\fmf{fermion,right=0.3}{v6,v5}
\fmf{fermion,right=0.3}{v5,v2}
\fmf{boson}{v6,v7}
\fmf{boson}{v8,v5}
\fmfv{decor.size=0, label=${\scs 1}$, l.dist=1mm, l.angle=-180}{v7}
\fmfv{decor.size=0, label=${\scs 2}$, l.dist=1mm, l.angle=0}{v8}
\fmfdot{v1,v2,v3,v4,v5,v6}
\end{fmfgraph*}
}}
%
%
\hspace*{0.3cm}+ \hspace*{0.3cm}
\parbox{19mm}{\centerline{
\begin{fmfgraph*}(8,8)
\setval
\fmfforce{0w,1/2h}{v1}
\fmfforce{1w,1/2h}{v2}
\fmfforce{3/4w,0.933h}{v3}
\fmfforce{1/4w,0.933h}{v4}
\fmfforce{3/4w,0.067h}{v5}
\fmfforce{1/4w,0.067h}{v6}
\fmfforce{-2.8/8w,0.067h}{v7}
\fmfforce{10.8/8w,0.067h}{v8}
\fmf{fermion,right=1}{v4,v3,v4}
\fmf{boson,right=0.3}{v2,v3}
\fmf{boson,right=0.3}{v4,v1}
\fmf{fermion}{v1,v2}
\fmf{fermion,left=0.3}{v6,v1}
\fmf{fermion,left=0.3}{v5,v6}
\fmf{fermion,left=0.3}{v2,v5}
\fmf{boson}{v6,v7}
\fmf{boson}{v8,v5}
\fmfv{decor.size=0, label=${\scs 1}$, l.dist=1mm, l.angle=-180}{v7}
\fmfv{decor.size=0, label=${\scs 2}$, l.dist=1mm, l.angle=0}{v8}
\fmfdot{v1,v2,v3,v4,v5,v6}
\end{fmfgraph*}
}}
\no \\%
&& \quad 
\no \\&& \quad 
\no \\
&& \hspace*{0.2cm}
%
%
%
+ \hspace*{0.3cm}
\parbox{28mm}{\centerline{
\begin{fmfgraph*}(25,10)
\setval
\fmfforce{0w,1/4h}{v1}
\fmfforce{1/5w,1/4h}{v2}
\fmfforce{2/5w,1/4h}{v3}
\fmfforce{3/5w,1/4h}{v4}
\fmfforce{4/5w,1/4h}{v5}
\fmfforce{5/5w,1/4h}{v6}
\fmfforce{3/5w,3/4h}{v7}
\fmfforce{4/5w,3/4h}{v8}
\fmf{boson}{v2,v1}
\fmf{fermion,right=1}{v2,v3,v2}
\fmf{boson}{v4,v3}
\fmf{fermion,right=0.4}{v4,v5}
\fmf{fermion,right=0.4}{v7,v4}
\fmf{fermion,right=0.4}{v8,v7}
\fmf{fermion,right=0.4}{v5,v8}
\fmf{boson}{v6,v5}
\fmf{boson,left=0.4}{v8,v7}
\fmfdot{v2,v3,v4,v5,v7,v8}
\fmfv{decor.size=0, label=${\scs 1}$, l.dist=1mm, l.angle=-180}{v1}
\fmfv{decor.size=0, label=${\scs 2}$, l.dist=1mm, l.angle=0}{v6}
\end{fmfgraph*}}}
%
%
\hspace*{0.3cm}+ \hspace*{0.3cm}
\parbox{28mm}{\centerline{
\begin{fmfgraph*}(25,10)
\setval
\fmfforce{0w,1/4h}{v1}
\fmfforce{1/5w,1/4h}{v2}
\fmfforce{2/5w,1/4h}{v3}
\fmfforce{3/5w,1/4h}{v4}
\fmfforce{4/5w,1/4h}{v5}
\fmfforce{5/5w,1/4h}{v6}
\fmfforce{3/5w,3/4h}{v7}
\fmfforce{4/5w,3/4h}{v8}
\fmf{boson}{v2,v1}
\fmf{fermion,right=1}{v2,v3,v2}
\fmf{boson}{v4,v3}
\fmf{fermion,left=0.4}{v5,v4}
\fmf{fermion,left=0.4}{v4,v7}
\fmf{fermion,left=0.4}{v7,v8}
\fmf{fermion,left=0.4}{v8,v5}
\fmf{boson}{v6,v5}
\fmf{boson,left=0.4}{v8,v7}
\fmfdot{v2,v3,v4,v5,v7,v8}
\fmfv{decor.size=0, label=${\scs 1}$, l.dist=1mm, l.angle=-180}{v1}
\fmfv{decor.size=0, label=${\scs 2}$, l.dist=1mm, l.angle=0}{v6}
\end{fmfgraph*}}}
%
%
 \hspace*{0.3cm}+ \hspace*{0.3cm}
\parbox{28mm}{\centerline{
\begin{fmfgraph*}(25,10)
\setval
\fmfforce{0w,1/4h}{v1}
\fmfforce{1/5w,1/4h}{v2}
\fmfforce{2/5w,1/4h}{v3}
\fmfforce{3/5w,1/4h}{v4}
\fmfforce{4/5w,1/4h}{v5}
\fmfforce{5/5w,1/4h}{v6}
\fmfforce{1/5w,3/4h}{v7}
\fmfforce{2/5w,3/4h}{v8}
\fmf{boson}{v1,v2}
\fmf{boson}{v3,v4}
\fmf{boson}{v6,v5}
\fmf{boson,right=0.4}{v7,v8}
\fmf{fermion,right=0.4}{v2,v3}
\fmf{fermion,right=0.4}{v7,v2}
\fmf{fermion,right=0.4}{v8,v7}
\fmf{fermion,right=0.4}{v3,v8}
\fmf{fermion,right=1}{v4,v5,v4}
\fmfdot{v2,v3,v4,v5,v7,v8}
\fmfv{decor.size=0, label=${\scs 1}$, l.dist=1mm, l.angle=-180}{v1}
\fmfv{decor.size=0, label=${\scs 2}$, l.dist=1mm, l.angle=0}{v6}
\end{fmfgraph*}}}
%
%
\hspace*{0.3cm}+ \hspace*{0.3cm}
\parbox{28mm}{\centerline{
\begin{fmfgraph*}(25,10)
\setval
\fmfforce{0w,1/4h}{v1}
\fmfforce{1/5w,1/4h}{v2}
\fmfforce{2/5w,1/4h}{v3}
\fmfforce{3/5w,1/4h}{v4}
\fmfforce{4/5w,1/4h}{v5}
\fmfforce{5/5w,1/4h}{v6}
\fmfforce{1/5w,3/4h}{v7}
\fmfforce{2/5w,3/4h}{v8}
\fmf{boson}{v1,v2}
\fmf{boson}{v3,v4}
\fmf{boson}{v6,v5}
\fmf{boson,right=0.4}{v7,v8}
\fmf{fermion,left=0.4}{v3,v2}
\fmf{fermion,left=0.4}{v2,v7}
\fmf{fermion,left=0.4}{v7,v8}
\fmf{fermion,left=0.4}{v8,v3}
\fmf{fermion,right=1}{v4,v5,v4}
\fmfdot{v2,v3,v4,v5,v7,v8}
\fmfv{decor.size=0, label=${\scs 1}$, l.dist=1mm, l.angle=-180}{v1}
\fmfv{decor.size=0, label=${\scs 2}$, l.dist=1mm, l.angle=0}{v6}
\end{fmfgraph*}}}
\no \\%
&& \quad 
\no \\%
&& \quad 
\no \\
&& \hspace*{0.2cm}
%
%
+ \hspace*{0.3cm}
\parbox{30mm}{\centerline{
\begin{fmfgraph*}(27,7)
\setval
\fmfforce{0w,1/2h}{v1}
\fmfforce{5/27w,1/2h}{v2}
\fmfforce{10/27w,1/2h}{v3}
\fmfforce{15/27w,1/2h}{v4}
\fmfforce{22/27w,1/2h}{v5}
\fmfforce{27/27w,1/2h}{v6}
\fmfforce{18.5/27w,0h}{v7}
\fmfforce{18.5/27w,1h}{v8}
\fmf{boson}{v1,v2}
\fmf{boson}{v6,v5}
\fmf{boson}{v7,v8}
\fmf{boson}{v4,v3}
\fmf{fermion,right=1}{v2,v3,v2}
\fmf{fermion,right=0.4}{v4,v7,v5,v8,v4}
\fmfdot{v2,v3,v4,v5,v7,v8}
\fmfv{decor.size=0, label=${\scs 1}$, l.dist=1mm, l.angle=-180}{v1}
\fmfv{decor.size=0, label=${\scs 2}$, l.dist=1mm, l.angle=0}{v6}
\end{fmfgraph*}}}
%
%
 \hspace*{0.3cm}+ \hspace*{0.3cm}
\parbox{30mm}{\centerline{
\begin{fmfgraph*}(27,7)
\setval
\fmfforce{0w,1/2h}{v1}
\fmfforce{5/27w,1/2h}{v2}
\fmfforce{10/27w,1/2h}{v3}
\fmfforce{15/27w,1/2h}{v4}
\fmfforce{22/27w,1/2h}{v5}
\fmfforce{27/27w,1/2h}{v6}
\fmfforce{18.5/27w,0h}{v7}
\fmfforce{18.5/27w,1h}{v8}
\fmf{boson}{v1,v2}
\fmf{boson}{v6,v5}
\fmf{boson}{v7,v8}
\fmf{boson}{v4,v3}
\fmf{fermion,right=1}{v2,v3,v2}
\fmf{fermion,left=0.4}{v4,v8,v5,v7,v4}
\fmfdot{v2,v3,v4,v5,v7,v8}
\fmfv{decor.size=0, label=${\scs 1}$, l.dist=1mm, l.angle=-180}{v1}
\fmfv{decor.size=0, label=${\scs 2}$, l.dist=1mm, l.angle=0}{v6}
\end{fmfgraph*}}}
%
%
\hspace*{0.3cm}  + \hspace*{0.3cm}
\parbox{38mm}{\centerline{
\begin{fmfgraph*}(35,5)
\setval
\fmfforce{0w,1/2h}{v1}
\fmfforce{1/7w,1/2h}{v2}
\fmfforce{2/7w,1/2h}{v3}
\fmfforce{3/7w,1/2h}{v4}
\fmfforce{4/7w,1/2h}{v5}
\fmfforce{5/7w,1/2h}{v6}
\fmfforce{6/7w,1/2h}{v7}
\fmfforce{7/7w,1/2h}{v8}
\fmf{boson}{v1,v2}
\fmf{boson}{v3,v4}
\fmf{boson}{v5,v6}
\fmf{boson}{v7,v8}
\fmf{fermion,right=1}{v2,v3,v2}
\fmf{fermion,right=1}{v4,v5,v4}
\fmf{fermion,right=1}{v6,v7,v6}
\fmfdot{v2,v3,v4,v5,v7,v6}
\fmfv{decor.size=0, label=${\scs 1}$, l.dist=1mm, l.angle=-180}{v1}
\fmfv{decor.size=0, label=${\scs 2}$, l.dist=1mm, l.angle=0}{v8}
\end{fmfgraph*}}}
\hspace*{0.4cm} . \la{PH3}
\eeq
\end{fmffile}
\begin{fmffile}{sd16}
\subsection{One-Particle Irreducible Vacuum Diagrams}
\la{CONNVAC2}
By analogy with the discussion of the connected vacuum diagrams in Subection
\r{CONNVAC}, also the one-particle irreducible vacuum diagrams of QED can
be generated together with their weights in two different ways.  
First, we show that they can be constructed from the diagrams of the
electron and photon connected two-point function as well as the
one-particle irreducible three-point function which have already been
determined in the previous subsection. Second, we derive from
the functional identities developed so far a nonlinear functional 
differential equation for the effective energy of the first kind and convert
it into a graphical recursion relation which directly generates the
one-particle irreducible vacuum diagrams.
\subsubsection{Relation to the Diagrams of the One-Particle Irreducible $n$-Point
Functions}
After having iteratively solved the closed set of graphical recursion 
relations (\r{R1})--(\r{R5}) for the electron and photon self-energy
and connected two-point function
as well as the one-particle irreducible three-point function, the 
corresponding loopwise one-particle irreducible vacuum diagrams can be
constructed as follows. Going back to the defining equations 
(\r{S3})--(\r{V3}), we obtain with (\r{DIV}), with the functional chain rule (\r{FCR}) and its analogue
\beq
\frac{\delta}{\delta D^{-1}_{12}} = - \frac{1}{2} \int_{34} \left\{ D_{13} D_{24} +  D_{14} D_{23} \right\} 
\frac{\delta}{\delta D_{34}} 
\eeq
three functional differential equations for the effective energy of the 
first kind:
\beq
\int_{12} S_{12} \frac{\delta \Gamma_1}{\delta S_{12}} & = &
\int_{12} S^{-1}_{21} \fulls_{12} \, ,  \la{VV1} \\
\int_{12} D_{12} \frac{\delta \Gamma_1}{\delta D_{12}} & = &
- \frac{1}{2} \int_{12} D^{-1}_{12} \fulld_{12} 
- \frac{1}{2} \int_{12} D_{12}^{-1} \fulla_1 \fulla_2  \, , \la{VV2} \\
\int_{123} V_{123} \frac{\delta \Gamma_1}{\delta V_{123}} & = &
\int_{123456} V_{123} \tau_{456} \fulls_{24} \fulls_{51} \fulld_{36}
- \int_{123} V_{123} \fulls_{21} \fulla_3 \, . \la{VV3}
\eeq
Their graphical representiations are
\beq
\parbox{8mm}{\begin{center}
\begin{fmfgraph*}(2.5,5)
\setval
\fmfstraight
\fmfforce{1w,0h}{v1}
\fmfforce{1w,1h}{v2}
\fmf{electron,left=1}{v1,v2}
\fmfv{decor.size=0, label=${\scs 2}$, l.dist=1mm, l.angle=0}{v1}
\fmfv{decor.size=0, label=${\scs 1}$, l.dist=1mm, l.angle=0}{v2}
\end{fmfgraph*}
\end{center}}
\hspace*{0.3cm} \dephi{-\Gamma_1}{1}{2} \quad & = & \quad - \, 
\parbox{8mm}{\centerline{
\begin{fmfgraph}(5,5)
\setval
\fmfforce{0w,0.5h}{v1}
\fmfforce{1w,0.5h}{v2}
\fmf{heavy,width=0.2mm,right=1}{v2,v1}
\fmf{fermion,right=1}{v1,v2}
\end{fmfgraph} }} 
\hspace*{0.4cm} , \la{IVAC1} \\
\parbox{8mm}{\begin{center}
\begin{fmfgraph*}(2.5,5)
\setval
\fmfstraight
\fmfforce{1w,0h}{v1}
\fmfforce{1w,1h}{v2}
\fmf{photon,left=1}{v1,v2}
\fmfv{decor.size=0, label=${\scs 2}$, l.dist=1mm, l.angle=0}{v1}
\fmfv{decor.size=0, label=${\scs 1}$, l.dist=1mm, l.angle=0}{v2}
\end{fmfgraph*}
\end{center}}
\hspace*{0.3cm} \dbphi{- \Gamma_1}{1}{2} \quad & = & \quad  
\frac{1}{2} \hspace*{0.2cm}
\parbox{8mm}{\centerline{
\begin{fmfgraph}(5,5)
\setval
\fmfforce{0w,0.5h}{v1}
\fmfforce{1w,0.5h}{v2}
\fmf{dbl_wiggly,width=0.2mm,right=1}{v2,v1}
\fmf{boson,right=1}{v1,v2}
\end{fmfgraph} }} 
\hspace*{0.2cm} + \hspace*{0.2cm} \frac{1}{2} \hspace*{0.2cm}
\parbox{8mm}{\centerline{
\begin{fmfgraph}(5,5)
\setval
\fmfforce{0w,0.5h}{v1}
\fmfforce{1w,0.5h}{v2}
\fmf{photon}{v1,v2}
\fmfdot{v1,v2}
\end{fmfgraph} }} 
\hspace*{0.4cm} , \la{IVAC2} \\
\parbox{5mm}{\begin{center}
\begin{fmfgraph*}(3,4)
\setval
\fmfstraight
\fmfforce{0w,1/2h}{v1}
\fmfforce{1w,1/2h}{v2}
\fmfforce{1w,1.25h}{v3}
\fmfforce{1w,-0.25h}{v4}
\fmf{photon}{v1,v2}
\fmf{electron}{v1,v3}
\fmf{electron}{v4,v1}
\fmfv{decor.size=0, label=${\scs 3}$, l.dist=1mm, l.angle=0}{v2}
\fmfv{decor.size=0, label=${\scs 1}$, l.dist=1mm, l.angle=0}{v3}
\fmfv{decor.size=0, label=${\scs 2}$, l.dist=1mm, l.angle=0}{v4}
\fmfdot{v1}
\end{fmfgraph*}
\end{center}}
\quad \dvertex{- \Gamma_1}{3}{2}{1} \quad & = & \quad - \,\,
\parbox{10mm}{\begin{center}
\begin{fmfgraph}(7,6)
\setval
\fmfforce{0w,1/2h}{v1}
\fmfforce{5/7w,1/2h}{v2}
\fmfforce{6/7w,1/2h}{v3}
\fmfforce{1w,1/2h}{v4}
\fmfforce{0.8w,0.67h}{v5}
\fmfforce{0.8w,0.33h}{v6}
\fmf{dbl_wiggly,width=0.2mm}{v1,v2}
\fmf{double,width=0.2mm,left=1}{v2,v4,v2}
\fmf{heavy,width=0.2mm,right=0.9}{v5,v1}
\fmf{heavy,width=0.2mm,right=0.9}{v1,v6}
\fmfdot{v1}
\end{fmfgraph}
\end{center} }
\quad - \quad
\parbox{13mm}{\centerline{
\begin{fmfgraph}(10,5)
\setval
\fmfforce{0w,0.5h}{v1}
\fmfforce{0.5w,0.5h}{v2}
\fmfforce{1w,0.5h}{v3}
\fmfforce{3/4w,0h}{v4}
\fmfforce{3/4w,1h}{v5}
\fmf{heavy,width=0.2mm,right=1}{v4,v5}
\fmf{double,width=0.2mm,right=1}{v5,v4}
\fmf{photon}{v1,v2}
\fmfdot{v2,v1}
\end{fmfgraph} }} 
\hspace*{0.4cm} . \la{IVAC3}
\eeq
They are based on counting the electron lines, the photon lines, and the
three-vertices of the one-particle irreducible vacuum diagrams. If the 
interaction $V$ vanishes, the Eqs. (\r{IVAC1})--(\r{IVAC3}) are solved by the free effective energy of the first kind
(\r{ZEGA}) due to (\r{SST1})--(\r{SST3}). For a non-vanishing interaction
$V$ the Eqs. (\r{IVAC1})--(\r{IVAC3}) produce corrections to (\r{ZEGA})
which we shall denote with $\Gamma_1^{({\rm int})}$. Thus the effective energy
of the first kind $\Gamma_1$ decomposes according to
\beq
\la{DCO}
\Gamma_1 = \Gamma^{({\rm free})}_1 + \Gamma_1^{({\rm int})} \, .
\eeq
In the following we recursively determine $\Gamma_1^{({\rm int})}$ 
in a graphical way 
for a vanishing field expectation value, so that we can neglect the 
last term in both (\r{IVAC2}) and (\r{IVAC3}). 
Performing a loopwise expansion of the interaction part of the
effective energy of the first kind
\beq
\la{GLOOP}
\Gamma_1^{({\rm int})} = \sum_{l=2}^{\infty} \Gamma_1^{(l)} \, , 
\eeq
we use the following eigenvalue problems for $l \ge 2$:
\beq
\parbox{8mm}{\begin{center}
\begin{fmfgraph*}(2.5,5)
\setval
\fmfstraight
\fmfforce{1w,0h}{v1}
\fmfforce{1w,1h}{v2}
\fmf{electron,left=1}{v1,v2}
\fmfv{decor.size=0, label=${\scs 2}$, l.dist=1mm, l.angle=0}{v1}
\fmfv{decor.size=0, label=${\scs 1}$, l.dist=1mm, l.angle=0}{v2}
\end{fmfgraph*}
\end{center}}
\hspace*{0.3cm} \dephi{\Gamma_1^{(l)}}{1}{2} 
\quad & = & \quad 2 (l-1) \,\,\Gamma_1^{(l)} 
\hspace*{0.4cm} , \la{EWW1} \\
\parbox{8mm}{\begin{center}
\begin{fmfgraph*}(2.5,5)
\setval
\fmfstraight
\fmfforce{1w,0h}{v1}
\fmfforce{1w,1h}{v2}
\fmf{photon,left=1}{v1,v2}
\fmfv{decor.size=0, label=${\scs 2}$, l.dist=1mm, l.angle=0}{v1}
\fmfv{decor.size=0, label=${\scs 1}$, l.dist=1mm, l.angle=0}{v2}
\end{fmfgraph*}
\end{center}}
\hspace*{0.3cm} \dbphi{\Gamma_1^{(l)}}{1}{2} \quad & = & \quad (l-1) \,\, 
\Gamma_1^{(l)}  \hspace*{0.4cm} ,  \la{EWW2} \\
\parbox{5mm}{\begin{center}
\begin{fmfgraph*}(3,4)
\setval
\fmfstraight
\fmfforce{0w,1/2h}{v1}
\fmfforce{1w,1/2h}{v2}
\fmfforce{1w,1.25h}{v3}
\fmfforce{1w,-0.25h}{v4}
\fmf{photon}{v1,v2}
\fmf{electron}{v1,v3}
\fmf{electron}{v4,v1}
\fmfv{decor.size=0, label=${\scs 3}$, l.dist=1mm, l.angle=0}{v2}
\fmfv{decor.size=0, label=${\scs 1}$, l.dist=1mm, l.angle=0}{v3}
\fmfv{decor.size=0, label=${\scs 2}$, l.dist=1mm, l.angle=0}{v4}
\fmfdot{v1}
\end{fmfgraph*}
\end{center}}
\quad \dvertex{\Gamma_1^{(l)}}{3}{2}{1} \quad & = & \quad 2 (l-1) \,\,
\Gamma_1^{(l)} \hspace*{0.4cm} . \la{EWW3}
\eeq
With these we explicitly solve (\r{IVAC1})--(\r{IVAC3})  for the expansion 
coefficients $\Gamma_1^{(l)}$ and obtain for $l \ge 2$
\beq
\la{VACC1}
- \Gamma^{(l)}_1 \quad & = & \quad - \frac{1}{2(l-1)} \hspace*{0.2cm}
\parbox{8mm}{\centerline{
\begin{fmfgraph*}(5,5)
\setval
\fmfforce{0w,0.5h}{v1}
\fmfforce{1w,0.5h}{v2}
\fmfforce{0.5w,1.1h}{v3}
\fmf{heavy,width=0.2mm,right=1}{v2,v1}
\fmf{fermion,right=1}{v1,v2}
\fmfv{decor.size=0, label=${\scs (l-1)}$, l.dist=1mm, l.angle=90}{v3}
\end{fmfgraph*} }} 
\hspace*{0.4cm} , \\ && \\
\la{VACC2}
-\Gamma_1^{(l)} \quad & = & \quad \frac{1}{2(l-1)} \hspace*{0.2cm} 
\parbox{8mm}{\centerline{
\begin{fmfgraph*}(5,5)
\setval
\fmfforce{0w,0.5h}{v1}
\fmfforce{1w,0.5h}{v2}
\fmfforce{0.5w,1.1h}{v3}
\fmf{dbl_wiggly,width=0.2mm,right=1}{v2,v1}
\fmf{boson,right=1}{v1,v2}
\fmfv{decor.size=0, label=${\scs (l-1)}$, l.dist=1mm, l.angle=90}{v3}
\end{fmfgraph*} }} 
\hspace*{0.4cm} , \\ && \\
\la{VACC3}
-\Gamma^{(l)}_1 \quad & = & \quad - \frac{1}{2(l-1)} 
\hspace*{0.2cm} \sum_{k_1=0}^l \sum_{k_2=0}^l 
\sum_{k_3=0}^{l-k_1-k_2-2} 
\hspace*{0.5cm} 
\parbox{14mm}{\centerline{
\begin{fmfgraph*}(11,6)
\setval
\fmfforce{-2/11w,0.3h}{v1}
\fmfforce{5/11w,0.3h}{v2}
\fmfforce{1w,0.3h}{v3}
\fmfforce{8/11w,0.3h}{v4}
\fmfforce{8/11w,-0.2h}{v5}
\fmfforce{8/11w,0.8h}{v6}
\fmfforce{7/11w,0.77h}{v7}
\fmfforce{7/11w,-0.197h}{v8}
\fmfforce{5.5/11w,1.6h}{v9}
\fmfforce{2.5/11w,0.7h}{v10}
\fmfforce{3/11w,-0.9h}{v11}
\fmf{double,width=0.2mm,left=1}{v2,v3,v2}
\fmf{dbl_wiggly,width=0.2mm}{v1,v2}
\fmf{heavy,width=0.2mm,right=0.8}{v1,v8}
\fmf{heavy,width=0.2mm,right=0.8}{v7,v1}
\fmfdot{v1}
\fmfv{decor.size=0, label=${\scs k_3}$, l.dist=0mm, l.angle=90}{v4}
\fmfv{decor.size=0, label=${\scs (l-k_1-k_2-k_3-2)}$, l.dist=0mm, l.angle=90}{v9}
\fmfv{decor.size=0, label=${\scs (k_2)}$, l.dist=0mm, l.angle=90}{v10}
\fmfv{decor.size=0, label=${\scs (k_1)}$, l.dist=0mm, l.angle=-90}{v11}
\end{fmfgraph*} }} 
\hspace*{0.4cm} . \\
&&\no 
\eeq
Inserting (\r{SST1})--(\r{PH3}) for the lower loop contributions to the 
connected electron and photon two-point function as well as the 
one-particle irreducible three-point function in one of the equations 
(\r{VACC1})--(\r{VACC3}), we get the effective energy of the first kind
for $l=2$ loops
\beq
\la{G2}
- \Gamma_1^{(2)} \quad = \quad 
%
%
- \hspace*{0.1cm} \frac{1}{2} \hspace*{0.3cm}  
\parbox{10mm}{\centerline{
\begin{fmfgraph}(6,6)
\setval
\fmfforce{0w,1/2h}{v1}
\fmfforce{1w,1/2h}{v2}
\fmf{boson}{v1,v2}
\fmf{fermion,right=1}{v1,v2}
\fmf{fermion,right=1}{v2,v1}
\fmfdot{v1,v2}
\end{fmfgraph}
}} 
\hspace*{0.4cm} , 
\eeq
for $l=3$ loops
\beq
\la{G3}
- \Gamma_1^{(3)} \quad = \quad 
%
%
- \hspace*{0.1cm} \frac{1}{4}\hspace*{0.2cm}
\parbox{8mm}{\begin{center}
\begin{fmfgraph}(5,5)
\setval
\fmfforce{0w,0h}{v1}
\fmfforce{0w,1h}{v2}
\fmfforce{1w,0h}{v3}
\fmfforce{1w,1h}{v4}
\fmf{fermion,right=0.4}{v2,v1}
\fmf{fermion,right=0.4}{v4,v2}
\fmf{fermion,right=0.4}{v3,v4}
\fmf{fermion,right=0.4}{v1,v3}
\fmf{boson}{v1,v4}
\fmf{boson}{v3,v2}
\fmfdot{v1,v2,v3,v4}
\end{fmfgraph}
\end{center}}
%
%
\hspace*{0.2cm}+ \hspace*{0.1cm} \frac{1}{4} \hspace*{0.2cm}
\parbox{14mm}{\begin{center}
\begin{fmfgraph}(12,12)
\setval
\fmfforce{3.5/12w,1/2h}{v1}
\fmfforce{8.5/12w,1/2h}{v2}
\fmfforce{1/2w,0h}{v3}
\fmfforce{1/2w,3.5/12h}{v4}
\fmfforce{1/2w,8.5/12h}{v5}
\fmfforce{1/2w,1h}{v6}
\fmf{fermion,right=1}{v3,v6,v3}
\fmf{boson}{v3,v4}
\fmf{boson}{v5,v6}
\fmf{fermion,left=1}{v4,v5,v4}
\fmfdot{v3,v4,v5,v6}
\end{fmfgraph}
\end{center}}
%
%
\hspace*{0.2cm} - \hspace*{0.1cm} \frac{1}{2}\hspace*{0.2cm}
\parbox{8mm}{\begin{center}
\begin{fmfgraph}(5,5)
\setval
\fmfforce{0w,0h}{v1}
\fmfforce{0w,1h}{v2}
\fmfforce{1w,0h}{v3}
\fmfforce{1w,1h}{v4}
\fmf{fermion,right=0.4}{v2,v1}
\fmf{fermion,right=0.4}{v4,v2}
\fmf{fermion,right=0.4}{v3,v4}
\fmf{fermion,right=0.4}{v1,v3}
\fmf{boson,right=0.4}{v1,v2}
\fmf{boson,right=0.4}{v4,v3}
\fmfdot{v1,v2,v3,v4}
\end{fmfgraph}
\end{center}}
\hspace*{0.4cm} ,
\eeq 
and for four loops
\beq
- \Gamma_1^{(4)} \hspace*{0.2cm} & = & \hspace*{0.2cm}
%
%
\frac{1}{6}\hspace*{0.1cm}
\parbox{16mm}{\begin{center}
\begin{fmfgraph}(12,12)
\setval
\fmfforce{3.5/12w,1/2h}{v1}
\fmfforce{8.5/12w,1/2h}{v2}
\fmfforce{1/2w,0h}{v3}
\fmfforce{1/2w,3.5/12h}{v4}
\fmfforce{1/2w,8.5/12h}{v5}
\fmfforce{1/2w,1h}{v6}
\fmfforce{8.165/12w,7.25/12h}{v7}
\fmfforce{3.835/12w,7.25/12h}{v8}
\fmfforce{11.196/12w,9/12h}{v9}
\fmfforce{0.834/12w,9/12h}{v10}
\fmf{fermion,right=0.6}{v3,v9,v10,v3}
\fmf{boson}{v7,v9}
\fmf{boson}{v3,v4}
\fmf{boson}{v8,v10}
\fmf{fermion,right=0.55}{v4,v7,v8,v4}
\fmfdot{v3,v4,v7,v8,v9,v10}
\end{fmfgraph}
\end{center}}
%
%
\hspace*{0.1cm} + \hspace*{0.05cm} \frac{1}{6}\hspace*{0.1cm}
\parbox{16mm}{\begin{center}
\begin{fmfgraph}(12,12)
\setval
\fmfforce{3.5/12w,1/2h}{v1}
\fmfforce{8.5/12w,1/2h}{v2}
\fmfforce{1/2w,0h}{v3}
\fmfforce{1/2w,3.5/12h}{v4}
\fmfforce{1/2w,8.5/12h}{v5}
\fmfforce{1/2w,1h}{v6}
\fmfforce{8.165/12w,7.25/12h}{v7}
\fmfforce{3.835/12w,7.25/12h}{v8}
\fmfforce{11.196/12w,9/12h}{v9}
\fmfforce{0.834/12w,9/12h}{v10}
\fmf{fermion,right=0.6}{v3,v9,v10,v3}
\fmf{boson}{v7,v9}
\fmf{boson}{v3,v4}
\fmf{boson}{v8,v10}
\fmf{fermion,left=0.55}{v4,v8,v7,v4}
\fmfdot{v3,v4,v7,v8,v9,v10}
\end{fmfgraph}
\end{center}}
%
%
\hspace*{0.3cm}- \hspace*{0.1cm} \frac{1}{6}\hspace*{0.3cm}
\parbox{11mm}{\begin{center}
\begin{fmfgraph}(8,8)
\setval
\fmfforce{0w,0.5h}{v1}
\fmfforce{0.25w,0.933h}{v2}
\fmfforce{0.75w,0.933h}{v3}
\fmfforce{1w,0.5h}{v4}
\fmfforce{0.75w,0.067h}{v5}
\fmfforce{0.25w,0.067h}{v6}
\fmf{fermion,right=0.3}{v1,v6,v5,v4,v3,v2,v1}
\fmf{boson}{v1,v4}
\fmf{boson}{v2,v5}
\fmf{boson}{v3,v6}
\fmfdot{v1,v2,v3,v4,v5,v6}
\end{fmfgraph}
\end{center}}
%
%
\hspace*{0.3cm}- \hspace*{0.1cm}\frac{1}{2}\hspace*{0.3cm}
\parbox{11mm}{\begin{center}
\begin{fmfgraph}(8,8)
\setval
\fmfforce{0w,0.5h}{v1}
\fmfforce{0.25w,0.933h}{v2}
\fmfforce{0.75w,0.933h}{v3}
\fmfforce{1w,0.5h}{v4}
\fmfforce{0.75w,0.067h}{v5}
\fmfforce{0.25w,0.067h}{v6}
\fmf{fermion,right=0.3}{v1,v6,v5,v4,v3,v2,v1}
\fmf{boson}{v1,v4}
\fmf{boson}{v2,v6}
\fmf{boson}{v3,v5}
\fmfdot{v1,v2,v3,v4,v5,v6}
\end{fmfgraph}
\end{center}}
%
%
\hspace*{0.3cm}+ \hspace*{0.1cm}\frac{1}{2}\hspace*{0.3cm}
\parbox{15mm}{\begin{center}
\begin{fmfgraph}(12,12)
\setval
\fmfforce{3.5/12w,1/2h}{v1}
\fmfforce{8.5/12w,1/2h}{v2}
\fmfforce{1/2w,0h}{v3}
\fmfforce{1/2w,3.5/12h}{v4}
\fmfforce{1/2w,8.5/12h}{v5}
\fmfforce{1/2w,1h}{v6}
\fmf{fermion,right=1}{v3,v6,v3}
\fmf{boson}{v1,v2}
\fmf{boson}{v3,v4}
\fmf{boson}{v5,v6}
\fmf{fermion,left=0.4}{v4,v1,v5,v2,v4}
\fmfdot{v1,v2,v3,v4,v5,v6}
\end{fmfgraph}
\end{center}}
\no \\
&& \hspace*{0.2cm}
%
%
- \hspace*{0.1cm}\frac{1}{6}\hspace*{0.3cm}
\parbox{13mm}{\begin{center}
\begin{fmfgraph}(10,10)
\setval
\fmfforce{0w,1/2h}{v1}
\fmfforce{1/4w,0h}{v2}
\fmfforce{3/4w,0h}{v3}
\fmfforce{1w,1/2h}{v4}
\fmfforce{3/4w,1h}{v5}
\fmfforce{1/4w,1h}{v6}
\fmfforce{1/2w,1/2h}{v7}
\fmf{fermion,right=1}{v6,v5}
\fmf{fermion,right=1}{v5,v6}
\fmf{fermion,right=1}{v1,v2}
\fmf{fermion,right=1}{v2,v1}
\fmf{fermion,right=1}{v3,v4}
\fmf{fermion,right=1}{v4,v3}
\fmf{boson}{v2,v3}
\fmf{boson}{v4,v5}
\fmf{boson}{v6,v1}
\fmfdot{v1,v2,v3,v4,v5,v6}
\end{fmfgraph}
\end{center}}
%
%
\hspace*{0.3cm}- \hspace*{0.1cm}\frac{1}{3}\hspace*{0.3cm}
\parbox{11mm}{\begin{center}
\begin{fmfgraph}(8,8)
\setval
\fmfforce{0w,0.5h}{v1}
\fmfforce{0.25w,0.933h}{v2}
\fmfforce{0.75w,0.933h}{v3}
\fmfforce{1w,0.5h}{v4}
\fmfforce{0.75w,0.067h}{v5}
\fmfforce{0.25w,0.067h}{v6}
\fmf{fermion,right=0.3}{v1,v6,v5,v4,v3,v2,v1}
\fmf{boson,right=0.7}{v2,v3}
\fmf{boson,right=0.7}{v4,v5}
\fmf{boson,right=0.7}{v6,v1}
\fmfdot{v1,v2,v3,v4,v5,v6}
\end{fmfgraph}
\end{center}} 
%
%
\hspace*{0.3cm}- \hspace*{0.1cm}\frac{1}{2}\hspace*{0.3cm}
\parbox{11mm}{\begin{center}
\begin{fmfgraph}(8,8)
\setval
\fmfforce{0w,0.5h}{v1}
\fmfforce{0.25w,0.933h}{v2}
\fmfforce{0.75w,0.933h}{v3}
\fmfforce{1w,0.5h}{v4}
\fmfforce{0.75w,0.067h}{v5}
\fmfforce{0.25w,0.067h}{v6}
\fmf{fermion,right=0.3}{v1,v6,v5,v4,v3,v2,v1}
\fmf{boson,right=0.7}{v2,v3}
\fmf{boson}{v1,v4}
\fmf{boson,right=0.7}{v5,v6}
\fmfdot{v1,v2,v3,v4,v5,v6}
\end{fmfgraph}
\end{center}}
%
%
\hspace*{0.3cm}- \hspace*{0.15cm}
\parbox{11mm}{\begin{center}
\begin{fmfgraph}(8,8)
\setval
\fmfforce{0w,0.5h}{v1}
\fmfforce{0.25w,0.933h}{v2}
\fmfforce{0.75w,0.933h}{v3}
\fmfforce{1w,0.5h}{v4}
\fmfforce{0.75w,0.067h}{v5}
\fmfforce{0.25w,0.067h}{v6}
\fmf{fermion,right=0.3}{v1,v6,v5,v4,v3,v2,v1}
\fmf{boson,right=0.7}{v2,v3}
\fmf{boson,right=0.2}{v4,v6}
\fmf{boson,right=0.2}{v5,v1}
\fmfdot{v1,v2,v3,v4,v5,v6}
\end{fmfgraph}
\end{center}}
%
%
\hspace*{0.3cm}+ \hspace*{0.3cm}
\parbox{14mm}{\begin{center}
\begin{fmfgraph}(12,12)
\setval
\fmfforce{3.5/12w,1/2h}{v1}
\fmfforce{8.5/12w,1/2h}{v2}
\fmfforce{1/2w,0h}{v3}
\fmfforce{1/2w,3.5/12h}{v4}
\fmfforce{1/2w,8.5/12h}{v5}
\fmfforce{1/2w,1h}{v6}
\fmfforce{0.7/12w,3.5/12h}{v7}
\fmfforce{0.7/12w,8.5/12h}{v8}
\fmf{fermion,right=1}{v3,v6}
\fmf{boson}{v3,v4}
\fmf{boson}{v5,v6}
\fmf{fermion,right=0.3}{v6,v8}
\fmf{fermion,right=0.25}{v8,v7}
\fmf{boson,left=0.25}{v8,v7}
\fmf{fermion,right=0.3}{v7,v3}
\fmf{fermion,right=1}{v5,v4,v5}
\fmfdot{v3,v4,v5,v6,v7,v8}
\end{fmfgraph}
\end{center}}
\hspace*{0.4cm}. \la{G4}
\eeq
A comparison with the corresponding results for the vacuum
energy in (\r{W2}), (\r{W3}), and (\r{W4}) shows that the effective energy
of the first kind contains precisely all one-particle irreducible vacuum diagrams.
\subsubsection{Graphical Recursion Relation}
Each of the three functional differential equations for the effective
energy of the first kind (\r{VV1})--(\r{VV3}) can be used to derive a
graphical recursion relation which directly generates the one-particle
irreducible vacuum diagrams. Here we restrict ourselves to the functional
differential equation (\r{VV1}) which is based on counting the number
of electron lines of the one-particle irreducible
vacuum diagrams. Inserting (\r{S3}),
(\r{DD3}), (\r{SE1}), (\r{CSD4}), and (\r{ZW2}) for the connected electron and
photon two-point function, the electron self-energy, and the
one-particle irreducible three-point function, we obtain from (\r{VV1}):
\beq
\delta_{11} \int_1 + 
\int_{12} S^{-1}_{12} \frac{\delta \Gamma_1}{\delta S_{12}^{-1}}
& = &  - \int_{123456} V_{123} V_{456} 
\frac{\delta^2 \Gamma_1}{\delta S_{12}^{-1} \delta S_{45}^{-1}}
\left\{ 2 \frac{\delta \Gamma_1}{\delta D_{36}^{-1}} - \fulla_3 \fulla_6
\right\}  - \int_{123} V_{123}
\frac{\delta \Gamma_1}{\delta S_{12}^{-1}} \fulla_3 \, .
\la{EFG}
\eeq
If the interaction $V$ vanishes, this is solved by the free effective energy of the first kind (\r{FEF}) which 
has the functional derivatives
\beq
\la{FU}
\frac{\delta \Gamma_1^{({\rm free})}}{\delta D^{-1}_{12}} = \frac{1}{2} D_{12} 
+ \frac{1}{2} \fulla_1 \fulla_2 \, , \hspace*{1cm}
\frac{\delta \Gamma_1^{({\rm free})}}{\delta S^{-1}_{12}} = - S_{21} \,, \hspace*{1cm}
\frac{\delta^2 \Gamma_1^{({\rm free})}}{\delta S^{-1}_{12} \delta S^{-1}_{34}}
= S_{23} S_{41} \, .
\eeq
For a non-vanishing interaction $V$, the right-hand side in (\r{EFG}) 
corrects (\r{FEF}) by the interaction part of the effective energy of the
first kind $\Gamma_1^{({\rm int})}$. 
Inserting the decomposition (\r{DCO})
into (\r{EFG}) and using (\r{FU}), we obtain together
with the functional chain rule the following functional differential 
equation for the interaction part of the effective energy of the first
kind $\Gamma_1^{({\rm int})}$:
\beq
&& \int_{12} S_{12} \frac{\delta \Gamma_1^{({\rm int})}}{\delta S_{12}} = 
\int_{123456} V_{123} V_{456} S_{24} S_{51} D_{36}
+ 2 \int_{12345678} V_{123} V_{456} D_{36} S_{51} S_{28} S_{74}
\frac{\delta \Gamma_1^{({\rm int})}}{\delta S_{78}} 
\no \\  && \hspace*{1cm}
- 2 \int_{12345678} V_{123} V_{456} S_{24} S_{51} D_{37} D_{68}
\frac{\delta \Gamma_1^{({\rm int})}}{\delta D_{78}} 
+ \int_{123456789\bar{1}} V_{123} V_{456} D_{36} S_{71} S_{28} S_{94}
S_{5\bar{1}} 
\frac{\delta^2 \Gamma_1^{({\rm int})}}{\delta S_{78} 
\delta S_{9\bar{1}}}
\no \\  && \hspace*{1cm}
- 2 \int_{123456789\bar{1}\bar{2} \bar{3}} V_{123}V_{456}
S_{71} S_{28} S_{94} S_{5\bar{1}} D_{3\bar{2}} D_{6\bar{3}}
\frac{\delta^2 \Gamma_1^{({\rm int})}}{\delta S_{78}
\delta S_{9\bar{1}}}
\frac{\delta \Gamma_1^{({\rm int})}}{\delta D_{\bar{2} \bar{3}}}
\no \\  && \hspace*{1cm}
- 4 \int_{123456789\bar{1}} V_{123} V_{456} 
S_{51} S_{74} S_{28} D_{39} D_{6\bar{1}}
\frac{\delta \Gamma_1^{({\rm int})}}{\delta S_{78}}
\frac{\delta \Gamma_1^{({\rm int})}}{\delta D_{9\bar{1}}}
- \int_{123} V_{123} S_{21} \fulla_3
- \int_{12345} V_{123} S_{41} S_{25} \fulla_3
\frac{\delta \Gamma_1^{({\rm int})}}{\delta S_{45}} \, .
\eeq
Its graphical representation reads:
\beq
\parbox{8mm}{\begin{center}
\begin{fmfgraph*}(2.5,5)
\setval
\fmfstraight
\fmfforce{1w,0h}{v1}
\fmfforce{1w,1h}{v2}
\fmf{electron,left=1}{v1,v2}
\fmfv{decor.size=0, label=${\scs 2}$, l.dist=1mm, l.angle=0}{v1}
\fmfv{decor.size=0, label=${\scs 1}$, l.dist=1mm, l.angle=0}{v2}
\end{fmfgraph*}
\end{center}}
\hspace*{0.2cm} \dephi{- \Gamma_1^{({\rm int})}}{1}{2} &\hspace*{0.2cm} = 
\hspace*{0.2cm} &
- \hspace*{0.2cm}  
\parbox{9mm}{\centerline{
\begin{fmfgraph}(6,6)
\setval
\fmfforce{0w,1/2h}{v1}
\fmfforce{1w,1/2h}{v2}
\fmf{boson}{v1,v2}
\fmf{fermion,right=1}{v1,v2}
\fmf{fermion,right=1}{v2,v1}
\fmfdot{v1,v2}
\end{fmfgraph}
}} 
\hspace*{0.2cm}+ \hspace*{0.1cm}2 \hspace*{0.2cm}
\parbox{10.5mm}{\begin{center}
\begin{fmfgraph*}(7.5,5)
\setval
\fmfstraight
\fmfforce{1/3w,0h}{v1}
\fmfforce{1/3w,1h}{v2}
\fmfforce{1w,0h}{v3}
\fmfforce{1w,1h}{v4}
\fmf{photon}{v1,v2}
\fmf{fermion}{v3,v1}
\fmf{fermion}{v2,v4}
\fmf{fermion,left=1}{v1,v2}
\fmfv{decor.size=0, label=${\scs 2}$, l.dist=1mm, l.angle=0}{v3}
\fmfv{decor.size=0, label=${\scs 1}$, l.dist=1mm, l.angle=0}{v4}
\fmfdot{v1,v2}
\end{fmfgraph*}
\end{center}}
\hspace*{0.2cm} \dephi{-\Gamma_1^{({\rm int})}}{1}{2}
\hspace*{0.2cm}- \hspace*{0.1cm}2 \hspace*{0.2cm}
\parbox{10.5mm}{\begin{center}
\begin{fmfgraph*}(7.5,5)
\setval
\fmfstraight
\fmfforce{1/3w,0h}{v1}
\fmfforce{1/3w,1h}{v2}
\fmfforce{1w,0h}{v3}
\fmfforce{1w,1h}{v4}
\fmf{photon}{v1,v3}
\fmf{photon}{v2,v4}
\fmf{fermion,right=1}{v1,v2,v1}
\fmfv{decor.size=0, label=${\scs 2}$, l.dist=1mm, l.angle=0}{v3}
\fmfv{decor.size=0, label=${\scs 1}$, l.dist=1mm, l.angle=0}{v4}
\fmfdot{v1,v2}
\end{fmfgraph*}
\end{center}}
\hspace*{0.2cm} \dbphi{-\Gamma_1^{({\rm int})}}{1}{2}
\hspace*{0.2cm}+ \hspace*{0.2cm}
\parbox{8mm}{\begin{center}
\begin{fmfgraph*}(5,9)
\setval
\fmfstraight
\fmfforce{0w,1/6h}{v1}
\fmfforce{0w,5/6h}{v2}
\fmfforce{1w,0h}{v3}
\fmfforce{1w,1/3h}{v4}
\fmfforce{1w,2/3h}{v5}
\fmfforce{1w,1h}{v6}
\fmf{photon}{v2,v1}
\fmf{fermion}{v3,v1}
\fmf{fermion}{v1,v4}
\fmf{fermion}{v5,v2}
\fmf{fermion}{v2,v6}
\fmfv{decor.size=0, label=${\scs 1}$, l.dist=1mm, l.angle=0}{v6}
\fmfv{decor.size=0, label=${\scs 2}$, l.dist=1mm, l.angle=0}{v5}
\fmfv{decor.size=0, label=${\scs 3}$, l.dist=1mm, l.angle=0}{v4}
\fmfv{decor.size=0, label=${\scs 4}$, l.dist=1mm, l.angle=0}{v3}
\fmfdot{v1,v2}
\end{fmfgraph*}
\end{center}}
\hspace*{0.3cm} \ddfermi{- \Gamma_1^{({\rm int})}}{1}{2}{3}{4}
\no \\ &&
+ \hspace*{0.1cm}
2 \hspace*{0.2cm} \dbphi{-\Gamma_1^{({\rm int})}}{1}{2}\hspace*{0.2cm} 
\parbox{14mm}{\begin{center}
\begin{fmfgraph*}(10,9)
\setval
\fmfstraight
\fmfforce{1/2w,1/6h}{v1}
\fmfforce{1/2w,5/6h}{v2}
\fmfforce{1w,0h}{v3}
\fmfforce{1w,1/3h}{v4}
\fmfforce{1w,2/3h}{v5}
\fmfforce{1w,1h}{v6}
\fmfforce{0w,1/6h}{v7}
\fmfforce{0w,5/6h}{v8}
\fmf{fermion}{v3,v1}
\fmf{fermion}{v1,v4}
\fmf{fermion}{v5,v2}
\fmf{fermion}{v2,v6}
\fmf{boson}{v1,v7}
\fmf{boson}{v2,v8}
\fmfv{decor.size=0, label=${\scs 3}$, l.dist=1mm, l.angle=0}{v6}
\fmfv{decor.size=0, label=${\scs 4}$, l.dist=1mm, l.angle=0}{v5}
\fmfv{decor.size=0, label=${\scs 5}$, l.dist=1mm, l.angle=0}{v4}
\fmfv{decor.size=0, label=${\scs 6}$, l.dist=1mm, l.angle=0}{v3}
\fmfv{decor.size=0, label=${\scs 2}$, l.dist=1mm, l.angle=-180}{v7}
\fmfv{decor.size=0, label=${\scs 1}$, l.dist=1mm, l.angle=-180}{v8}
\fmfdot{v1,v2}
\end{fmfgraph*}
\end{center}}
\hspace*{0.2cm} \ddfermi{- \Gamma_1^{({\rm int})}}{3}{4}{5}{6}
\hspace*{0.2cm}+ \hspace*{0.1cm}
4 \hspace*{0.2cm}\dbphi{-\Gamma_1^{({\rm int})}}{1}{2}
\hspace*{0.2cm}
\parbox{14mm}{\begin{center}
\begin{fmfgraph*}(10,5)
\setval
\fmfstraight
\fmfforce{0w,0h}{v1}
\fmfforce{0w,1h}{v2}
\fmfforce{1/2w,0h}{v3}
\fmfforce{1/2w,1h}{v4}
\fmfforce{1w,0h}{v5}
\fmfforce{1w,1h}{v6}
\fmf{photon}{v1,v3}
\fmf{photon}{v2,v4}
\fmf{fermion}{v3,v4}
\fmf{fermion}{v4,v6}
\fmf{fermion}{v5,v3}
\fmfv{decor.size=0, label=${\scs 1}$, l.dist=1mm, l.angle=-180}{v2}
\fmfv{decor.size=0, label=${\scs 2}$, l.dist=1mm, l.angle=-180}{v1}
\fmfv{decor.size=0, label=${\scs 3}$, l.dist=1mm, l.angle=0}{v6}
\fmfv{decor.size=0, label=${\scs 4}$, l.dist=1mm, l.angle=0}{v5}
\fmfdot{v3,v4}
\end{fmfgraph*}
\end{center}}
\hspace*{0.2cm} \dephi{-\Gamma_1^{({\rm int})}}{3}{4}
\no \\ &&
- \hspace*{0.2cm}  
\parbox{13mm}{\centerline{
\begin{fmfgraph}(10,5)
\setval
\fmfforce{0w,0.5h}{v1}
\fmfforce{0.5w,0.5h}{v2}
\fmfforce{1w,0.5h}{v3}
\fmfforce{3/4w,0h}{v4}
\fmfforce{3/4w,1h}{v5}
\fmf{fermion,right=1}{v4,v5}
\fmf{plain,right=1}{v5,v4}
\fmf{photon}{v1,v2}
\fmfdot{v2,v1}
\end{fmfgraph} }} 
\hspace*{0.1cm}+\hspace*{0.1cm}
\parbox{11.5mm}{\begin{center}
\begin{fmfgraph*}(8.5,7)
\setval
\fmfstraight
\fmfforce{0w,1/2h}{v1}
\fmfforce{5/8.5w,1/2h}{v2}
\fmfforce{1w,0h}{v3}
\fmfforce{1w,1h}{v4}
\fmf{photon}{v1,v2}
\fmf{fermion}{v3,v2}
\fmf{fermion}{v2,v4}
\fmfv{decor.size=0, label=${\scs 2}$, l.dist=1mm, l.angle=0}{v3}
\fmfv{decor.size=0, label=${\scs 1}$, l.dist=1mm, l.angle=0}{v4}
\fmfdot{v1,v2}
\end{fmfgraph*}
\end{center}}
\hspace*{0.2cm} \dephi{- \Gamma_1^{({\rm int})}}{1}{2} \hspace*{0.4cm}.
\la{GRGR}
\eeq
Again, we illustrate the graphical recursive solution of (\r{GRGR})
only for a vanishing field expectation value, so that we can drop the last two terms. Inserting the loop expansion (\r{GLOOP})
and taking into account the eigenvalue problem (\r{EWW1}), we obtain a 
graphical recursion relation for the expansion coefficients $\Gamma_1^{(l)}$
of the effective energy of the first kind for $l \ge 3$:
\beq
-\Gamma_1^{(l)} & \hspace*{0.2cm} = \hspace*{0.2cm} & \frac{1}{l-1} \left\{
\parbox{10.5mm}{\begin{center}
\begin{fmfgraph*}(7.5,5)
\setval
\fmfstraight
\fmfforce{1/3w,0h}{v1}
\fmfforce{1/3w,1h}{v2}
\fmfforce{1w,0h}{v3}
\fmfforce{1w,1h}{v4}
\fmf{photon}{v1,v2}
\fmf{fermion}{v3,v1}
\fmf{fermion}{v2,v4}
\fmf{fermion,left=1}{v1,v2}
\fmfv{decor.size=0, label=${\scs 2}$, l.dist=1mm, l.angle=0}{v3}
\fmfv{decor.size=0, label=${\scs 1}$, l.dist=1mm, l.angle=0}{v4}
\fmfdot{v1,v2}
\end{fmfgraph*}
\end{center}}
\hspace*{0.2cm} \dephi{-\Gamma_1^{(l-1)}}{1}{2}
\hspace*{0.2cm}- \hspace*{0.2cm}
\parbox{10.5mm}{\begin{center}
\begin{fmfgraph*}(7.5,5)
\setval
\fmfstraight
\fmfforce{1/3w,0h}{v1}
\fmfforce{1/3w,1h}{v2}
\fmfforce{1w,0h}{v3}
\fmfforce{1w,1h}{v4}
\fmf{photon}{v1,v3}
\fmf{photon}{v2,v4}
\fmf{fermion,right=1}{v1,v2,v1}
\fmfv{decor.size=0, label=${\scs 2}$, l.dist=1mm, l.angle=0}{v3}
\fmfv{decor.size=0, label=${\scs 1}$, l.dist=1mm, l.angle=0}{v4}
\fmfdot{v1,v2}
\end{fmfgraph*}
\end{center}}
\hspace*{0.2cm} \dbphi{-\Gamma_1^{(l-1)}}{1}{2}
\hspace*{0.2cm}+ \frac{1}{2} \hspace*{0.2cm}
\parbox{8mm}{\begin{center}
\begin{fmfgraph*}(5,9)
\setval
\fmfstraight
\fmfforce{0w,1/6h}{v1}
\fmfforce{0w,5/6h}{v2}
\fmfforce{1w,0h}{v3}
\fmfforce{1w,1/3h}{v4}
\fmfforce{1w,2/3h}{v5}
\fmfforce{1w,1h}{v6}
\fmf{photon}{v2,v1}
\fmf{fermion}{v3,v1}
\fmf{fermion}{v1,v4}
\fmf{fermion}{v5,v2}
\fmf{fermion}{v2,v6}
\fmfv{decor.size=0, label=${\scs 1}$, l.dist=1mm, l.angle=0}{v6}
\fmfv{decor.size=0, label=${\scs 2}$, l.dist=1mm, l.angle=0}{v5}
\fmfv{decor.size=0, label=${\scs 3}$, l.dist=1mm, l.angle=0}{v4}
\fmfv{decor.size=0, label=${\scs 4}$, l.dist=1mm, l.angle=0}{v3}
\fmfdot{v1,v2}
\end{fmfgraph*}
\end{center}}
\hspace*{0.3cm} \ddfermi{- \Gamma_1^{(l-1)}}{1}{2}{3}{4}
\right. \no \\
&& \left. 
+ \sum_{k=2}^{l-2} \left[ \, \dbphi{-\Gamma_1^{(k)}}{1}{2}\hspace*{0.2cm} 
\parbox{14mm}{\begin{center}
\begin{fmfgraph*}(10,9)
\setval
\fmfstraight
\fmfforce{1/2w,1/6h}{v1}
\fmfforce{1/2w,5/6h}{v2}
\fmfforce{1w,0h}{v3}
\fmfforce{1w,1/3h}{v4}
\fmfforce{1w,2/3h}{v5}
\fmfforce{1w,1h}{v6}
\fmfforce{0w,1/6h}{v7}
\fmfforce{0w,5/6h}{v8}
\fmf{fermion}{v3,v1}
\fmf{fermion}{v1,v4}
\fmf{fermion}{v5,v2}
\fmf{fermion}{v2,v6}
\fmf{boson}{v1,v7}
\fmf{boson}{v2,v8}
\fmfv{decor.size=0, label=${\scs 3}$, l.dist=1mm, l.angle=0}{v6}
\fmfv{decor.size=0, label=${\scs 4}$, l.dist=1mm, l.angle=0}{v5}
\fmfv{decor.size=0, label=${\scs 5}$, l.dist=1mm, l.angle=0}{v4}
\fmfv{decor.size=0, label=${\scs 6}$, l.dist=1mm, l.angle=0}{v3}
\fmfv{decor.size=0, label=${\scs 2}$, l.dist=1mm, l.angle=-180}{v7}
\fmfv{decor.size=0, label=${\scs 1}$, l.dist=1mm, l.angle=-180}{v8}
\fmfdot{v1,v2}
\end{fmfgraph*}
\end{center}}
\hspace*{0.2cm} \ddfermi{- \Gamma_1^{(l-k)}}{3}{4}{5}{6}
\hspace*{0.2cm}+ \hspace*{0.1cm}
2 \hspace*{0.2cm}\dbphi{-\Gamma_1^{(k)}}{1}{2}
\hspace*{0.2cm}
\parbox{14mm}{\begin{center}
\begin{fmfgraph*}(10,5)
\setval
\fmfstraight
\fmfforce{0w,0h}{v1}
\fmfforce{0w,1h}{v2}
\fmfforce{1/2w,0h}{v3}
\fmfforce{1/2w,1h}{v4}
\fmfforce{1w,0h}{v5}
\fmfforce{1w,1h}{v6}
\fmf{photon}{v1,v3}
\fmf{photon}{v2,v4}
\fmf{fermion}{v3,v4}
\fmf{fermion}{v4,v6}
\fmf{fermion}{v5,v3}
\fmfv{decor.size=0, label=${\scs 1}$, l.dist=1mm, l.angle=-180}{v2}
\fmfv{decor.size=0, label=${\scs 2}$, l.dist=1mm, l.angle=-180}{v1}
\fmfv{decor.size=0, label=${\scs 3}$, l.dist=1mm, l.angle=0}{v6}
\fmfv{decor.size=0, label=${\scs 4}$, l.dist=1mm, l.angle=0}{v5}
\fmfdot{v3,v4}
\end{fmfgraph*}
\end{center}}
\hspace*{0.2cm}\dephi{-\Gamma_1^{(l-k)}}{3}{4}
\right] \right\} \hspace*{0.4cm}.
\la{GRGRL}
\eeq
It is solved starting from $\Gamma_1^{(2)}$ in (\r{G2}). With the
line amputations
\beq
\dbphi{-\Gamma_1^{(2)}}{1}{2} \hspace*{0.2cm}= - \frac{1}{2} \hspace*{0.2cm}
\parbox{14mm}{\begin{center}
\begin{fmfgraph*}(11,5)
\setval
\fmfforce{0w,0.5h}{v1}
\fmfforce{3/11w,0.5h}{v2}
\fmfforce{8/11w,0.5h}{v3}
\fmfforce{1w,0.5h}{v4}
\fmf{boson}{v1,v2}
\fmf{fermion,right}{v2,v3,v2}
\fmf{boson}{v3,v4}
\fmfdot{v2,v3}
\fmfv{decor.size=0, label=${\scs 1}$, l.dist=1mm, l.angle=-180}{v1}
\fmfv{decor.size=0, label=${\scs 2}$, l.dist=1mm, l.angle=0}{v4}
\end{fmfgraph*}
\end{center}}
\hspace*{0.2cm}, \hspace*{0.5cm}
\dephi{-\Gamma_1^{(2)}}{1}{2} \hspace*{0.2cm}= - \hspace*{0.2cm}
\parbox{14mm}{\begin{center}
\begin{fmfgraph*}(11,5)
\setval
\fmfforce{0w,0.5h}{v1}
\fmfforce{3/11w,0.5h}{v2}
\fmfforce{8/11w,0.5h}{v3}
\fmfforce{1w,0.5h}{v4}
\fmf{fermion}{v4,v3,v2,v1}
\fmf{boson,left=1}{v2,v3}
\fmfdot{v2,v3}
\fmfv{decor.size=0, label=${\scs 1}$, l.dist=1mm, l.angle=-180}{v1}
\fmfv{decor.size=0, label=${\scs 2}$, l.dist=1mm, l.angle=0}{v4}
\end{fmfgraph*}
\end{center} }
\hspace*{0.2cm}, \hspace*{0.5cm}
\ddfermi{-\Gamma_1^{(2)}}{1}{2}{3}{4}\hspace*{0.2cm} = - \hspace*{0.2cm} 
\parbox{12mm}{\begin{center}
\begin{fmfgraph*}(9.243,4.243)
\setval
\fmfforce{0w,0h}{v1}
\fmfforce{0w,1h}{v2}
\fmfforce{2.12/9.243w,1/2h}{v3}
\fmfforce{7.12/9.243w,1/2h}{v4}
\fmfforce{1w,0h}{v5}
\fmfforce{1w,1h}{v6}
\fmf{boson}{v3,v4}
\fmf{fermion}{v3,v1}
\fmf{fermion}{v2,v3}
\fmf{fermion}{v5,v4}
\fmf{fermion}{v4,v6}
\fmfv{decor.size=0, label=${\scs 1}$, l.dist=1mm, l.angle=-180}{v1}
\fmfv{decor.size=0, label=${\scs 2}$, l.dist=1mm, l.angle=0}{v5}
\fmfv{decor.size=0, label=${\scs 3}$, l.dist=1mm, l.angle=-180}{v2}
\fmfv{decor.size=0, label=${\scs 4}$, l.dist=1mm, l.angle=0}{v6}
\fmfdot{v3,v4}
\end{fmfgraph*}
\end{center} }
\hspace*{0.4cm} ,
\eeq
we obtain from (\r{GRGRL}) the three-loop result $\Gamma_1^{(3)}$ in (\r{G3}). Performing the amputation of one photon line
\beq
\dbphi{-\Gamma_1^{(3)}}{1}{2} \hspace*{0.2cm}= 
%
%
\hspace*{0.2cm} - \hspace*{0.1cm} \frac{1}{2} \hspace*{0.3cm} 
\parbox{16mm}{\begin{center}
\begin{fmfgraph*}(13,7)
\setval
\fmfforce{0w,1/2h}{v1}
\fmfforce{3/13w,1/2h}{v2}
\fmfforce{10/13w,1/2h}{v3}
\fmfforce{1w,1/2h}{v4}
\fmfforce{1/2w,0h}{v5}
\fmfforce{1/2w,1h}{v6}
\fmf{boson}{v1,v2}
\fmf{boson}{v3,v4}
\fmf{boson}{v5,v6}
\fmf{fermion,right=0.4}{v2,v5}
\fmf{fermion,right=0.4}{v5,v3}
\fmf{fermion,right=0.4}{v3,v6}
\fmf{fermion,right=0.4}{v6,v2}
\fmfdot{v2,v3,v5,v6}
\fmfv{decor.size=0, label=${\scs 1}$, l.dist=1mm, l.angle=-180}{v1}
\fmfv{decor.size=0, label=${\scs 2}$, l.dist=1mm, l.angle=0}{v4}
\end{fmfgraph*}
\end{center} }
%
%
\hspace*{0.3cm} - \hspace*{0.1cm} \frac{1}{2} \hspace*{0.3cm} 
\parbox{14mm}{\begin{center}
\begin{fmfgraph*}(11,5)
\setval
\fmfforce{0w,0h}{v1}
\fmfforce{3/11w,0h}{v2}
\fmfforce{8/11w,0h}{v3}
\fmfforce{1w,0h}{v4}
\fmfforce{3/11w,1h}{v5}
\fmfforce{8/11w,1h}{v6}
\fmf{boson}{v1,v2}
\fmf{boson}{v3,v4}
\fmf{boson,right=0.4}{v5,v6}
\fmf{fermion,right=0.4}{v5,v2}
\fmf{fermion,right=0.4}{v6,v5}
\fmf{fermion,right=0.4}{v3,v6}
\fmf{fermion,right=0.4}{v2,v3}
\fmfdot{v2,v3,v5,v6}
\fmfv{decor.size=0, label=${\scs 1}$, l.dist=1mm, l.angle=-180}{v1}
\fmfv{decor.size=0, label=${\scs 2}$, l.dist=1mm, l.angle=0}{v4}
\end{fmfgraph*}
\end{center} }
%
%
\hspace*{0.3cm} - \hspace*{0.1cm} \frac{1}{2} \hspace*{0.3cm} 
\parbox{14mm}{\begin{center}
\begin{fmfgraph*}(11,5)
\setval
\fmfforce{0w,0h}{v1}
\fmfforce{3/11w,0h}{v2}
\fmfforce{8/11w,0h}{v3}
\fmfforce{1w,0h}{v4}
\fmfforce{3/11w,1h}{v5}
\fmfforce{8/11w,1h}{v6}
\fmf{boson}{v1,v2}
\fmf{boson}{v3,v4}
\fmf{boson,right=0.4}{v5,v6}
\fmf{fermion,left=0.4}{v2,v5}
\fmf{fermion,left=0.4}{v5,v6}
\fmf{fermion,left=0.4}{v6,v3}
\fmf{fermion,left=0.4}{v3,v2}
\fmfdot{v2,v3,v5,v6}
\fmfv{decor.size=0, label=${\scs 1}$, l.dist=1mm, l.angle=-180}{v1}
\fmfv{decor.size=0, label=${\scs 2}$, l.dist=1mm, l.angle=0}{v4}
\end{fmfgraph*}
\end{center} }
%
%
\hspace*{0.3cm} + \hspace*{0.1cm} \frac{1}{2}\hspace*{0.3cm}
\parbox{24mm}{\begin{center}
\begin{fmfgraph*}(21,5)
\setval
\fmfforce{0w,1/2h}{v1}
\fmfforce{3/21w,1/2h}{v2}
\fmfforce{8/21w,1/2h}{v3}
\fmfforce{13/21w,1/2h}{v4}
\fmfforce{18/21w,1/2h}{v5}
\fmfforce{21/21w,1/2h}{v6}
\fmf{boson}{v1,v2}
\fmf{fermion,right=1}{v2,v3,v2}
\fmf{boson}{v3,v4}
\fmf{fermion,right=1}{v4,v5,v4}
\fmf{boson}{v5,v6}
\fmfdot{v2,v3,v4,v5}
\fmfv{decor.size=0, label=${\scs 1}$, l.dist=1mm, l.angle=-180}{v1}
\fmfv{decor.size=0, label=${\scs 2}$, l.dist=1mm, l.angle=0}{v6}
\end{fmfgraph*}
\end{center} }
\hspace*{0.4cm} ,
\eeq
one electron line
\beq
\dephi{-\Gamma_1^{(3)}}{1}{2} \hspace*{0.2cm}= - \hspace*{0.2cm}
%
%
\parbox{24mm}{\begin{center}
\begin{fmfgraph*}(21,5)
\setval
\fmfforce{0w,0h}{v1}
\fmfforce{3/21w,0h}{v2}
\fmfforce{8/21w,0h}{v3}
\fmfforce{13/21w,0h}{v4}
\fmfforce{18/21w,0h}{v5}
\fmfforce{1w,0h}{v6}
\fmf{fermion}{v6,v5,v4,v3,v2,v1}
\fmf{boson,left=0.75}{v2,v5}
\fmf{boson,left=1}{v3,v4}
\fmfdot{v2,v3,v4,v5}
\fmfv{decor.size=0, label=${\scs 1}$, l.dist=1mm, l.angle=-180}{v1}
\fmfv{decor.size=0, label=${\scs 2}$, l.dist=1mm, l.angle=0}{v6}
\end{fmfgraph*}
\end{center} }
%
%
\hspace*{0.3cm} - \hspace*{0.3cm} 
\parbox{24mm}{\begin{center}
\begin{fmfgraph*}(21,10)
\setval
\fmfforce{0w,1/2h}{v1}
\fmfforce{3/21w,1/2h}{v2}
\fmfforce{8/21w,1/2h}{v3}
\fmfforce{13/21w,1/2h}{v4}
\fmfforce{18/21w,1/2h}{v5}
\fmfforce{1w,1/2h}{v6}
\fmf{fermion}{v6,v5,v4,v3,v2,v1}
\fmf{boson,left=0.75}{v2,v4}
\fmf{boson,right=0.75}{v3,v5}
\fmfdot{v2,v3,v4,v5}
\fmfv{decor.size=0, label=${\scs 1}$, l.dist=1mm, l.angle=-180}{v1}
\fmfv{decor.size=0, label=${\scs 2}$, l.dist=1mm, l.angle=0}{v6}
\end{fmfgraph*}
\end{center} }
%
%
\hspace*{0.3cm} + \hspace*{0.3cm}  
\parbox{14mm}{\begin{center}
\begin{fmfgraph*}(11,7.5)
\setval
\fmfforce{0w,0h}{v1}
\fmfforce{3/11w,0h}{v2}
\fmfforce{8/11w,0h}{v3}
\fmfforce{3/3w,0h}{v4}
\fmfforce{3/11w,2/3h}{v5}
\fmfforce{8/11w,2/3h}{v6}
\fmf{fermion}{v4,v3,v2,v1}
\fmf{fermion,right=1}{v5,v6}
\fmf{fermion,right=1}{v6,v5}
\fmf{boson}{v2,v5}
\fmf{boson}{v3,v6}
\fmfdot{v2,v3,v5,v6}
\fmfv{decor.size=0, label=${\scs 1}$, l.dist=1mm, l.angle=-180}{v1}
\fmfv{decor.size=0, label=${\scs 2}$, l.dist=1mm, l.angle=0}{v4}
\end{fmfgraph*}
\end{center} }
%
%
\hspace*{0.3cm} - \hspace*{0.3cm} 
\parbox{24mm}{\begin{center}
\begin{fmfgraph*}(21,2.5)
\setval
\fmfforce{0w,0h}{v1}
\fmfforce{3/21w,0h}{v2}
\fmfforce{8/21w,0h}{v3}
\fmfforce{13/21w,0h}{v4}
\fmfforce{18/21w,0h}{v5}
\fmfforce{21/21w,0h}{v6}
\fmf{fermion}{v6,v5,v4,v3,v2,v1}
\fmf{boson,left=1}{v2,v3}
\fmf{boson,left=1}{v4,v5}
\fmfdot{v2,v3,v4,v5}
\fmfv{decor.size=0, label=${\scs 1}$, l.dist=1mm, l.angle=-180}{v1}
\fmfv{decor.size=0, label=${\scs 2}$, l.dist=1mm, l.angle=0}{v6}
\end{fmfgraph*}
\end{center} }
\hspace*{0.4cm} , 
\eeq
and two electron lines
\beq
\ddfermi{- \Gamma_1^{(3)}}{1}{2}{3}{4} &\quad = \quad &
%
%
- \hspace*{0.3cm}
\parbox{26.39mm}{\centerline{
\begin{fmfgraph*}(19.24,5)
\setval
\fmfforce{0w,0.38/5h}{v1}
\fmfforce{2.12/19.24w,1/2h}{v2}
\fmfforce{7.12/19.24w,1/2h}{v3}
\fmfforce{12.12/19.24w,1/2h}{v4}
\fmfforce{17.12/19.24w,1/2h}{v5}
\fmfforce{1w,0.38/5h}{v6}
\fmfforce{0w,4.62/5h}{v7}
\fmfforce{1w,4.62/5h}{v8}
\fmf{electron}{v2,v1}
\fmf{electron}{v6,v5}
\fmf{electron}{v7,v2}
\fmf{electron}{v5,v8}
\fmf{boson}{v2,v3}
\fmf{boson}{v4,v5}
\fmf{electron,right=1}{v3,v4,v3}
\fmfv{decor.size=0, label=${\scs 1}$, l.dist=1mm, l.angle=-180}{v1}
\fmfv{decor.size=0, label=${\scs 2}$, l.dist=1mm, l.angle=0}{v6}
\fmfv{decor.size=0, label=${\scs 3}$, l.dist=1mm, l.angle=-180}{v7}
\fmfv{decor.size=0, label=${\scs 4}$, l.dist=1mm, l.angle=0}{v8}
\fmfdot{v2,v3,v4,v5}
\end{fmfgraph*} 
}}
%
%
\hspace*{0.3cm}- \hspace*{0.3cm}
\parbox{14mm}{\centerline{
\begin{fmfgraph*}(11,5)
\setval
\fmfforce{0w,0h}{v1}
\fmfforce{3/11w,0h}{v2}
\fmfforce{8/11w,0h}{v3}
\fmfforce{1w,0h}{v4}
\fmfforce{0w,1h}{v5}
\fmfforce{3/11w,1h}{v6}
\fmfforce{8/11w,1h}{v7}
\fmfforce{1w,1h}{v8}
\fmf{electron}{v4,v3}
\fmf{electron}{v2,v1}
\fmf{electron,left=0.4}{v3,v2}
\fmf{electron}{v8,v7}
\fmf{electron,right=0.4}{v7,v6}
\fmf{electron}{v6,v5}
\fmf{boson,left=0.4}{v2,v6}
\fmf{boson,right=0.4}{v3,v7}
\fmfv{decor.size=0, label=${\scs 1}$, l.dist=1mm, l.angle=-180}{v1}
\fmfv{decor.size=0, label=${\scs 2}$, l.dist=1mm, l.angle=0}{v4}
\fmfv{decor.size=0, label=${\scs 4}$, l.dist=1mm, l.angle=-180}{v5}
\fmfv{decor.size=0, label=${\scs 3}$, l.dist=1mm, l.angle=0}{v8}
\fmfdot{v2,v3,v6,v7}
\end{fmfgraph*} 
}}
%
%
\hspace*{0.3cm}- \hspace*{0.3cm}
\parbox{14mm}{\centerline{
\begin{fmfgraph*}(11,5)
\setval
\fmfforce{0w,0h}{v1}
\fmfforce{3/11w,0h}{v2}
\fmfforce{8/11w,0h}{v3}
\fmfforce{1w,0h}{v4}
\fmfforce{0w,1h}{v5}
\fmfforce{3/11w,1h}{v6}
\fmfforce{8/11w,1h}{v7}
\fmfforce{1w,1h}{v8}
\fmf{electron}{v4,v3}
\fmf{electron}{v2,v1}
\fmf{electron,left=0.4}{v3,v2}
\fmf{electron}{v7,v8}
\fmf{electron,left=0.4}{v6,v7}
\fmf{electron}{v5,v6}
\fmf{boson,left=0.4}{v2,v6}
\fmf{boson,right=0.4}{v3,v7}
\fmfv{decor.size=0, label=${\scs 1}$, l.dist=1mm, l.angle=-180}{v1}
\fmfv{decor.size=0, label=${\scs 2}$, l.dist=1mm, l.angle=0}{v4}
\fmfv{decor.size=0, label=${\scs 3}$, l.dist=1mm, l.angle=-180}{v5}
\fmfv{decor.size=0, label=${\scs 4}$, l.dist=1mm, l.angle=0}{v8}
\fmfdot{v2,v3,v6,v7}
\end{fmfgraph*} 
}}
%
%
\hspace*{0.3cm}+ \hspace*{0.3cm}
\parbox{14mm}{\centerline{
\begin{fmfgraph*}(11,5)
\setval
\fmfforce{0w,0h}{v1}
\fmfforce{3/11w,0h}{v2}
\fmfforce{8/11w,0h}{v3}
\fmfforce{1w,0h}{v4}
\fmfforce{0w,1h}{v5}
\fmfforce{3/11w,1h}{v6}
\fmfforce{8/11w,1h}{v7}
\fmfforce{1w,1h}{v8}
\fmf{electron}{v4,v3}
\fmf{electron}{v2,v1}
\fmf{boson,left=0.4}{v3,v2}
\fmf{electron}{v7,v8}
\fmf{boson,left=0.4}{v6,v7}
\fmf{electron}{v5,v6}
\fmf{fermion,right=0.4}{v6,v2}
\fmf{fermion,right=0.4}{v3,v7}
\fmfv{decor.size=0, label=${\scs 1}$, l.dist=1mm, l.angle=-180}{v1}
\fmfv{decor.size=0, label=${\scs 2}$, l.dist=1mm, l.angle=0}{v4}
\fmfv{decor.size=0, label=${\scs 3}$, l.dist=1mm, l.angle=-180}{v5}
\fmfv{decor.size=0, label=${\scs 4}$, l.dist=1mm, l.angle=0}{v8}
\fmfdot{v2,v3,v6,v7}
\end{fmfgraph*} 
}}
%
%
\hspace*{0.3cm}+ \hspace*{0.3cm}
\parbox{14mm}{\centerline{
\begin{fmfgraph*}(11,5)
\setval
\fmfforce{0w,0h}{v1}
\fmfforce{3/11w,0h}{v2}
\fmfforce{8/11w,0h}{v3}
\fmfforce{1w,0h}{v4}
\fmfforce{0w,1h}{v5}
\fmfforce{3/11w,1h}{v6}
\fmfforce{8/11w,1h}{v7}
\fmfforce{1w,1h}{v8}
\fmf{electron}{v4,v3}
\fmf{electron}{v2,v1}
\fmf{boson}{v7,v2}
\fmf{electron}{v7,v8}
\fmf{boson}{v6,v3}
\fmf{electron}{v5,v6}
\fmf{fermion,right=0.4}{v6,v2}
\fmf{fermion,right=0.4}{v3,v7}
\fmfv{decor.size=0, label=${\scs 1}$, l.dist=1mm, l.angle=-180}{v1}
\fmfv{decor.size=0, label=${\scs 2}$, l.dist=1mm, l.angle=0}{v4}
\fmfv{decor.size=0, label=${\scs 3}$, l.dist=1mm, l.angle=-180}{v5}
\fmfv{decor.size=0, label=${\scs 4}$, l.dist=1mm, l.angle=0}{v8}
\fmfdot{v2,v3,v6,v7}
\end{fmfgraph*} 
}}
\no \\
&& \quad 
\no \\
&& \quad 
\no \\
&&
%
%
+ \hspace*{0.3cm}
\parbox{14mm}{\centerline{
\begin{fmfgraph*}(10.12,12.12)
\setval
\fmfforce{0w,2.12/12.12h}{v1}
\fmfforce{3/10.12w,2.12/12.12h}{v2}
\fmfforce{8/10.12w,2.12/12.12h}{v3}
\fmfforce{1w,0h}{v4}
\fmfforce{0w,1h}{v5}
\fmfforce{3/10.12w,7.12/12.12h}{v6}
\fmfforce{3/10.12w,1h}{v7}
\fmfforce{1w,4.24/12.12h}{v8}
\fmf{electron}{v2,v1}
\fmf{electron}{v6,v2}
\fmf{electron}{v7,v6}
\fmf{electron}{v5,v7}
\fmf{electron}{v3,v8}
\fmf{electron}{v4,v3}
\fmf{boson}{v3,v2}
\fmf{boson,right=1}{v6,v7}
\fmfv{decor.size=0, label=${\scs 1}$, l.dist=1mm, l.angle=-180}{v1}
\fmfv{decor.size=0, label=${\scs 2}$, l.dist=1mm, l.angle=0}{v4}
\fmfv{decor.size=0, label=${\scs 3}$, l.dist=1mm, l.angle=-180}{v5}
\fmfv{decor.size=0, label=${\scs 4}$, l.dist=1mm, l.angle=0}{v8}
\fmfdot{v2,v3,v6,v7}
\end{fmfgraph*} 
}}
%
%
\hspace*{0.3cm}+ \hspace*{0.3cm}
\parbox{14mm}{\centerline{
\begin{fmfgraph*}(10.12,12.12)
\setval
\fmfforce{0w,0h}{v1}
\fmfforce{2.12/10.12w,2.12/12.12h}{v2}
\fmfforce{7.12/10.12w,2.12/12.12h}{v3}
\fmfforce{1w,2.12/12.12h}{v4}
\fmfforce{0w,4.24/12.12h}{v5}
\fmfforce{7.12/10.12w,7.12/12.12h}{v6}
\fmfforce{7.12/10.12w,1h}{v7}
\fmfforce{1w,1h}{v8}
\fmf{electron}{v2,v1}
\fmf{electron}{v5,v2}
\fmf{electron}{v4,v3}
\fmf{electron}{v3,v6}
\fmf{electron}{v6,v7}
\fmf{electron}{v7,v8}
\fmf{boson}{v3,v2}
\fmf{boson,left=1}{v6,v7}
\fmfv{decor.size=0, label=${\scs 1}$, l.dist=1mm, l.angle=-180}{v1}
\fmfv{decor.size=0, label=${\scs 2}$, l.dist=1mm, l.angle=0}{v4}
\fmfv{decor.size=0, label=${\scs 3}$, l.dist=1mm, l.angle=-180}{v5}
\fmfv{decor.size=0, label=${\scs 4}$, l.dist=1mm, l.angle=0}{v8}
\fmfdot{v2,v3,v6,v7}
\end{fmfgraph*} 
}}
\hspace*{0.3cm}+ \hspace*{0.3cm}
%
%
\parbox{13.12mm}{\centerline{
\begin{fmfgraph*}(10.12,10)
\setval
\fmfforce{0w,0h}{v1}
\fmfforce{3/10.12w,0h}{v2}
\fmfforce{3/10.12w,5/10h}{v3}
\fmfforce{3/11w,10/10h}{v4}
\fmfforce{0w,1h}{v5}
\fmfforce{8/10.12w,5/10h}{v6}
\fmfforce{1w,2.88/10h}{v7}
\fmfforce{1w,7.12/10h}{v8}
\fmf{electron}{v2,v1}
\fmf{electron}{v3,v2}
\fmf{electron}{v4,v3}
\fmf{electron}{v5,v4}
\fmf{electron}{v7,v6}
\fmf{electron}{v6,v8}
\fmf{boson}{v3,v6}
\fmf{boson,left=0.6}{v2,v4}
\fmfv{decor.size=0, label=${\scs 1}$, l.dist=1mm, l.angle=-180}{v1}
\fmfv{decor.size=0, label=${\scs 2}$, l.dist=1mm, l.angle=0}{v7}
\fmfv{decor.size=0, label=${\scs 3}$, l.dist=1mm, l.angle=-180}{v5}
\fmfv{decor.size=0, label=${\scs 4}$, l.dist=1mm, l.angle=0}{v8}
\fmfdot{v2,v3,v6,v4}
\end{fmfgraph*} 
}}
%
%
\hspace*{0.3cm}+ \hspace*{0.3cm}
\parbox{13.12mm}{\centerline{
\begin{fmfgraph*}(10.12,10)
\setval
\fmfforce{0w,2.88/10h}{v1}
\fmfforce{2.12/10.12w,1/2h}{v2}
\fmfforce{0w,7.12/10h}{v3}
\fmfforce{7.12/10.12w,0h}{v4}
\fmfforce{7.12/10.12w,5/10h}{v5}
\fmfforce{7.12/10.12w,1h}{v6}
\fmfforce{1w,1h}{v7}
\fmfforce{1w,0h}{v8}
\fmf{electron}{v2,v1}
\fmf{electron}{v3,v2}
\fmf{electron}{v4,v5}
\fmf{electron}{v5,v6}
\fmf{electron}{v6,v7}
\fmf{electron}{v8,v4}
\fmf{boson}{v2,v5}
\fmf{boson,left=0.6}{v6,v4}
\fmfv{decor.size=0, label=${\scs 1}$, l.dist=1mm, l.angle=-180}{v1}
\fmfv{decor.size=0, label=${\scs 2}$, l.dist=1mm, l.angle=0}{v8}
\fmfv{decor.size=0, label=${\scs 3}$, l.dist=1mm, l.angle=-180}{v3}
\fmfv{decor.size=0, label=${\scs 4}$, l.dist=1mm, l.angle=0}{v7}
\fmfdot{v2,v5,v6,v4}
\end{fmfgraph*} 
}}
\no \\ &&
\no \\ &&
%
%
- \hspace*{0.3cm} 
\parbox{14mm}{\begin{center}
\begin{fmfgraph*}(11,5)
\setval
\fmfforce{0w,0h}{v1}
\fmfforce{3/11w,0h}{v2}
\fmfforce{8/11w,0h}{v3}
\fmfforce{11/11w,0h}{v4}
\fmfforce{0w,1h}{v5}
\fmfforce{3/11w,1h}{v6}
\fmfforce{8/11w,1h}{v7}
\fmfforce{11/11w,1h}{v8}
\fmf{fermion}{v8,v7,v6,v5}
\fmf{fermion}{v4,v3,v2,v1}
\fmf{boson,right=1}{v2,v3}
\fmf{boson,left=1}{v6,v7}
\fmfdot{v2,v3,v6,v7}
\fmfv{decor.size=0, label=${\scs 1}$, l.dist=1mm, l.angle=-180}{v5}
\fmfv{decor.size=0, label=${\scs 2}$, l.dist=1mm, l.angle=0}{v4}
\fmfv{decor.size=0, label=${\scs 3}$, l.dist=1mm, l.angle=-180}{v1}
\fmfv{decor.size=0, label=${\scs 4}$, l.dist=1mm, l.angle=0}{v8}
\end{fmfgraph*}
\end{center} }
%
%
\hspace*{0.3cm} - \hspace*{0.3cm} 
\parbox{23mm}{\begin{center}
\begin{fmfgraph*}(20,5)
\setval
\fmfforce{0w,1/2h}{v1}
\fmfforce{3/20w,1/2h}{v2}
\fmfforce{8/20w,1/2h}{v3}
\fmfforce{13/20w,1/2h}{v4}
\fmfforce{13/20w,5.5/5h}{v5}
\fmfforce{18/20w,1/2h}{v6}
\fmfforce{20/20w,0.5/5h}{v7}
\fmfforce{20/20w,4.5/5h}{v8}
\fmf{fermion}{v4,v3,v2,v1}
\fmf{fermion}{v6,v8}
\fmf{fermion}{v7,v6}
\fmf{fermion}{v5,v4}
\fmf{boson,left=1}{v2,v3}
\fmf{boson}{v4,v6}
\fmfdot{v2,v3,v4,v6}
\fmfv{decor.size=0, label=${\scs 1}$, l.dist=1mm, l.angle=-180}{v1}
\fmfv{decor.size=0, label=${\scs 2}$, l.dist=1mm, l.angle=0}{v7}
\fmfv{decor.size=0, label=${\scs 3}$, l.dist=1mm, l.angle=0}{v8}
\fmfv{decor.size=0, label=${\scs 4}$, l.dist=1mm, l.angle=90}{v5}
\end{fmfgraph*}
\end{center} }
%
%
\hspace*{0.3cm} - \hspace*{0.3cm} 
\parbox{23mm}{\begin{center}
\begin{fmfgraph*}(20,5)
\setval
\fmfforce{0w,0.5/5h}{v1}
\fmfforce{0w,4.5/5h}{v2}
\fmfforce{3/20w,1/2h}{v3}
\fmfforce{8/20w,1/2h}{v4}
\fmfforce{8/20w,5.5/5h}{v5}
\fmfforce{13/20w,1/2h}{v6}
\fmfforce{18/20w,1/2h}{v7}
\fmfforce{20/20w,1/2h}{v8}
\fmf{fermion}{v8,v7,v6,v4}
\fmf{fermion}{v1,v3}
\fmf{fermion}{v3,v2}
\fmf{fermion}{v4,v5}
\fmf{boson,left=1}{v6,v7}
\fmf{boson}{v3,v4}
\fmfdot{v3,v4,v5,v6}
\fmfv{decor.size=0, label=${\scs 1}$, l.dist=1mm, l.angle=90}{v5}
\fmfv{decor.size=0, label=${\scs 2}$, l.dist=1mm, l.angle=-180}{v1}
\fmfv{decor.size=0, label=${\scs 3}$, l.dist=1mm, l.angle=-180}{v2}
\fmfv{decor.size=0, label=${\scs 4}$, l.dist=1mm, l.angle=0}{v8}
\end{fmfgraph*}
\end{center} }
\hspace*{0.4cm} , 
\eeq
we obtain from (\r{GRGRL}) the four-loop result $\Gamma_1^{(4)}$ in (\r{G4}).
\section{Summary and Outlook}
We have derived a closed set of Schwinger-Dyson equations in QED by using functional analytic methods developed in Refs. 
\cite{Kleinert1,Vasiliev}. Their conversion to graphical recursion relations allows us to systematically generate connected and
one-particle irreducible Feynman diagrams for $n$-point functions. In the subsequent paper \cite{skeleton} we show that corrections
of the electron and the photon propogator as well as the vertex can be iteratively eliminated by introducing higher Legendre
transformations \cite{Kleinert1,Vasiliev}. This will lead to graphical recursion relations for all skeleton Feynman diagrams
in QED \cite{Drell}.
\section*{Acknowledgement}
We are grateful to Konstantin Glaum for carefully reading our manuscript.
\end{fmffile}
\end{document}